\documentclass[12pt]{article}

\usepackage[default]{jasa_harvard}    
\usepackage{JASA_manu}

\newcommand {\ctn}{\citeasnoun} 
\usepackage{graphicx,subfigure,amsmath,latexsym,amssymb}
\usepackage{float,epsfig,multirow,rotating,times}
\usepackage{upgreek,wrapfig}
\usepackage{comment}

\newcommand{\btheta}{\boldsymbol{\theta}}

\newcommand{\bbeta}{\boldsymbol{\beta}}

\newcommand{\bSigma}{\boldsymbol{\Sigma}}

\newcommand{\bmu}{\boldsymbol{\mu}}

\newcommand{\bD}{\boldsymbol{D}}

\newcommand{\bG}{\boldsymbol{G}}
\newcommand{\bh}{\boldsymbol{h}}
\newcommand{\bH}{\boldsymbol{H}}

\newcommand{\bI}{\boldsymbol{I}}

\newcommand{\bA}{\boldsymbol{A}}

\newcommand{\bR}{\boldsymbol{R}}

\newcommand{\bs}{\boldsymbol{s}}

\begin{document}

\normalsize

\title{\vspace{-0.8in}
Bayesian Inference in Nonparametric Dynamic State-Space Models}
\author{Anurag Ghosh, Soumalya Mukhopadhyay, Sandipan Roy and Sourabh Bhattacharya\thanks{
Anurag Ghosh is a PhD student in Department of Statistical Science, Duke University, 
Soumalya Mukhopadhyay is a PhD student in Agricultural and Ecological Research Unit, 
Indian Statistical Institute, Sandipan Roy 
is a PhD student in Department of Statistics, University of Michigan, Ann Arbor, and Sourabh Bhattacharya 
is an Assistant Professor in
Bayesian and Interdisciplinary Research Unit, Indian Statistical
Institute, 203, B. T. Road, Kolkata 700108.
Corresponding e-mail: sourabh@isical.ac.in.}}
\date{\vspace{-0.5in}}
\maketitle

\begin{abstract}
We introduce state-space models where the functionals of the observational
and the evolutionary equations are unknown, and treated as random functions evolving with time.
Thus, our model is nonparametric and generalizes the traditional parametric state-space models.
This random function approach also frees us from the restrictive
assumption that the functional forms, although time-dependent, are of fixed forms.
The traditional approach of assuming known, parametric functional forms is
questionable, particularly in state-space models, since the validation of the assumptions 
require data on both the observed time series and the latent states; however, data on the
latter are not available in state-space models. 

We specify Gaussian processes as priors of the random functions and exploit
the ``look-up table approach" of \ctn{Bhattacharya07} to efficiently handle the dynamic
structure of the model. We consider both univariate and multivariate situations, using 
the Markov chain Monte Carlo (MCMC) approach for studying the posterior distributions of interest.
We illustrate our methods with simulated data sets, in both univariate
and multivariate situations.
Moreover, using our Gaussian process approach we analyse a real data set, which has also
been analysed by \ctn{Shumway82} and \ctn{Carlin92} using the linearity assumption. Interestingly, 
our analyses indicate that towards the end of the time series, the linearity assumption  
is perhaps questionable. 
\\[2mm]
\end{abstract}

\section{Introduction}
\label{sec:intro}

The state-space models play important role in dealing with dynamic systems that arise in various disciplines such as finance, 
engineering, ecology, medicine, and statistics. 
The time-varying regression structure and the flexibility inherent in the sequential nature of state-space models make them very suitable 
for analysis and prediction of dynamic data. Indeed, as is well-known, most time series models of interest are 
expressible as state-space models; see \ctn{Durbin01} and \ctn{Shumway11} for details.
However, till date, the state-space models have considered only known forms of the equations, typically linear. 
But testing the parametric assumptions 
require data on both the observed time series and the unobserved states; unfortunately, data on the
latter are not available in state-space models. 
Moreover, the regression structures of the state-space models may 
evolve with time, changing from linear to non-linear, and even the non-linear structure may also evolve
with time, yielding further different non-linear structures. 
We are not aware of any nonparametric state-space approach in the statistical literature that can 
handle unknown functional forms, which may
or may not be evolving with time.
Another criticism of the existing state space models is the assumption that the (unobserved) states satisfy the Markov property.
Although such Markov models have been useful in many situations where there are natural laws supporting such conditional
independence, in general such assumption is not expected to hold.
These arguments point towards the need for developing general, nonparametric, approaches to state-space models,
and this indeed, is our aim in this article. We adopt the Bayesian paradigm for its inherent flexibility.

In a nutshell, in this work, adopting a nonparametric Bayesian framework, we treat the regression structures as 
unknown and model these as Gaussian processes, 
and develop the consequent theory 
in the Bayesian framework, considering both univariate and multivariate situations. Our Gaussian process approach of
viewing the unknown functional forms allows very flexible modeling of the unknown structures, even though they might
evolve with time. Also, as we discusss in Section \ref{subsec:non_Markovian}, as a consequence of our nonparametric 
approach, the unobserved state variables do not follow any Markov model. Thus our approach provides a realistic dependence structure
between the state variables.  
We also develop efficient MCMC-based methods for simulating from the resulting posterior distributions. 
We demonstrate our methods in the case of both univariate and
multivariate situations using simulated data.
Application of our ideas to a real data set which has been analysed
by \ctn{Shumway82} and \ctn{Carlin92} assuming linearity, provided an interesting insight that, although
the linearity assumption may not be unreasonable for most part of the time series, the assumption may be
called in question towards the end of the time series. This vindicates that our approach is indeed 
capable of modeling
unknown functions even if the forms are changing with time, without requiring any change point analysis
and specification of functional forms before and after change points. 

Before introducing our approach, we provide a brief overview of state-space models.

\section{Overview of state-space models}
\label{sec:overview}

Generally, state-space models are of the following form: for $t=1,2,\ldots$,
\begin{eqnarray}
y_t&=&f_t(x_t)+\epsilon_t\label{eq:obs1}\\
x_t&=&g_t(x_{t-1})+\eta_t\label{eq:evo1}
\end{eqnarray}
In the above, $f_t$ and $g_t$ are assumed to be functions of known forms which may or may not explicitly depend upon $t$; $\eta_t,\epsilon_t$ are usually assumed to be zero mean $iid$ normal variates.
The choice $f_t(x_{t})=F_tx_{t}$ and $g_t(x_{t-1})=G_tx_{t-1}$, assuming known $F_t,G_t$, have found very wide use in the literature.
Obviously, $x_t,y_t$ may be univariate or multivariate. Matrix-variate  dynamic linear models have been considered by \ctn{Quintana87}
and \ctn{West97} (see also \ctn{Carvalho07}). 
Equation (\ref{eq:obs1}) is called the observational equation, while (\ref{eq:evo1}) is known as the evolutionary equation.
Letting $\bD_T=(y_1,\ldots,y_T)'$ denote the available data, 
the goal is to obtain inferences about $y_{T+1}$ (single-step
forecast), $y_{T+k}$ ($k$-step forecast), $x_{T+1}$ conditional on $y_{T+1}$ (filtering), $x_{T-k}$ (retrospection). In the Bayesian paradigm, 
the interests center upon analyzing the corresponding  posteriors $[y_{T+1}\mid \bD_T]$, $[y_{T+k}\mid \bD_T]$, $[x_{T+1}\mid \bD_T,y_{T+1}]$ 
(also, $[x_{T+1}\mid \bD_T]$) and $[x_{T-k}\mid \bD_T]$.

In the non-Bayesian framework,
 solutions to dynamic systems are quite
generally available via the well-known Kalman filter. However, the performance of Kalman filter is heavily dependent on the assumption of Gaussian errors and linearity of the functions in the observation and the evolution equations. 
In the case of non-linear dynamic models, various linearization techniques are used to obtain approximate solutions. For details on these issues, see \ctn{Brockwell87}, \ctn{Meinhold89}, \ctn{West97} and the references therein.
The Bayesian paradigm frees the investigator from restrictions of linear functions or Gaussian errors, and allows for very general dynamic model building through coherent combination of prior and the available time series data,
and using Markov chain Monte Carlo (MCMC) for inference. Bayesian non-linear dynamic models with non-Gaussian errors, in conjunction with the Gibbs sampling approach
for inference, have been considered in \ctn{Carlin92}. 
For general details on non-linear and non-Gaussian approches to state space models, see \ctn{Durbin01}.

However, even non-linear and non-Gaussian approaches assume that there is an underlying known natural phenomenon supporting
some standard parametric model. Except in well-studied scientific contexts such assumptions are not unquestionable. 
In this work, we particularly concern ourselves with situations where parametric models are not established for the underlying
scientific study. A case in point may be the context of climate change dynamics, where observed climate depends upon various
factors in the forms of latent states, but in our knowledge, no clear parametric model is available for this extremely 
important and challenging problem. 
In medicine, growth of cancer cells at any time point may depend upon various unobserved factors (states), but no clear parametric model
is available, in our knowledge. Similar challenges exist in sociology, in studies of dynamic social networks; in 
astrophysics, associated with the study of the evolution of the universe; in computer science, for target tracking
and data-driven computer animation, and in various other fields.
Thus, we expect our nonparametric approach to be quite relevant and useful for these investigations.

For the sake of clarity, in this main article, we consider only one-dimensional $x_t$ and $y_t$, 
but we provide additional details of the univariate cases, and generalize our approach to accommodate
multivariate situations in the supplement \ctn{Ghosh13supp}, whose sections, figures and tables have 
the prefix ``S-" when referred to in this paper. A brief description of the contents of the supplement
can be found at the end of this article.

In Section \ref{sec:univariate} we introduce our novel nonparametric dynamic model where we use Gaussian processes
to model the unknown functions $f_t$ and $g_t$. Assuming the functions to be random allows us to accommodate
even those functions the forms of which are changing with time. In order to describe the distribution of the
unobserved states, we adopt the ``look-up table" idea of \ctn{Bhattacharya07}. Since this idea is
an integral part of the development of our methodology, we devote Section \ref{sec:lookup_details} to 
its detailed discussion. In Section \ref{sec:prior_univariate} we provide the forms of the prior distributions 
of the hyperparameters of our model
and build an MCMC based methodology for Bayesian inference.
In Section \ref{sec:simstudies} we include a brief discussion of two simulation studies,
the details of which are reported in Section S-3 of the supplement.
In Section \ref{sec:cps_realdata} we consider application to a real, univariate data set.
We present a summary of the current work, along with discussion of further work in Section \ref{sec:conclusions}.

\section{Nonparametric dynamic model: univariate case}
\label{sec:univariate}

In this section we model the unknown observational and the evolutionary functions using Gaussian processes
assuming that the true functions are evolving with time $t$; our approach includes the time-invariant 
situation as a simple special case; see Section S-3.2. 


In 
(\ref{eq:obs1}) and (\ref{eq:evo1})
we now assume $f_t$ and $g_t$ to be of unknown functional forms varying with time $t$. 
For convenience, we denote $f_t(x_t)$ and $g_t(x_{t-1})$ as $f(t,x_t)$ and $g(t,x_{t-1})$, respectively.
That is, we treat time $t$ as an input to both the observational and
evolutionary functions, in addition to the other relevant inputs $x_t$ and $x_{t-1}$. 
With this understanding, we re-write (\ref{eq:obs1}) and (\ref{eq:evo1}) as
\begin{eqnarray}
y_t&=&f(t,x_t)+\epsilon_t,\hspace{2mm}\epsilon_t\sim N(0,\sigma^2_{\epsilon}),\label{eq:obs2}\\
x_t&=&g(t,x_{t-1})+\eta_t,\hspace{2mm}\eta_t\sim N(0,\sigma^2_{\eta}). \label{eq:evo2}
\end{eqnarray}
We assume that $x_0\sim N(\mu_{x_0},\sigma^2_{x_0})$; $\mu_{x_0},\sigma^2_{x_0}$ being known.
Crucially, we allow $f(\cdot,\cdot)$ and $g(\cdot,\cdot)$ to be of unknown functional forms,
which we model as two independent Gaussian processes.
To present the details of the Gaussian processes, we find it convenient to use the notation
$x^*_{t,t}=(t,x_t)'$, and $x^*_{t,t-1}=(t,x_{t-1})'$ so that (\ref{eq:obs2}) and (\ref{eq:evo2})
can be re-written as
\begin{eqnarray}
y_t&=&f(x^*_{t,t})+\epsilon_t,\hspace{2mm}\epsilon_t\sim N(0,\sigma^2_{\epsilon}),\label{eq:obs3}\\
x_t&=&g(x^*_{t,t-1})+\eta_t,\hspace{2mm}\eta_t\sim N(0,\sigma^2_{\eta}); \label{eq:evo3}
\end{eqnarray}
in fact, more generally, we use the notation $x^*_{t,u}=(t,x_u)$.
This general notation will be convenient for describing theoretical and computational details. 
Next, we provide details of the independent Gaussian processes used to model $f$ and $g$.

\subsection{Modeling the unknown observational and evolutionary time-varying functions
using independent Gaussian processes}
\label{subsec:gp_models}

The functions $f$ and $g$ are modeled
as independent Gaussian processes with mean functions $\mu_f(\cdot)=\bh(\cdot)'\bbeta_f$ and 
$\mu_g(\cdot)=\bh(\cdot)'\bbeta_g$ with $\bh(x^*)=(1,x^*)'$ for any $x^*$,
and covariance
functions of the form $\sigma^2_f c_f(\cdot,\cdot)$ and $\sigma^2_g c_g(\cdot,\cdot)$, respectively. 
The process variances are $\sigma^2_f$ and $\sigma^2_g$ and $c_f, c_g$ are the correlation functions.
Typically, for any $z_1,z_2$, 
$c_f(z_1,z_2)=\exp\{-(z_1-z_2)'\bR_f(z_1-z_2)\}$ and $c_g(z_1,z_2)=
\exp\{-(z_1-z_2)'\bR_g(z_1-z_2)\}$, where $\bR_f$ and $\bR_g$ 
are $2\times 2$-dimensional diagonal matrices consisting of respective smoothness (or, roughness) parameters
$\{r_{1,f},r_{2,f}\}$ and $\{r_{1,g},r_{2,g}\}$, which are responsible for the smoothness of the process 
realizations. These choices of the correlation functions imply that the functions,
modeled by the process realizations,
are infinitely smooth. 


The sets of parameters $\btheta_f=(\bbeta_f,\sigma^2_f,\bR_f)$ and $\btheta_g=(\bbeta_g,\sigma^2_g,\bR_g)$ are assumed to be 
independent {\it a priori}. 
We consider the following form of prior distribution of the parameters:
$[\bbeta_f,\sigma^2_f,\bR_f,\bbeta_g,\sigma^2_g,R_g,\sigma^2_{\epsilon},\sigma^2_{\eta}]=[\bbeta_f,\sigma^2_f,\bR_f][\bbeta_g,\sigma^2_g,\bR_g][\sigma^2_{\epsilon},\sigma^2_{\eta}]$.

\subsection{Hierarchical structure induced by our Gaussian process approach}
\label{subsec:hierarchy}
In summary, our approach can be described in the following hierarchical form:
\begin{align}
[y_t|f,\btheta_f,x_t]&\sim N\left(f(x^*_{t,t}),\sigma^2_{\epsilon}\right);~t=1,\ldots,T,\label{eq:y_t_dist}\\
[x_t|g,\btheta_g,x_{t-1}]&\sim N\left(g(x^*_{t,t-1}),\sigma^2_{\eta}\right);~t=1,\ldots,T,\label{eq:x_t_dist}\\
[x_0]&\sim N\left(\mu_{x_0},\sigma^2_{x_0}\right),\label{eq:x_0_dist}\\
[f(\cdot)|\btheta_f]&\sim GP\left(\bh'(\cdot)'\bbeta_f,\sigma^2_fc_f(\cdot,\cdot)\right),\label{eq:gp_f}\\
[g(\cdot)|\btheta_g]&\sim GP\left(\bh'(\cdot)\bbeta_g,\sigma^2_gc_g(\cdot,\cdot)\right),\label{eq:gp_g}\\
[\bbeta_f,\sigma^2_f,\bR_f,\bbeta_g,\sigma^2_g,R_g,\sigma^2_{\epsilon},\sigma^2_{\eta}]
&=[\bbeta_f,\sigma^2_f,\bR_f][\bbeta_g,\sigma^2_g,\bR_g][\sigma^2_{\epsilon},\sigma^2_{\eta}].\label{eq:theta_prior}
\end{align}
In the above, GP stands for ``Gaussian Process". 
Forms of the prior distributions in (\ref{eq:theta_prior}) are provided in (\ref{sec:prior_univariate})
and specific details are provided in the relevant applications.

\subsection{Conditional distribution of the observed data induced by the Gaussian process prior on
the observational function and a brief discussion of the difficulty of obtaining the joint
distribution of the state variables}
\label{subsec:data_dist}

It follows from the Gaussian process prior assumption on the unknown observational function $f$ 
that the distribution of $\bD_T$, conditional on $x^*_{1,1},\ldots,x^*_{T,T}$
(equivalently, conditional on $x_1,\ldots,x_T$), and the other parameters is multivariate normal: 
\begin{equation}
\bD_T\sim N_T(\bH_{D_T}\bbeta_f,\sigma_f^2\bA_{f,D_T}+\sigma^2_{\epsilon}\bI_T),
\label{eq:data_distribution}
\end{equation}
where $\bH^\prime_{D_T}=\left[\bh(x^*_{1,1}),\ldots,\bh(x^*_{T,T})\right]$, $\bA_{f,D_T}$ is a $T\times T$ 
matrix with $(i,j)$-th element $c_f(x^*_{i,i},x^*_{j,j})$; $(i,j)=1,\ldots,T$, and $\bI_T$ is the $T$-th 
order identity matrix.

The joint distribution of the state variables $(x_1,\ldots,x_T)$, however, is much less straightforward. 
Observe that, although we have
$[x_0]\sim N(\mu_{x_0},\sigma^2_{x_0})$, $[x_1\mid x_0]\sim N(\bh(x_0)'\bbeta_g,\sigma^2_g+\sigma^2_{\eta})$, 
but $[x_2\mid x_1,x_0]$$=$$[g(2,x_1)+\eta_2\mid x_1,x_0]$$=$$[g(2,g(1,x_0)+\eta_1)+\eta_2\mid g(1,x_0)+\eta_1,x_0]$, the rightmost expression suggesting that special techniques may
be necessary to get hold of the conditional distribution. We adopt the procedure introduced by \ctn{Bhattacharya07} to deal with this problem. 
The idea is to conceptually simulate the entire function $g$ modeled by the Gaussian process, and use the simulated process as a look-up table to
obtain the conditional distributions of $\{x_i;i\geq 2\}$. 
The intuition behind the look-up table concept is briefly dicussed in the next subsection, while the detailed
procedure of approximating the joint distribution of the state variables is provided in Section  \ref{sec:lookup_details}.

\subsection{Intuition behind the look-up table idea for approximating the joint distribution of the state variables 
}
\label{subsec:lookup}

For the purpose of illustration only let us assume that $\epsilon_t=0$ for all $t$, yielding the model
$x_t=g(x^*_{t,t-1})$. The concept of look-up table in this problem can be briefly explained as follows. 
Let us first assume that the entire process $g(\cdot)$ is available. This means that for every input $z$, 
the corresponding $g(z)$ is available, thus constituting a look-up table, with the first column representing
$z$ and the second column representing the corresponding $g(z)$.
Conditional on $x^*_{t,t-1}$ (equivalently, conditional on $x_{t-1}$), $x_t=g(x^*_{t,t-1})$ can be obtained 
by simply picking the input $x^*_{t,t-1}$ from the first column of the look-up table and 
reporting the corresponding output value $g(x^*_{t,t-1})$, located in the second column of the look-up table. 
This hypothetical look-up table concept suggests that conditional on 
the simulated process $g$, it can be safely assumed that $x_t$ depends only upon $x^*_{t,t-1}$ via
$x_t=g(x^*_{t,t-1})$. Thus, if for all possible inputs, a simulation of the entire random function
$g$, following the Gaussian process, is available, then for any input $x^*_{t,t-1}$, we only need
to identify the corresponding $g(x^*_{t,t-1})$ in the look-up table. In practice, we can have a simulation
of the Gaussian process $g$ on a fine enough grid of inputs. Given this simulation on a fine grid,
we can simulate from the conditional distribution of $g(x^*_{t,t-1})$, fixing $x^*_{t,t-1}$ as given. 
This simulation from the conditional distribution of $g(x^*_{t,t-1})$ will 
approximate $x_t$ as accurately
as we desire by making the grid as fine as required. By repeating this procedure for each $t$, we
can approximate the joint distribution of the state variables as closely as we desire. In the next section we provide details
regarding this approach.

\section{Detailed procedure of approximating the joint distribution of the state variable using the look-up table concept}
\label{sec:lookup_details}

\subsection{Distribution of $x_1$}
\label{subsec:dist_x1}
Note that given $x_0$ we can simulate $x_1=g(x^*_{1,0})\sim N(\bh(x^*_{1,0})'\bbeta_g,\sigma^2_g)$, 
which is the marginal
distribution of the Gaussian process prior. Thus, $x_1$ is simulated without resorting to any approximation. 
It then remains to simulate the rest of the dynamic sequence, for which we need to simulate the
rest of the process $\{g(x^*);x^*\neq x^*_{1,0}\}$.

\subsection{Introduction of a set of auxiliary variables to act as proxy to the Gaussian process $g$}
\label{subsec:auxiliary}
In practice, it is not possible to have a simulation of this entire set $\{g(x^*);x^*\neq x^*_{1,0}\}$..  
We only have available a set of grid points $\bG_n=\{z_1,\ldots,z_n\}$ obtained, perhaps, 
by Latin hypercube sampling 
(see, for example, \ctn{Santner03})
and a corresponding
simulation of $g$, given by $\bD^*_n=\{g(z_1),\ldots,g(z_n)\}$, the latter having a joint
multivariate normal distribution with mean 
\begin{equation}
E\left[\bD^*_n\mid\btheta_g\right]=\bH_{D^*_n}\bbeta_g\label{eq:mean1}
\end{equation}
and covariance matrix 
\begin{equation}
V\left[\bD^*_n\mid\btheta_g\right]=\sigma^2_g\bA_{g,D^*_n},
\label{eq:var1}
\end{equation}
where
$\bH^{\prime}_{D^*_n}$=$[\bh(z_1),\ldots,\bh(z_n)]$ and $\bA_{g,D^*_n}$ is a 
correlation matrix with the $(i,j)$-th element $c_g(z_i,z_j)$. 
and $\bs_{g,D^*_n}(\cdot)=\left(c_g(\cdot,z_{1}),\ldots,c_g(\cdot,z_{n})\right)'$. 

Given $(x_0,g(x^*_{1,0}))$, we simulate $\bD^*_n$ from $[\bD^*_n\mid\btheta_g, g(x^*_{1,0}),x_{0}]$.
Since the joint distribution of $\left[\bD^*_n,g(x^*_{1,0})\mid x_0\right]$ is multivariate normal with mean vector
$\left( \begin{array}{c}
\bH_{D^*_n}\bbeta_g \\
\bh(x_0)'\bbeta_g
\end{array}
\right)$ and covariance matrix $\sigma^2_g\bA_{D^*_n,g(x^*_{1,0})}$ where 
\begin{eqnarray}
\bA_{D^*_n,g(x^*_{1,0})}&=&\left( \begin{array}{cc}\bA_{g,D^*_n} &  \bs_{g,D^*_n}(x^*_{1,0})\\
\bs_{g,D^*_n}(x^*_{1,0})' & 1\end{array}\right),
\end{eqnarray}
it follows that the conditional $[\bD^*_n\mid g(x^*_{1,0}),x^*_{1,0}]$
has an $n$-variate normal distribution with mean vector 
\begin{equation}
E[\bD^*_n\mid g(x^*_{1,0}),x_0,\btheta_g]=\bmu_{g,D^*_n}=\bH_{D^*_n}\bbeta_g+\bs_{g,D^*_n}(x^*_{1,0})(g(x^*_{1,0})-\bh(x^*_{1,0})'\bbeta_g) \label{eq:mean2}
\end{equation}
and covariance matrix
\begin{equation}
V[\bD^*_n\mid g(x^*_{1,0}),x_0,\btheta_g]=\sigma^2_g\bSigma_{g,D^*_n},
\label{eq:var2}
\end{equation}
where
\begin{equation}
\bSigma_{g,D^*_n}=\bA_{g,D^*_n}-\bs_{g,D^*_n}(x^*_{1,0})\bs_{g,D^*_n}(x^*_{1,0})'.
\label{eq:Sigma_g_Dz}
\end{equation}


\subsection{Distribution of each state variable conditional on the look-up table proxy $\bD^*_n$}
\label{subsec:dist_x_t}

We now seek the conditional distribution $[x_t=g(x^*_{t,t-1})\mid \bD^*_n,x_{t-1},x_{t-2},\ldots,x_1]$. 
To notationally distinguish between the conditional distribution of $g(\cdot)$ from the elements of the
set $\bD^*_n$, we henceforth denote the elements of $\bD^*_n$ as $g_{true}(\cdot)$. In other words,
we henceforth write $\bD^*_n=\{g_{true}(z_1),\ldots,g_{true}(z_n)\}$.

Recall that
the look-up table idea supports conditional independence, that is, given a simulation of the entire
random function $g$, $x_t$ depends only upon $x_{t-1}$ via $x_t=g(x^*_{t,t-1})$, so that given $g$, $x_t$
is conditionally independent of $\{x_k;~k<t-1\}$. Indeed, given a fine enough grid $\bG_n$,
$\bD^*_n$ approximates the random function $g$, which contains all information regarding
the conditioned state variables $x_1,\ldots,x_{t-1}$. Hence,  
\begin{equation}
[g(x^*_{t,t-1})\mid \bD^*_n,x_{t-1},x_{t-2},\ldots,x_1]\approx [g(x^*_{t,t-1})\mid \bD^*_n,x_{t-1}],
\label{eq:gp_approx}
\end{equation}
and this approximation can be made arbitrarily accurate by making the grid $\bG_n$ as fine as desired.
Hence, it is sufficient for our purpose to deal with the conditional distribution of 
$[g(x^*_{t,t-1})\mid \bD^*_n,x_{t-1}]$. This is easy to obtain: since given $x_{t-1}$,
$(g(x^*_{t,t-1}), \bD^*_n)$ is jointly multivariate normal, it is easily seen that 
$[g(x^*_{t,t-1})\mid \bD^*_n,x_{t-1}]$ is
normal with mean
\begin{equation}
\mu_t=\bh(x^*_{t,t-1})^\prime \bbeta_g+\bs_{g,D^*_n}(x^*_{t,t-1})^\prime \bA_{g,D^*_n}^{-1}(\bD^*_n-\bH_{D^*_n}\bbeta_g)
\label{eq:cond_mean_xt}
\end{equation}
and variance 
\begin{equation} \sigma_t^2=\sigma^2_g\left\{1-\bs_{g,D^*_n}(x^*_{t,t-1})^\prime \bA_{g,D^*_n}^{-1}\bs_{g,D^*_n}(x^*_{t,t-1})\right\}.
\label{eq:cond_var_xt}
\end{equation}

One subtlety involved in the assumption of conditional independence is that, conditional on $x^*_{1,0}$, $\bG_n$ must not contain $x^*_{1,0}$; otherwise
$\bD^*_n$ would contain $g(x^*_{1,0})$ implying that $[g(x^*_{t,t-1})\mid\btheta_g,\bD^*_n,x_{t}]$ is dependent on $x_1=g(x^*_{1,0})$, violating the conditional
independence assumption. 

\subsection{Accuracy of the Markov approximation of the distributions of the state
variables conditional on $\bD^*_n$}
\label{subsec:accuracy}

To first heuristically understand how the approximation (\ref{eq:gp_approx}) can be made arbitrarily accurate, note that 
thanks to the Gaussian process assumption,
conditioning on $\bD^*_n$ forces the random function $g(\cdot)$ to pass through the points in $(\bG_n,\bD^*_n)$ 
since the conditional $[g(x)\mid\btheta_g, \bD^*_n,x]$ has zero variance if $x\in\bG_n$ (see, for example, \ctn{Bhattacharya07} and the references therein).  In other words,
if $x^*_{t,t-1}\in\bG_n$, then $\sigma^2_t=0$ so that
\begin{equation}
[g(x^*_{t,t-1})\mid\btheta_g, \bD^*_n,x_{t-1}]=\delta_{g_{true}(x^*_{t,t-1})},
\label{eq:cond_delta}
\end{equation}
where $\delta_z$ denotes point mass at $z$. This property of the conditional associated with Gaussian process is in keeping with the insight gained from
the discussion related to look-up table associated with prediction of the outputs of deterministic function having dynamic behaviour. However, $x^*_{t,t-1}\notin\bG_n$ with
probability 1 and the conditional $[g(x^*_{t,t-1})\mid\btheta_g, \bD^*_n,x_{t-1}]$ provides spatial interpolation within $(\bG_n,\bD^*_n)$ (see, for example, \ctn{Cressie93}, \ctn{Stein99}).
Finer the set $\bG_n$, closer is $[g(x^*_{t,t-1})\mid\btheta_g, \bD^*_n,x_{t-1}]$ to $\delta_{g(x^*_{t,t-1})}$. The conditional independence assumption of 
$g(x^*_{t,t-1})$ of all $\{x^*_{t,k};k<(t-1)\}$
given $(\bG_n,\bD^*_n)$ is in accordance with the motivation provided by the deterministic sequence and here $\bD^*_n$ acts as a set of auxiliary variables,
greatly simplifying computation, while not compromising on accuracy. 

In Section S-1 of the supplement we formally prove a theorem stating that, given a particular design $\bG_n$, 
the order of approximation of $g_{true}(\cdot)$ by the 
conditional distribution of $g(\cdot)$ given $\bD^*_n$, within a finite region (but as large as required
for all practical purposes), is $O\left(n^{-1}\right)$. Hence, given a judiciously chosen sufficiently fine grid $\bG_n$ and
corresponding $\bD^*_n$, the conditioned state variables $\{x_k;~k<t\}$ do not provide any extra information to $x_t=g(x^*_{t,t-1})$
regarding $g_{true}(x^*_{t,t-1})$ in an asymptotic sense with respect to $\bG_n$. Hence, the Markov approximation (\ref{eq:gp_approx})
is valid for appropriate $\bG_n$, and the accuracy of the approximation can be improved arbitrarily.

\subsection{Summary of the look-up table procedure for obtaining the joint distribution of the state variables}
\label{subsec:lookup_summary}

To summarize the ideas, let $x_0\sim\pi$, where $\pi$ is some appropriate prior distribution of $x_0$. 
The entire dynamic sequence $\{x_0,x_1=g(x^*_{1,0}),x_2=g(x^*_{2,1}),\ldots\}$ can then be simulated using the
following steps sequentially:
\begin{itemize}
\item[(1)] Draw $x_0\sim\pi$.
\item[(2)] Given $x_0$, draw $x_1=g(x^*_{1,0})\sim N(\bh(x^*_{1,0})'\bbeta_g,\sigma^2_g)$.
\item[(3)] Given $x_0$, and $x_1=g(x^*_{1,0})$, draw $\bD^*_n\sim [\bD^*_n\mid\btheta_g, g(x^*_{1,0}),x_{0}]$.
\item[(4)] For $t=2,3,\ldots$, draw $x_t\sim [x_t=g(x^*_{t,t-1})\mid\btheta_g,\bD^*_n,x_{t-1}]$.
\end{itemize}
Step (1) is a simulation of $x_0$ from its prior, step (2) is simply drawn from the known marginal distribution of 
$g(x^*_{1,0})$ given $x_0$.
In step (3) $\bD^*_n$ is drawn conditional on $g(x^*_{1,0})$ (and $x^*_{1,0}$), conceptually implying that the rest of the process
$\{g(x^*);x\neq x^*_{1,0}\}$ is drawn once $g(x^*_{1,0})$ is known. 
Step (4) then uses this simulated $\bD^*_n$ to obtain
the rest of the dynamic sequence, using the assumed conditional independence structure.



\subsection{Explicit form of the look-up table induced joint distribution of $\{x_0,x_1,\ldots,x_{T+1},\bD^*_n\}$} 
\label{subsec:lookup_prior_conditionals}

Once $\bG_n$ and $\bD^*_n$ are available, we write down the joint distribution
of $(\bD^*_n,x_0,x_1,\ldots,x_T)$ conditional on the other parameters as
\\[2mm]
$[x_0,x_1,\ldots,x_{T+1},\bD^*_n\mid\btheta_f,\btheta_g,\sigma^2_{\epsilon},\sigma^2_{\eta}]$
\begin{eqnarray}
&&=[x_0][x_1=g(x^*_{1,0})+\eta_1\mid x_0,\sigma^2_{\eta}]
[\bD^*_n\mid\btheta_g]\nonumber\\
&& \ \ \prod_{t=1}^{T} [x_{t+1}=g(x^*_{t+1,t})+\eta_{t+1}\mid \bD^*_n,x_{t},\btheta_g,\sigma^2_{\eta}]
\label{eq:x_joint}
\end{eqnarray}
Recall that $[x_0]\sim N(\mu_{x_0},\sigma^2_{x_0})$, 
$[x_1=g(x^*_{1,0})+\eta_1\mid x^*_{1,0},\sigma^2_{\eta}]$$\sim N(\bh(x^*_{1,0})'\bbeta_g,\sigma^2_g+\sigma^2_{\eta})$
and the distribution of $\bD^*_n$ is multivariate normal with mean and variance given by (\ref{eq:mean1}) 
and (\ref{eq:var1}).
The conditional distribution $[x_{t+1}=g(x^*_{t+1,t})+\eta_{t+1}\mid \bD^*_n,x_t,\btheta_g,\sigma^2_{\eta}]$
is normal with mean
\begin{equation}
\mu_{x_t}=\bh(x^*_{t+1,t})^\prime \bbeta_g+\bs_{g,\bD^*_n}(x^*_{t+1,t})^\prime \bA_{g,D^*_n}^{-1}(\bD^*_n-\bH_{D^*_n}\bbeta_g)
\label{eq:cond_mean_xt1}
\end{equation}
and variance 
\begin{equation} \sigma^2_{x_t}=\sigma^2_{\eta}+\sigma^2_g\left\{1-\bs_{g,\bD^*_n}(x^*_{t+1,t})^\prime \bA_{g,\bD^*_n}^{-1}\bs_{g,\bD^*_n}(x^*_{t+1,t})\right\}
\label{eq:cond_var_xt1}
\end{equation}
Observe that in this case even if $x^*_{t+1,t}\in\bG_n$, due to the presence of the additive error term 
$\eta_{t+1}$, 
the conditional variance of $x_{t+1}$ is non-zero, equalling $\sigma^2_{x_t}=\sigma^2_{\eta}$,
the error variance.

\subsection{Non-Markovian dependence structure of the marginalized joint distribution of
the state variables $(x_0,x_1,\ldots,x_{T+1})$}
\label{subsec:non_Markovian}

As in \ctn{Bhattacharya07}, here also it is possible to marginalize out $\bD^*_n$ from (\ref{eq:x_joint}) 
to obtain the approximate joint distribution of
$(x_0,x_1,\ldots,x_T)$. 
However, if $\bD^*_n$ is integrated out, it is clear that
the conditional indpendepence (Markov) property of $x_t$'s given $\bD^*_n$, as seen in 
(\ref{eq:x_joint}), will be lost. Thus, the marginalized conditional distribution of $x_{t+1}$ depends
upon $\{x_k;~k<{t+1}\}$, that is, the set of all the past state variables, unlike the non-marginalized case, where
conditionally on $\bD^*_n$, $x_{t+1}$ depends only upon $x_{t}$. 
This makes it clear that even though for fixed (known) evolutionary function $g$
the corresponding equation (\ref{eq:evo1}) satisfies the Markov property, such Markov
property is lost when the function is modeled as Gaussian processes. Hence, in our approach
based on Gaussian process, the state variables are non-Markovian.

\subsection{To marginalize or not to marginalize with respect to $\bD^*_n$?}
\label{subsec:marginalization}

As discussed in detail in \ctn{Bhattacharya07}, the complicated dependence structure associated with the
marginalized joint distribution of $(x_0,x_1,\ldots,x_{T+1})$ is also the root of all numerical instabilities
associated with MCMC implementation of our model. 
To understand this heuristically, note that evaluations of the conditionals 
$[x_{t+1}|x_{t},\ldots,x_0,\btheta_g,\sigma^2_g]$ are required for MCMC implementation, but 
evaluations of the conditionals require inversions of covariance matrices involving the random states 
$x_0,x_1,\ldots,x_{t-1}$ in the correlation terms. By sample path continuity of the underlying
Gaussian process, the sampled states $x_0,x_1,\ldots,x_{t-1}$ will be often close to each other with high probability,
particularly if $\sigma^2_g$ and $\sigma^2_{\eta}$ are small,
rendering the correlation matrix almost singular. Moreover, inversion of such correlation matrices
at every iteration of MCMC is also very costly computationally. The problems are much aggravated for large $t$. 

On the other hand, if $\bD^*_n$ is retained, then such problem is avoided, since in that case we only
need to compute the conditionals $[x_{t+1}|D^*_n,x_t,\btheta_g,\sigma^2_g]$, which involves
inversion of the correlation matrix $\bA_{g,D^*_n}$, which has $(i,j)$-th element 
of the form $c(z_i,z_j)$, where $z_1,\ldots,z_n$ are fixed constants selected by the user.
Also, quite importantly, $\bA^{-1}_{g,D^*_n}$ can be computed even before beginning MCMC simulations, and it remains
fixed thereafter, saving a lot of computational time, in addition to providing protection against
numerical instability. For further details, see \ctn{Bhattacharya07}.
 
Hence, we retain the set of auxiliary variables $\bD^*_n$ in (\ref{eq:x_joint})
for implementation of our MCMC methods, and finally discard them from the resultant MCMC samples to infer
about the quantities of our interest.


In the next section we complete specification of our fully Bayesian model by 
choosing appropriate prior distributions of the parameters.

\section{Prior specifications and Bayesian inference using MCMC}
\label{sec:prior_univariate}
We assume the following prior distributions:
\begin{align}
[\log(r_{i,f})]&\sim N\left(\mu_{r_{i,f}},\sigma^2_{r_{i,f}}\right);\ \ \mbox{for}\ \ i=1,2.
\label{eq:rf_prior}\\
[\log(r_{i,g})]&\sim N\left(\mu_{r_{i,g}},\sigma^2_{r_{i,g}}\right);\ \ \mbox{for}\ \ i=1,2.
\label{eq:rg_prior}\\
[\sigma^2_{\epsilon}]&\propto\left(\sigma^2_{\epsilon}\right)^{-\left(\frac{\alpha_{\epsilon}+2}{2}\right)}\exp\left\{-\frac{\gamma_{\epsilon}}{2\sigma^2_{\epsilon}}\right\};
\alpha_{\epsilon},\gamma_{\epsilon}>0
\label{eq:sigma_epsilon_prior}\\
[\sigma^2_{\eta}]&\propto\left(\sigma^2_{\eta}\right)^{-\left(\frac{\alpha_{\eta}+2}{2}\right)}\exp\left\{-\frac{\gamma_{\eta}}{2\sigma^2_{\eta}}\right\};
\alpha_{\eta},\gamma_{\eta}>0
\label{eq:sigma_eta_prior}\\
[\sigma^2_f]&\propto\left(\sigma^2_f\right)^{-\left(\frac{\alpha_f+2}{2}\right)}\exp\left\{-\frac{\gamma_f}{2\sigma^2_f}\right\};
\alpha_f,\gamma_f>0
\label{eq:sigma_f_prior}\\
[\sigma^2_g]&\propto\left(\sigma^2_g\right)^{-\left(\frac{\alpha_g+2}{2}\right)}\exp\left\{-\frac{\gamma_g}{2\sigma^2_g}\right\};
\alpha_g,\gamma_g>0
\label{eq:sigma_g_prior}\\
[\bbeta_f]&\sim N_m\left(\bbeta_{f,0},\bSigma_{\beta_{f,0}}\right)
\label{eq:betaf_prior}\\
[\bbeta_g]&\sim N_m\left(\bbeta_{g,0},\bSigma_{\beta_{g,0}}\right)
\label{eq:betag_prior}
\end{align}
All the prior parameters are assumed to be known. Now we discuss our approach to
selecting the prior parameters for our application our Bayesian model in simulation studies and
real data application in the univariate situations.

In order to choose the parameters of the log-normal priors of the smoothness parameters, we
set the mean of the log-normal prior with parameters $\mu$ and $\sigma^2$,
given by $\exp(\mu+\sigma^2/2)$, to 1. This yields $\mu= -\sigma^2/2$. Since the variance
of this log-normal prior is given by $(\exp(\sigma^2)-1)\exp(2\mu+\sigma^2)$, the relation
$\mu=-\sigma^2/2$ implies that the variance is $\exp(\sigma^2)-1=\exp(-2\mu)-1$. We set
$\sigma^2=1$, so that $\mu=-0.5$. This implies that the mean is 1 and the variance is approximately 2,
for the priors of each smoothness parameter $r_{i,f}$ and $r_{i,g}$; $i=1,2$.

For the choice of the parameters of the priors of $\sigma^2_f$, $\sigma^2_g$, $\sigma^2_{\epsilon}$
and $\sigma^2_{\eta}$,
we first note that the mean is of the form $\gamma/(\alpha-2)$ and the variance is of the form
$2\gamma^2/\{(\alpha-2)^2(\alpha-4)\}$. Thus, if we set $\gamma/(\alpha-2)=a$, then the variance becomes
$2a^2/(\alpha-4)$. Here we set $a=0.5,0.5,0.1,0.1$, respectively, for $\sigma^2_f$, $\sigma^2_g$,
$\sigma^2_{\epsilon}$, $\sigma^2_{\eta}$. For each of these priors we set $\alpha=4.01$, so that the 
variance is of the form $200a^2$. 

We set the priors of $\bbeta_f$ and $\bbeta_g$ to be trivariate normal with zero mean and
the identity matrix as the variance.



\subsection{MCMC-based Bayesian inference}

In this section we begin with the problem of forecasting $y_{T+1}$, given the data set $\bD_T$.
Interestingly, our approach to this problem provides an MCMC methodology which generates inference
about all the posterior distributions required, either as by-products or by simple generalization
of this MCMC approach using an augmentation scheme. Details follow.

The posterior predictive distribution of $y_{T+1}$ given $\bD_T$ is
\begin{eqnarray}
[y_{T+1}\mid \bD_T]&=&\int [y_{T+1}\mid \bD_T,x_0,x_1,\ldots,x_{T+1},\bbeta_f,\bR_f,\sigma^2_{\epsilon}]\nonumber\\
\ \ \ &&\times[x_0,x_1,\ldots,x_{T+1},\bbeta_f,\bbeta_g,\bR_f,\bR_g,\sigma^2_{\epsilon}\sigma^2_{\eta}\mid\bD_T]d\btheta_f d\btheta_g d\sigma^2_{\epsilon},\sigma^2_{\eta}
dx_0dx_1,\ldots,dx_{t+1}.\nonumber\\
\label{eq:posterior_predictive}
\end{eqnarray}
This posterior is not available analytically, and so simulation methods are necessary to make inferences.
In particular, once a sample is available from the posterior $[x_0,x_1,\ldots,x_{t+1},\bbeta_f,\bbeta_g,\bR_f,\bR_g,\sigma^2_{\epsilon}\sigma^2_{\eta}\mid\bD_T]$,
the corresponding samples drawn from $[y_{T+1}\mid \bD_T,x_0,x_1,\ldots,x_{T+1},\bbeta_f,\bR_f,\sigma^2_{\epsilon}]$ 
are from the posterior predictive (\ref{eq:posterior_predictive}), using which required posterior summaries can be obtained.
Note that the conditional distribution $[y_{T+1}=f(x^*_{T+1,T+1})+\epsilon_{T+1}\mid \bD_T,x_0,x_1,\ldots,x_{T+1}\bbeta_f,\sigma^2_f,\bR_f,\sigma^2_{\epsilon}]$
is normal with mean
\begin{equation}
\mu_{y_{T+1}}=\bh(x^*_{T+1,T+1})^\prime \bbeta_f+\bs_{f,\bD_T}(x^*_{T+1,T+1})^\prime \bA_{f,D_T}^{-1}(\bD_T-\bH_{D_T}\bbeta_f)
\label{eq:cond_mean_y}
\end{equation}
and variance 
\begin{equation} \sigma^2_{y_{T+1}}=\sigma^2_{\epsilon}+\sigma^2_f\left\{1-\bs_{f,\bD_T}(x^*_{T+1,T+1})^\prime \bA_{f,\bD_T}^{-1}\bs_{f,\bD_T}(x^*_{T+1,T+1})\right\}
\label{eq:cond_var_y}
\end{equation}

Using $\bD^*_n$, the conditional posterior $[x_0,x_1,\ldots,x_{T+1},\bbeta_f,\bbeta_g,\bR_f,\bR_g,\sigma^2_{\epsilon},\sigma^2_{\eta}\mid\bD_T]$ can be
written as
\\[2mm]
$[x_0,x_1,\ldots,x_{T+1},\btheta_f,\btheta_g,\sigma^2_{\epsilon},\sigma^2_{\eta}\mid\bD_T]$
\begin{eqnarray} 
&=&\int [x_0,x_1,\ldots,x_{T+1},\btheta_f,\btheta_g,\sigma^2_{\epsilon},\sigma^2_{\eta},g(x^*_{1,0}),\bD^*_n\mid\bD_T]dg(x^*_{1,0})d\bD^*_n 
\label{eq:state_posterior1}\\
&\propto &\int [x_0,x_1,\ldots,x_{T+1},\btheta_f,\btheta_g,\sigma^2_{\epsilon},\sigma^2_{\eta},g(x^*_{1,0}),\bD^*_n,\bD_T]dg(x_0)d\bD^*_n 
\label{eq:state_posterior2}\\
&=&\int [\btheta_f,\btheta_g,\sigma^2_{\epsilon},\sigma^2_{\eta}][x_0][g(x^*_{1,0})\mid x^*_{1,0}][\bD^*_n\mid g(x^*_{1,0}),x_{0},\btheta_g]\nonumber\\
&&\ \ \ \ \times [x_1=g(x^*_{1,0})+\eta_1\mid g(x^*_{1,0}),x_0,\bbeta_g,\sigma^2_g,\sigma^2_{\eta}]\nonumber\\
&&\ \ \ \ \times[x_{T+1}=g(x^*_{T+1,T})+\eta_T \mid\btheta_g,\sigma^2_{\eta},\bD^*_n,x_T]\nonumber\\
&&\ \ \ \ \times\prod_{t=1}^{T-1} [x_{t+1}\mid\btheta_g,\sigma^2_{\eta},\bD^*_n,x_t]
[\bD_T\mid x_1,\ldots,x_{T},\btheta_f,\sigma^2_{\epsilon}]dg(x^*_{1,0})d\bD^*_n
\label{eq:state_posterior3}
\end{eqnarray}
In the above, $[x_1=g(x^*_{1,0})+\eta_1\mid g(x^*_{1,0}),\bbeta_g,\sigma^2_g,\sigma^2_{\eta}]\sim N\left(g(x^*_{1,0}),\sigma^2_{\eta}\right)$.
Although the analytic form of (\ref{eq:state_posterior3}) is not available, MCMC simulation from 
$[x_0$$,x_1,$$\ldots,$$x_{T+1},$$\btheta_f,$$\btheta_g,$$\sigma^2_{\epsilon},$$\sigma^2_{\eta},$$g(x^*_{1,0}),$$\bD^*_n$$\mid$$\bD_T]$,
which is proportional to the integrand in (\ref{eq:state_posterior3}),
is possible. Ignoring $g(x^*_{1,0})$ and $\bD^*_n$ in these MCMC simulations yields the desired samples from
$[x_0,x_1,\ldots,x_{T+1},\btheta_f,\btheta_g,\sigma^2_{\epsilon},\sigma^2_{\eta}\mid\bD_T]$.
The details of the MCMC are provided in Section S-2.

\subsection{MCMC-based single and mutiple step forecasts, filtering and retrospection}
\label{subsubsec:forecasting}

Observe that one can readily study the posteriors $[y_{T+1}\mid\bD_T]$, $[x_{T+1}\mid\bD_T]$ and $[x_{T-k}\mid\bD_T]$
using the readily available MCMC samples of $\{y_{T+1},x_{T+1},x_{T-k}\}$ after ignoring the samples corresponding to the
rest of the unknowns.
To study the posterior $[x_{T+1}\mid\bD_T,y_{T+1}]$, we only need to augment $y_{T+1}$ to $\bD_T$ to create
$\bD_{T+1}=(\bD'_T,y_{T+1})'$. Then our methodology can be followed exactly to generate samples from
$[x_{T+1}\mid\bD_{T+1}]$. Sample generation from $[y_{T+k}\mid\bD_T]$ requires a slight generalization of this
augmentation strategy. Here
we use successive augmentation, adding
each simulated $y_{T+j}$ to the previous $\bD_{T+j-1}$ to create $\bD_{T+j}=(\bD'_{T+j-1},y_{T+j})$; $j=1,2,\ldots,k$.
Then our MCMC methodology can be implemented successively to generate samples from $y_{T+j+1}$ and all other variables.
This implies that at each augmentation stage we need to draw a single MCMC sample from
$[x_0,x_1,\ldots,x_{T+j+1},\bbeta_f,\bbeta_g,\bR_f,\bR_g,\sigma^2_f,\sigma^2_g,\sigma^2_{\epsilon},\sigma^2_{\eta}\mid\bD_{T+j}]$.
Once this sample is generated, we can draw a single realization from $y_{T+j+1}$ by
drawing from $y_{T+j+1}\sim N\left(\mu_{y_{T+j+1}},\sigma^2_{y_{T+j+1}}\right)$, where, analogous to (\ref{eq:cond_mean_y})
and (\ref{eq:cond_var_y}),
\begin{equation}
\mu_{y_{T+j+1}}=\bh(x^*_{T+j+1,T+j+1})^\prime \bbeta_f+\bs_{f,\bD_{T+j}}(x^*_{T+j+1,T+j+1})^\prime \bA_{f,D_{T+j}}^{-1}(\bD_{T+j}-\bH_{D_{T+j}}\bbeta_f)
\label{eq:cond_mean_yj}
\end{equation}
and variance 
\begin{equation} \sigma^2_{y_{T+j+1}}=\sigma^2_{\epsilon}+\sigma^2_f\left\{1-\bs_{f,\bD_{T+j}}(x^*_{T+j+1,T+j+1})^\prime 
\bA_{f,\bD_{T+j}}^{-1}\bs_{f,\bD_{T+j}}(x^*_{T+j+1,T+j+1})\right\}.
\label{eq:cond_var_yj}
\end{equation}

\section{Brief discussion on simulation studies}
\label{sec:simstudies}

In Section S-3 of the supplement we present two detailed simulation experiments. In the first experiment,
both the true observational and the evolutionary functions are of linear forms. In the second study, both
these true functions are non-linear. In both the cases we fitted our Gaussian process based nonparametric
model to the data generated from the true models. Whenever the true parameters are comparable to the parameters
associated with our model, the posteriors of the parameters of our model successfully captured them. 
The true time series $\{x_0,\ldots,x_T\}$ fell well within their respective 95\% 
Bayesian credible intervals in both the simulation studies. 
However, the lengths of the credible regions of the states seem to be somewhat larger than desired.
Indeed, since the posterior distribution of the states depend upon the observational and evolutionary functions,
both of which are treated as unknown, somewhat larger credible intervals only reflect the uncertainty
associated with these unknown functions. In the context of our simulation studies where the true functions
are known, the lengths of our credible intervals may not seem to be particularly encouraging, but as we  
already mentioned in the Section \ref{sec:intro}, we have in mind complex, realistic problems, where
true observational and evolutionary functions are extremely difficult to ascertain. In such problems,
there exist large amounts of uncertainties regarding these functions, which would be coherently modeled by our
Gaussian process approach, and relatively large Bayesian credible regions, that would arise as a result of acknowledging
such uncertainties, would be coherent and make good practical sense.

We now apply our ideas based on Gaussian processes to a real data set.

\section{Application to a real data set}
\label{sec:cps_realdata}

Assuming a parametric, dynamic, linear model set up, 
\ctn{Shumway82} and \ctn{Carlin92} analysed a data set consisting of estimated total
physician expenditures by year ($y_t;t=1,\ldots,25$) as measured by the Social Security Administration.
The unobserved true annual physician expenditures are denoted by $x_t;t=1,\ldots,25$. \ctn{Shumway82}
used a maximum likelihood approach based on the EM algorithm, while \ctn{Carlin92} considered a Gibbs sampling
based Bayesian approach. 

We apply our Bayesian nonparametric Gaussian process based model and methodology on the same data set and
check if the results of the former authors who analysed this data based on the linearity assumption,
agree with those obtained by our far more general analysis.

\subsection{Choice of prior parameters and MCMC implementation}
\label{subsec:cps_realdata_prior_choice}

We use the same prior structures as detailed in Section \ref{sec:prior_univariate}. 
We set $\alpha=4.01$ so that the prior variance is of the form $200a^2$; we 
set $a=0.5, 0.5, 100, 100$, respectively, for $\sigma^2_f$, $\sigma^2_g$, $\sigma^2_{\epsilon}$
and $\sigma^2_{\eta}$. These imply that the prior expectations of the variances are 0.5, 0.5, 100, and 100,
respectively, along with the aforementioned variances. The choices of high values of the prior expectations of $\sigma^2_{\epsilon}$
and $\sigma_{\eta}$ are motivated by \ctn{Carlin92}, while the small variabilities of $\sigma^2_f$ and $\sigma^2_g$
reflects the belief that the true functional forms are perhaps not very different from linearity, the belief being motivated
by the assumption of linearity by both \ctn{Shumway82} and \ctn{Carlin92}. Also motivated by the latter we chose
$x_0\sim N(2500,100^2)$ to be the prior distribution of $x_0$. We set the priors of $\bbeta_f$ and $\bbeta_g$
to be trivariate normal distributions with zero mean and the identity matrix as the covariance matrix.
For the prior parameters of the smoothness parameters we set $\sigma^2=100$ and $\mu=-50$ in the log-normal
prior distributions, so that the means are 1 and the variances are $\exp(100)-1$.
Here we set high variance
for the smoothness parameters to account for much higher degree of uncertainty about smoothness
in this real data situation compared to the simulation experiment.

To set up the grid $\bG_n$ we noted that the MLEs of the time series obtained by \ctn{Shumway82} using
linear state space model are contained in $[2000,30000]$.
We divide this interval into 200 sub-intervals of equal length, and select
a value randomly from each such sub-interval, obtaining values of the second component of the
two-dimensional grid $\bG_n$. For the first component, we generate a number uniformly from each of the
200 sub-intervals $[i,i+1]$; $i=0,\ldots,199$. 

We discard the first 10,000 MCMC iterations as burn-in and store the next 50,000 iterations for 
inference. We used the normal random walk proposal with variance 0.05 for updating $\sigma_f$, $\sigma_g$,
$\sigma_{\epsilon}$ and $\sigma_{\eta}$ and the normal random walk proposal with variance 10 for updating
$\{x_0,x_1,\ldots,x_T\}$. These choices of the variances are based on pilot runs of our MCMC algorithm. 
As before, informal diagnostics indicated good convergence properties of our MCMC algorithm. 
It took around 17 hours to implement this application in an ordinary laptop machine.

\subsection{Results of model-fitting}
\label{subsec:cps_realdata_results}

\begin{figure}
\centering
  \subfigure{ \label{fig:beta_f_1_realdata_cps_hist}
   \includegraphics[width=5cm,height=4cm]{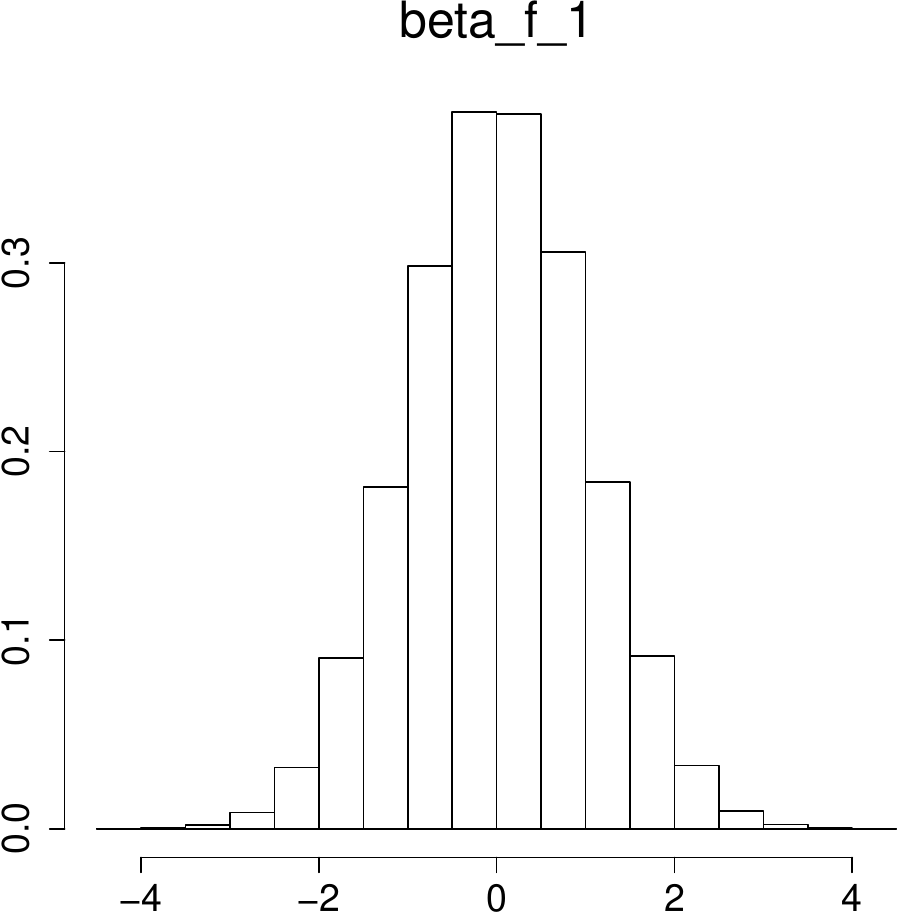}}
   \hspace{2mm}
  \subfigure{ \label{fig:beta_f_2_realdata_cps_hist} 
  \includegraphics[width=5cm,height=4cm]{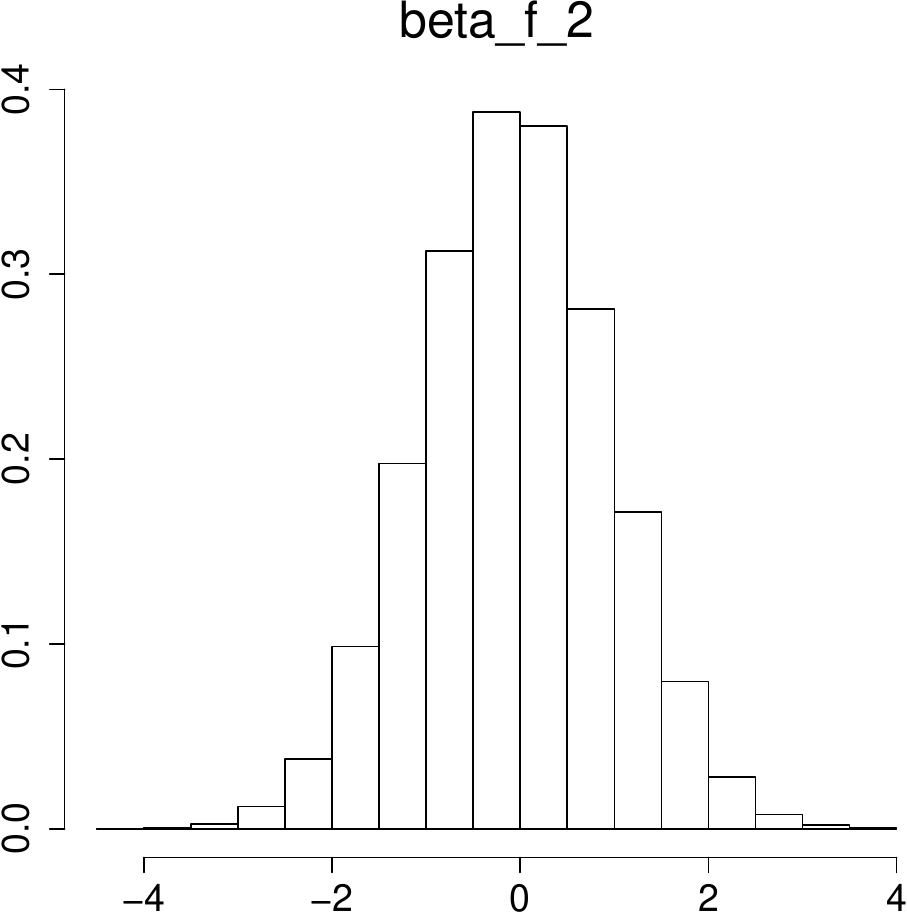}}
   \hspace{2mm}
  \subfigure{ \label{fig:beta_f_3_realdata_cps_hist} 
  \includegraphics[width=5cm,height=4cm]{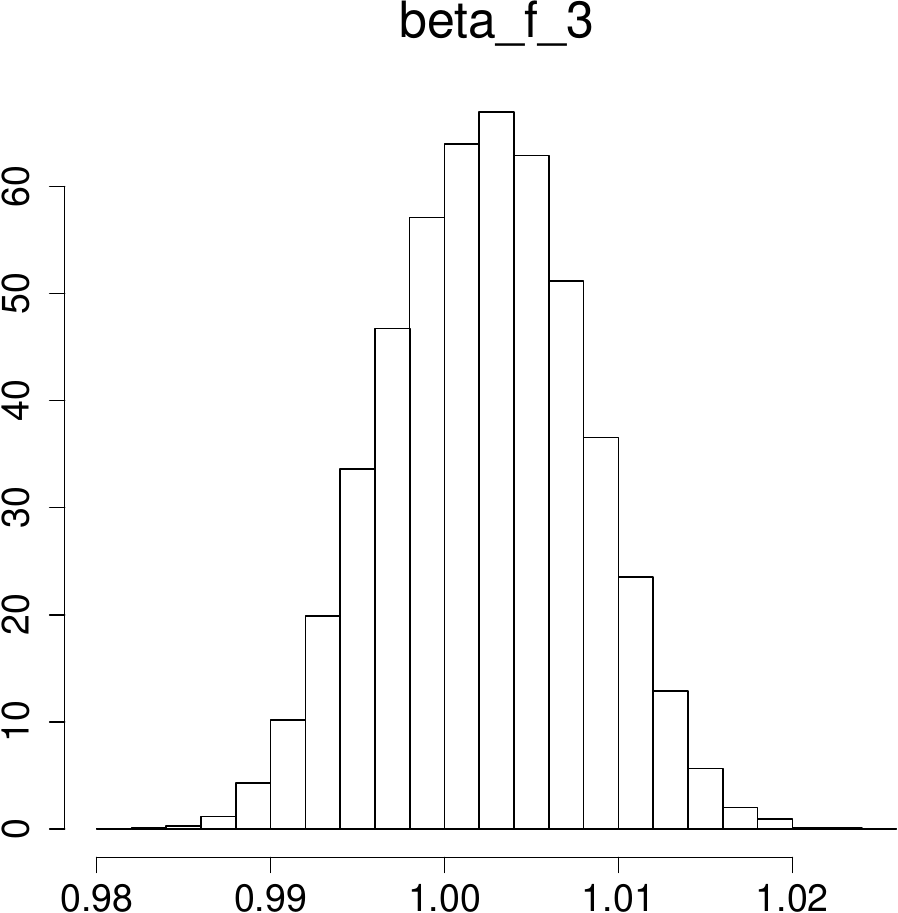}}\\
  \vspace{2mm}
  \subfigure{ \label{fig:beta_g_1_realdata_cps_hist}
   \includegraphics[width=5cm,height=4cm]{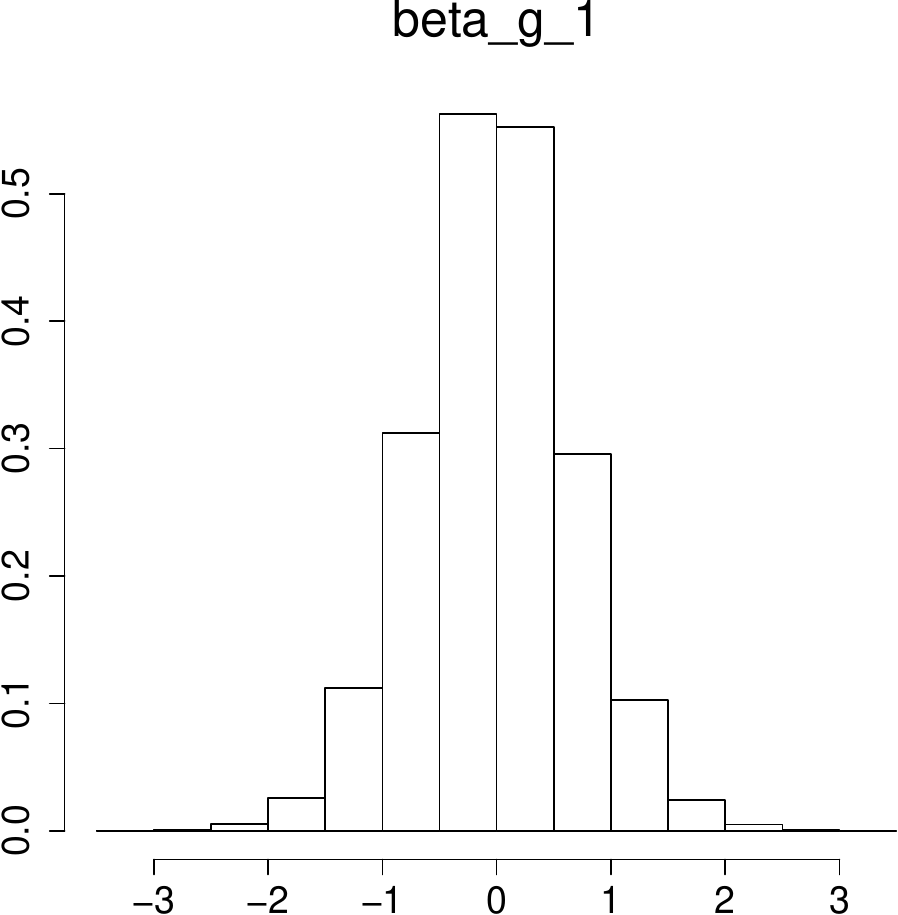}}
   \hspace{2mm}
  \subfigure{ \label{fig:beta_g_2_realdata_cps_hist} 
  \includegraphics[width=5cm,height=4cm]{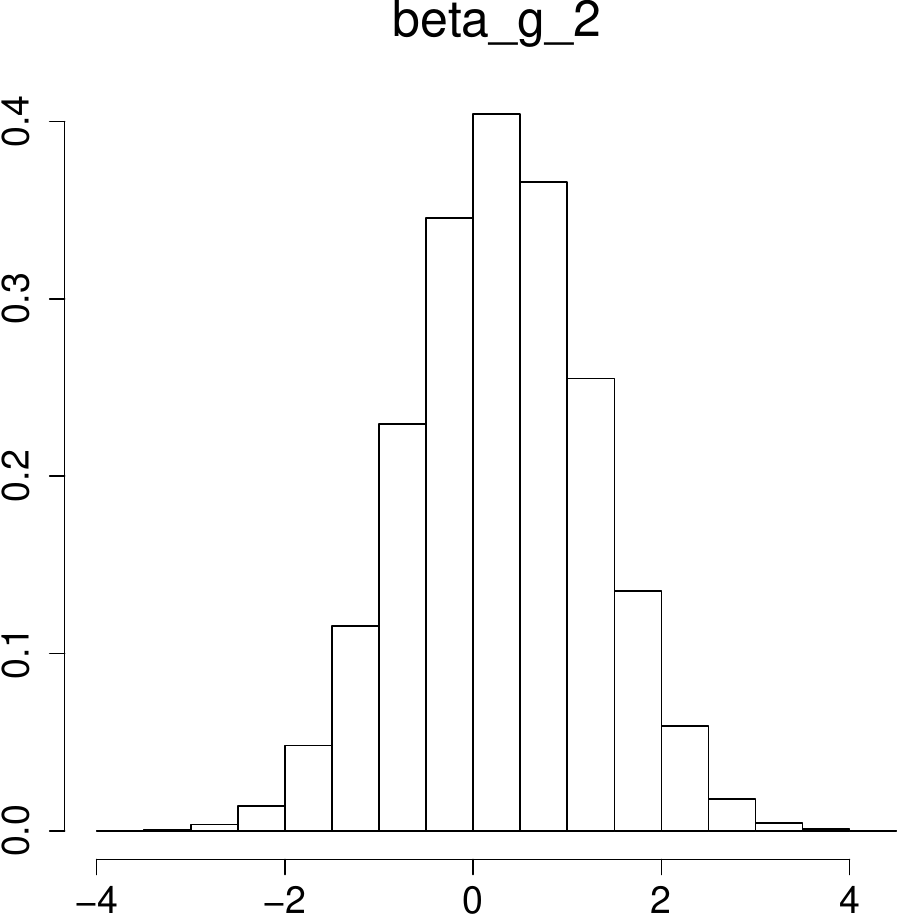}}
   \hspace{2mm}
  \subfigure{ \label{fig:beta_g_3_realdata_cps_hist} 
  \includegraphics[width=5cm,height=4cm]{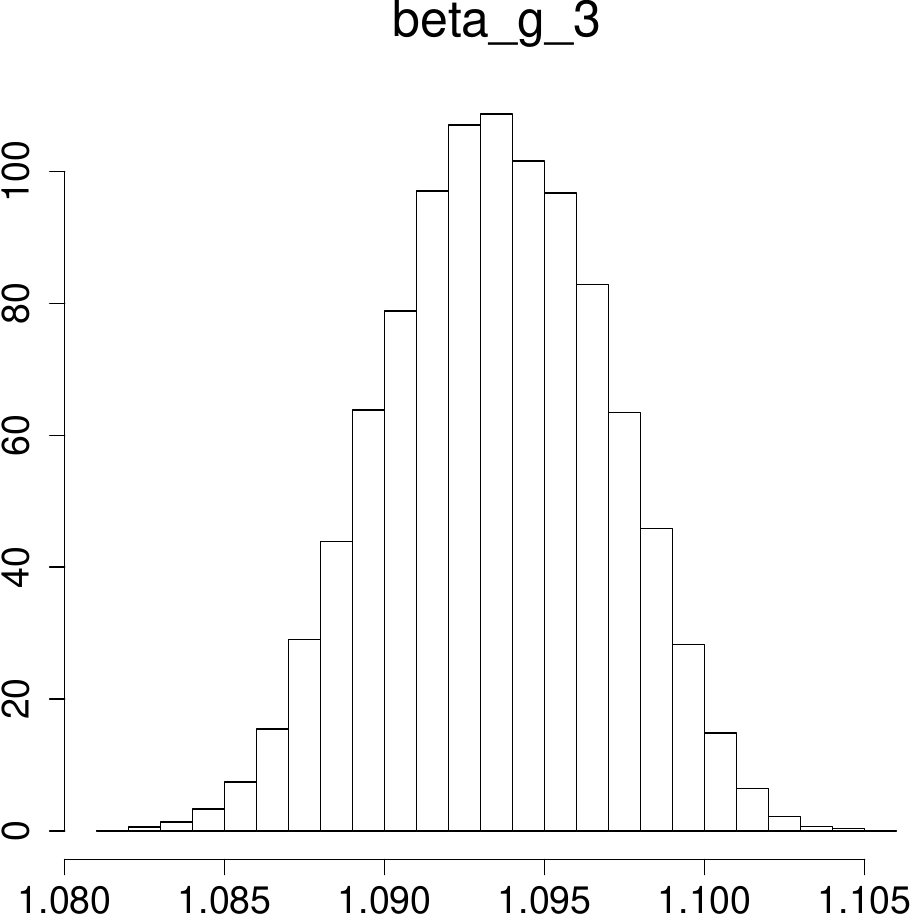}}\\
  \vspace{2mm}
  \subfigure{ \label{fig:sigma_f_realdata_cps_hist}
   \includegraphics[width=5cm,height=4cm]{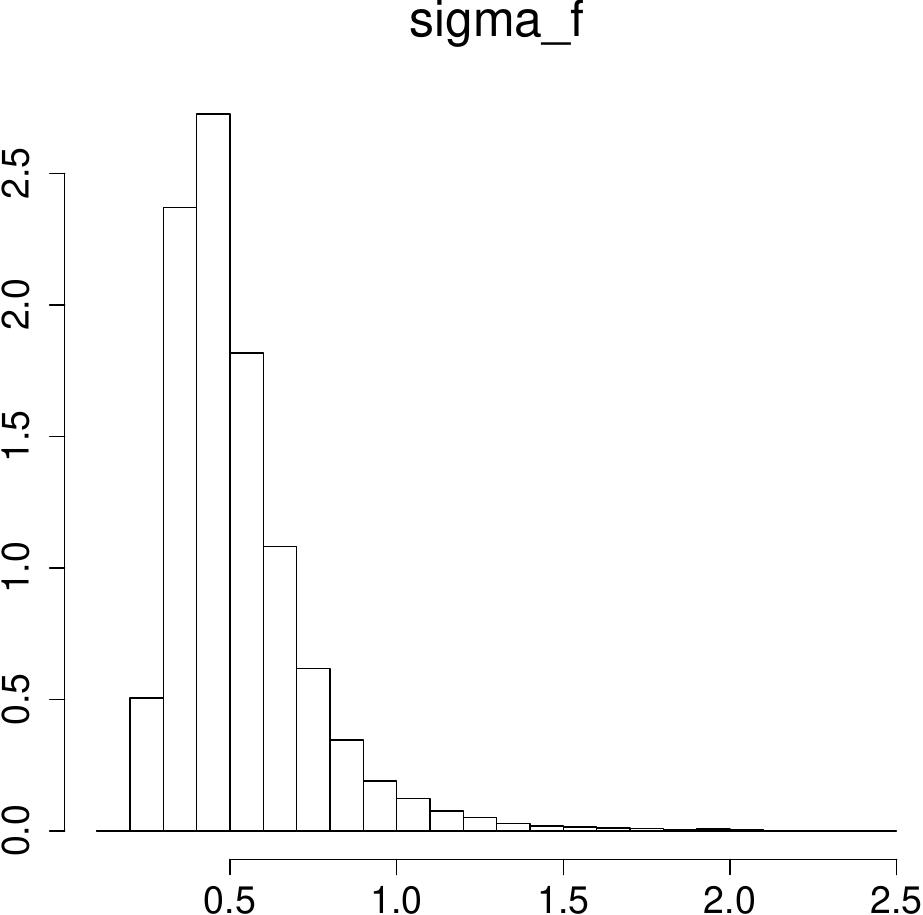}}
   \hspace{2mm}
  \subfigure{ \label{fig:sigma_g_realdata_cps_hist} 
  \includegraphics[width=5cm,height=4cm]{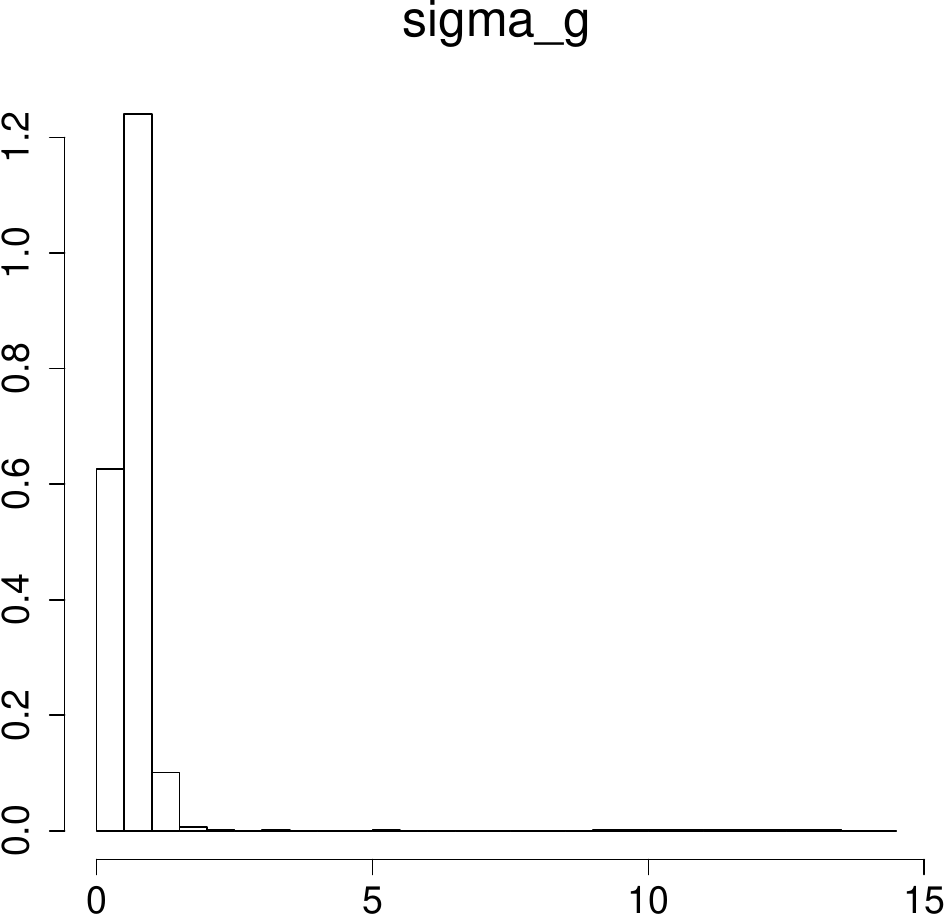}}
   \hspace{2mm}
  \subfigure{ \label{fig:sigma_e_realdata_cps_hist} 
  \includegraphics[width=5cm,height=4cm]{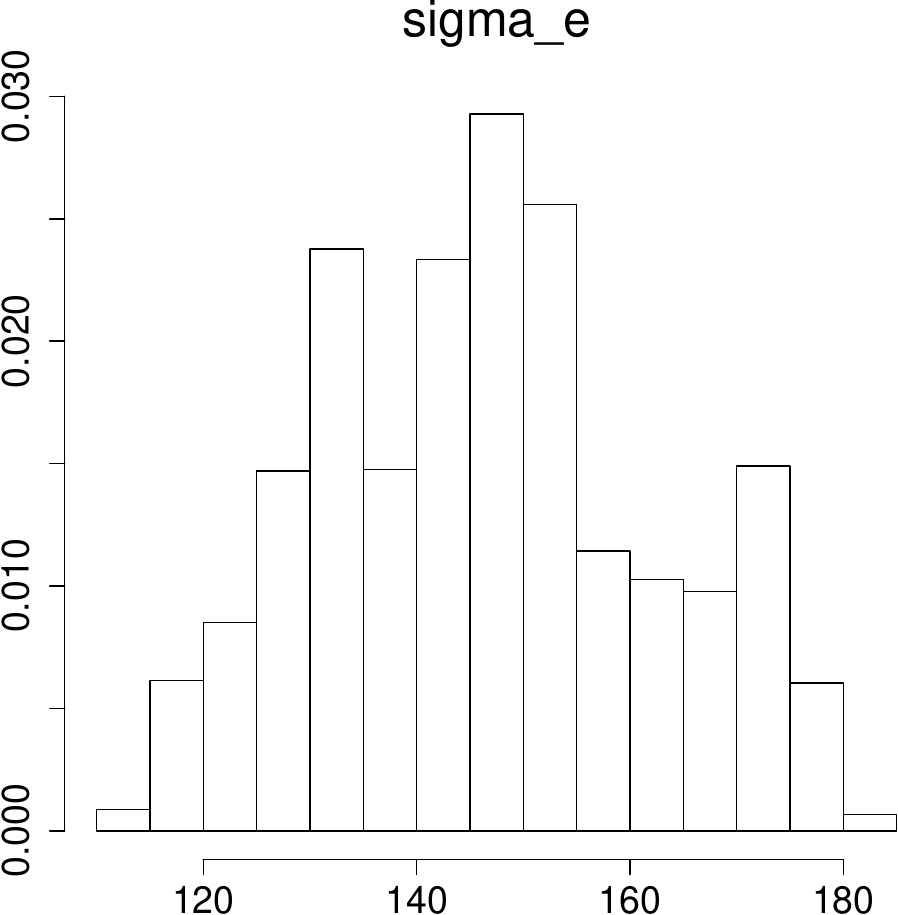}}
 \caption{{\bf Real data analysis}: 
 Posterior densities of $\beta_{0,f}$, $\beta_{1,f}$, $\beta_{2,f}$, $\beta_{0,g}$, $\beta_{1,g}$,
 $\beta_{2,g}$, $\sigma_f$, $\sigma_g$, and $\sigma_{\epsilon}$.} 
 \label{fig:realdata_cps_hist}
 \end{figure}


\begin{figure}
\centering
  \subfigure{ \label{fig:sigma_eta_realdata_cps_hist}
   \includegraphics[width=5cm,height=4cm]{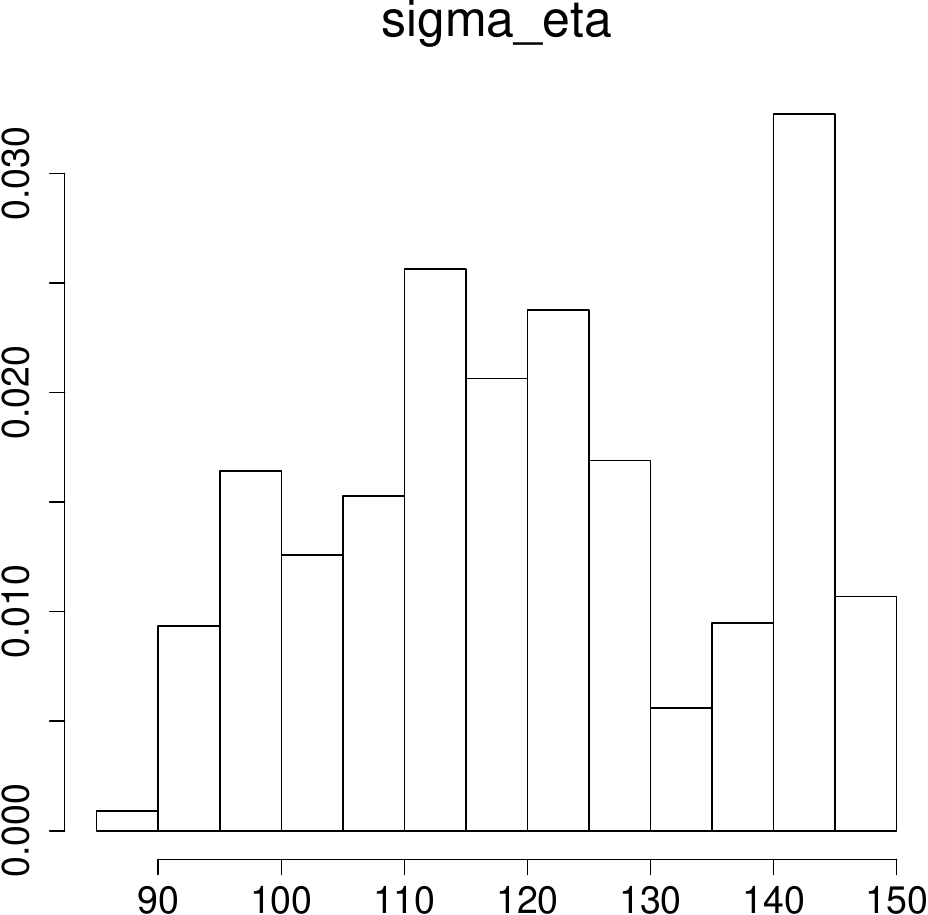}}
   \hspace{2mm}
  \subfigure{ \label{fig:r_f_1_realdata_cps_hist} 
  \includegraphics[width=5cm,height=4cm]{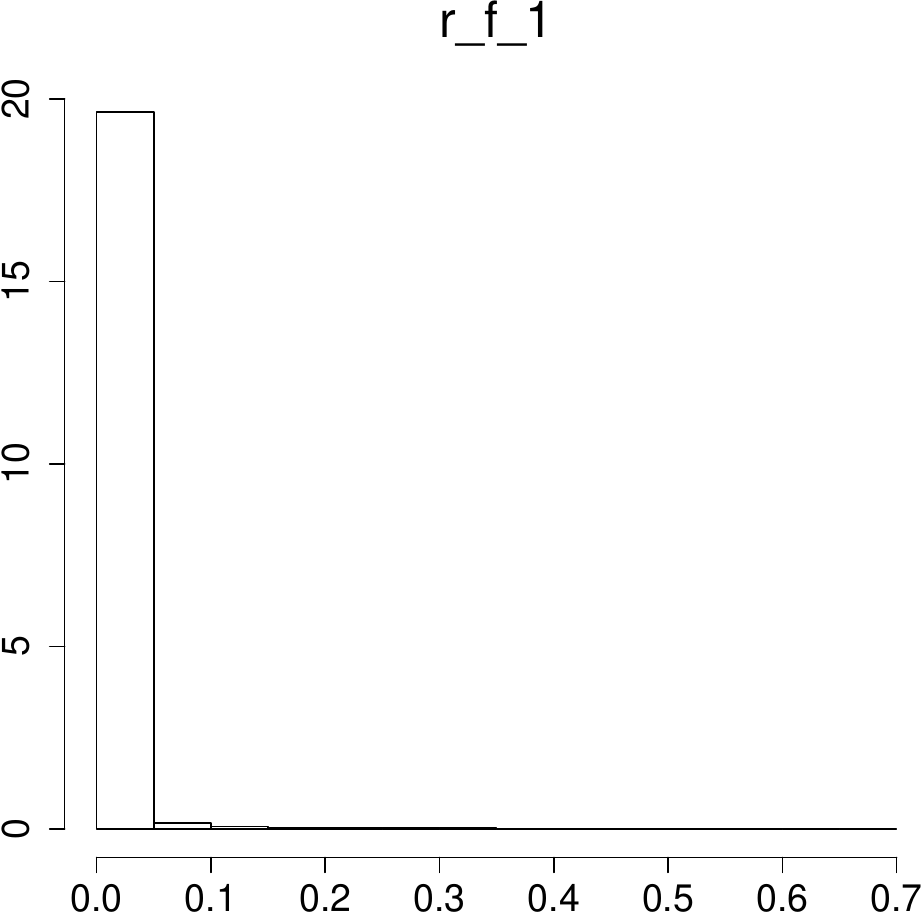}}
   \hspace{2mm}
  \subfigure{ \label{fig:r_f_2_realdata-cps} 
  \includegraphics[width=5cm,height=4cm]{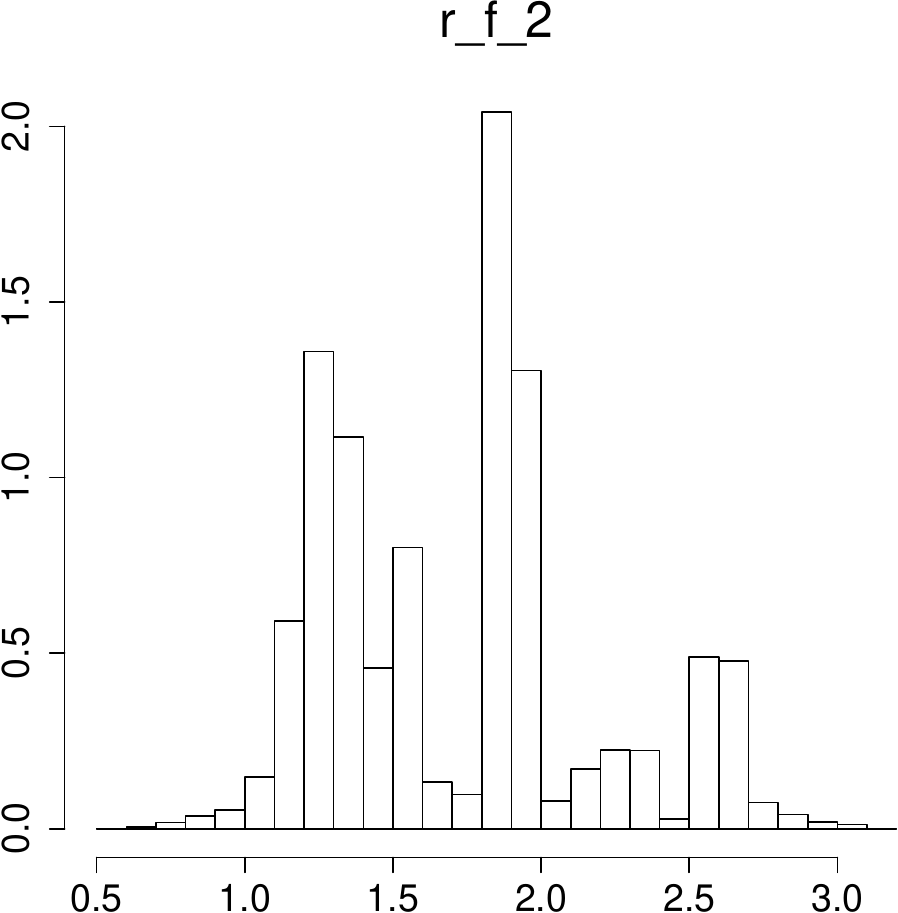}}\\
  \vspace{2mm}
  \subfigure{ \label{fig:r_g_1_realdata_cps_hist}
   \includegraphics[width=5cm,height=4cm]{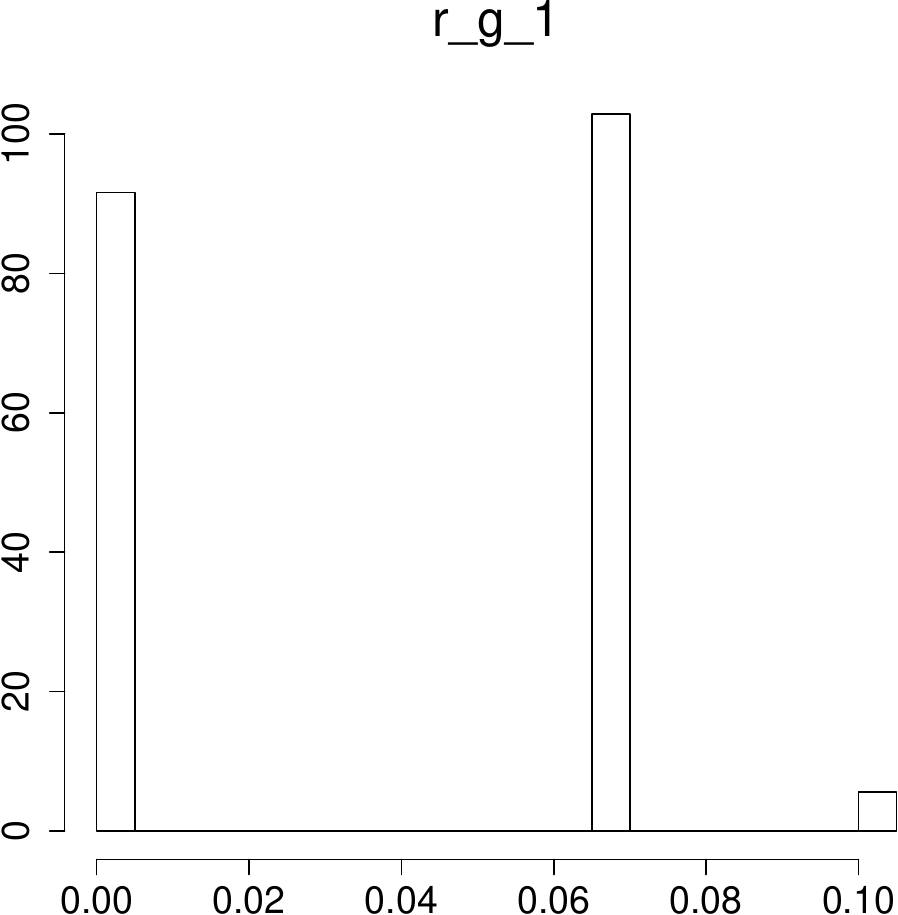}}
   \hspace{2mm}
  \subfigure{ \label{fig:r_g_2_realdata_cps_hist} 
  \includegraphics[width=5cm,height=4cm]{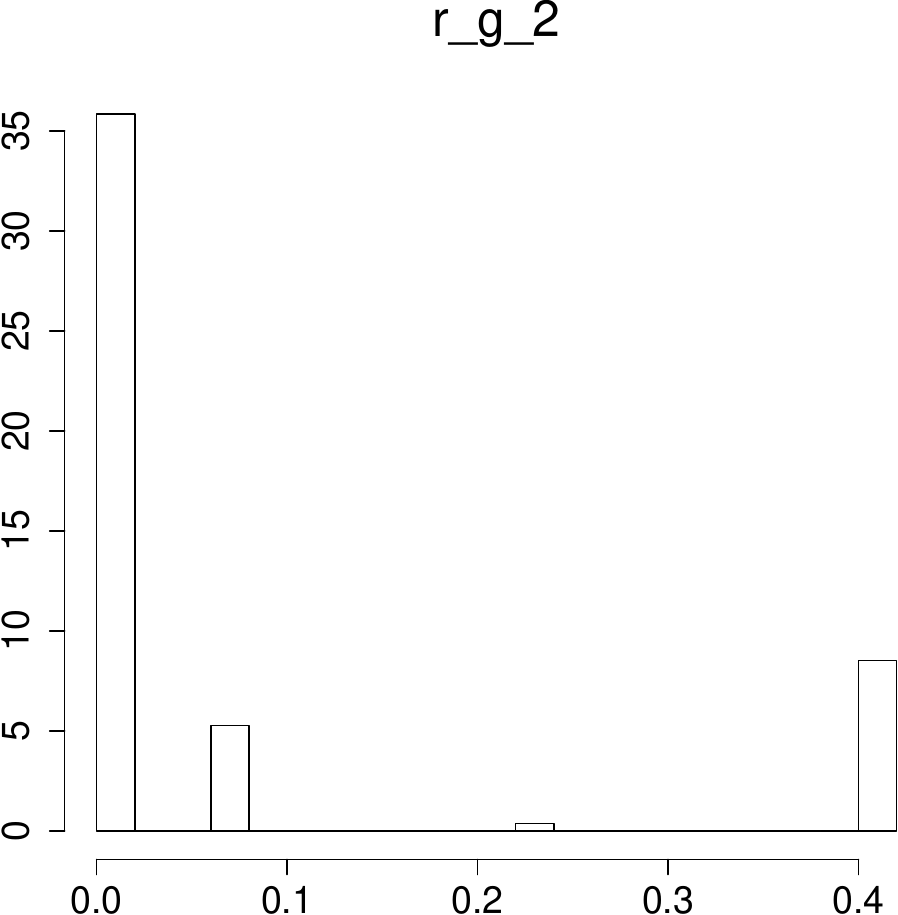}}
   \hspace{2mm}
  \subfigure{ \label{fig:x0_realdata_cps_hist} 
  \includegraphics[width=5cm,height=4cm]{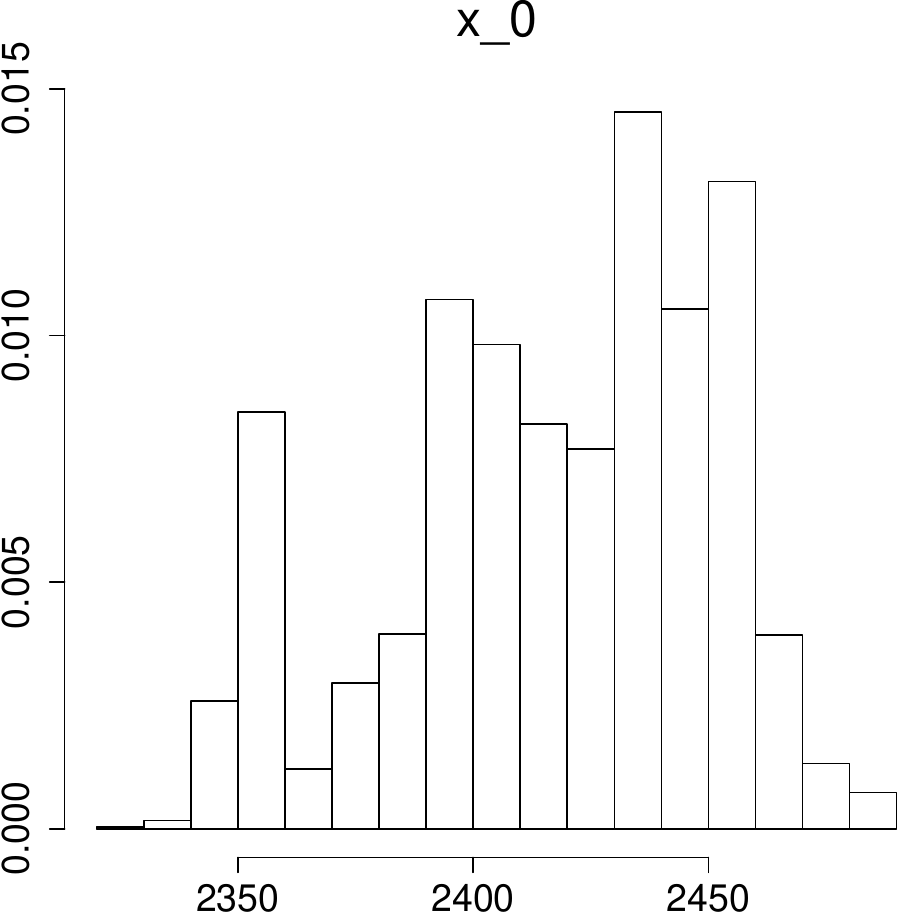}}\\
  \vspace{2mm}
  \subfigure{ \label{fig:forecasted_x_realdata_cps_hist}
   \includegraphics[width=5cm,height=4cm]{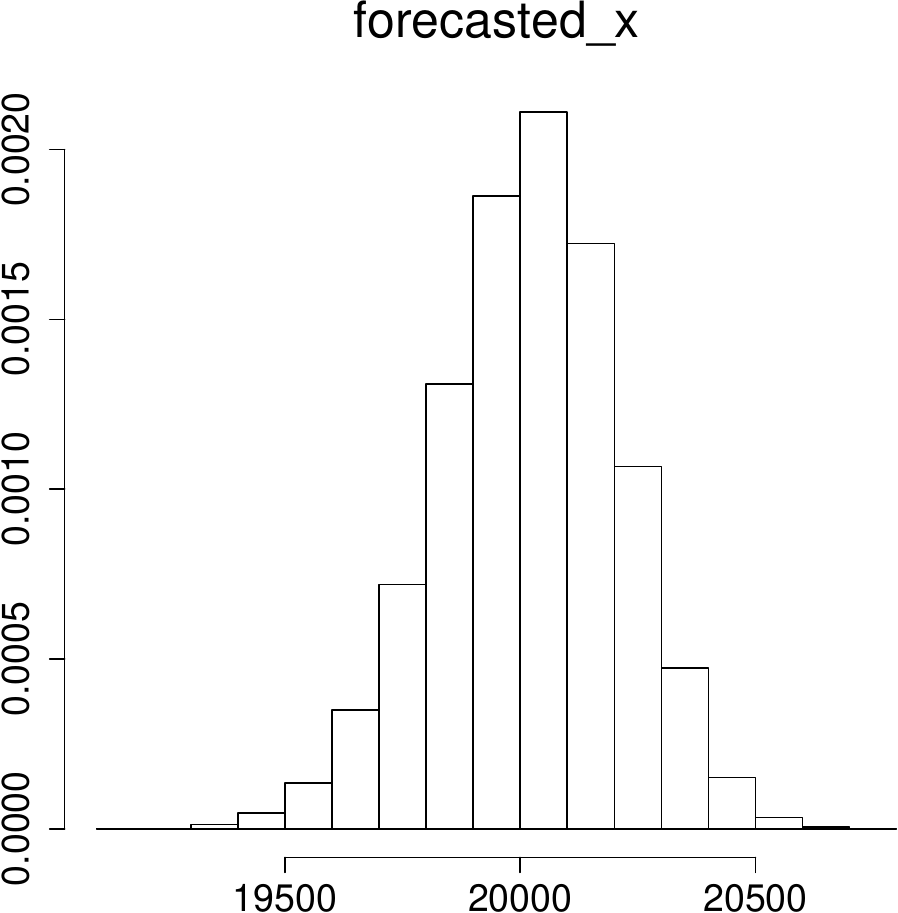}}
   \hspace{2mm}
  \subfigure{ \label{fig:forecasted_y_realdata_cps_hist} 
  \includegraphics[width=5cm,height=4cm]{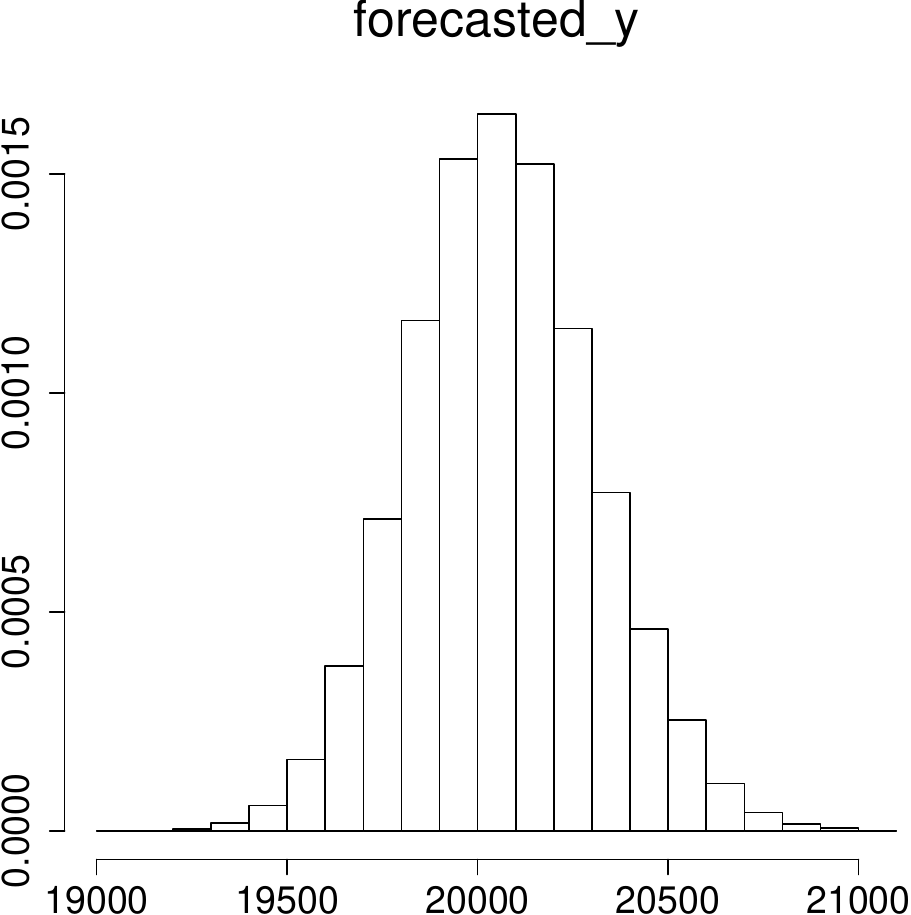}}
 \caption{{\bf Real data analysis}: 
 Posterior densities of $\sigma_{\eta}$, $r_{1,f}$, $r_{2,f}$, $r_{1,g}$, $r_{2,g}$,
 $x_0$, $x_{T+1}$ (one-step forecasted $x$), and $y_{T+1}$ (one-step forecasted $y$).} 
 \label{fig:realdata_cps_hist2}
 \end{figure}


\begin{figure}
\centering
  \subfigure{ \label{fig:x_time_series_realdata_cps_hist}
   \includegraphics[width=15cm,height=10cm]{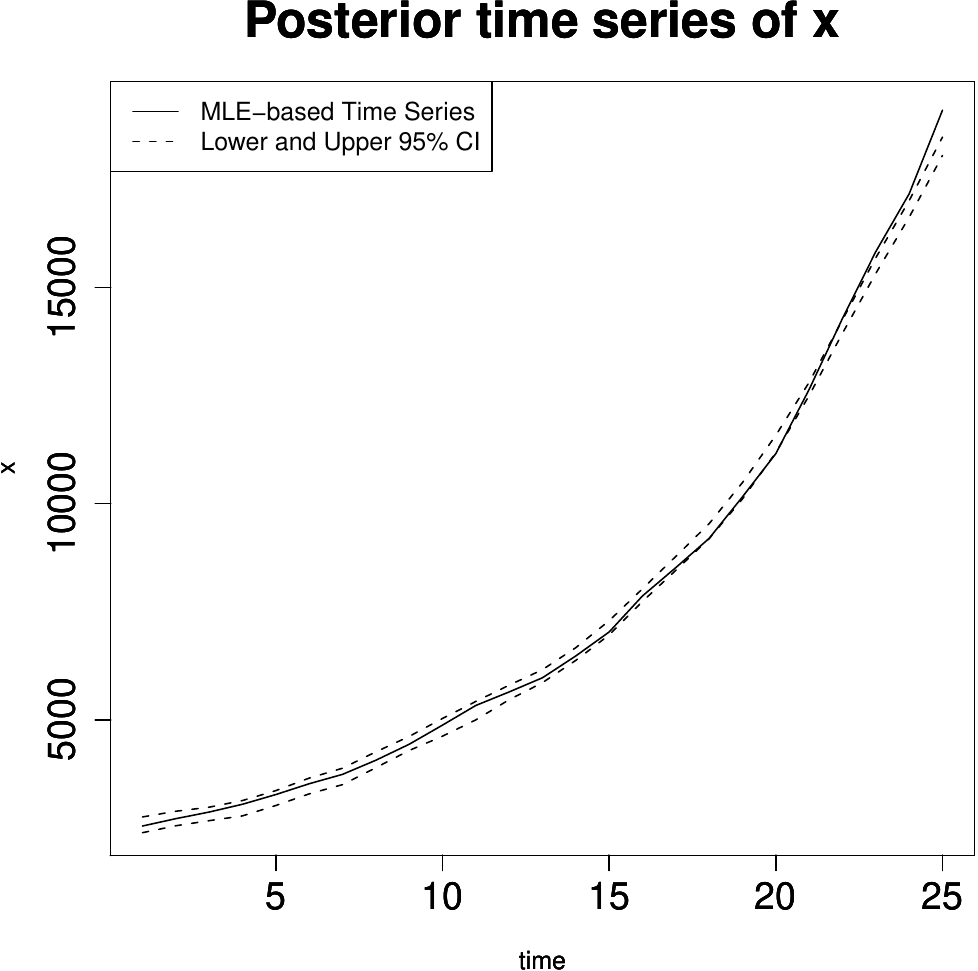}}
 \caption{{\bf Real data analysis}: 
 95\% highest posterior density credible intervals of the time series $x_1,\ldots,x_T$. 
 The solid line stands for the MLE time series obtained by Shumway and Stoffer (1982).}
 \label{fig:realdata_cps_hist3}
 \end{figure}

Figures \ref{fig:realdata_cps_hist}, \ref{fig:realdata_cps_hist2} and \ref{fig:realdata_cps_hist3} show 
the posterior distributions of the unknowns. Our posterior time series of $x_t$ has relatively
narrow 95\% highest posterior density credible intervals, vindicating that the linearity
assumption is not unreasonable as claimed by \ctn{Shumway82} and \ctn{Carlin92}.
Indeed, for such linear relationships, it is well-known that the Gaussian process priors 
that we adopted are expected to perform very well. 

However, perhaps not all is well with the aforementioned linearity assumption of 
\ctn{Shumway82} and \ctn{Carlin92}.
Note that although the MLE time series obtained by \ctn{Shumway82} fall mostly within our Bayesian
95\% highest posterior density credible intervals of $x_t$, five observations towards the
end of the time series fall outside. These observations correspond to the years $1968$,
$1970$, $1971$, $1972$, and $1973$. The values of $y_t$ in these years are $11,099$, $14,306$,
$15,835$, $16,916$, and $18,200$. We suspect that linearity breaks down towards the end of the
time series, an issue which is perhaps overlooked by the linear model based approaches of \ctn{Shumway82} and \ctn{Carlin92}.
On the other hand, without any change point analysis, our flexible nonparametric model, 
based on Gaussian processes, is able to 
accommodate changes in the regression structures.

\section{Conclusions and future work}
\label{sec:conclusions}
In this article, using Gaussian process priors and the ``look-up table" idea
of \ctn{Bhattacharya07} we proposed a novel methodology for Bayesian inference in nonparametric
state space models, in both univariate and multivariate cases. The Gaussian process priors on the unknown 
functional forms of the observational
and the evolutionary equations allow for very flexible modeling of time-varying random functions,
where even the functional forms may change over time, without requiring any change point analysis. 
We have vindicated the effectiveness of our model and methodology with simulation experiments,
and a real data analysis which provided interesting insight
into nonlinearity of the underlying time series towards its end.

For our current purpose, we have assumed $iid$ Gaussian noises $\epsilon_t$ and $\eta_t$; however,
it is straightforward to generalize these to other parametric distributions (thick-tailed or otherwise). 
It may, however, be quite interesting to consider nonparametric error distributions, for example, mixtures
of Dirichlet processes; such a work in the linear cases has been undertaken by \ctn{Caron07}. 
We shall also consider matrix-variate extensions to our current work, in addition to nonparametric
error distributions.

\section*{Acknowledgment} We are extremely grateful to an anonymous reviewer whose suggestions resulted in
improved presentation of our ideas.


\section*{Description of the supplement}
In Section S-1 of the supplement we state and prove results on the accuracy of the look-up table idea
and in Section S-2 we provide details of MCMC sampling in the univariate situation.
In Section S-3 two simulation studies are considered in details.
We provide extension of our Gaussian process based methodology for univariate cases to multivariate
situations in Section S-4; followed by details of MCMC sampling in the multivariate set-up in
Section S-5. In Section S-6 we conduct a detailed simulation study to illustrate our methodology in 
the multivariate set-up. Finally, in Section S-7 we provide detailed discussion of posteriors
of the composite observational and evolutionary functions and their comparisons with the respective
true composite functions; in particular, we provide the comparisons in the case of the three simulation
studies (two univariate cases, and one multivariate case). 


\newpage

\renewcommand\baselinestretch{1.3}
\normalsize
\bibliographystyle{ECA_jasa}
\bibliography{irmcmc}

\end{document}


\normalsize

\title{\vspace{-0.8in}
Supplement to ``Bayesian Inference in Nonparametric Dynamic State-Space Models"}
\author{Anurag Ghosh, Soumalya Mukhopadhyay, Sandipan Roy and Sourabh Bhattacharya\thanks{
Anurag Ghosh is a PhD student in Department of Statistical Science, Duke University, 
Soumalya Mukhopadhyay is a PhD student in Agricultural and Ecological Research Unit, 
Indian Statistical Institute, Sandipan Roy 
is a PhD student in Department of Statistics, University of Michigan, Ann Arbor, and Sourabh Bhattacharya 
is an Assistant Professor in
Bayesian and Interdisciplinary Research Unit, Indian Statistical
Institute, 203, B. T. Road, Kolkata 700108.
Corresponding e-mail: sourabh@isical.ac.in.}}
\date{\vspace{-0.5in}}
\maketitle

Throughout, we refer to our main paper \ctn{Ghosh13} as GMRB.

\section{Result on the accuracy of the Markov approximation of the conditional distribution of $x_t$
given $\bD^*_n$}
\label{sec:accuracy_markov}
\begin{theorem}
Let $g^{(t)}(\cdot)$ denote composition applied to $g(\cdot)$
with itself $t~(t\geq 1)$ times, where $g^{(1)}(\cdot)=g(\cdot)$. 
Let $g^{(t)}_{true}(\cdot)$ be defined analogously. Let the inputs $\{z_1,\ldots,z_n\}$ satisfy
$z_i-z_{i-1}=(b-a)/n$ for $i=2,3,\ldots,n$, 
where $a,b$ are finite constants such that $a<\min\{z_1,\ldots,z_n\}<\min\{z_1,\ldots,z_n\}<b$.
Also assume that $\bG_n$ is of the form $\{z_1,\ldots,z_n,g_{true}(z_1),\ldots,g_{true}(z_n),
g^{(2)}_{true}(z_1),\\ \ldots,g^{(2)}_{true}(z_n),\ldots,g^{(T-1)}_{true}(z_1), \ldots,g^{(T-1)}_{true}(z_n)\}$, so that $\bD^*_n
=\{g(z_1),\ldots,g(z_n),g(g_{true}(z_1)),\ldots,\\ g(g_{true}(z_n)),
g(g^{(2)}_{true}(z_1)),\ldots, g(g^{(2)}_{true}(z_n)),\ldots,g(g^{(T-1)}_{true}(z_1)),\ldots,g(g^{(T-1)}_{true}(z_n))\}$.
Then, for $t=1,\ldots,T$, for $r>0$,
\begin{equation}
\int_a^b\left|E[g^{(t)}(z)|\bD^*_n,\btheta_g]-g^{(t)}_{true}(z)\right|^rdz=O\left(n^{-1}\right). 
\label{eq:L_r}
\end{equation}
Also, for any $z^*\in [a,b]$,
\begin{equation}
\left|E[g^{(t)}(z^*)|\bD^*_n,\btheta_g]-g^{(t)}_{true}(z^*)\right|=O\left(n^{-1}\right),
\label{eq:pointwise}
\end{equation}
and 
\begin{equation}
Var[g^{(t)}(z^*)|\bD^*_n,\btheta_g]=O\left(n^{-1}\right).
\label{eq:pointwise_var}
\end{equation}
\end{theorem}
{\it Proof:}\\
Note that for any $r>0$, we have
\begin{align}
&\int_a^b\left|E[g^{(t)}(z)|\bD^*_n,\btheta_g]-g^{(t)}_{true}(z)\right|^rdz
=\frac{b-a}{n}\sum_{i=1}^n\left|E[g^{(t)}(z_i)|\bD^*_n,\btheta_g]-g^{(t)}_{true}(z_i)\right|^r+O\left(n^{-1}\right),\notag
\end{align}
by Riemann sum approximation. But by the choice of $\bG_n$ and the interpolation property of Gaussian processes,
it follows that, given $\bD^*_n$, $g(z_i)=g_{true}(z_i)$ with probability one. Hence, given $\bD^*_n$,
$g(g(z_i))=g(g_{true}(z_i))=g_{true}(g_{true}(z_i))$ with probability 1 since $g_{true}(z_i)$ is in $\bG_n$, and so on.
In general, given $\bD^*_n$, $g^{(t)}(z_i)=g^{(t)}_{true}(z_i)$ with probability 1.
Hence, we have
$\left|E[g^{(t)}(z_i)|\bD^*_n,\btheta_g]-g^{(t)}_{true}(z_i)\right|=0$ for $k=1,\ldots,n$.
The result (\ref{eq:L_r}) thus follows.

To see (\ref{eq:pointwise}) note that if $z^*\in\bG_n$, then $\left|E[g^{(t)}(z^*)|\bD^*_n,\btheta_g]-g^{(t)}_{true}(z^*)\right|=0$
by the interpolation property of Gaussian processes, so that (\ref{eq:pointwise}) trivially holds. If $z^*\in[a,b]$ is not
a design point in $\bG_n$, then there exists $z_i\in\bG_n$ such that $z^*\in [z_i,z_{i+1}]$. Hence, $z^*=z_i+h$, where
$h<(b-a)/n$. Then, letting $\nu(\cdot)=E[g^{(t)}(\cdot)|\bD^*_n,\btheta_g]$ and $\gamma(\cdot)=g^{(t)}_{true}(\cdot)$ we have
\begin{align}
\nu(z_i+h)&=\nu(z_i)+h\nu'(\xi_1)\quad\mbox{and}\label{eq:nu}\\
\gamma(z_i+h)&=\gamma(z_i)+h\gamma'(\xi_2),\label{eq:gamma}
\end{align}
where $\xi_1$ and $\xi_2$ lie between $z_i$ and $z_i+h$. Since $\nu(z_i)=\gamma(z_i)$ and since
$\nu'(\cdot)$ and $\gamma'(\cdot)$ are bounded on $[a,b]$ (as the exponential correlation function of our Gaussian process
ensures that $\nu$ and $\gamma$ are continuously differentiable), we have, using (\ref{eq:nu}) and (\ref{eq:gamma}),
\begin{equation*}
\left|E[g^{(t)}(z^*)|\bD^*_n,\btheta_g]-g^{(t)}_{true}(z^*)\right|=\left|\nu(z_i+h)-\gamma(z_i+h)\right|=O\left(h\right).
\end{equation*}
Since $h<(b-a)/n$, result (\ref{eq:pointwise}) follows.

To prove (\ref{eq:pointwise_var}), let $\varphi(\cdot)= Var[g^{(t)}(\cdot)|\bD^*_n,\btheta_g]$.
Then
\begin{align}
\varphi(z_i+h)&=\varphi(z_i)+h\varphi'(\xi_3),\notag
\end{align}
where $\xi_3$ lies between $z_i$ and $z_i+h$. But $\varphi(z_i)=0$ due to the interpolation property of Gaussian processes,
and $\varphi'(\cdot)$ is bounded on $[a,b]$ (again, the exponential correlation function ensuring that $\varphi$ is
continuously differentiable). Moreover, since $h<(b-a)/n$, the result (\ref{eq:pointwise_var}) follows.
\hfill$\square$

\section{Details of MCMC sampling in the univariate situation}
\label{sec:fullconds}

Let $\bG_{n+1}=\bG_n\cup\{x^*_{1,0}\}$, $\bD^*_{n+1}=\left({\bD^*_n}',g(x^*_{1,0})\right)'$, 
$\bA_{g,n+1}=\left(\begin{array}{cc}\bA_{g,D^*_n} & \bs_{g,D^*_n}(x^*_{1,0})\\ \bs_{g,D^*_n}(x^*_{1,0})' & 1\end{array}\right)$
and $\bH_{g,D^*_{n+1}}'=[\bH_{g,D^*_n}', \bh(x^*_{1,0})]$. Clearly, $[\bD^*_{n+1}\mid x^*_{1,0},\btheta_g]=[g(x^*_{1,0})\mid x_{0},\btheta_g][\bD^*_n\mid g(x^*_{1,0}),x_{0},\btheta_g]$.
With these definitions, the forms of the full conditional distributions of the unknowns are provided below.
\begin{align}
[\bbeta_f\mid \cdots]&\propto [\bbeta_f][\bD_T\mid x_1,\ldots,x_{T},\btheta_f,\sigma^2_{\epsilon}]\label{eq:fullcond_betaf}\\
[\bbeta_g\mid \cdots]&\propto [\bbeta_g]
[\bD^*_{n+1}\mid x^*_{1,0},\btheta_g]
\prod_{t=1}^{T}[x_{t+1} \mid\btheta_g,\sigma^2_{\eta},\bD^*_n,x_t]\label{eq:fullcond_betag}\\
[\sigma^2_f\mid \cdots]&\propto [\sigma^2_f][\bD_T\mid x_1,\ldots,x_{T},\btheta_f,\sigma^2_{\epsilon}]\label{eq:fullcond_sigmasqf}\\
[\sigma^2_g\mid \cdots]&\propto [\sigma^2_g][\bD^*_{n+1}\mid x^*_{1,0},\btheta_g]
\prod_{t=1}^{T}[x_{t+1} \mid\btheta_g,\sigma^2_{\eta},\bD^*_n,x_t]\label{eq:fullcond_sigmasqg}\\
[\sigma^2_{\epsilon}\mid \cdots]&\propto [\sigma^2_{\epsilon}][\bD_T\mid x_1,\ldots,x_{T},\btheta_f,\sigma^2_{\epsilon}]\label{eq:fullcond_sigmasq_epsilon}\\
[\sigma^2_{\eta}\mid \cdots]&\propto [\sigma^2_{\eta}][x_1\mid g(x^*_{1,0}),x_0,\btheta_g,\sigma^2_{\eta}]\prod_{t=1}^{T}[x_{t+1} \mid\btheta_g,\sigma^2_{\eta},\bD^*_n,x_t]
\label{eq:fullcond_sigmasq_eta}\\
[r_{i,f}\mid\cdots]&\propto [r_{i,f}][\bD_T\mid x_1,\ldots,x_{T},\btheta_f,\sigma^2_{\epsilon}];\ \ i=1,2\label{eq:fullcond_bf}\\
[r_{i,g}\mid\cdots]&\propto [r_{i,g}][\bD^*_n\mid g(x^*_{1,0}),x_0,\btheta_g]
\prod_{t=1}^{T}[x_{t+1} \mid\btheta_g,\sigma^2_{\eta},\bD^*_n,x_t];\ \ i=1,2\label{eq:fullcond_bg}\\
[g(x^*_{1,0})\mid\cdots]&\propto [g(x^*_{1,0})\mid x_0,\bbeta_g,\sigma^2_g][\bD^*_n\mid g(x^*_{1,0}),x_0,\btheta_g][x_1\mid g(x^*_{1,0}),x_0,\sigma^2_{\eta}]
\label{eq:fullcond_g}\\
[\bD^*_n\mid\cdots]&\propto \prod_{t=1}^{T} [x_{t+1}\mid\btheta_g,\sigma^2_{\eta},\bD^*_n,x_t][\bD^*_n\mid g(x^*_{1,0}),x_0,\btheta_g]\label{eq:fullcond_Dz}\\
[x_0\mid\cdots]&\propto[x_0]
[\bD^*_{n+1}\mid x_0,\btheta_g]\label{eq:fullcond_x0}\\
[x_1\mid\cdots]&\propto [x_1\mid g(x^*_{1,0}),x_0,\bbeta_g,\sigma^2_g,\sigma^2_{\eta}][x_2 \mid\btheta_g,\sigma^2_{\eta},\bD^*_n,x_1]\nonumber\\
& \ \ \ \ \times[\bD_T\mid x_1,\ldots,x_{T},\btheta_f,\sigma^2_{\epsilon}]\label{eq:fullcond_x1}\\
[x_{t+1}\mid\cdots]&\propto [x_{t+1} \mid\btheta_g,\sigma^2_{\eta},\bD^*_n,x_t][x_{t+2} \mid\btheta_g,\sigma^2_{\eta},\bD^*_n,x_{t+1}]\nonumber\\
& \ \ \ \ \times[\bD_T\mid x_1,\ldots,x_{T},\btheta_f,\sigma^2_{\epsilon}]; \ \ \ \ t=1,\ldots,T-1\label{eq:fullcond_xt}\\
[x_{T+1}\mid\cdots]&\propto [x_{T+1} \mid\btheta_g,\sigma^2_{\eta},\bD^*_n,x_T]\label{eq:fullcond_xlast}
\end{align}
Although some of the full conditionals are of standard forms permitting Gibbs sampling steps, others are non-standard
and sampling requires Metropolis-Hastings (MH) steps in those situations. We describe below the Gibbs steps and 
construct proposal distributions when MH steps are needed.

\subsection{Updating $\bbeta_f$ using Gibbs step}
\label{subsec:fullcond_beta_f}

The full conditional of $\bbeta_f$ 
is $m$-variate normal with mean
\begin{align}
E\left[\bbeta_f\mid\cdots\right]&=\left\{\bH'_{D_T}\left(\sigma^2_f\bA_{f,D_T}+\sigma^2_{\epsilon}\bI\right)^{-1}\bH_{D_T}+\bSigma^{-1}_{\beta_f,0}\right\}^{-1}\nonumber\\
&\ \ \ \ \times\left\{\bH'_{D_T}\left(\sigma^2_f\bA_{f,D_T}+\sigma^2_{\epsilon}\bI\right)^{-1}\bD_T+\bSigma^{-1}_{\beta_f,0}\bbeta_{f,0}\right\}.
\label{eq:condmean_betaf}
\end{align}
and variance
\begin{align}
V\left[\bbeta_f\mid\cdots\right]&=\left\{\bH'_{D_T}\left(\sigma^2_f\bA_{f,D_T}+\sigma^2_{\epsilon}\bI\right)^{-1}\bH_{D_T}+\bSigma^{-1}_{\beta_f,0}\right\}^{-1}.
\label{eq:condvar_betaf}
\end{align}

\subsection{Updating $\bbeta_g$ using Gibbs step}
\label{subsec:fullcond_beta_g}

The conditional distribution of $\bbeta_g$ 
is $m$-variate normal with mean
\\[2mm]
$E\left[\bbeta_g\mid\cdots\right]$
\begin{align}
&=\left\{\bSigma^{-1}_{\beta_g,0}+\frac{\bH_{D^*_{n+1}}'\bA^{-1}_{g,D^*_{n+1}}\bH_{D^*_{n+1}}}{\sigma^2_g}\right.\nonumber\\
&\ \ \ \ \left.+\sum_{t=1}^{T}\frac{\left(\bH_{D^*_n}'\bA_{g,D^*_n}^{-1}\bs_{g,D^*_n}(x^*_{t+1,t})-\bh(x^*_{t+1,t})\right)\left(\bH_{D^*_n}'\bA_{g,D^*_n}^{-1}\bs_{g,D^*_n}(x^*_{t+1,t})-\bh(x^*_{t+1,t})\right)'}{\sigma^2_{g,\eta,t}}\right\}^{-1}\nonumber\\
&\ \ \ \ \times\left\{\bSigma^{-1}_{\beta_g,0}\bbeta_{g,0}+\frac{\bH_{D^*_{n+1}}'\bA^{-1}_{g,D^*_{n+1}}\bD^*_{n+1}}{\sigma^2_g}\right.\nonumber\\ 
&\left. \ \ \ \ +\sum_{t=1}^{T}\frac{\left(x_{t+1}-\bs_{g,D^*_n}(x^*_{t+1,t})'\bA^{-1}_{g,D^*_n}\bD^*_n\right)\left(\bh(x^*_{t+1,t})-\bH_{D^*_n}'\bA^{-1}_{g,D^*_n}\bs_{g,D^*_n}(x^*_{t+1,t})\right)}{\sigma^2_{g,\eta,t}}\right\}.
\label{eq:condmean_betag}
\end{align}
and variance
\\[2mm]
$V\left[\bbeta_g\mid\cdots\right]$
\begin{align}
&=\left\{\bSigma^{-1}_{\beta_g,0}+\frac{\bH_{D^*_{n+1}}'\bA^{-1}_{g,D^*_{n+1}}\bH_{D^*_{n+1}}}{\sigma^2_g}\right.\nonumber\\
&\ \ \ \ \left.+\sum_{t=1}^{T}\frac{\left(\bH_{D^*_n}'\bA_{g,D^*_n}^{-1}\bs_{g,D^*_n}(x^*_{t+1,t})-\bh(x^*_{t+1,t})\right)\left(\bH_{D^*_n}'\bA_{g,D^*_n}^{-1}\bs_{g,D^*_n}(x^*_{t+1,t})-\bh(x^*_{t+1,t})\right)'}{\sigma^2_{g,\eta,t}}\right\}^{-1}
\label{eq:condvar_betag}
\end{align}
In (\ref{eq:condmean_betag}) and (\ref{eq:condvar_betag}),
\begin{equation}
\sigma^2_{g,\eta,t}=\sigma^2_g\left\{1-\bs_{g,D^*_n}(x^*_{t+1,t})'\bA^{-1}_{g,D^*_n}\bs_{g,D^*_n}(x^*_{t+1,t})\right\}+\sigma^2_{\eta}.
\label{eq:betag_error}
\end{equation}


\subsection{Updating $\sigma^2_f$ and $\sigma^2_g$ using MH steps}
\label{subsec:fullcond_sigma_f_g}

The full conditionals of $\sigma^2_f$ and $\sigma^2_g$ 
are not available in closed forms
and MH steps are necessary here. For proposal
distributions we first construct a new model by setting $\sigma^2_{\epsilon}=\sigma^2_f$
and $\sigma^2_{\eta}=\sigma^2_g$.
For this model the full conditional distributions
of $\sigma^2_f$ and $\sigma^2_g$ are inverse Gamma distributions, given by
\begin{align}
q_{\sigma^2_f}(\sigma^2_f)&\propto\left(\sigma^2_f\right)^{-\left(\frac{T+\alpha_f+2}{2}\right)}\exp\left[-\frac{1}{2\sigma^2_f}\left\{\gamma_f+\left(\bD_T-\bH_{D_T}\bbeta_f\right)'\left(\bA_{f,D_T}+\bI_T\right)^{-1}\left(\bD_T-\bH_{D_T}\bbeta_f\right)\right\}\right]
\label{eq:proposal_sigmasqf}
\end{align}
and
\begin{align}
q_{\sigma^2_g}(\sigma^2_g)&\propto\left(\sigma^2_g\right)^{-\left(\frac{\alpha_g+4+n+T}{2}\right)}
\exp\left[-\frac{1}{2\sigma^2_g}\left\{\gamma_g+(x_1-g(x^*_{1,0}))^2\right.\right.\nonumber\\
&\ \ \ \ \left.\left.+\sum_{t=1}^{T}\frac{\left\{x_{t+1}-\bh(x^*_{t+1,t})'\bbeta_g-\bs_{g,D^*_n}(x^*_{t+1,t})'\bA^{-1}_{g,D^*_n}\left(\bD^*_n-\bH_{D^*_n}\bbeta_g\right)\right\}^2}{\sigma^2_{g,t}}\right.\right.\nonumber\\
&\ \ \ \ \left.\left.+\left(\bD^*_{n+1}-\bH_{D^*_{n+1}}\bbeta_g\right)'\bA^{-1}_{g,D^*_{n+1}}\left(\bD^*_{n+1}-\bH_{D^*_{n+1}}\bbeta_g\right)\right\}\right]\nonumber\\
\label{eq:proposal_sigmasqg}
\end{align}
In (\ref{eq:proposal_sigmasqg}),
\begin{equation}
\sigma^2_{g,t}
=2-\bs_{g,D^*_n}(x^*_{t+1,t})'\bA^{-1}_{g,D^*_n}\bs_{g,D^*_n}(x^*_{t+1,t}).
\label{eq:betag_lesserror}
\end{equation}
In (\ref{eq:proposal_sigmasqg}) the term $\left(\sigma^2_g\right)^{-1/2}\exp\left\{-\frac{1}{2\sigma^2_g}(x_1-g(x^*_{1,0}))^2\right\}$
is also taken into account since, given $\sigma^2_{\eta}=\sigma^2_g$, $[x_1\mid g(x^*_{1,0}),x_0]\sim N(g(x^*_{1,0}),\sigma^2_g)$.
It is useful to remark here that unless $\sigma_f\approx\sigma_{\epsilon}$ and $\sigma_g\approx\sigma_{\eta}$ these proposal mechanisms
may not be efficient.
We shall discuss other proposal distributions in the context of applications.


\subsection{Updating $\sigma^2_{\epsilon}$ and $\sigma^2_{\eta}$ using MH steps}
\label{subsec:fullcond_sigma_e_eta}

As before, the full conditionals of $\sigma^2_{\epsilon}$ and $\sigma^2_{\eta}$ 
are not available in closed forms. For MH steps, we construct proposal distributions 
obtained by setting $\sigma^2_f=\sigma^2_{\epsilon}$ and $\sigma^2_g=\sigma^2_{\eta}$.
The proposal distributions are given by
\begin{align}
q_{\sigma^2_{\epsilon}}(\sigma^2_{\epsilon})&\propto\left(\sigma^2_{\epsilon}\right)^{-\left(\frac{T+\alpha_{\epsilon}+2}{2}\right)}
\exp\left[-\frac{1}{2\sigma^2_{\epsilon}}
\left\{\gamma_{\epsilon}+\left(\bD_T-\bH_{D_T}\bbeta_f\right)'\left(\bA_{f,D_T}+\bI_T\right)^{-1}\left(\bD_T-\bH_{D_T}\bbeta_f\right)\right\}\right] 
\label{eq:proposal_sigmasq_epsilon}\\
q_{\sigma^2_{\eta}}(\sigma^2_{\eta})&\propto\left(\sigma^2_{\eta}\right)^{-\left(\frac{\alpha_{\eta}+4+n+T}{2}\right)}
\exp\left[-\frac{1}{2\sigma^2_{\eta}}\left\{\gamma_{\eta}+(x_1-g(x^*_{1,0}))^2\right.\right.\nonumber\\
&\ \ \ \ \left.\left.+\left(\bD^*_{n+1}-\bH_{D^*_{n+1}}\bbeta_g\right)'\bA^{-1}_{g,D^*_{n+1}}\left(\bD^*_{n+1}-\bH_{D^*_{n+1}}\bbeta_g\right)\right.\right.\nonumber\\
&\ \ \ \ \left.\left.+\sum_{t=1}^{T}\frac{\left\{x_{t+1}-\bh(x^*_{t+1,t})'\bbeta_g-\bs_{g,D^*_n}(x^*_{t+1,t})'\bA^{-1}_{g,D^*_n}\left(\bD^*_n-\bH_{D^*_n}\bbeta_g\right)\right\}^2}{\sigma^2_{g,t}}\right\}\right]\nonumber\\
\label{eq:proposal_sigmasq_eta}
\end{align}
In (\ref{eq:proposal_sigmasq_eta}) the term $\left(\sigma^2_{\eta}\right)^{-(n+1)/2}
\exp\left\{-\frac{1}{2\sigma^2_{\eta}}\left(\bD^*_{n+1}-\bH_{D^*_{n+1}}\bbeta_g\right)'\bA^{-1}_{g,D^*_{n+1}}\left(\bD^*_{n+1}-\bH_{D^*_{n+1}}\bbeta_g\right)\right\}$
occurs since, given $\sigma^2_g=\sigma^2_{\eta}$, $[\bD^*_{n+1}\mid\bbeta_g,\sigma^2_g,\bR_g]\sim N_{n+1}\left(\bH_{D^*_{n+1}}\bbeta_g,\sigma^2_{\eta}\bA_{g,D^*_{n+1}}\right)$.
Again, these proposals need not be efficient unless $\sigma_f\approx\sigma_{\epsilon}$ and $\sigma_g\approx\sigma_{\eta}$.
In the context of specific applications we shall discuss other proposal distributions.

\subsection{Updating $\bR_f$ and $\bR_g$ using MH steps}
\label{subsec:fullcond_smoothness}

For $i=1,2$, the full conditionals of the smoothness parameters $r_{i,f}$ and $r_{i,g}$
are not available in closed forms, and we suggest MH steps
with normal random walk proposals with adequately optimized variances. 

\subsection{Updating $g(x^*_{1,0})$ using Gibbs step}
\label{subsec:fullcond_g}

The full conditional of $g(x^*_{1,0})$ is univariate normal with mean and variance given, respectively, by
\begin{align}
E[g(x^*_{1,0})\mid\cdots]
&=\left\{\frac{1}{\sigma^2_{\eta}}+\frac{1+\bs_{g,D^*_n}(x^*_{1,0})'\bSigma^{-1}_{g,D^*_n}\bs_{g,D^*_n}(x^*_{1,0})}{\sigma^2_g}\right\}^{-1}\nonumber\\
& \ \ \ \ \times\left\{\frac{x_1}{\sigma^2_{\eta}}+\frac{\bh(x^*_{1,0})'\bbeta_g+\bs_{g,D^*_n}(x^*_{1,0})'\bSigma^{-1}_{g,D^*_n}\bD^*_z}{\sigma^2_g}\right\}\nonumber\\
\label{eq:condmean_g}
\end{align}
and
\begin{align}
V[g(x^*_{1,0})\mid\cdots]
&=\left\{\frac{1}{\sigma^2_{\eta}}+\frac{1+\bs_{g,D^*_n}(x^*_{1,0})'\bSigma^{-1}_{g,D^*_n}\bs_{g,D^*_n}(x^*_{1,0})}{\sigma^2_g}\right\}^{-1}\nonumber\\
\label{eq:condvar_g}
\end{align}
In (\ref{eq:condmean_g}),
\begin{equation}
\bD^*_z=\bD^*_n-\bH_{D^*_n}\bbeta_g+\bs_{g,D^*_n}(x^*_{1,0})\bh(x^*_{1,0})'\bbeta_g
\label{eq:D_star_z}
\end{equation}

\subsection{Updating $\bD^*_n$ using Gibbs step}
\label{subsec:fullcond_D^*_n}

The full conditional distribution of $\bD^*_n$ 
is $n$-variate normal with mean
\begin{align}
E\left[\bD^*_n\mid\cdots\right]
&=\left\{\frac{\bSigma^{-1}_{g,D^*_n}}{\sigma^2_g}
+\bA^{-1}_{g,D^*_n}\left(\sum_{t=1}^{T}\frac{\bs_{g,D^*_n}(x^*_{t+1,t})\bs_{g,D^*_n}(x^*_{t+1,t})'}{\sigma^2_{g,\eta,t}}\right)\bA^{-1}_{g,D^*_n}\right\}^{-1}\nonumber\\
&\times\left\{\frac{\bSigma^{-1}_{g,D^*_n}\bmu_{g,D^*_n}}{\sigma^2_g}
+\bA^{-1}_{g,D^*_n}\sum_{t=1}^{T}\frac{\bs_{g,D^*_n}(x^*_{t+1,t})\{x_{t+1}-\bbeta_g'(\bh(x^*_{t+1,t})-\bH'_{D^*_n}\bA^{-1}_{g,D^*_n}\bs_{g,D^*_n}(x^*_{t+1,t}))\}}{\sigma^2_{g,\eta,t}}\right\}\nonumber\\
\label{eq:condmean_Dz}
\end{align}
and variance
\begin{align}
V\left[\bD^*_n\mid\cdots\right]&=\left\{\frac{\bSigma^{-1}_{g,D^*_n}}{\sigma^2_g}+\bA^{-1}_{g,D^*_n}\left(\sum_{t=1}^{T}\frac{\bs_{g,D^*_n}(x^*_{t+1,t})\bs_{g,D^*_n}(x^*_{t+1,t})'}{\sigma^2_{g,\eta,t}}\right)\bA^{-1}_{g,D^*_n}\right\}^{-1}
\label{eq:condvar_Dz}
\end{align}
In (\ref{eq:condmean_Dz}) and (\ref{eq:condvar_Dz}), $\bmu_{g,D^*_n}$ and $\bSigma_{g,D^*_n}$ are given by
(9) and (10), respectively, of GMRB.

\subsection{Updating $x_0$ using MH step}
\label{subsec:fullcond_x_0}

Let $\bbeta_f=(\bbeta_{0,f}^\prime,\beta_{1,f})'$ and $\bbeta_g=(\bbeta_{0,g}^\prime,\beta_{1,g})'$,
where $\bbeta_{0,f}$ and $\bbeta_{0,g}$ are two-component vectors.
Assuming the prior of $x_0$ to be normal with mean $\mu_{x_0}$ and
variance $\sigma^2_{x_0}$, and setting $\sigma^2_g=0$, the full conditional distribution of $x_1$
is univariate normal, with mean and variance given, respectively, by
\begin{align}
\xi_0&=\left(\frac{1}{\sigma^2_{x_0}}+\frac{\beta^2_{1,g}}{\sigma^2_{\eta}}\right)^{-1}
\left\{\frac{\mu_{x_0}}{\sigma^2_{x_0}}+\frac{(x_1-\bk_1'\bbeta_{0,g})\beta_{1,g}}{\sigma^2_{\eta}}\right\},
\label{eq:proposal_mean_x0}\\
\phi^2_0&=\left(\frac{1}{\sigma^2_{x_0}}+\frac{\beta^2_{1,g}}{\sigma^2_{\eta}}\right)^{-1}.
\label{eq:proposal_var_x0}
\end{align}
In (\ref{eq:proposal_mean_x0}), for any $t$, $\bk_t=(1,t)'$.
We use $q_{x_0}(x_0)\equiv N\left(x_0:\xi_0,\phi^2_0\right)$ as the proposal distribution for updating $x_0$
using MH step.
Observe that, under $\sigma^2_g=0$, $g(x^*_{1,0})=\bh(x^*_{1,0})'\bbeta_g$ with probability one; hence
$[x_1\mid g(x^*_{1,0}),x_0,\bbeta_g,\sigma^2_{\eta}]\sim N(\bh(x^*_{1,0})'\bbeta_g,\sigma^2_{\eta})$, which has
been taken into account while constructing the above proposal distribution.
Note that this proposal will only be efficient when the $g(\cdot,\cdot)$ is close to linear.
As a result, for non-linear applications, we shall often use other proposal mechanisms, such as the normal
random walk.

\subsection{Updating $\{x_1,\ldots,x_T\}$ using MH steps}
\label{subsec:fullcond_x_t}

We construct proposal distributions 
for simulating $\{x_1,\ldots,x_T\}$
based on linear observational and evolutionary equations,
setting $\sigma^2_f=\sigma^2_g=0$. Thus, for $t=0,\ldots,T-1$, the proposal distributions of $x_{t+1}$ are of the form
$q_{x_{t+1}}(x_{t+1})\equiv N\left(x_{t+1}:\xi_{t+1},\phi_{t+1}\right)$, that is, a normal distribution
with mean $\xi_{t+1}$ and variance $\phi^2_{t+1}$, where the latter quantities are given by
\begin{align}
\xi_{t+1}&=\left(\frac{1+\beta^2_{1,g}}{\sigma^2_{\eta}}+\frac{\beta^2_{1,f}}{\sigma^2_{\epsilon}}\right)^{-1}
\left\{\frac{\bk_{t+1}'\bbeta_{0,g}+\beta_{1,g}x_t+(x_{t+2}-\bk_{t+2}'\bbeta_{0,g})\beta_{1,g}}{\sigma^2_{\eta}}
+\frac{(y_{t+1}-\bk_{t+1}'\bbeta_{0,f})\beta_{1,f}}{\sigma^2_{\epsilon}}\right\}
\label{eq:proposal_mean_xt}\\
\phi^2_{t+1}&=\left(\frac{1+\beta^2_{1,g}}{\sigma^2_{\eta}}+\frac{\beta^2_{1,f}}{\sigma^2_{\epsilon}}\right)^{-1}
\label{eq:proposal_var_xt}
\end{align}
These proposal mechanisms will be efficient only if both $f(\cdot,\cdot)$ and $g(\cdot,\cdot)$ are close to linear.
Hence, for non-linear situations, we shall consider other proposal distributions.

\subsection{Updating $x_{T+1}$ using Gibbs step}
\label{subsec:fullcond_x_last}

The full conditional distribution of $x_{T+1}$ 
is normal with mean and variance given, respectively, by
\begin{equation}
E\left[x_{T+1}\mid\cdots\right]=\bh(x^*_{T+1,T})^\prime \bbeta_g+\bs_{g,\bD^*_n}(x^*_{T+1,T})^\prime \bA_{g,D^*_n}^{-1}(\bD^*_n-\bH_{D^*_n}\bbeta_g)
\label{eq:fullcond_mean_xt1}
\end{equation}
and variance 
\begin{equation} V\left[x_{T+1}\mid\cdots\right]=\sigma^2_{\eta}+\sigma^2_g\left\{1-\bs_{g,\bD^*_n}(x^*_{T+1,T})^\prime \bA_{g,\bD^*_n}^{-1}\bs_{g,\bD^*_n}(x^*_{T+1,T})\right\}
\label{eq:fullcond_var_xt1}
\end{equation}

\section{Simulation studies in the univariate situations}
\label{sec:simstudies}

In this section we consider two simulation studies: in the first study 
we consider univariate data generated from linear observational and evolutionary
equations, the linear models being invariant with respect to time--we fit our Gaussian process based model on this
linear data.
In the second case we generate data from the parametric, univariate
growth model of \ctn{Carlin92}
and fit our nonparametric model to the data, validating our model and methodology in the process when the data
are governed by non-linear observational and evolutionary equations. 

\subsection{Data generated from a univariate linear model}
\label{subsec:linear_data_generation}

We generate data from the following linear model:
\begin{align}
x_t&=\beta_{x,0}+\beta_{x,1} x_{t-1}+u_t\label{eq:linear1}\\
y_t&=\beta_{y,0}+\beta_{y,1} x_t+v_t, \ \ \ t=1,\ldots,100,\label{eq:linear2}
\end{align}
where $x_0=0$. As before, we assume
$u_t\sim N(0,\sigma^2_{\epsilon})$, $v_t\sim N(0,\sigma^2_{\eta})$, independently for all $t$, and set
$\sigma_{\epsilon}=0.1=\sigma_{\eta}$. We fix
the true values of $(\beta_{x,0},\beta_{x,1},\beta_{y,0},\beta_{y,1})$ to be $(1.0,0.1,8.0,0.05)$, respectively.
We then generated the data set consisting of 101 data points using these true values.
As before, we set aside the last data point for the purpose of forecasting.

Note that under the above true model, the coefficient of time $t$ is zero for both the observational and 
the evolutionary equations. In other words, the functions are not time-variant. Thus, while fitting our
Gaussian process based model, we expect the corresponding coefficients in the mean functions $\beta_{2,f}$,
$\beta_{2,g}$
and the corresponding smoothness parameters $r_{1,f}$ and $r_{1,g}$ to have
large posterior probabilities around zero. The results of our model implementation show that it is
indeed the case.

\subsubsection{Choice of grid and MCMC implementation}
\label{subsubsec:linear_data_prior_choice}

To set up the grid $\bG_n$ we first note the interval containing the entire true time series; in this
example, the entire true time series fall within $[0.8,1.5]$.
We then consider
a much larger interval, $[-30,30]$, containing the aforementioned interval, and divide the larger interval into 100
sub-intervals of equal length. Then we select a value randomly from each sub-interval. 
This yields values of the second component of the
two-dimensional grid $\bG_n$. 
It is worth mentioning that we experimented with other reasonable grid choices, but the results
suggested considerable robustness with respect to the grid choices.
For the first component we generate a number uniformly from each of the
100 sub-intervals $[i,i+1]$; $i=0,\ldots,99$. 

We discarded the first 10,000 MCMC iterations as burn-in and stored the next 50,000 iterations for
inference. 
For updating $\{x_0,\ldots,x_T\}$ we used the 
linear model based proposals detailed in Sections S-2.8 and S-2.9. Since the true models are also linear,
the performance of these proposal distributions, as measured by informal MCMC diagnostics, turned out to be adequate.
However, updating $\{\sigma_f,\sigma_g,\sigma_{\epsilon},\sigma_{\eta}\}$ using the linear model based proposals
detailed in Sections S-2.3 and S-2.4 did not perform satisfactorily since the assumptions $\sigma_f\approx\sigma_{\epsilon}$
and $\sigma_g\approx\sigma_{\eta}$ do not hold here (clearly, since the true models are linear, $\sigma_f=\sigma_g=0$,
whereas $\sigma_{\epsilon}$ and $\sigma_{\eta}$ are positive). So, to update the variance parameters we used the normal
random walk proposal with variance 0.05, which worked adequately.
We note, however, that it is possible to modify the proposals provided in Sections S-2.3 and S-2.4 so that
the fact $\sigma_f=\sigma_g=0$ is taken account of, but since our random walk proposals performed adequately
here we refrained from further experiments on proposal distributions.
It took around 15 hours in an ordinary laptop to implement this experiment.

\subsubsection{Results of model-fitting}
\label{subsubsec:linear_data_results}

\begin{figure}
\centering
\subfigure{ \label{fig:beta_f_1_simdata_linear_hist}
\includegraphics[width=5cm,height=4cm]{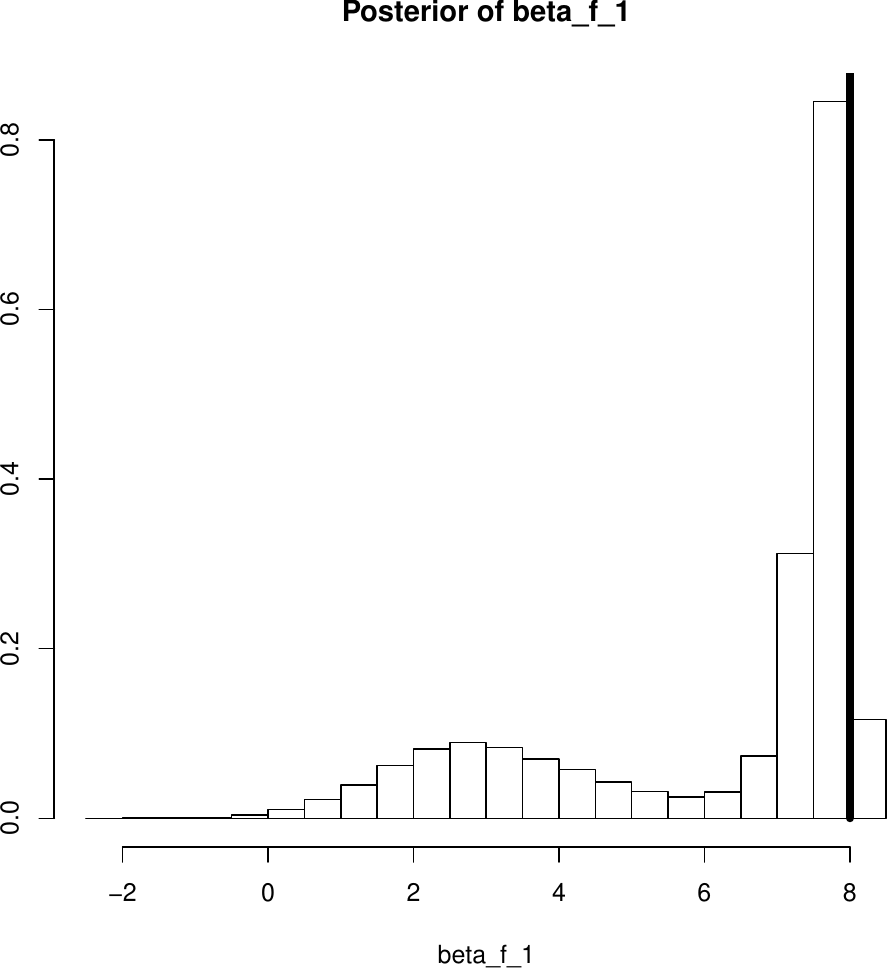}}
\hspace{2mm}
\subfigure{ \label{fig:beta_f_2_simdata_linear_hist}
\includegraphics[width=5cm,height=4cm]{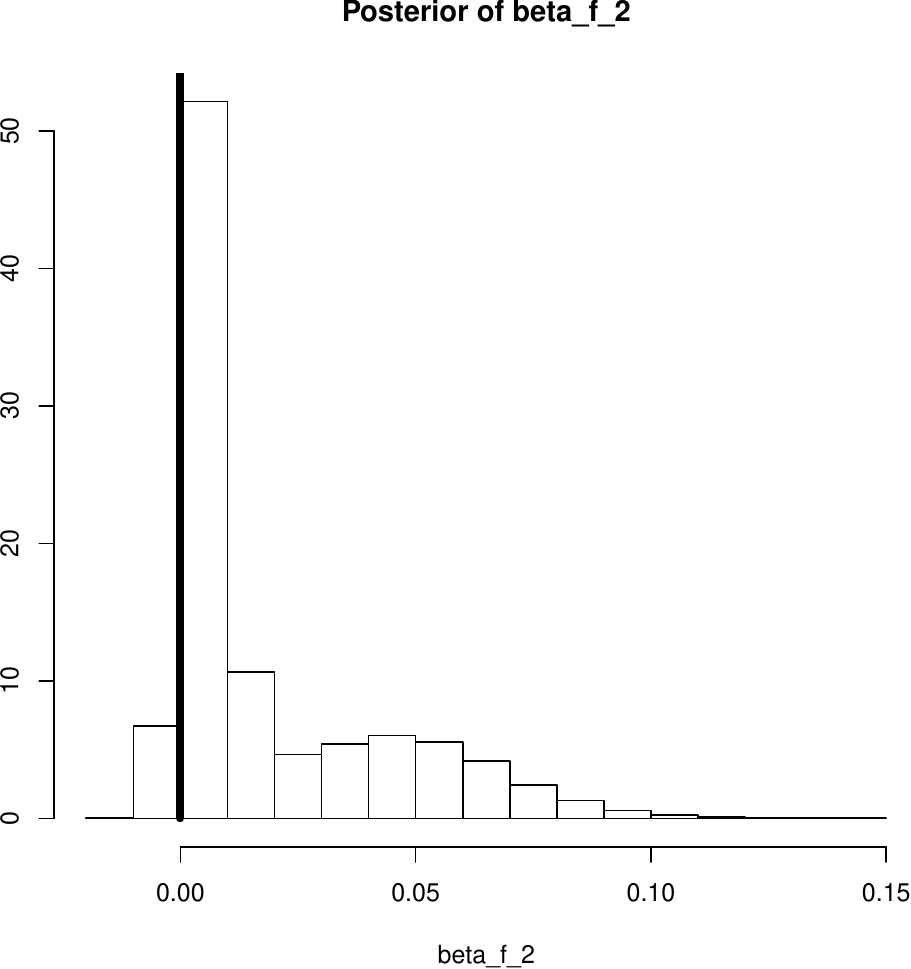}}
\hspace{2mm}
\subfigure{ \label{fig:beta_f_3_simdata_linear_hist}
\includegraphics[width=5cm,height=4cm]{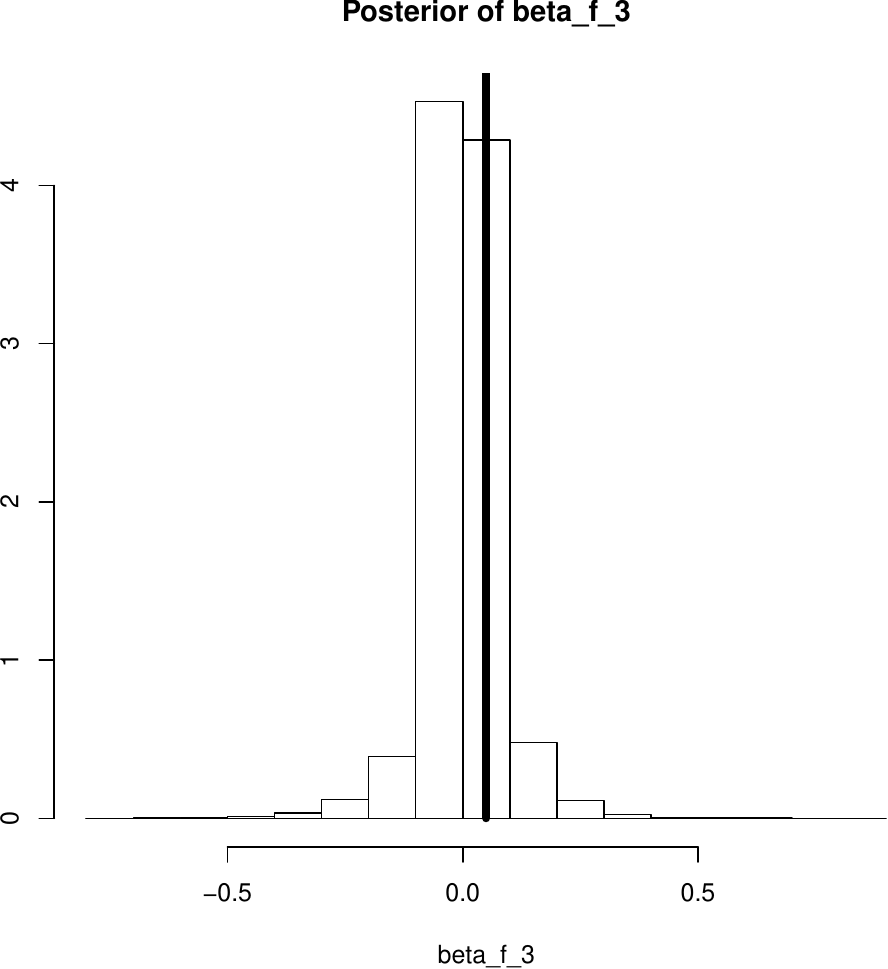}}\\
\vspace{2mm}
\subfigure{ \label{fig:beta_g_1_simdata_linear_hist}
\includegraphics[width=5cm,height=4cm]{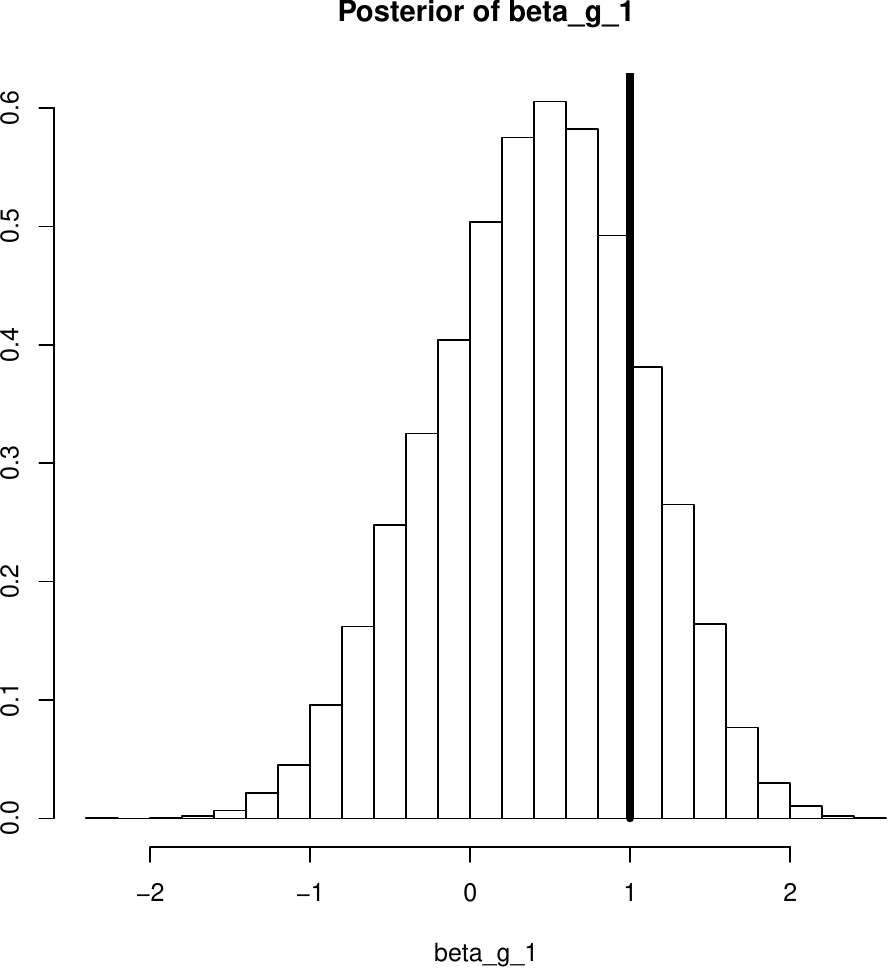}}
\hspace{2mm}
\subfigure{ \label{fig:beta_g_2_simdata_linear_hist}
\includegraphics[width=5cm,height=4cm]{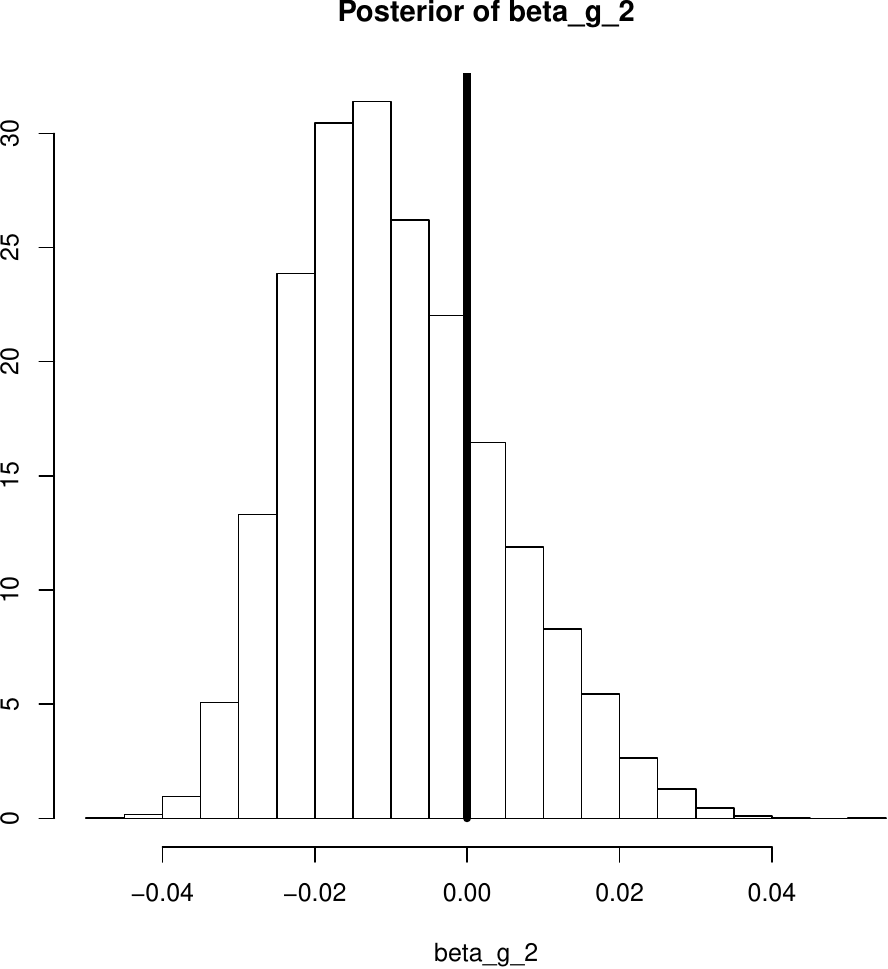}}
\hspace{2mm}
\subfigure{ \label{fig:beta_g_3_simdata_linear_hist}
\includegraphics[width=5cm,height=4cm]{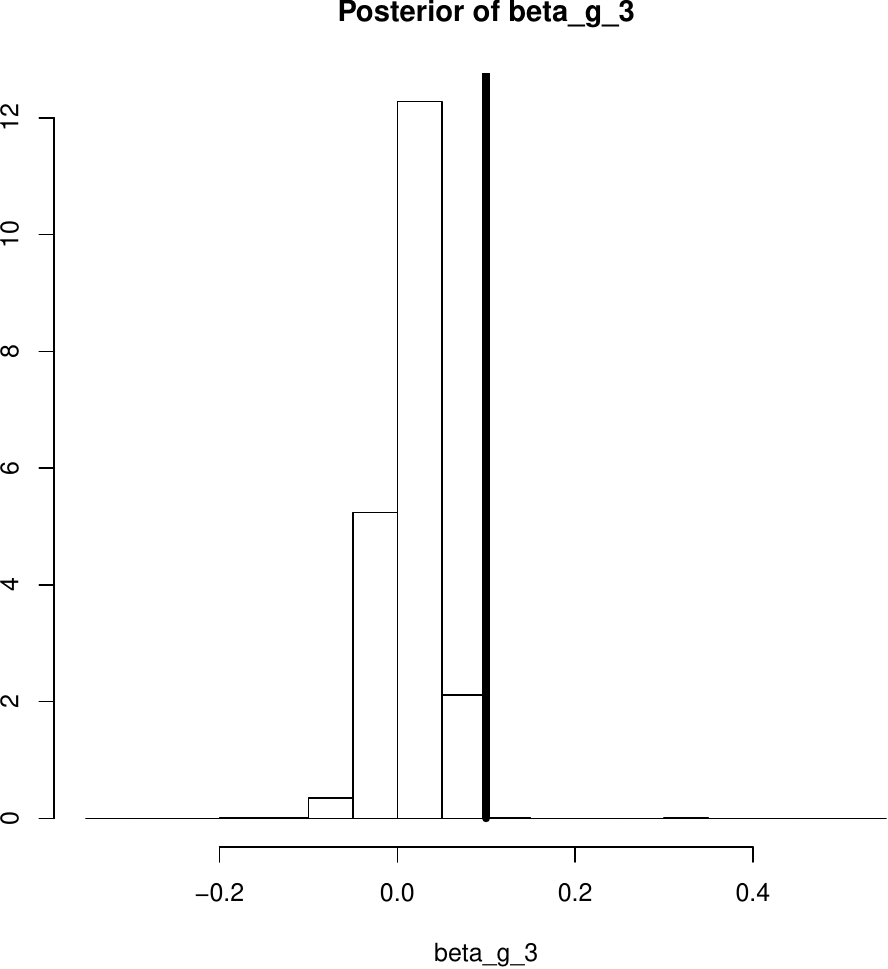}}\\
\vspace{2mm}
\subfigure{ \label{fig:sigma_f_simdata_linear_hist}
\includegraphics[width=5cm,height=4cm]{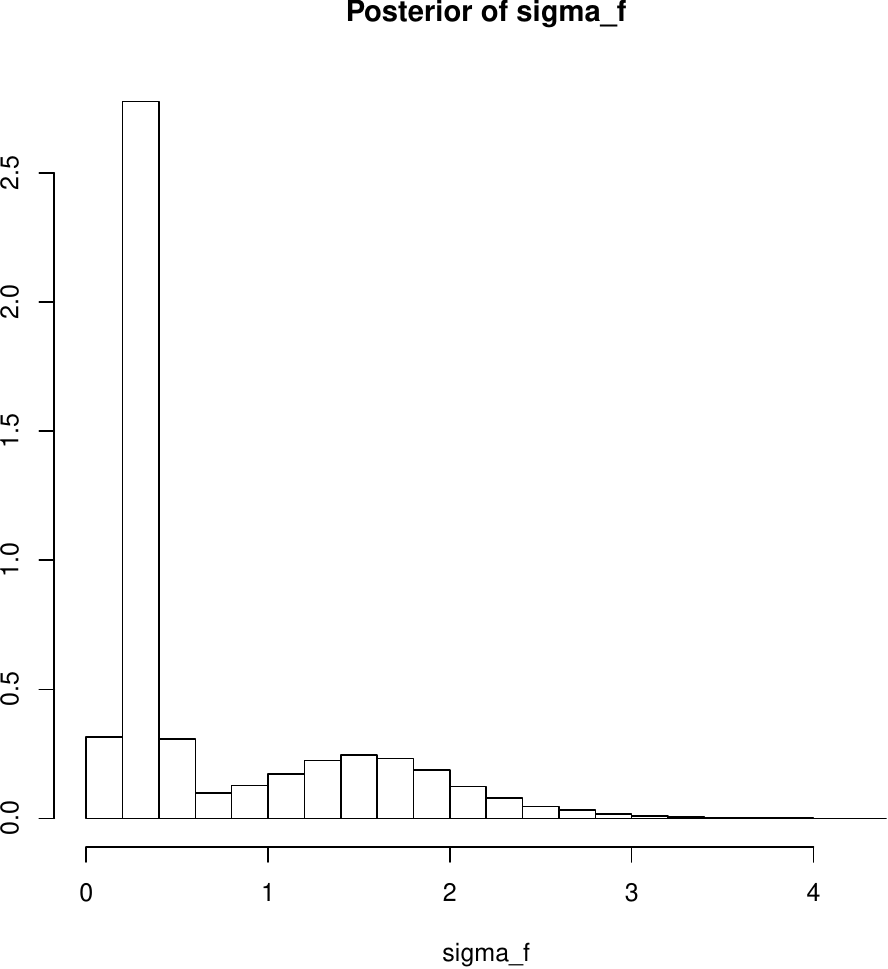}}
\hspace{2mm}
\subfigure{ \label{fig:sigma_g_simdata_linear_hist}
\includegraphics[width=5cm,height=4cm]{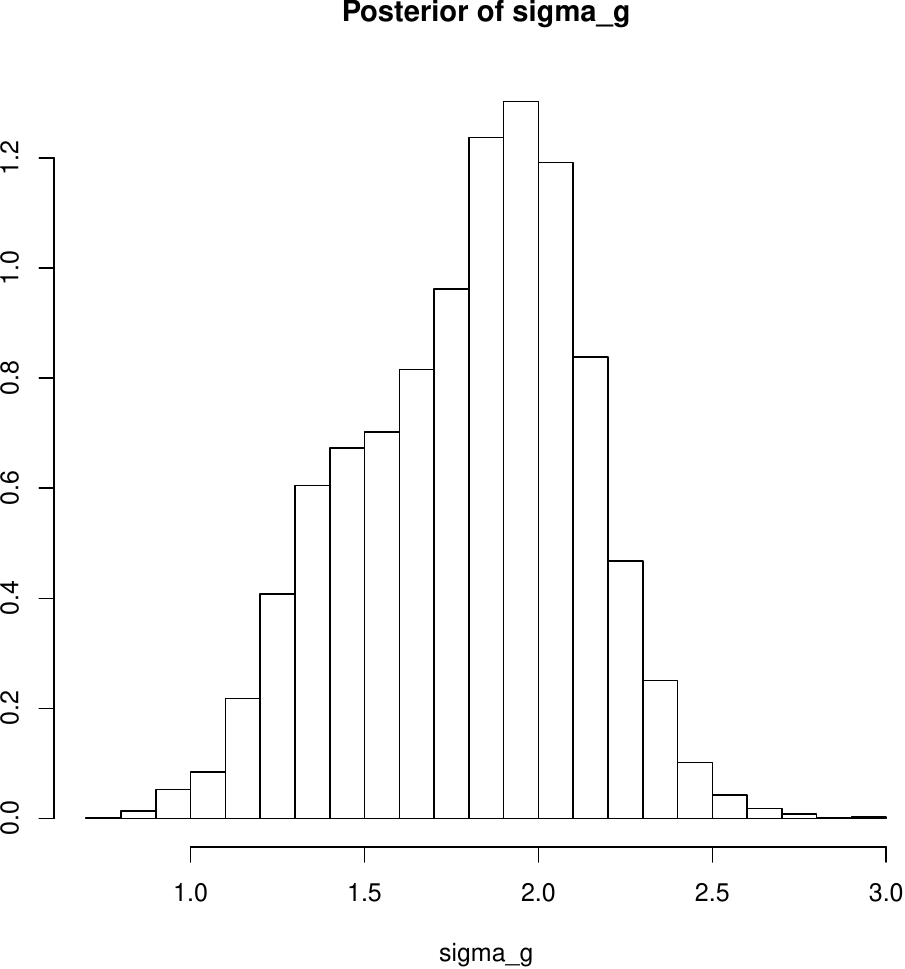}}
\hspace{2mm}
\subfigure{ \label{fig:sigma_e_simdata_linear_hist}
\includegraphics[width=5cm,height=4cm]{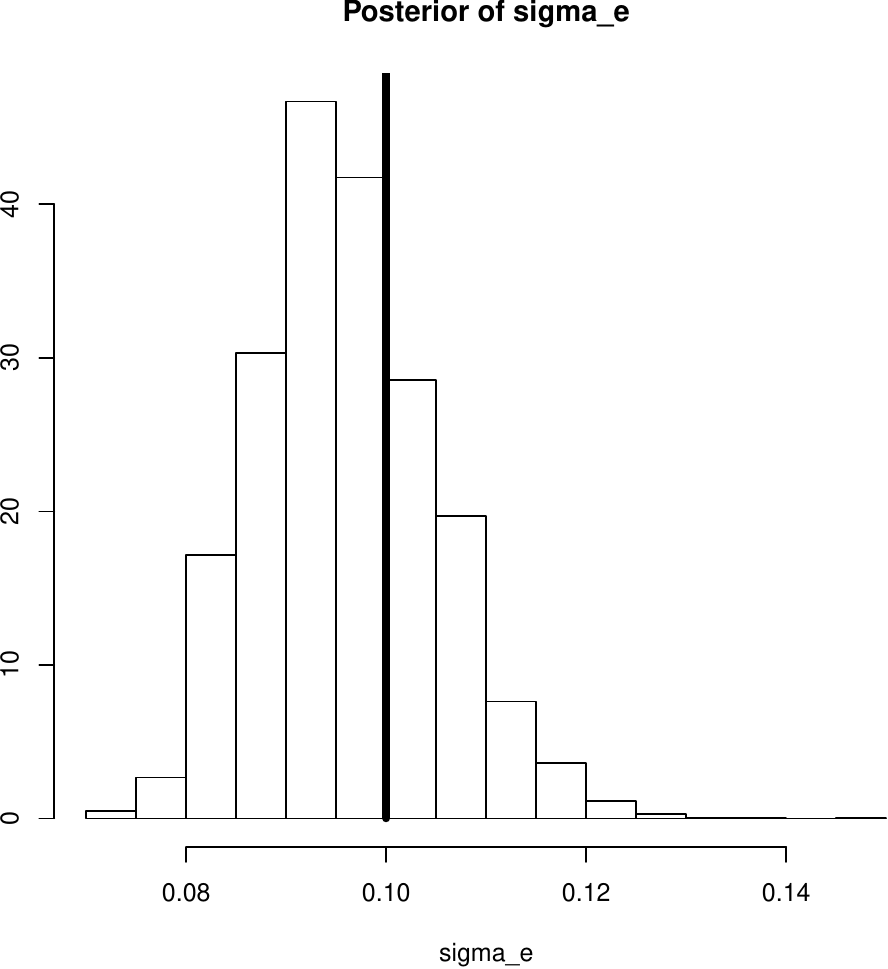}}
\caption{{\bf Simulation study with data generated from the univariate linear model}:
Posterior densities of $\beta_{0,f}$, $\beta_{1,f}$, $\beta_{2,f}$, $\beta_{0,g}$, $\beta_{1,g}$,
$\beta_{2,g}$, $\sigma_f$, $\sigma_g$, and $\sigma_{\epsilon}$.
The solid line stands for the true values of the respective parameters.}
\label{fig:simdata_linear_hist}
\end{figure}


\begin{figure}
\centering
\subfigure{ \label{fig:sigma_eta_simdata_linear_hist}
\includegraphics[width=5cm,height=4cm]{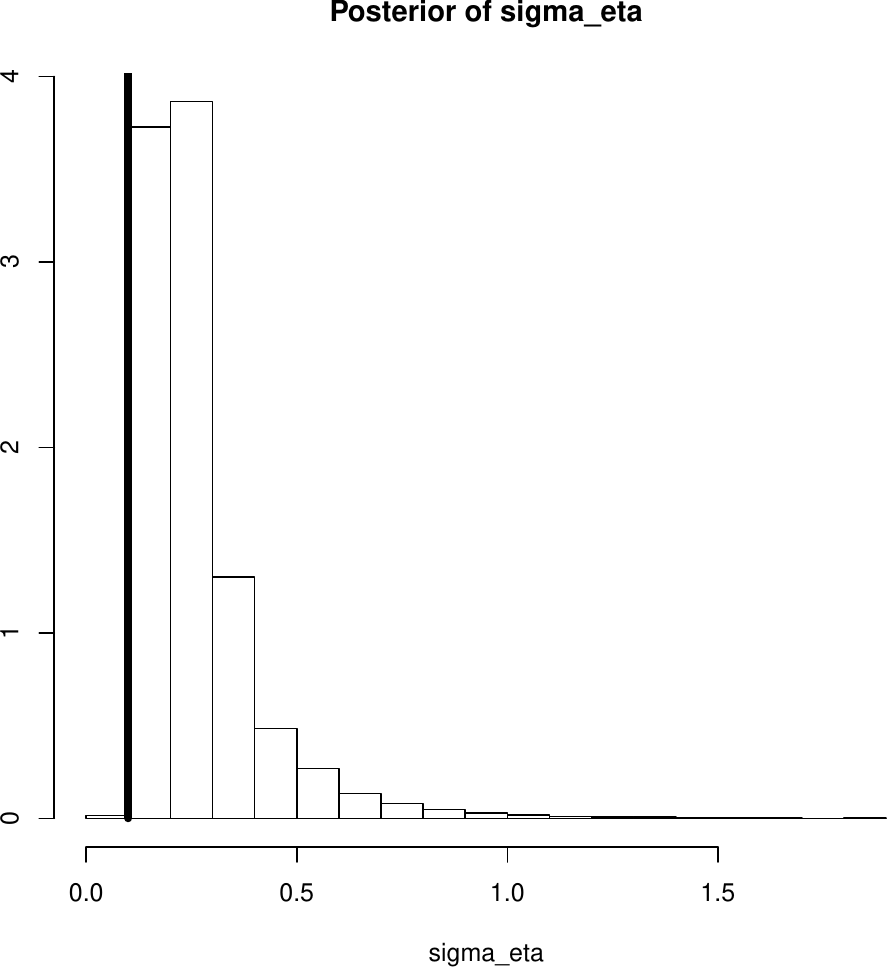}}
\hspace{2mm}
\subfigure{ \label{fig:r_f_1_simdata_linear_hist}
\includegraphics[width=5cm,height=4cm]{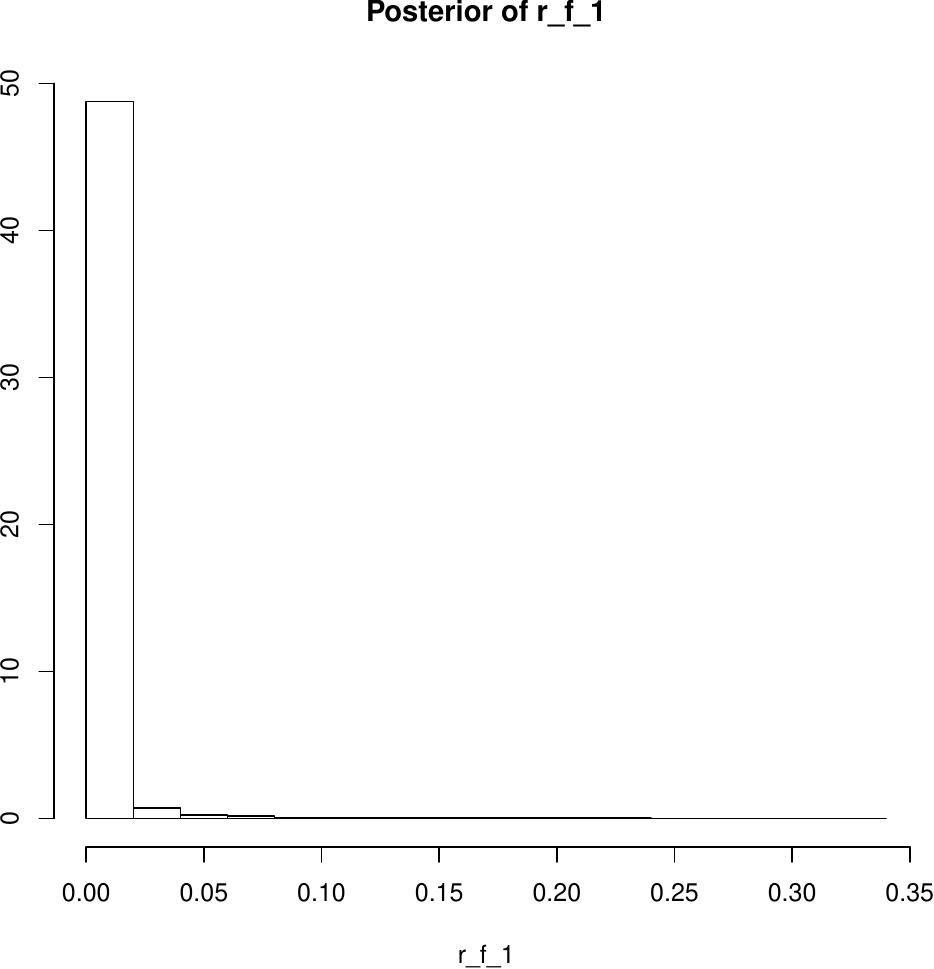}}
\hspace{2mm}
\subfigure{ \label{fig:r_f_2_simdata-linear}
\includegraphics[width=5cm,height=4cm]{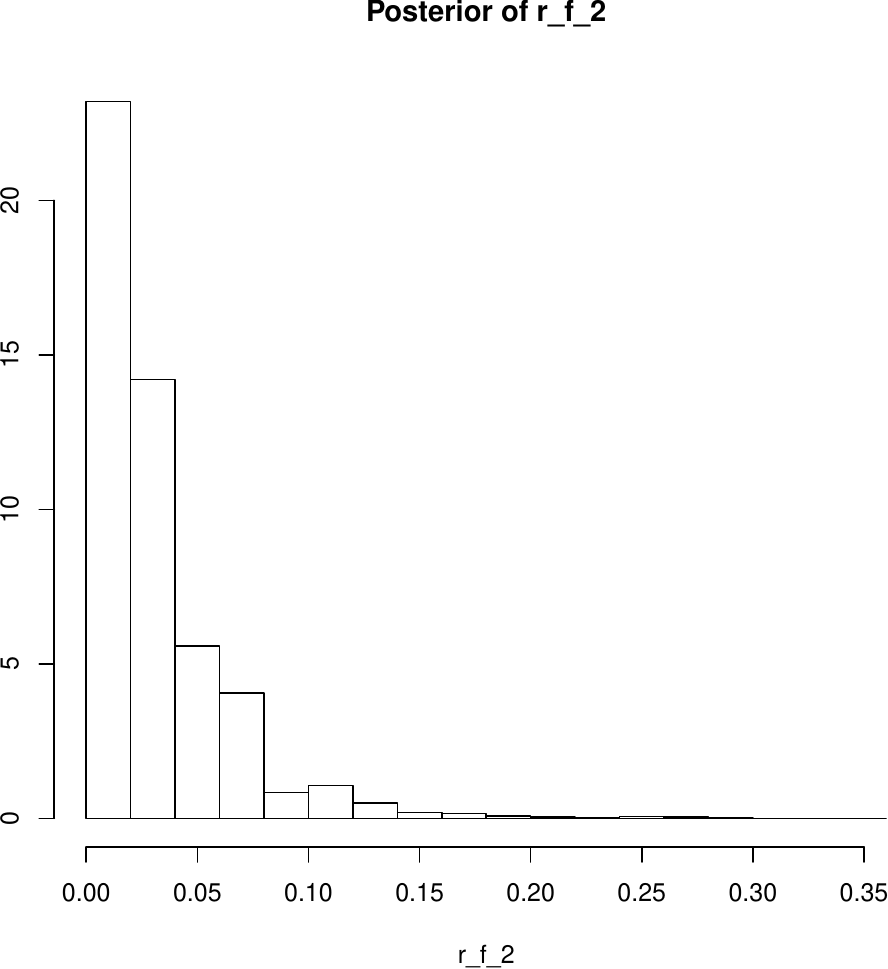}}\\
\vspace{2mm}
\subfigure{ \label{fig:r_g_1_simdata_linear_hist}
\includegraphics[width=5cm,height=4cm]{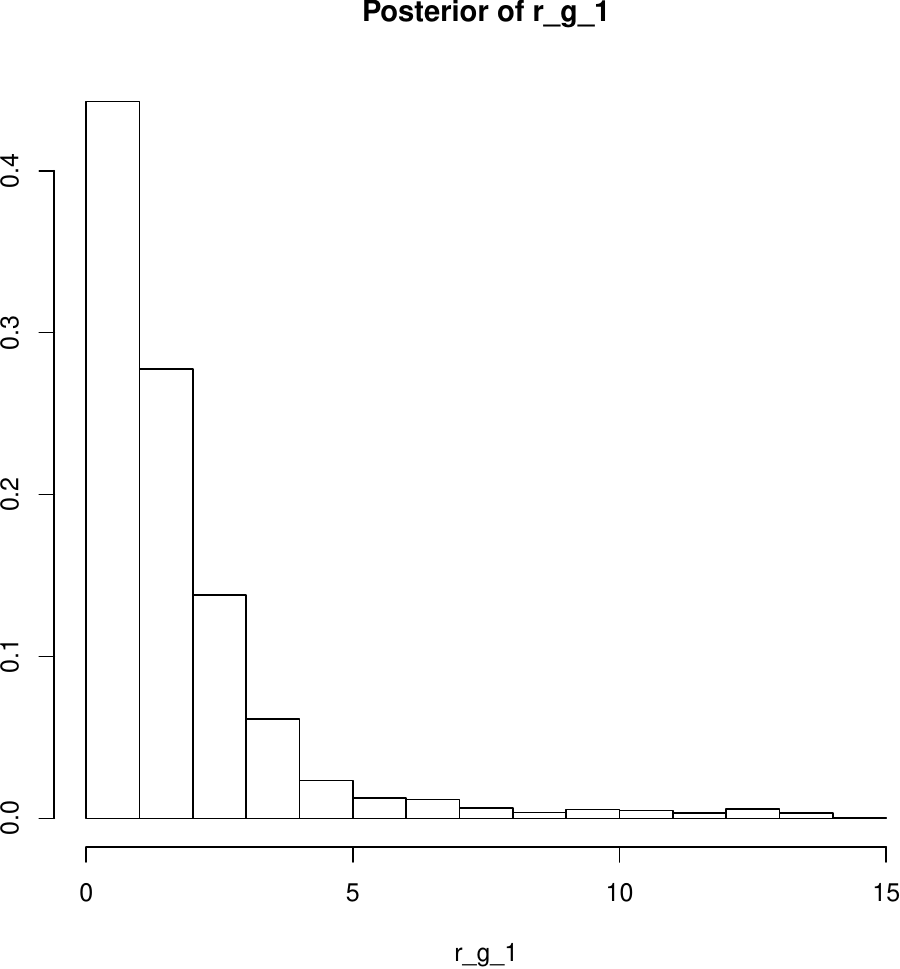}}
\hspace{2mm}
\subfigure{ \label{fig:r_g_2_simdata_linear_hist}
\includegraphics[width=5cm,height=4cm]{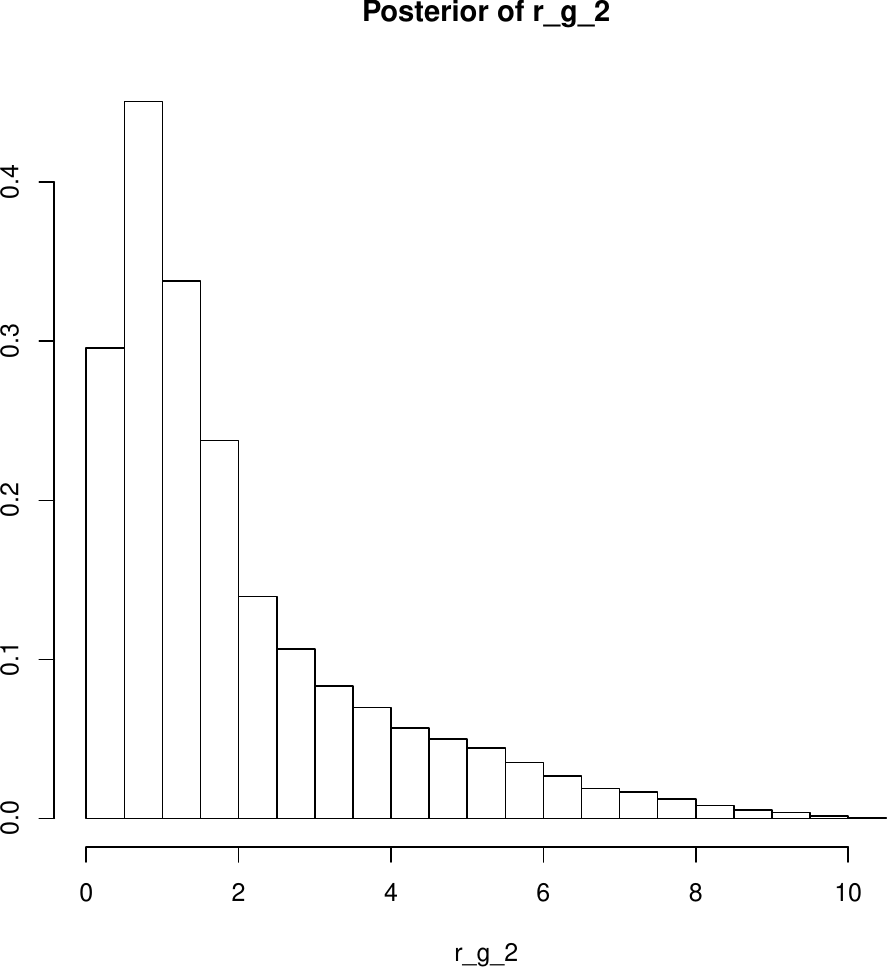}}
\hspace{2mm}
\subfigure{ \label{fig:x0_simdata_linear_hist}
\includegraphics[width=5cm,height=4cm]{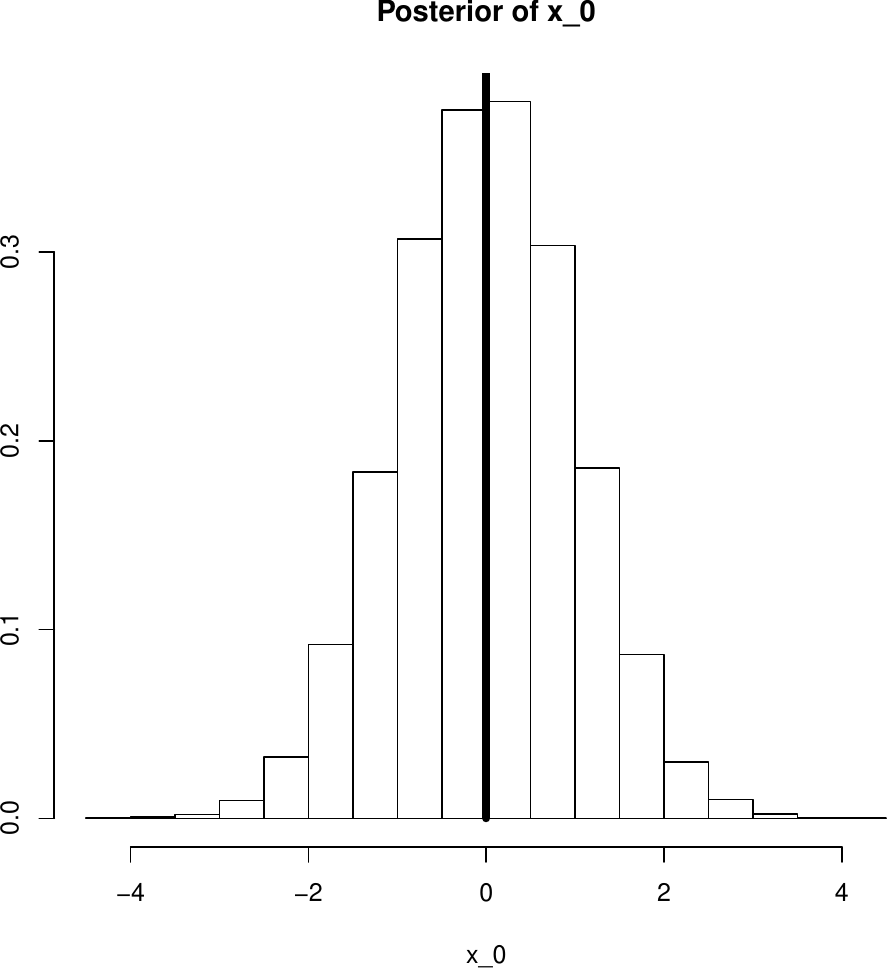}}\\
\vspace{2mm}
\subfigure{ \label{fig:forecasted_x_simdata_linear_hist}
\includegraphics[width=5cm,height=4cm]{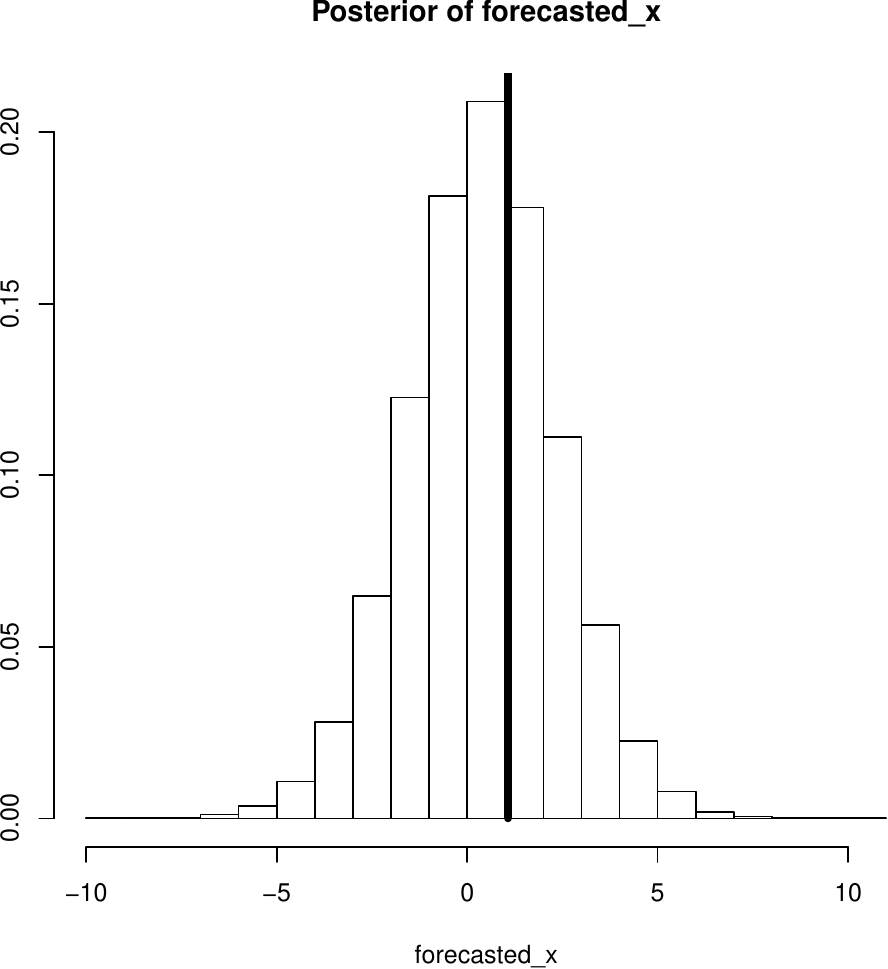}}
\hspace{2mm}
\subfigure{ \label{fig:forecasted_y_simdata_linear_hist}
\includegraphics[width=5cm,height=4cm]{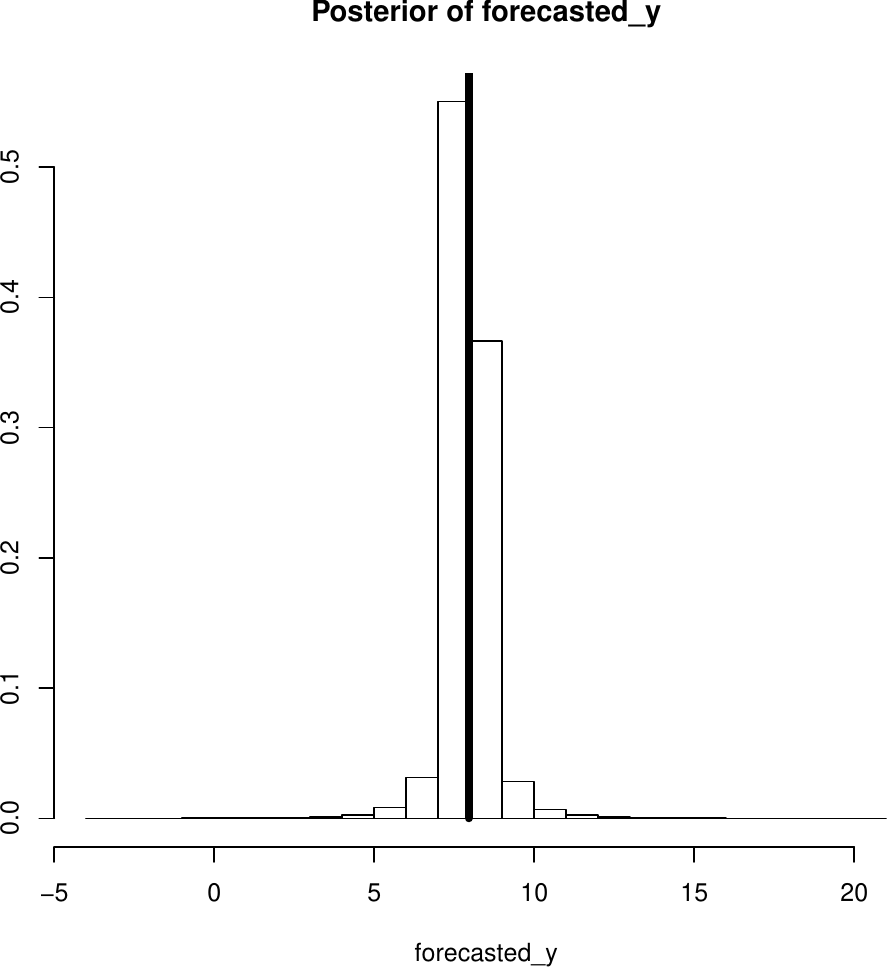}}
\caption{{\bf Simulation study with data generated from the univariate linear model}:
Posterior densities of $\sigma_{\eta}$, $r_{1,f}$, $r_{2,f}$, $r_{1,g}$, $r_{2,g}$,
$x_0$, $x_{T+1}$ (one-step forecasted $x$), and $y_{T+1}$ (one-step forecasted $y$).
The solid line stands for the true values of the respective parameters.}
\label{fig:simdata_linear_hist2}
\end{figure}


\begin{figure}
\centering
\subfigure{ \label{fig:x_time_series_simdata_linear_hist}
\includegraphics[width=15cm,height=10cm]{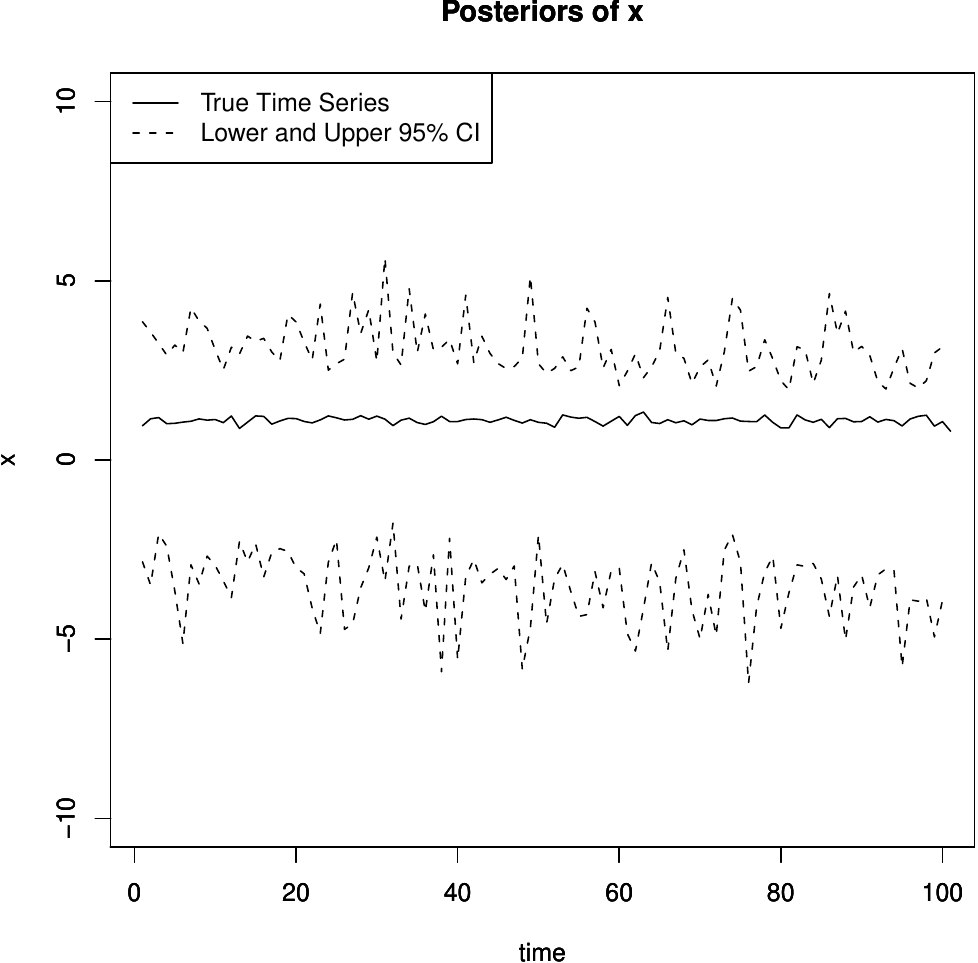}}
\caption{{\bf Simulation study with data generated from the univariate linear model}:
95\% highest posterior density credible intervals of the time series $x_1,\ldots,x_T$.
The solid line stands for the true time series.}
\label{fig:simdata_linear_hist3}
\end{figure}

Figures \ref{fig:simdata_linear_hist}, \ref{fig:simdata_linear_hist2} and \ref{fig:simdata_linear_hist3} show
the posterior distributions of the unknowns. Note that whenever applicable, the true value
is well-captured by the posteriors in question. Also, as expected, the posteriors of $\beta_{2,f}$,
$\beta_{2,g}$, $r_{1,f}$ and $r_{1,g}$, have high posterior probabilities around zero, supporting
the time-invariant nature of the true observational and the evolutionary equations.
The true values of $\sigma_{\epsilon}$, $\sigma_{\eta}$ and $x_0$ also fell well within
the highest density regions of their respective posterior distributions. The true forecasted
values of $x_{101}$ and $y_{101}$ are well-captured by their respective forecast posteriors.
The true time series $\{x_1,\ldots,x_{100}\}$ again
fell well within the lower and the upper 95\% highest posterior density credible limits, but
the credible regions seem to be wider than the usual parametric approaches. 
This is a consequence of 
acknowledging uncertainty about the functional forms of both $f$ and $g$.
Particularly because the posterior distribution of the state variables depends upon 
both the unknown functions $f$ and $g$, it is perhaps not surprising that the credible intervals 
reflect the effect of uncertainties about the random functions.   
In light of the knowledge
of the true time series in this simple simulated example, which has a linear structure, the relatively wide credible intervals
may not seem to indicate very encouraging performance, but in complex, realistic situations,
where any structure of the true time series is impossible to guess, the relatively large credible
intervals make good practical sense.

\subsection{Data generated from a univariate parametric growth model}
\label{subsec:cps_data_generation}

Following \ctn{Carlin92} we generate data from the following model:
\begin{align}
x_t&=\alpha x_{t-1}+\beta x_{t-1}/(1+x^2_{t-1})+\gamma\cos(1.2 (t-1))+u_t\label{eq:cps1}\\
y_t&=x^2_t/20+v_t, \ \ \ t=1,\ldots,100,\label{eq:cps2}
\end{align}
where $x_0=0$. 
Here we assume
$u_t\sim N(0,\sigma^2_{\epsilon})$ and $v_t\sim N(0,\sigma^2_{\eta})$, independently for all $t$.
We set $\sigma_{\epsilon}=0.1=\sigma_{\eta}$ and fix
the true values of $(\alpha,\beta,\gamma)$ to be $(0.05,0.1,0.2)$, respectively.
We then generated the data set consisting of 101 data points using these true values.
As before, we set aside the last data point for the purpose of forecasting.

\subsubsection{Choice of grid and MCMC implementation}
\label{subsubsec:cps_data_prior_choice}

In this example the entire true time series falls within the space $[-0.33,0.32]$.
As before, we consider the much larger space, $[-30,30]$, and divide it into 100 sub-intervals of equal length, 
and then select
a value randomly from each such sub-interval. This yields values of the second component of the
two-dimensional grid $\bG_n$. As in the linear model experiment, here also we experimented with 
other reasonable choices and even in this non-linear experiment, the results remained
exhibited considerable robustness. 
For the first component we generate a number uniformly from each of the
100 sub-intervals $[i,i+1]$; $i=0,\ldots,99$. 

We discarded the first 10000 MCMC iterations as burn-in and stored the next 50,000 iterations for
inference. We used the normal random walk proposal with variance 0.05 for updating $\sigma_f$, $\sigma_g$,
$\sigma_{\epsilon}$ and $\sigma_{\eta}$. For updating $x_0$ we used the normal random walk proposal with variance 1,
but for updating $x_t;t=1,\ldots,T$, we found in this example, using informal convergence diagnostics,
that the random walk proposal with variance $t$ provided
much better mixing than those with constant variances.
These proposals also performed much better than the linear model based proposals detailed in Sections S-2.8 and S-2.9.
It took around 15 hours in an ordinary laptop to implement this experiment.

\subsubsection{Results of model-fitting}
\label{subsubsec:cps_data_results}

\begin{figure}
\centering
\subfigure{ \label{fig:beta_f_1_simdata_cps_hist}
\includegraphics[width=5cm,height=4cm]{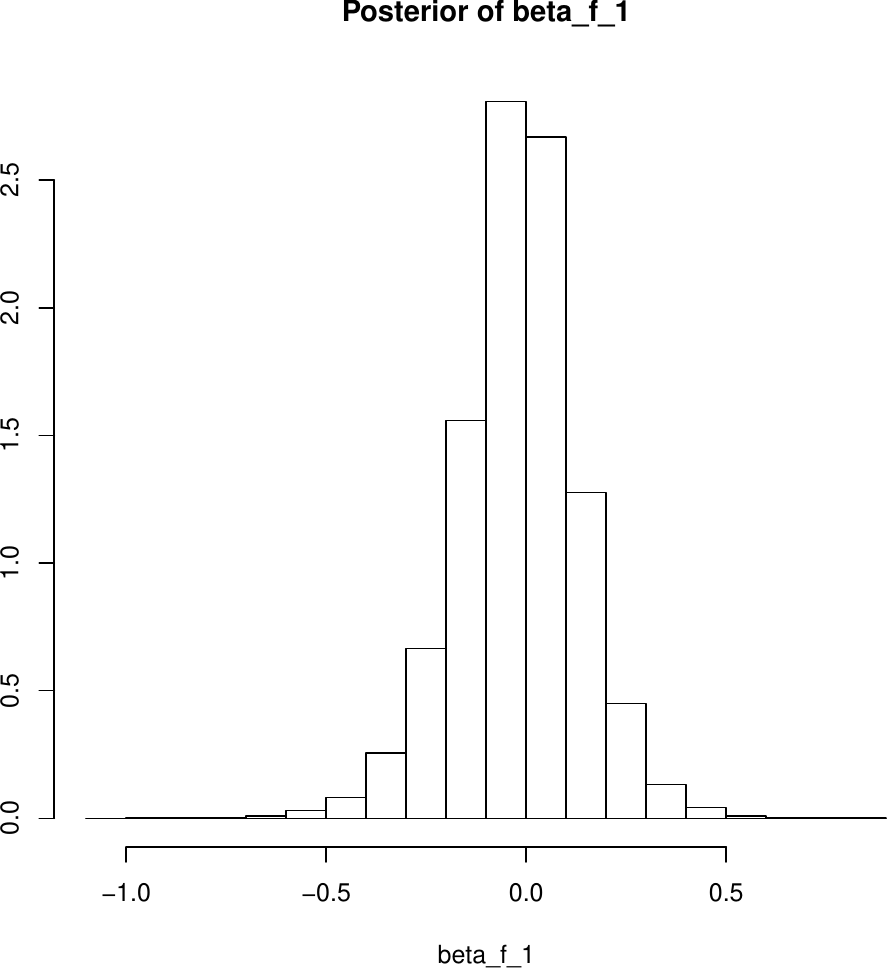}}
\hspace{2mm}
\subfigure{ \label{fig:beta_f_2_simdata_cps_hist}
\includegraphics[width=5cm,height=4cm]{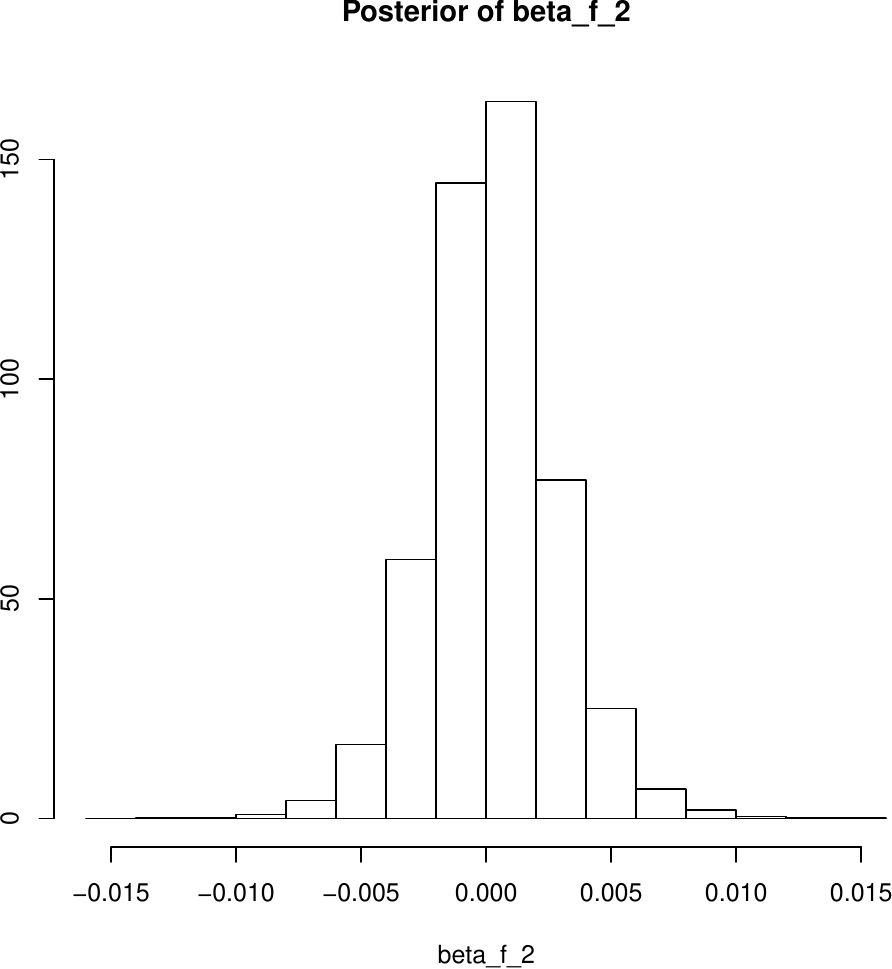}}
\hspace{2mm}
\subfigure{ \label{fig:beta_f_3_simdata_cps_hist}
\includegraphics[width=5cm,height=4cm]{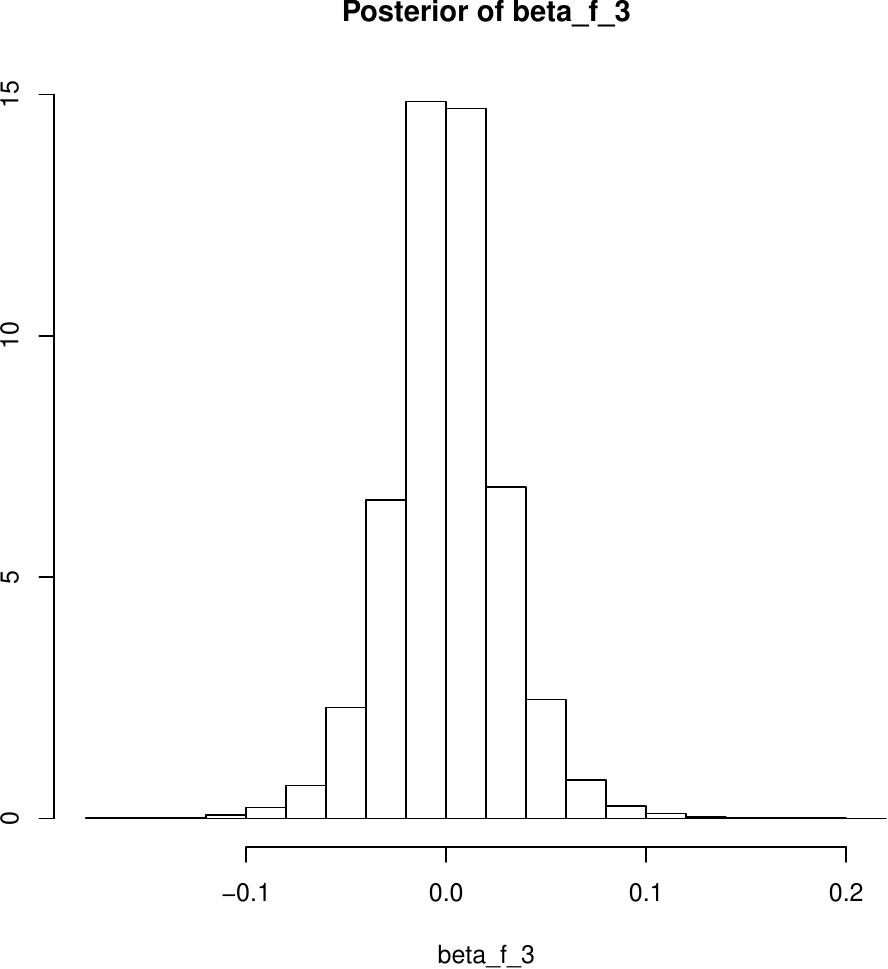}}\\
\vspace{2mm}
\subfigure{ \label{fig:beta_g_1_simdata_cps_hist}
\includegraphics[width=5cm,height=4cm]{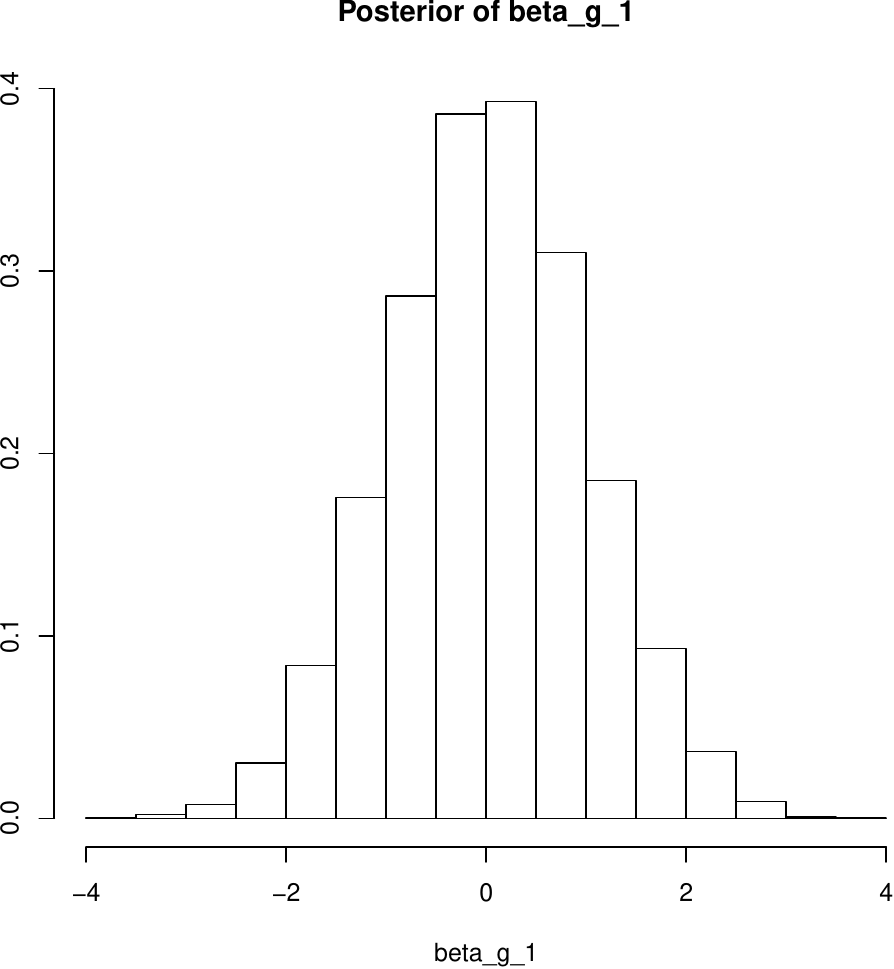}}
\hspace{2mm}
\subfigure{ \label{fig:beta_g_2_simdata_cps_hist}
\includegraphics[width=5cm,height=4cm]{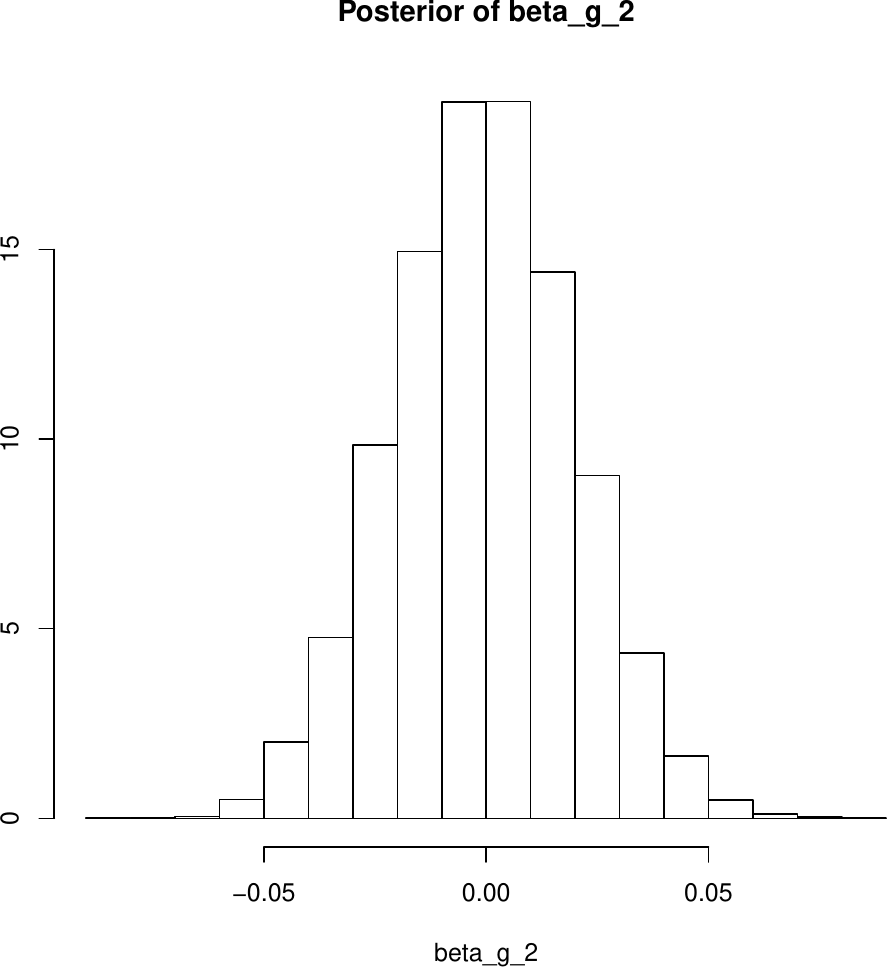}}
\hspace{2mm}
\subfigure{ \label{fig:beta_g_3_simdata_cps_hist}
\includegraphics[width=5cm,height=4cm]{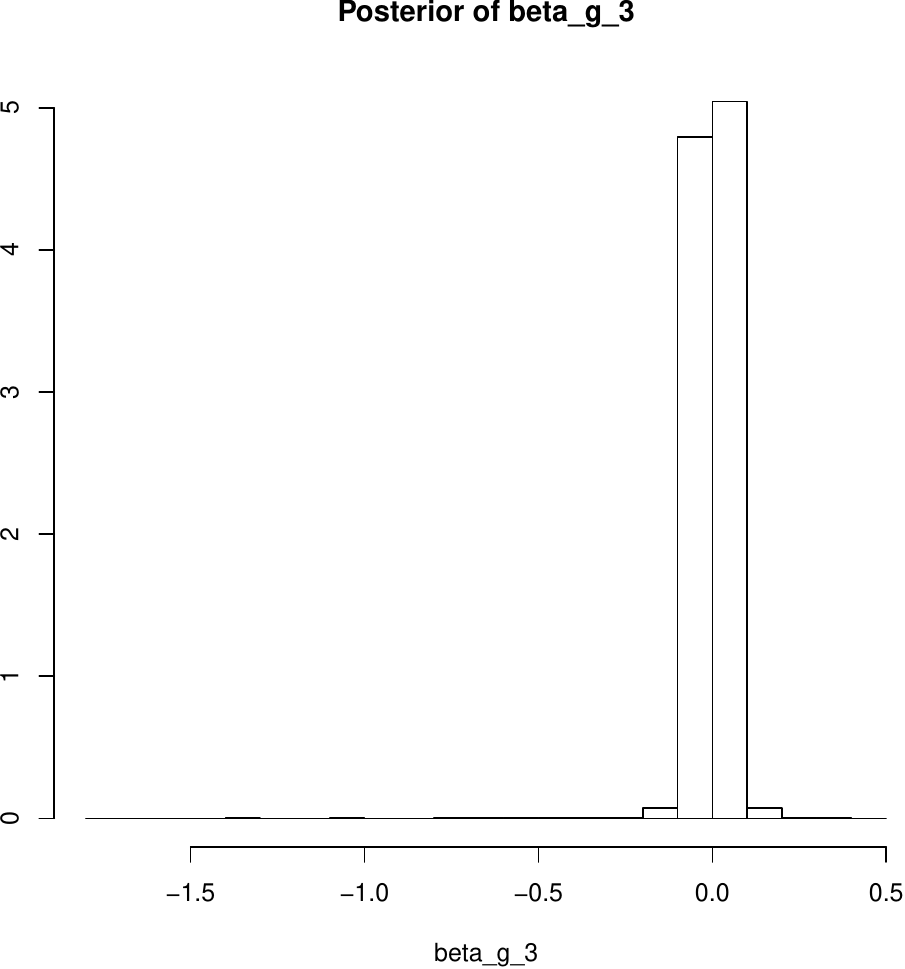}}\\
\vspace{2mm}
\subfigure{ \label{fig:sigma_f_simdata_cps_hist}
\includegraphics[width=5cm,height=4cm]{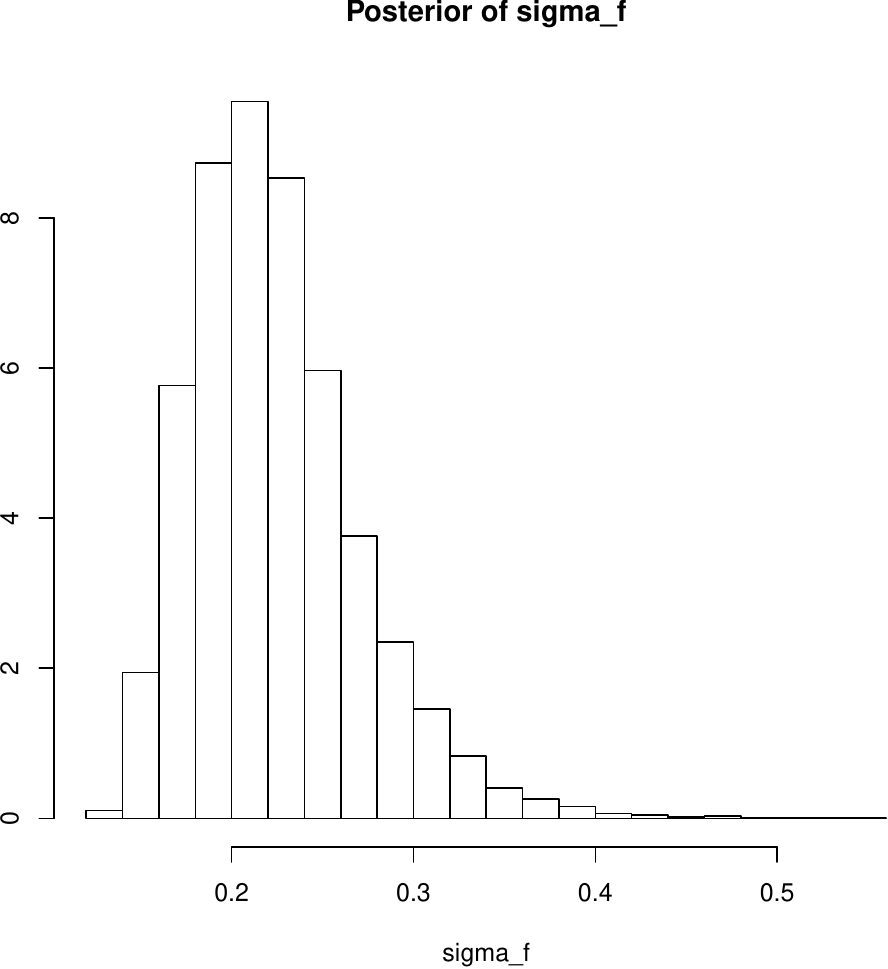}}
\hspace{2mm}
\subfigure{ \label{fig:sigma_g_simdata_cps_hist}
\includegraphics[width=5cm,height=4cm]{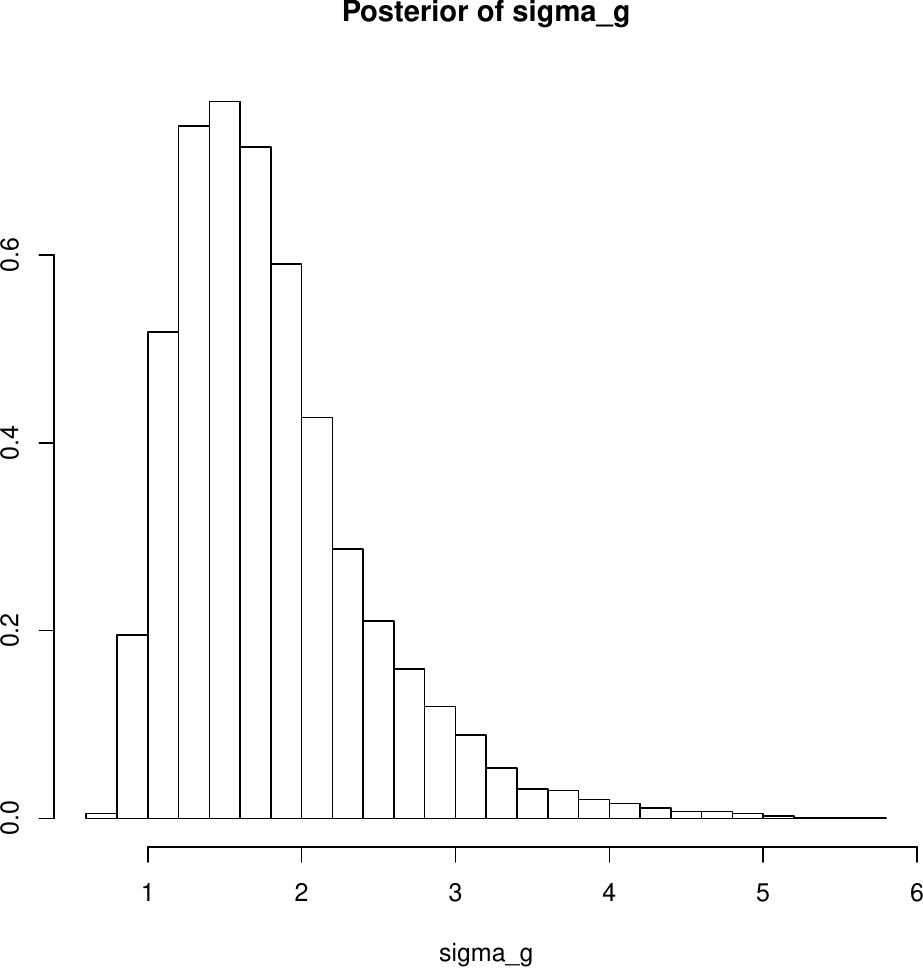}}
\hspace{2mm}
\subfigure{ \label{fig:sigma_e_simdata_cps_hist}
\includegraphics[width=5cm,height=4cm]{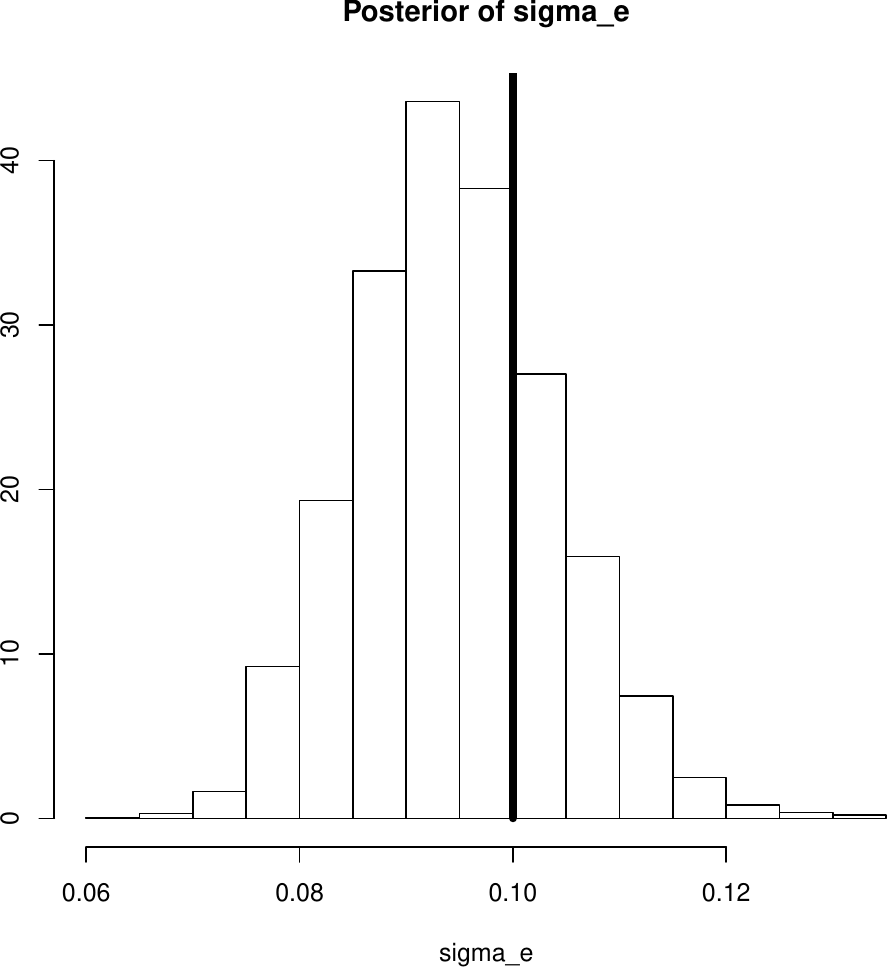}}
\caption{{\bf Simulation study with data generated from the univariate growth model
of CPS}:
Posterior densities of $\beta_{0,f}$, $\beta_{1,f}$, $\beta_{2,f}$, $\beta_{0,g}$, $\beta_{1,g}$,
$\beta_{2,g}$, $\sigma_f$, $\sigma_g$, and $\sigma_{\epsilon}$.
The solid line stands for the true values of the respective parameters.}
\label{fig:simdata_cps_hist}
\end{figure}


\begin{figure}
\centering
\subfigure{ \label{fig:sigma_eta_simdata_cps_hist}
\includegraphics[width=5cm,height=4cm]{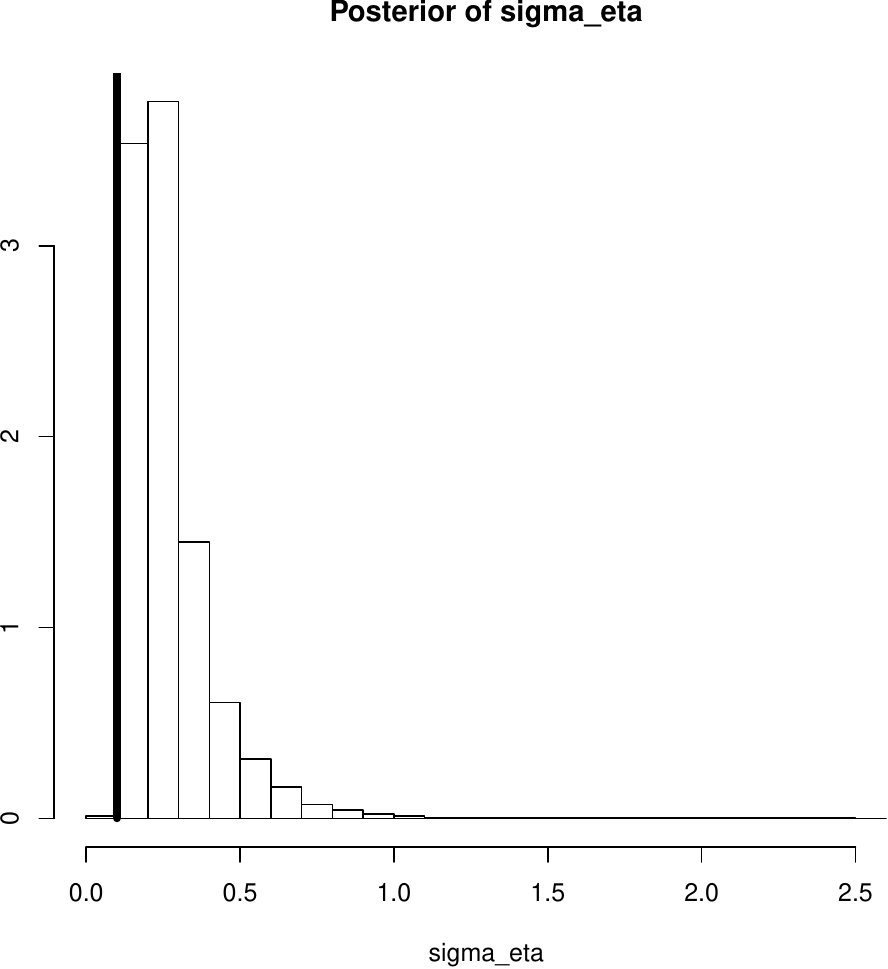}}
\hspace{2mm}
\subfigure{ \label{fig:r_f_1_simdata_cps_hist}
\includegraphics[width=5cm,height=4cm]{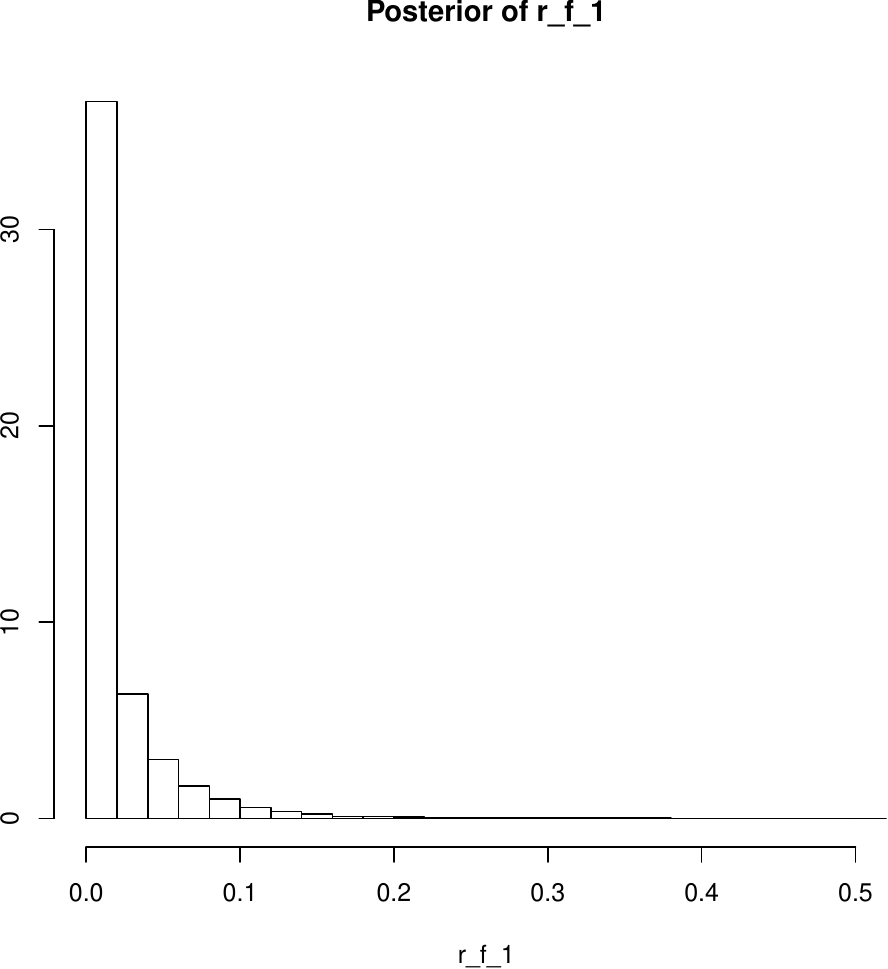}}
\hspace{2mm}
\subfigure{ \label{fig:r_f_2_simdata-cps}
\includegraphics[width=5cm,height=4cm]{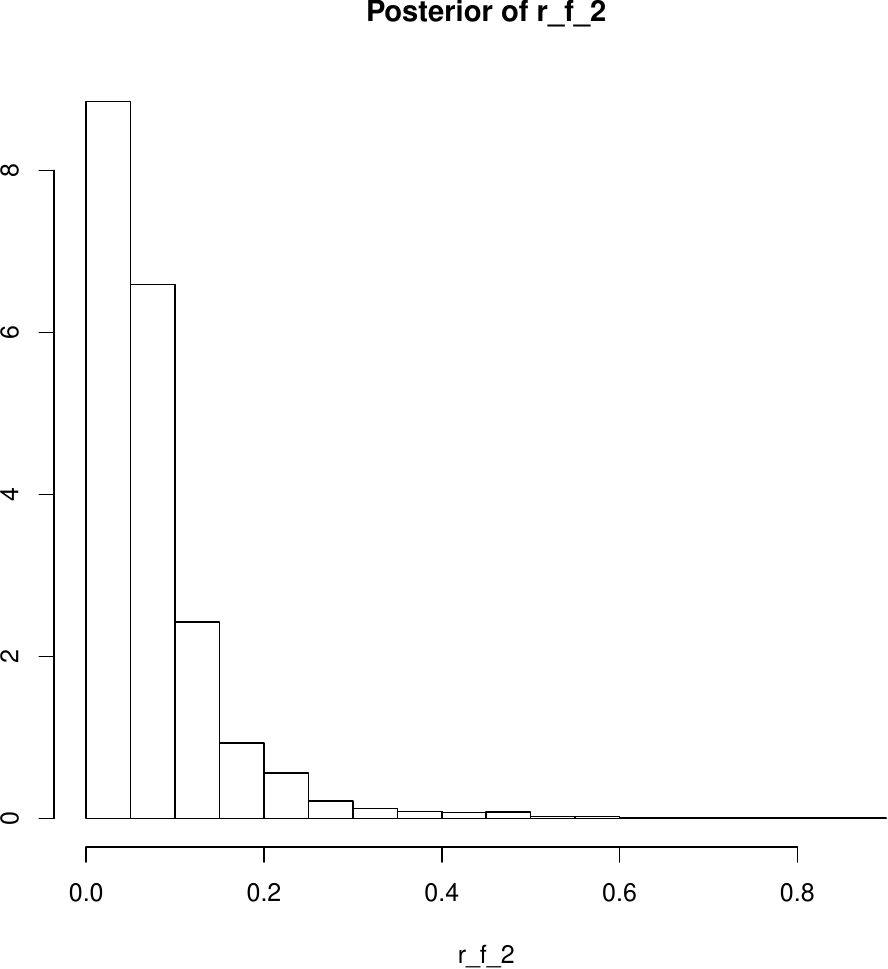}}\\
\vspace{2mm}
\subfigure{ \label{fig:r_g_1_simdata_cps_hist}
\includegraphics[width=5cm,height=4cm]{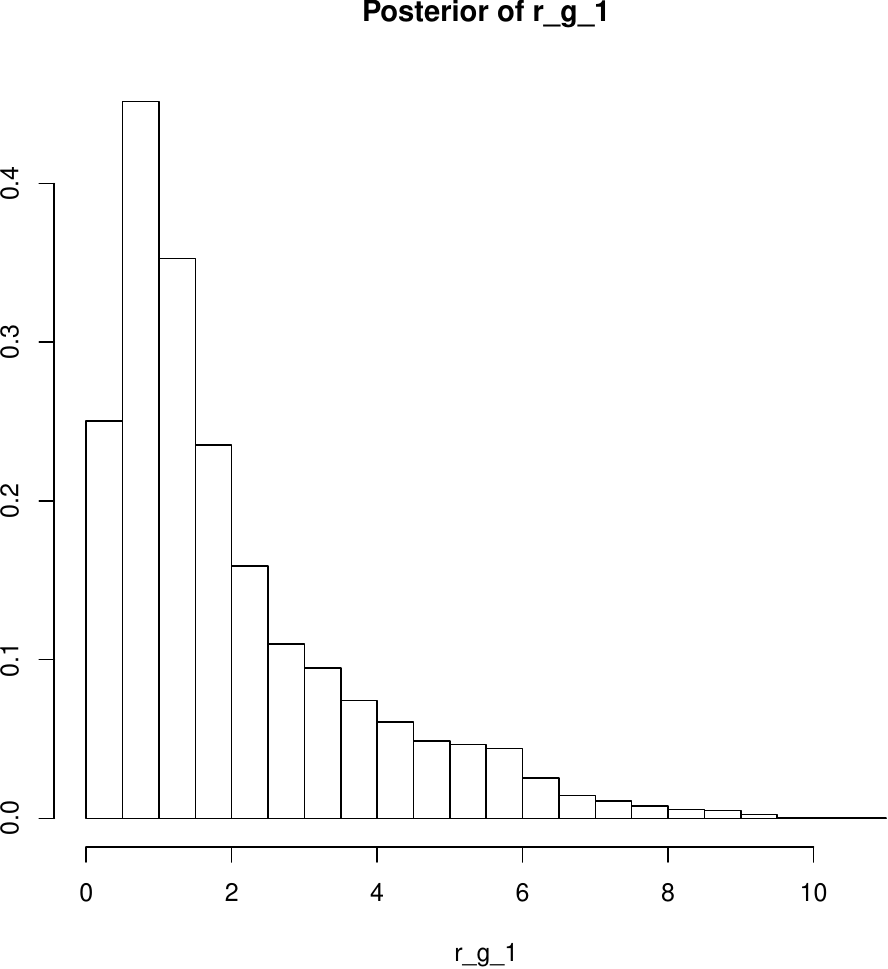}}
\hspace{2mm}
\subfigure{ \label{fig:r_g_2_simdata_cps_hist}
\includegraphics[width=5cm,height=4cm]{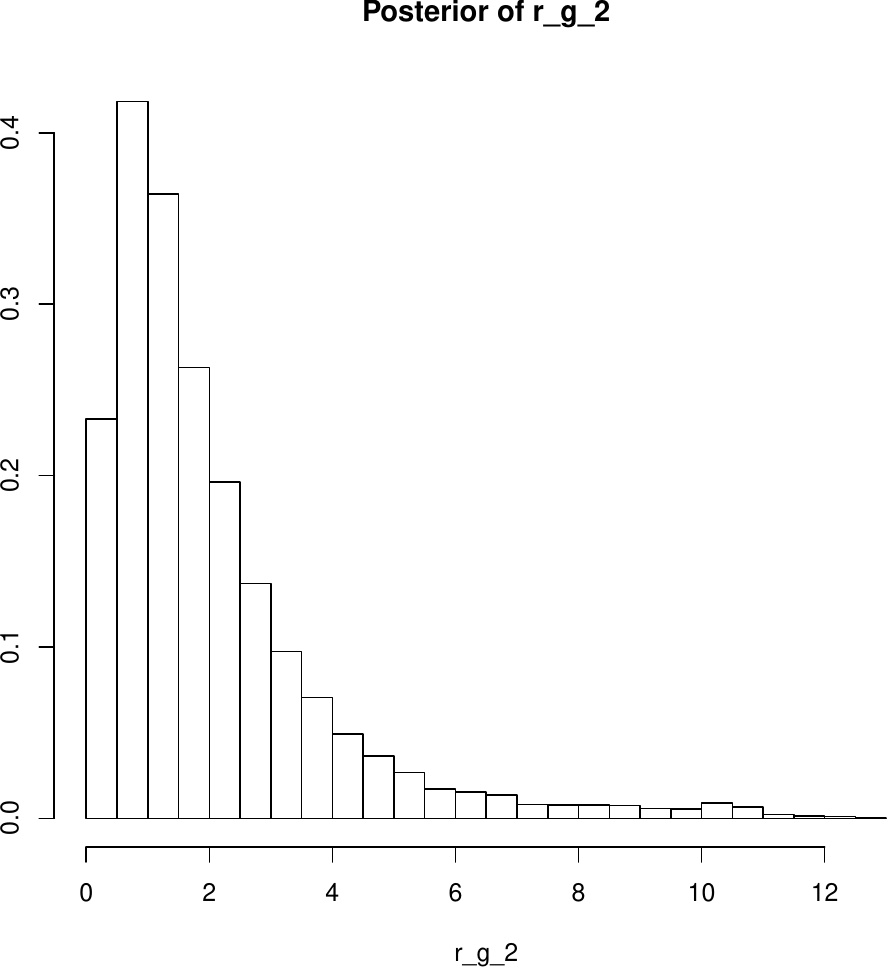}}
\hspace{2mm}
\subfigure{ \label{fig:x0_simdata_cps_hist}
\includegraphics[width=5cm,height=4cm]{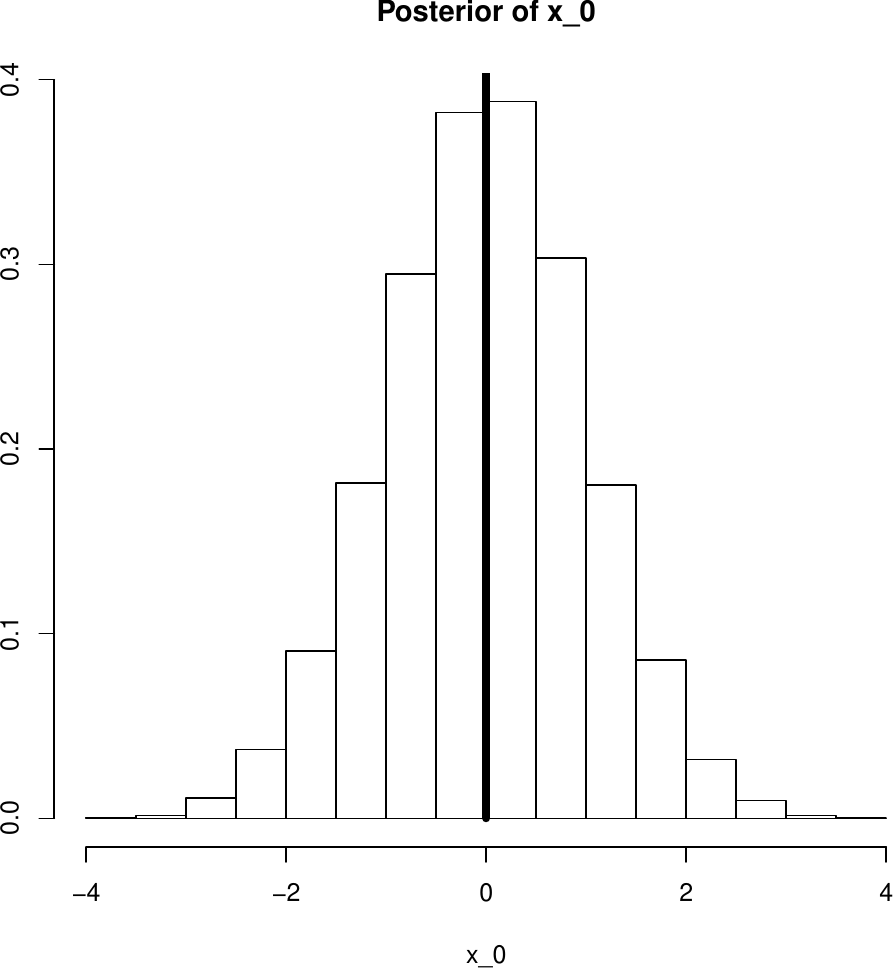}}\\
\vspace{2mm}
\subfigure{ \label{fig:forecasted_x_simdata_cps_hist}
\includegraphics[width=5cm,height=4cm]{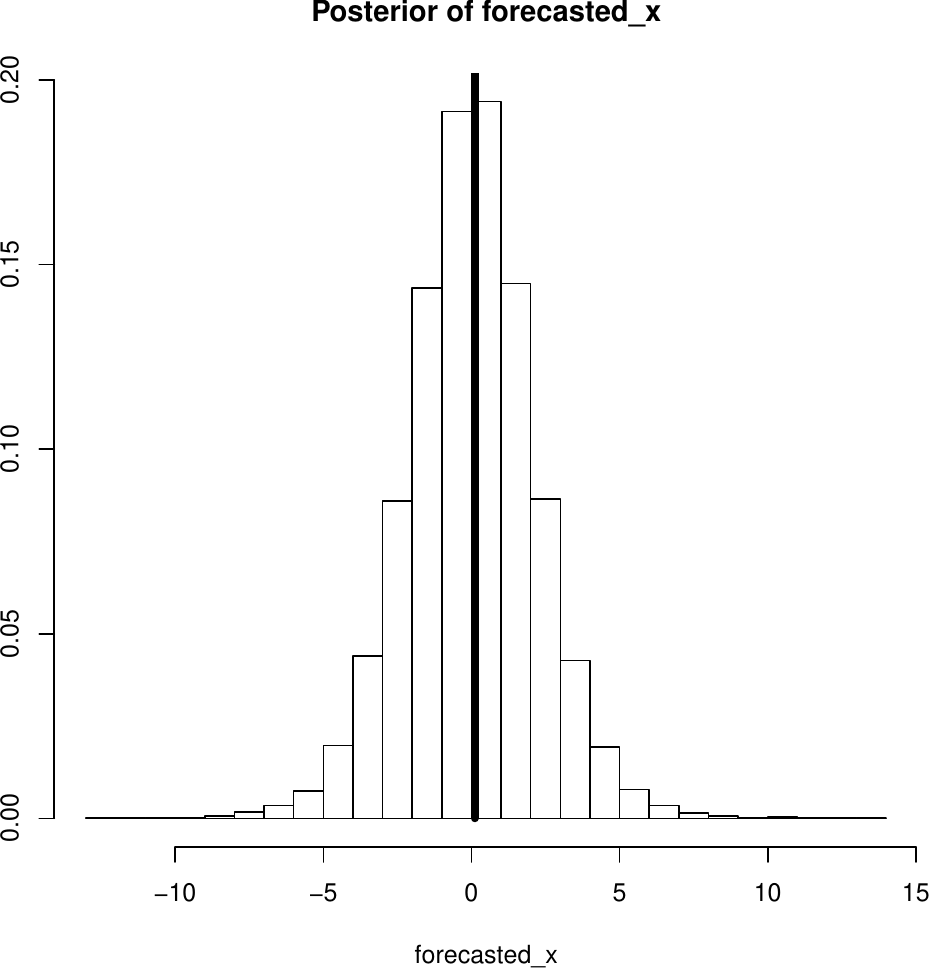}}
\hspace{2mm}
\subfigure{ \label{fig:forecasted_y_simdata_cps_hist}
\includegraphics[width=5cm,height=4cm]{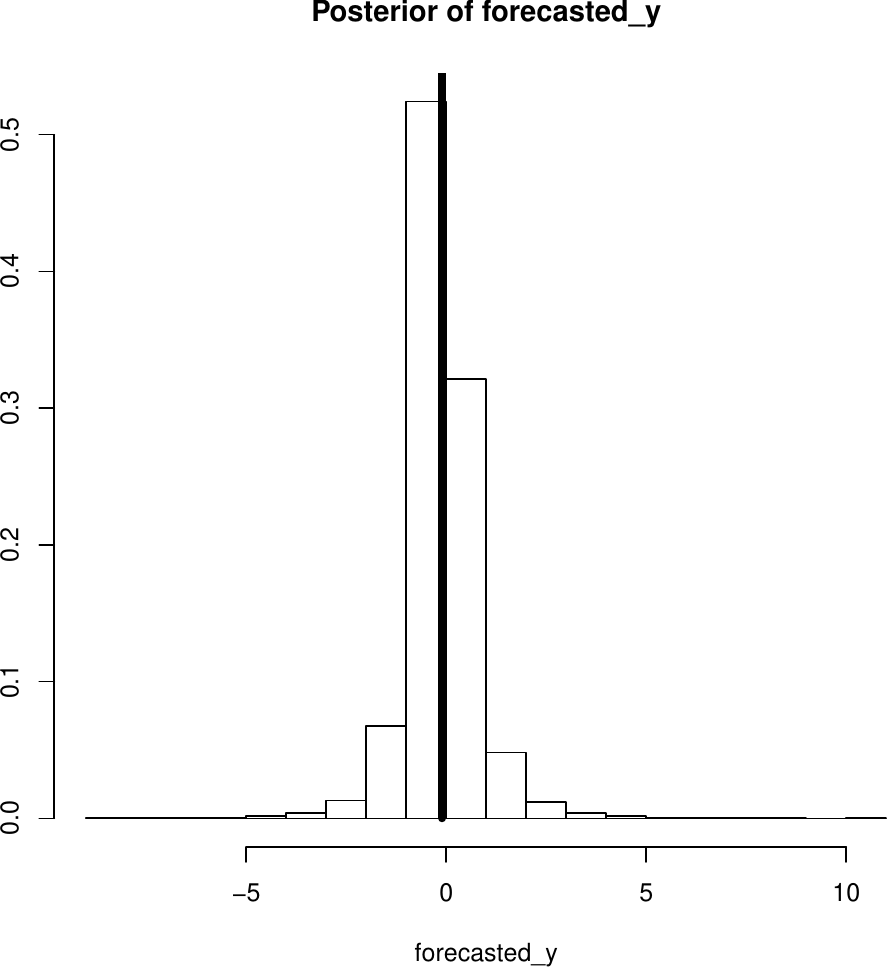}}
\caption{{\bf Simulation study with data generated from the univariate 
growth model of CPS}:
Posterior densities of $\sigma_{\eta}$, $r_{1,f}$, $r_{2,f}$, $r_{1,g}$, $r_{2,g}$,
$x_0$, $x_{T+1}$ (one-step forecasted $x$), and $y_{T+1}$ (one-step forecasted $y$).
The solid line stands for the true values of the respective parameters.}
\label{fig:simdata_cps_hist2}
\end{figure}


\begin{figure}
\centering
\subfigure{ \label{fig:x_time_series_simdata_cps_hist}
\includegraphics[width=15cm,height=10cm]{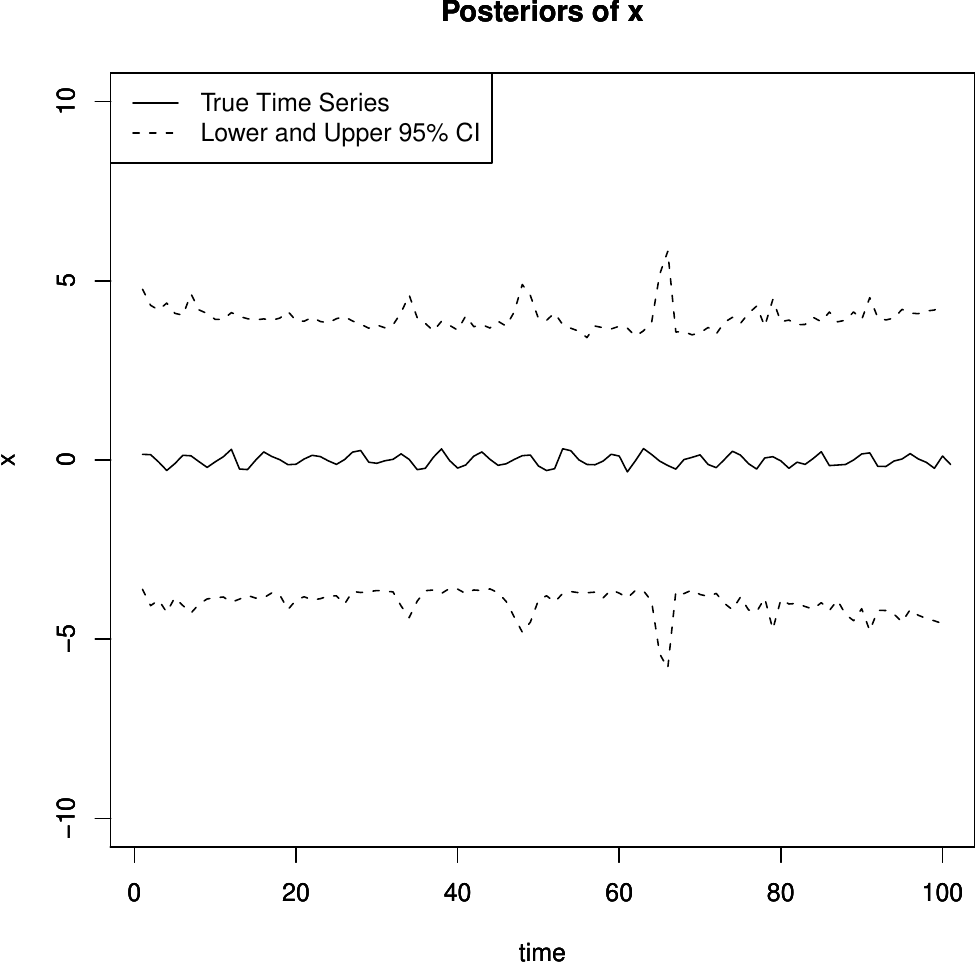}}
\caption{{\bf Simulation study with data generated from the univariate 
growth model of CPS}:
95\% highest posterior density credible intervals of the time series $x_1,\ldots,x_T$.
The solid line stands for the true time series.}
\label{fig:simdata_cps_hist3}
\end{figure}

Figures \ref{fig:simdata_cps_hist}, \ref{fig:simdata_cps_hist2} and \ref{fig:simdata_cps_hist3} show
the posterior distributions of the unknowns. Note that whenever applicable, the true value
is well-captured by the posteriors in question. To avoid any confusion, we mention that the true values
of $(\alpha,\beta,\gamma)$ associated with the true observational equation (\ref{eq:cps1}), for example, 
are not comparable with the parameters $(\beta_{0,f},\beta_{1,f},\beta_{2,f})$ associated with our Gaussian 
process model. Hence, the figures do not show any true value associated with the posteriors of the Gaussian process
parameters. Note that this is in contrast with the previous example where the true model has a linear state space
structure. There, the true functional forms are comparable with the linear mean functions of the underlying Gaussian 
processes, and that is the reason why the figures of the posterior distributions of the Gaussian process parameters
also consisted of the vertical line denoting the true values of the parameters.

As before, the true values of $x_{101}$ and $y_{101}$ are captured very well
by their respective posteriors. The true time series $\{x_1,\ldots,x_{100}\}$ again
fell well within the lower and the upper 95\% highest posterior density credible limits.
However, note that the credible intervals in this case are somewhat wider than the corresponding linear
model, which is to be expected.
The true values of $\sigma_{\epsilon}$, $\sigma_{\eta}$ and $x_0$ also fell well within
the highest density regions of their respective posterior distributions.
That our Gaussian process approach fits the data so well despite the fact that the
functions $f(\cdot,\cdot)$ and $g(\cdot,\cdot)$ are highly non-linear, is quite encouraging.

\section{Extension of our Gaussian process based approach to the multivariate situation}
\label{sec:multivariate}

Now we extend our nonparametric dynamic model to the case where both $\by_t$ and $\bx_t$ are multivariate.
In particular, we assume that they are $p$-component and $q$-component vectors, respectively.
Then our multivariate model is of the form 
\begin{eqnarray}
\by_t&=&\bof(\bx^*_{t,t})+\bepsilon_t,\hspace{2mm}\bepsilon_t\sim N_p(\bzero,\bSigma_{\epsilon}),\label{eq:obs2_mult}\\
\bx_t&=&\bg(\bx^*_{t,t-1})+\boeta_t,\hspace{2mm}\boeta_t\sim N_q(\bx_0,\bSigma_{\eta}), \label{eq:evo2_mult}
\end{eqnarray}
where $\bx_0\sim N_q(\bmu_{x_0},\bSigma_{x_0})$; $\bmu_{x_0},\bSigma_{x_0}$ assumed known.
In the above, $\bof(\cdot)=(f_1(\cdot),\ldots,f_p(\cdot))'$ is a function with $p$ components and 
$\bg(\cdot)=(g_1(\cdot),\ldots,g_q(\cdot))'$ is a function consisting of
$q$ components.
We assume that $\bof(\cdot)$ is a $p$-variate Gaussian process with 
mean $E[\bof(\cdot)]=\bB'_f\bh(\cdot)$ 
and with covariance function $cov (\bof(\bz_1),\bof(\bz_2))=c_f(\bz_1,\bz_2)\bSigma_f$, 
for any $q$-dimensional inputs $\bz_1,\bz_2$. Here 
$\bh(\cdot)=(h_1(\cdot),\ldots,h_m(\cdot))'$ and
$\bB_f=(\bbeta_{1,f},\ldots,\bbeta_{p,f})$, where,
for $j=1,\ldots,p$, $\bbeta_{j,f}$ are $m$-dimensional column vectors; 
clearly, $m=q+2$.
Also,
$c_f(\bz_1,\bz_2)=\exp\left\{-(\bz_1-\bz_2)'\bR_f(\bz_1-\bz_2)\right\}$,
where $\bR_f$ is a diagonal matrix consisting of $(q+1)$ smoothness parameters, denoted by 
$\{r_{1,f},\ldots,r_{(q+1),f}\}$.
Similarly, we assume that $\bg(\cdot)$ is a Gaussian process with mean 
$E[\bg(\cdot)]=\bB'_g\bh(\cdot)$ and covariance function $c_g(\bz_1,\bz_2)\bSigma_g$
$=\exp\left\{-(\bz_1-\bz_2)'\bR_g(\bz_1-\bz_2)\right\}$, the notation used being analogous
to those used for description of the Gaussian process $\bof(\cdot)$.

\subsection{Multivariate data and its distribution}

The multivariate data is given by the following $T\times p$ matrix:
$\bD_T=\left(\by_1,\by_2,\ldots,\by_T\right)'.$
Writing $\bD_T$ vectorically as a $Tp$-vector for convenience, as
$\bD_{Tp}=\left(\by'_1,\by'_2,\ldots,\by'_T\right)'$,
it follows that $[\bD_{Tp}\mid\bB_f,\bSigma_f,\bR_f,\bSigma_{\epsilon}]$ is a $Tp$-variate normal with mean vector
\begin{equation}
E[\bD_{Tp}\mid\bB_f,\bSigma_f,\bR_f,\bSigma_{\epsilon}]=\left(\begin{array}{c}\bB'_f\bh(\bx^*_{1,1})\\\bB'_f\bh(\bx^*_{2,2})\\\vdots\\\bB'_f\bh(\bx^*_{T,T})\end{array}\right)=\bmu_{D_{Tp}}\hspace{2mm}\mbox{(say)}
\label{eq:mult_data_mean}
\end{equation}
and covariance matrix 
\begin{equation}
V[\bD_{Tp}\mid\bB_f,\bSigma_f,\bR_f,\bSigma_{\epsilon}]=\bA_{f,D_T}\otimes\bSigma_f+\bI_T\otimes\bSigma_{\epsilon}=\bSigma_{D_{Tp}}\hspace{2mm}\mbox{(say)}.
\label{eq:cov_data}
\end{equation}
Since (\ref{eq:cov_data}) does not admit a Kronecker product form, the distribution of $\bD_{Tp}$ can not be written as matrix normal, which requires right and left
covariance matrices forming Kronecker product in the corresponding multivariate distribution of the vector.
It follows that
$[\by_{T+1}=\bof(\bx^*_{T+1,T+1})+\bepsilon_{T+1}\mid \bD_T,\bx_0,\bx_1,\ldots,\bx_{T+1}\bB_f,\bSigma_f,\bR_f,\bSigma_{\epsilon}]$
is $p$-variate normal with mean
\begin{equation}
\bmu_{y_{T+1}}=\bB'_f\bh(\bx^*_{T+1,T+1})+(\bD_T-\bH_{D_T}\bB_f)'\bA^{-1}_{f,D_T}\bs_{f,\bD_T}(\bx^*_{T+1,T+1})
\label{eq:cond_mean_y_mult}
\end{equation}
and variance 
\begin{equation} \bSigma_{y_{T+1}}=\left\{1-\bs_{f,\bD_T}(\bx^*_{T+1,T+1})^\prime \bA_{f,\bD_T}^{-1}\bs_{f,\bD_T}(\bx^*_{T+1,T+1})\right\}\bSigma_f+\bSigma_{\epsilon}
\label{eq:cond_var_y_mult}
\end{equation}

\subsection{Distributions of $\bg(\bx^*_{1,0})$ and $\bD^*_n$}

Conditional on $\bx_0$, $\bg(\bx^*_{1,0})$ is $q$-variate normal with mean $\bB_g'\bh(\bx^*_{1,0})$ and covariance matrix $\bSigma_g$. 

In contrast with the distribution of $\bD_T$, $\bD_{z,nq}=\left(\bg'(\bz_1),\bg'(\bz_2),\ldots,\bg'(\bz_n)\right)'$ has an $nq$-variate normal distribution with mean
\begin{equation}
E[\bD_{z,nq}\mid\bB_g,\bSigma_g,\bR_g]=\left(\begin{array}{c}\bB'_g\bh(\bz_1)\\\bB'_g\bh(\bz_2)\\\vdots\\\bB'_g\bh(\bz_n)\end{array}\right)=\bmu_{D_{z,nq}}\hspace{2mm}\mbox{(say)}
\label{eq:mult_Dz_mean}
\end{equation}
and covariance matrix 
\begin{equation}
V[\bD_{z,nq}\mid\bB_g,\bSigma_g,\bR_g]=\bA_{g,D^*_n}\otimes\bSigma_g=\bSigma_{D_{z,nq}}\hspace{2mm}\mbox{(say)}.
\label{eq:cov_Dz_vector}
\end{equation}
Hence, the distribution of the $n\times q$-dimensional matrix $\bD^*_n=\left(\bg(\bz_1),\bg(\bz_2),\ldots,\bg(\bz_n)\right)'$ is matrix normal:
\begin{equation}
[\bD^*_n\mid\bB_g,\bSigma_g,\bR_g]\sim\mathcal N_{n,q}\left(\bH_z\bB_g,\bA_{g,D^*_n},\bSigma_g\right)
\label{eq:Dz_matrix_normal}
\end{equation}
Conditionally on $(\bx_0,\bg(\bx^*_{1,0}))$,
it follows that $\bD^*_n$ is $n\times q$-dimensional matrix-normal:
\begin{equation}
[\bD^*_n\mid\bg(\bx^*_{1,0}),\bx_0,\bB_g,\bSigma_g,\bR_g,\bSigma_{\eta}]\sim\mathcal N_{n,q}\left(\bmu_{g,D^*_n},\bSigma_{g,D^*_n},\bSigma_g\right)
\label{eq:Dz_matrix_normal_conditional}
\end{equation}
In (\ref{eq:Dz_matrix_normal_conditional}) $\bmu_{g,D^*_n}$ is the mean matrix, given by
\begin{equation}
\bmu_{g,D^*_n}=\bH_{D^*_n}\bB_g+\bs_{g,D^*_n}(\bx^*_{1,0})(\bg(\bx^*_{1,0})'-\bh(\bx^*_{1,0})'\bB_g), \label{eq:mean1_mult}
\end{equation}
and 
\begin{equation}
\bSigma_{g,D^*_n}=\bA_{g,D^*_n}-\bs_{g,D^*_n}(\bx^*_{1,0})\bs_{g,D^*_n}(\bx^*_{1,0}).'
\label{eq:var1_mult}
\end{equation}
Here we slightly abuse notation to denote both univariate and multivariate versions of the mean matrix and the right
covariance matrix by $\bmu_{g,D^*_n}$ and $\bSigma_{g,D^*_n}$, respectively 
(see (9) and (11) of GMRB).

\subsection{Joint distribution of $\{\bx_0,\ldots,\bx_{T+1},\bD^*_n\}$}

Note that
\begin{align}
[\bx_1\mid \bg(\bx_0),\bx_0,\bB_g,\bSigma_g]&\sim N_q\left(\bg(\bx^*_{1,0}),\bSigma_{\eta}\right),\label{eq:dist_x1_mult}
\end{align}
and for $t=1,\ldots,T$, the conditional distribution $[\bx_{t+1}=\bg(\bx^*_{t+1,t})+\boeta_{t+1}\mid \bD^*_n,\bx_t,\bB_g,\bSigma_g,\bR_g,\bSigma_{\eta}]$
is $q$-variate normal with mean
\begin{equation}
\bmu_{x_t}=\bB_g'\bh(\bx^*_{t+1,t})+(\bD^*_n-\bH_{D^*_n}\bB_g)'\bA_{g,D^*_n}^{-1}\bs_{g,\bD^*_n}(\bx^*_{t+1,t})
\label{eq:cond_mean_xt1_mult}
\end{equation}
and variance 
\begin{equation} \bSigma_{x_t}=\left\{1-\bs_{g,\bD^*_n}(\bx^*_{t+1,t})^\prime \bA_{g,\bD^*_n}^{-1}\bs_{g,\bD^*_n}(\bx^*_{t+1,t})\right\}\bSigma_g+\bSigma_{\eta}.
\label{eq:cond_var_xt1_mult}
\end{equation}
Since $[\bx_0]\sim N_p\left(\bmu_{x_0},\bSigma_{x_0}\right)$ and the distribution of $\bD^*_n$ is given by (\ref{eq:Dz_matrix_normal})
the joint distribution is obtained by taking products of the individual distributions.

\subsection{Prior distributions}
We assume the following prior distributions:
\begin{align}
\mbox{For}\ \ i=1,\ldots,(q+1),\nonumber\\
[\log(r_{i,f})]&\stackrel{iid}{\sim} N\left(\mu_{R_f},\sigma^2_{R_f}\right)
\label{eq:Rf_prior}\\
\mbox{For}\ \ i=1,\ldots,(p+1),\nonumber\\
[\log(r_{i,g})]&\stackrel{iid}{\sim} N\left(\mu_{R_f},\sigma^2_{R_g}\right)
\label{eq:Rg_prior}\\
[\bSigma_{\epsilon}]&\propto\left|\bSigma_{\epsilon}\right|^{-\frac{\nu_{\epsilon}+p+1}{2}}\exp\left[-\frac{1}{2}tr\left(\bSigma^{-1}_{\epsilon}\bSigma_{\epsilon,0}\right)\right];\ \
\nu_{\epsilon}>p-1
\label{eq:Sigma_epsilon_prior}\\
[\bSigma_{\eta}]&\propto\left|\bSigma_{\eta}\right|^{-\frac{\nu_{\eta}+q+1}{2}}\exp\left[-\frac{1}{2}tr\left(\bSigma^{-1}_{\eta}\bSigma_{\eta,0}\right)\right];\ \
\nu_{\eta}>q-1
\label{eq:Sigma_eta_prior}\\
[\bSigma_f]&\propto\left|\bSigma_{f}\right|^{-\frac{\nu_f+p+1}{2}}\exp\left[-\frac{1}{2}tr\left(\bSigma^{-1}_f\bSigma_{f,0}\right)\right];\ \
\nu_f>p-1
\label{eq:Sigmaf_prior}\\
[\bSigma_g]&\propto\left|\bSigma_{g}\right|^{-\frac{\nu_g+q+1}{2}}\exp\left[-\frac{1}{2}tr\left(\bSigma^{-1}_g\bSigma_{g,0}\right)\right];\ \
\nu_g>q-1
\label{eq:Sigmag_prior}\\
[\bB_f\mid\bSigma_f]&\sim \mathcal N_{m,p}\left(\bB_{f,0},\bSigma_{B_f,0},\psi\bSigma_f\right)
\label{eq:Bf_prior}\\
[\bB_g\mid\bSigma_g]&\sim \mathcal N_{m,q}\left(\bB_{g,0},\bSigma_{B_g,0},\psi\bSigma_g\right)
\label{eq:Bg_prior}
\end{align}
All the prior parameters are assumed to be known; and their choices will be discussed in the specific applications
to be considered.

The full conditionals are of analogous forms as those in the univariate case, provided by equations from 
(\ref{eq:fullcond_betaf}) to (\ref{eq:fullcond_xlast}), but now
the univariate distributions must be replaced with multivariate distributions and multivariate 
distributions with matrix-variate distributions, and interestingly in some cases the full conditionals
are not available in closed form although these were available in the one-dimensional version of the problem.
For example, although the full conditional of $\bbeta_f$ and $\bbeta_g$ were available in the one-dimensional
problem, the corresponding distributions of $\bB_f$ and $\bB_g$ are no longer available in closed forms.
The problem of obtaining the closed form of the full conditional of $\bB_f$ can be attributed to the fact that 
(\ref{eq:cov_data}) can not be represented as a single
Kronecker product. The non-availability of the closed form in the case of $\bB_g$
is due to fact that the covariance matrices of $\bx_t$ are additive. In fact, it turns out that only
the full conditional of $\bx_{T+1}$ is of standard form.
Details of our MCMC methods for the multivariate case are provided in the next section.

\section{Details of MCMC sampling in the mulivariate situation}
\label{sec:mult_fullcond}


To obtain good proposal distributions,
we use our old strategy of ignoring the error terms $\bepsilon_t$ and $\boeta_t$. 
Below we provide details
of the proposal distributions used in each case. For our purpose, we abuse notation slightly
as in Section 4 of GMRB, that is, 
we denote by $\bG_{n+1}$, $\bD^*_{n+1}$, $\bA_{g,n+1}$ and $\bH_{g,D^*_{n+1}}$
the multivariate analogues of the quantities corresponding to the univariate situation described in 
Section 4 of GMRB. 
In other words, let $\bG_{n+1}=\bG\cup\{\bx^*_{1,0}\}$, $\bD^*_{n+1}=\left({\bD^*_n}',\bg(\bx^*_{1,0})\right)'$, 
$\bA_{g,n+1}=\left(\begin{array}{cc}\bA_{g,D^*_n} & \bs_{g,D^*_n}(\bx^*_{1,0})\\ \bs_{g,D^*_n}(\bx^*_{1,0})' & 1\end{array}\right)$
and $\bH_{g,D^*_{n+1}}'=[\bH_{g,D^*_n}', \bh(\bx^*_{1,0})]$. 

\subsection{Proposal distribution for updating $\bB_f$}
\label{subsec:mult_proposal_b_f}

Assuming $\bSigma_{\epsilon}=\bSigma_f$ in our multivariate dynamic state-space model, 
the full conditional of $\bB_f$ is $m\times p$-variate matrix-normal:
\begin{equation}
\bB_f\sim \mathcal N_{m,p}\left(\bmu_{B_f},\bSigma_{B_f},\bSigma_f\right), 
\label{eq:Bf_proposal}
\end{equation}
where
\begin{align}
\bmu_{B_f}&=\left(\bH'_{D_T}\left(\bA_{f,D_T}+\bI_T\right)^{-1}\bH_{D_T}+\psi^{-1}\bSigma^{-1}_{B_f,0}\right)^{-1}\nonumber\\
& \ \ \ \ \times \left(\bH'_{D_T}\left(\bA_{f,D_T}+\bI_T\right)^{-1}\bD_T+\psi^{-1}\bSigma^{-1}_{B_f,0}\bB_{f,0}\right),
\label{eq:condmean_betaf_mult}
\end{align}
and
\begin{align}
\bSigma_{B_f}&=\left(\bH'_{D_T}\left(\bA_{f,D_T}+\bI_T\right)^{-1}\bH_{D_T}+\psi^{-1}\bSigma^{-1}_{B_f,0}\right)^{-1}.
\label{eq:condvar_betaf_mult}
\end{align}

\subsection{Proposal distribution for $\bB_g$}
\label{subsec:mult_proposal_b_g}

Assuming $\bSigma_{\eta}=\bSigma_g$, the conditional distribution of $\bB_g$ is $m\times q$-variate matrix:
\begin{equation}
\bB_g\sim \mathcal N_{m,q}\left(\bmu_{B_g},\bSigma_{B_g},\bSigma_g\right), 
\label{eq:Bg_proposal}
\end{equation}
where
\begin{align}
\bmu_{B_g}
&=\left\{\psi^{-1}\bSigma^{-1}_{\bB_g,0}+\bH_{D^*_{n+1}}'\bA^{-1}_{g,D^*_{n+1}}\bH_{D^*_{n+1}}\right.\nonumber\\
&\ \ \ \ \left.+\sum_{t=1}^{T}\frac{\left(\bH_{D^*_n}'\bA_{g,D^*_n}^{-1}\bs_{g,D^*_n}(\bx^*_{t+1,t})-\bh(\bx^*_{t+1,t})\right)\left(\bH_{D^*_n}'\bA_{g,D^*_n}^{-1}\bs_{g,D^*_n}(\bx^*_{t+1,t})-\bh(\bx^*_{t+1,t})\right)'}{\sigma^2_{g,t}}\right\}^{-1}\nonumber\\
&\ \ \ \ \times\left\{\psi^{-1}\bSigma^{-1}_{\bB_g,0}\bB_{g,0}+\bH_{D^*_{n+1}}'\bA^{-1}_{g,D^*_{n+1}}\bD^*_{n+1} \right.\nonumber\\
&\left. \ \ \ \ +\sum_{t=1}^{T}
\frac{\left(\bh(\bx^*_{t+1,t})-\bH_{D^*_n}'\bA^{-1}_{g,D^*_n}\bs_{g,D^*_n}(\bx^*_{t+1,t})\right)
\left(\bx_{t+1}-\bD'_z\bA^{-1}_{g,D^*_n}\bs_{g,D^*_n}(\bx^*_{t+1,t})\right)'}{\sigma^2_{g,t}}\right\}.
\label{eq:condmean_bg_mult}
\end{align}
and 
\\[2mm]
\begin{align}
\bSigma_{B_g}
&=\left\{\psi^{-1}\bSigma^{-1}_{\bB_g,0}+\bH_{D^*_{n+1}}'\bA^{-1}_{g,D^*_{n+1}}\bH_{D^*_{n+1}}\right.\nonumber\\
&\ \ \ \ \left.+\sum_{t=1}^{T}\frac{\left(\bH_{D^*_n}'\bA_{g,D^*_n}^{-1}\bs_{g,D^*_n}(\bx^*_{t+1,t})-\bh(\bx^*_{t+1,t})\right)\left(\bH_{D^*_n}'\bA_{g,D^*_n}^{-1}\bs_{g,D^*_n}(\bx^*_{t+1,t})-\bh(\bx^*_{t+1,t})\right)'}{\sigma^2_{g,t}}\right\}^{-1}
\label{eq:condvar_bg_mult}
\end{align}
In (\ref{eq:condmean_bg_mult}) and (\ref{eq:condvar_bg_mult}),
$\sigma^2_{g,t}$ is given by (\ref{eq:betag_lesserror}), with obvious notational change from the univariate $x_t$ 
to the multivariate $\bx_t$. The corresponding changes from $x^*$ to $\bx^*$ are also self-explanatory.

\subsection{Proposal distributions for updating $\bSigma_f$ and $\bSigma_g$}
\label{subsec:mult_proposal_sigma_f_g}

Setting $\bSigma_{\epsilon}=\bSigma_f$ and $\bSigma_{\eta}=\bSigma_g$, we obtain the following inverse Wishart proposal
distributions of $\bSigma_f$ and $\bSigma_g$:
\begin{align}
q_{\bSigma_f}(\bSigma_f)
&\propto\left|\bSigma_f\right|^{-\left(\frac{\nu_f+p+1+T+m}{2}\right)}
\exp\left[-\frac{1}{2}tr\left\{\bSigma^{-1}_f\left(\bSigma_{f,0}+\psi^{-1}(\bB_f-\bB_{f,0})'\bSigma^{-1}_{B_f,0}(\bB_f-\bB_{f,0})\right.\right.\right.\nonumber\\
&\left.\left.\left.+\left(\bD_T-\bH_{D_T}\bB_f\right)'\left(\bA_{f,D_T}+\bI_T\right)^{-1}\left(\bD_T-\bH_{D_T}\bB_f\right)\right)\right\}\right]
\label{eq:proposal_sigmasqf_mult}
\end{align}
and
\begin{align}
q_{\bSigma_g}(\bSigma_g)&\propto\left|\bSigma_g\right|^{-\left(\frac{\nu_g+q+2+m+n+T}{2}\right)}
\exp\left[-\frac{1}{2}tr\bSigma^{-1}_g\left\{\bSigma_{g,0}+\psi^{-1}(\bB_g-\bB_{g,0})'\bSigma^{-1}_{B_g,0}(\bB_g-\bB_{g,0})\right.\right.\nonumber\\
&\left.\left.\ \ \ \ +(\bx_1-\bg(\bx^*_{1,0}))(\bx_1-\bg(\bx^*_{1,0}))'\right.\right.\nonumber\\
&\ \ \ \ \left.\left.+\sum_{t=1}^{T}\frac{1}{\sigma^2_{g,t}}
\left\{\bx_{t+1}-\bB'_g\bh(\bx^*_{t+1,t})-\left(\bD^*_n-\bH_{D^*_n}\bB_g\right)'\bA^{-1}_{g,D^*_n}\bs_{g,D^*_n}(\bx^*_{t+1,t})\right\}\right.\right.\nonumber\\
&\left.\left.\ \ \ \ \times\left\{\bx_{t+1}-\bB'_g\bh(\bx^*_{t+1,t})-\left(\bD^*_n-\bH_{D^*_n}\bB_g\right)'\bA^{-1}_{g,D^*_n}\bs_{g,D^*_n}(\bx^*_{t+1,t})\right\}'\right.\right.\nonumber\\
&\ \ \ \ \left.\left.+\left(\bD^*_{n+1}-\bH_{D^*_{n+1}}\bB_g\right)'\bA^{-1}_{g,D^*_{n+1}}\left(\bD^*_{n+1}-\bH_{D^*_{n+1}}\bB_g\right)\right\}\right]
\label{eq:proposal_sigmasqg_mult}
\end{align}

\subsection{Proposal distributions for updating $\bSigma_{\epsilon}$, $\bSigma_{\eta}$, $\bR_f$ and $\bR_g$}
\label{subsec:mult_proposal_sigma_smoothness}

As in the univariate case, here we set $\bSigma_f=\bSigma_{\epsilon}$ and $\bSigma_g=\bSigma_{\eta}$. Then the proposal
distributions of $\bSigma_{\epsilon}$ and $\bSigma_{\eta}$ are given by the following:
\begin{align}
q_{\bSigma_{\epsilon}}(\bSigma_{\epsilon})&\propto\left|\bSigma_{\epsilon}\right|^{-\left(\frac{\nu_{\epsilon}+p+1+T}{2}\right)}\nonumber\\
& \ \ \ \ \times \exp\left[-\frac{1}{2}tr\bSigma^{-1}_{\epsilon}
\left\{\bSigma_{\epsilon,0}+\left(\bD_T-\bH_{D_T}\bB_f\right)'\left(\bA_{f,D_T}+\bI_T\right)^{-1}\left(\bD_T-\bH_{D_T}\bB_f\right)\right\}\right]
\label{eq:proposal_sigmasq_epsilon_mult}\\
q_{\bSigma_{\eta}}(\bSigma_{\eta})&\propto\left|\bSigma_{\eta}\right|^{-\left(\frac{\nu_{\eta}+q+3+n+T}{2}\right)}
\exp\left[-\frac{1}{2}tr\bSigma^{-1}_{\eta}\left\{\bSigma_{\eta,0}+(\bx_1-\bg(\bx^*_{1,0}))(\bx_1-\bh(\bx^*_{1,0}))'\right.\right.\nonumber\\
&\ \ \ \ \left.\left.+\left(\bD^*_{n+1}-\bH_{D^*_{n+1}}\bB_g\right)'\bA^{-1}_{g,D^*_{n+1}}\left(\bD^*_{n+1}-\bH_{D^*_{n+1}}\bB_g\right)\right.\right.\nonumber\\
&\ \ \ \ \left.\left.+\sum_{t=1}^{T}\frac{1}{\sigma^2_{g,t}}
\left\{\bx_{t+1}-\bB_g'\bh(\bx^*_{t+1,t})-\left(\bD^*_n-\bH_{D^*_n}\bB_g\right)'\bA^{-1}_{g,D^*_n}\bs_{g,D^*_n}(\bx^*_{t+1,t})\right\}\right.\right.\nonumber\\
& \ \ \ \ \times \left.\left.\left\{\bx_{t+1}-\bB_g'\bh(\bx^*_{t+1,t})-\left(\bD^*_n-\bH_{D^*_n}\bB_g\right)'\bA^{-1}_{g,D^*_n}\bs_{g,D^*_n}(\bx^*_{t+1,t})\right\}'\right\}\right]\nonumber\\
\label{eq:proposal_sigmasq_eta_mult}
\end{align}
As in the case of the corresponding univariate proposals, in (\ref{eq:proposal_sigmasq_eta_mult}) 
the factor $$\left|\bSigma_{\eta}\right|^{-\left(\frac{n}{2}\right)}\exp\left[-\frac{1}{2}tr\bSigma^{-1}_{\eta}
\left(\bD^*_{n+1}-\bH_{D^*_{n+1}}\bB_g\right)'\bA^{-1}_{g,D^*_{n+1}}\left(\bD^*_{n+1}-\bH_{D^*_{n+1}}\bB_g\right)\right]$$ occurs because under 
$\bSigma_g=\bSigma_{\eta}$,
$[\bD^*_{n+1}\mid\bB_g,\bSigma_g,\bR_g]\sim\mathcal N_{n,q}\left(\bH_{D^*_{n+1}}\bB_g,\bA_{g,D^*_{n+1}},\bSigma_{\eta}\right)$.

The full conditionals of the smoothness parameters in $\bR_f$ and $\bR_g$ 
are not available in closed forms, and as before we suggest Metropolis-Hastings steps
with normal random walk proposals with adequately optimized variances. 

\subsection{Proposal distribution for updating $\bg(\bx^*_{1,0})$}
\label{subsec:mult_proposal_d_g_new}

Assuming $\bSigma_{\eta}=\bSigma_g$, the full conditional of $\bg(\bx^*_{1,0})$ is $q$-variate normal with mean and variance given, respectively, by
\begin{align}
\bmu^*_g=E[\bg(\bx^*_{1,0})\mid\cdots]
&=\{2+\bs_{g,D^*_n}(\bx^*_{1,0})'\bSigma^{-1}_{g,D^*_n}\bs_{g,D^*_n}(\bx^*_{1,0})\}^{-1}\bSigma_g\nonumber\\
& \ \ \ \ \times\left\{\bx_1+\bB_g'\bh(\bx^*_{1,0})+{\bD^*}'\bSigma^{-1}_{g,D^*_n}\bs_{g,D^*_n}(\bx^*_{1,0})\right\}\nonumber\\
\label{eq:condmean_mult_g}
\end{align}
and
\begin{align}
\bSigma^*_g=V[\bg(\bx^*_{1,0})\mid\cdots]
&=\{2+\bs_{g,D^*_n}(\bx^*_{1,0})'\bSigma^{-1}_{g,D^*_n}\bs_{g,D^*_n}(\bx^*_{1,0})\}^{-1}\bSigma_g\nonumber\\
\label{eq:condvar_mult_g}
\end{align}
In the above, $\bSigma_{g,D^*_n}$ is given by (59) of GMRB, 
and $\bD^*_z$ is given by
\begin{equation}
\bD^*_z=\bD^*_n-\bH_{D^*_n}\bB_g+\bs_{g,D^*_n}(\bx^*_{1,0})\bh(\bx^*_{1,0})'\bB_g
\label{eq:D_star_mult_z}
\end{equation}
We consider $q_{g(x^*_{1,0})}(\bg(\bx^*_{1,0}))\equiv N_q\left(\bg(\bx^*_{1,0}):\bmu^*_g,\bSigma^*_g\right)$
as the proposal distribution for updating $\bg(\bx^*_{1,0})$.

\subsection{Proposal distribution for updating $\bD^*_n$}
\label{subsec:mult_proposal_d_g}

The full conditional distribution of $\bD^*_n$ in our dynamic model after setting $\bSigma_{\eta}=\bSigma_g$, is matrix-normal:
\begin{equation}
[\bD^*_n]\sim\mathcal N_{n,q}\left(\bmu^*_{D^*_n},\bSigma^*_{D^*_n},\bSigma_g\right),
\label{eq:Dz_cond_mult}
\end{equation}
where
\begin{align}
\bmu^*_{D^*_n}
&=\left\{\bSigma^{-1}_{g,D^*_n}
+\bA^{-1}_{g,D^*_n}\left(\sum_{t=1}^{T}\frac{\bs_{g,D^*_n}(\bx^*_{t+1,t})\bs_{g,D^*_n}(\bx^*_{t+1,t})'}{\sigma^2_{g,t}}\right)\bA^{-1}_{g,D^*_n}\right\}^{-1}\nonumber\\
&\times\left\{\bSigma^{-1}_{g,D^*_n}\bmu_{g,D^*_n}
+\bA^{-1}_{g,D^*_n}\sum_{t=1}^{T}\frac{\bs_{g,D^*_n}(\bx^*_{t+1,t})\{\bx_{t+1}'-(\bh(\bx^*_{t+1,t})'-\bs_{g,D^*_n}(\bx^*_{t+1,t})'\bA^{-1}_{g,D^*_n}\bH_{D^*_n})\bB_g\}}{\sigma^2_{g,t}}\right\}\nonumber\\
\label{eq:condmean_Dz_mult}
\end{align}
and 
\begin{align}
\bSigma^*_{D^*_n}&=\left\{\bSigma^{-1}_{g,D^*_n}+\bA^{-1}_{g,D^*_n}\left(\sum_{t=1}^{T}\frac{\bs_{g,D^*_n}(\bx^*_{t+1,t})\bs_{g,D^*_n}(\bx^*_{t+1,t})'}{\sigma^2_{g,t}}\right)\bA^{-1}_{g,D^*_n}\right\}^{-1}
\label{eq:condvar_Dz_mult}
\end{align}
In the above, $\bmu_{g,D^*_n}$ and $\bSigma_{g,D^*_n}$ are given by
(58) and (59) of GMRB, respectively.
The distribution (\ref{eq:Dz_cond_mult}) will be used as proposal distribution to update $\bD^*_n$ using Metropolis-Hastings
step.

\subsection{MH proposal for updating $\bx_0$}
\label{subsec:mult_proposal_x_0}

For $j=1,\ldots,p$, let $\bbeta_{j,f}=(\bbeta'_{j,f,1},\bbeta'_{j,f,2})'$. Likewise, for $j=1,\ldots,q$, 
let $\bbeta_{j,g}=(\bbeta'_{j,g,1},\bbeta'_{j,g,2})'$.
Here $\bbeta_{j,f,1}$ and $\bbeta_{j,g,1}$ are two-dimensional and
$\bbeta_{j,f,2}$ and $\bbeta_{j,g,2}$ are $q$-dimensional vectors.
Also let $\ba_f=(\bbeta_{1,f,1},\ldots,\bbeta_{p,f,1})'$ and $\bb_f=(\bbeta_{1,f,2},\ldots,\bbeta_{p,f,2})'$.
Similarly,
$\ba_g=(\bbeta_{1,g,1},\ldots,\bbeta_{q,g,1})'$ and $\bb_g=(\bbeta_{1,g,2},\ldots,\bbeta_{q,g,2})'$.
Thus, $\ba_f$ and $\ba_g$ are $p\times 2$ and $q\times 2$-dimensional matrices respectively, 
while $\bb_f$ and $\bb_g$ are, 
respectively, $p\times q$ and $q\times q$ dimensional matrices.
Then, $\bB'_f\bh(\bx^*)=\ba_f\bk_1+\bb_f\bx$ and $\bB'_g\bh(\bx^*)=\ba_g\bk_1+\bb_g\bx$, for any $q$-dimensional $\bx$.
With these representations, and setting $\bSigma_g=0$, 
the full conditional distribution of $\bx_0$ is
$q$-variate normal with mean and variance given, respectively, by
\begin{align}
E[\bx_0\mid\cdots]&=\left\{\bSigma^{-1}_{x_0}+\bb'_g{\bSigma_{\eta}}^{-1}\bb_g\right\}^{-1}
\left\{\bSigma^{-1}_{x_0}\bmu_{x_0}+\bb'_g{\bSigma_{\eta}}^{-1}\left(\bx_1-\ba_g\bk_1\right)\right\}
\label{eq:mean_proposal_x0_mult}\\
V[\bx_0\mid\cdots]&=\left\{\bSigma^{-1}_{x_0}+\bb'_g{\bSigma_{\eta}}^{-1}\bb_g\right\}^{-1}.
\label{eq:var_proposal_x0_mult}
\end{align}

\subsection{MH proposals for $\{\bx_1,\ldots,\bx_T\}$}
\label{subsec:mult_proposal_x_t}
Ignoring the errors of the functions $\bof$ and $\bg$, that is, setting $\bSigma_f=\bSigma_g=0$,
it turns out that the proposal distribution of 
$\bx_{t+1}$, for $t=0,\ldots,T-1$,
can be taken as a $q$-variate normal distribution
with mean and variance given, respectively, by
\begin{align}
E[\bx_{t+1}\mid\cdots]&=\left(\bSigma^{-1}_{\eta}+\bb'_g\bSigma^{-1}_{\eta}\bb'_g+\bb'_f\bSigma^{-1}_{\epsilon}\bb_f\right)^{-1}\nonumber\\
&\times\left\{\bSigma^{-1}_{\eta}\left(\ba_g\bk_{t+1}+\bb_g\bx_{t}\right)+\bb'_g\bSigma^{-1}_{\eta}\left(\bx_{t+2}-\ba_g\bk_{t+2}\right)
+\bb'_f\bSigma^{-1}_{\epsilon}(\by_{t+1}-\ba_f\bk_{t+1})\right\}
\label{eq:mean_proposal_xt_mult}\\
V[\bx_{t+1}\mid\cdots]&=\left(\bSigma^{-1}_{\eta}+\bb'_g\bSigma^{-1}_{\eta}\bb'_g+\bb'_f\bSigma^{-1}_{\epsilon}\bb_f\right)^{-1}.
\label{eq:var_proposal_xt_mult}
\end{align}

\subsection{Gibbs step for updating $\bx_{T+1}$}
\label{subsec:mult_proposal_x_last}
The full conditional distribution of $\bx_{T+1}$ 
is $q$-variate normal with mean
\begin{equation}
E[\bx_{T+1}\mid\cdots]=\bB'_g\bh(\bx^*_{T+1,T})+(\bD^*_n-\bH_{D^*_n}\bB_g)'\bA^{-1}_{g,D^*_n}\bs_{g,\bD^*_n}(\bx^*_{T+1,T})
\label{eq:fullcond_mean_xt1_mult}
\end{equation}
and variance 
\begin{equation} 
V[\bx_{T+1}\mid\cdots]=\bSigma_{\eta}+\bSigma_g\left\{1-\bs_{g,\bD^*_n}(\bx^*_{T+1,T})^\prime \bA_{g,\bD^*_n}^{-1}\bs_{g,\bD^*_n}(\bx^*_{T+1,T})\right\}.
\label{eq:fullcond_var_xt1_mult}
\end{equation}

The caveat for using these proposal distributions is that if the underlying assumptions $\Sigma_f=\Sigma_{\epsilon}$,
$\Sigma_g=\Sigma_{\eta}$ and $\Sigma_g=\bzero$ do not hold, then the proposal mechanisms will turn out to be
much less effective, and more so compared to the univariate cases, due to the curse of dimensionality. So,
we shall also consider other appproaches to these updating procedures, to be discussed in the context of the 
specific applications.

\section{Simulation study: multivariate case}
\label{subsec:simstudy_multivariate}

We consider a simulation study where we generate the data from a 4-variate model, constructed using the
univariate nonstationary growth model of \ctn{Carlin92}. To reduce computational burden in this multi-dimensional
set up we generated 50 data points using the following model:
 for $t=1,\ldots,50$,
 \begin{align}
 x_{t,1}&= \alpha x_{t-1,1}+\beta x_{t-1,1}/(1+x^2_{t-1,1})+\gamma\cos(1.2(t-1))+\eta_{t,1}\label{eq:mult1}\\
 x_{t,2}&= \alpha x_{t-1,2}+\beta x_{t-1,2}/(1+x^2_{t-1,2})+\eta_{t,2}\label{eq:mult2}\\   
 x_{t,3}&=\alpha+\beta x_{t-1,3}+\eta_{t,3}\label{eq:mult3}\\
 x_{t,4}&=\gamma\cos(1.2(t-1))+\eta_{t,4}\label{eq:mult4}
 \\[2mm]
 y_{t,i}&=x^2_{t,i}/20+\epsilon_{t,i};\ \  i=1,\ldots,4,
 \end{align}
 where $\epsilon_{t,i}$ and $\eta_{t,i}$ are $iid$ zero-mean normal random variables with variances
 0.1. We set $\alpha = 0.05$, $\beta=0.1$ and $\gamma=0.2$, and $\bx_0=\bzero$.

\subsection{Choice of prior parameters}
\label{subsubsec:mult_data_prior_choice}

For the priors of $\bB_f$ and $\bB_g$, we set $\psi=1$, $\bB_{f,0}$ and $\bB_{g,0}$ to be null matrices,
and $\bSigma_{B_f,0}$ and $\bSigma_{B_g,0}$ to be identity matrices. For the priors of $\bSigma_{\epsilon}$,
$\bSigma_{\eta}$, $\bSigma_f$, and $\bSigma_g$ we set $\nu_\epsilon=p$, $\nu_{\eta}=q$, $\nu_f=p$,
and $\nu_g=q$. We chose $\bSigma_{\epsilon}=\bSigma_{\eta}=0.1\bI$, and $\bSigma_f=\bSigma_g=0.5\bI$,
where $\bI$ denotes the identity matrix. 
For the log-normal priors of the smoothness parameters
we set $\mu_{R_f}=\mu_{R_g}=-0.5$ and $\sigma^2_{R_f}=\sigma^2_{R_g}=1$. 
The choices imply that the prior mean and the prior variance of each of the smoothness
parameters are, respectively, 1 and 2 (approximately). In our simulation experiment 
these choices seemed adequate. 


\subsection{MCMC implementation}
\label{subsubsec:mult_data_mcmc}

For updating the smoothness parameters we used normal random walk proposals with variances 0.005.
To set up the grid $\bG_z$ for the model-fitting purpose, we considered $[-30,30]^4$ to be a grid space for the 4-dimensional
variable $\bx$. Indeed, this grid-space is much larger than, and contains the region where the 
entire true 4-dimensional time series lie. We divide $[-30,30]$ into 100 equal sub-intervals
and chose a point randomly from each of 100 sub-intervals, in each dimension, yielding $n=100$ 4-dimensional points corresponding to $\bx$.
As before, we chose the first component of the grid (corresponding to the time
component) by uniformly drawing from each subinterval $[i,i+1]$; $i=0,\ldots,99$.

The block updating proposal for updating $\bB_f$, described in 
Section S-2.2 
worked quite well,
but those for block updating $\bB_g$, $\bg(\bx^*_{1,0})$, $\bD^*_n$, and those for the covariance matrices, as decribed in Section S-2 
yielded
poor acceptance rates. In order to update the covariance matrices $\bSigma_{\epsilon}$, $\bSigma_{\eta}$, $\bSigma_g$
and $\bSigma_f$, we considered the following strategy: we re-wrote the matrices in the form $\bC\bC'$, where $\bC$ is a lower triangular matrix, and used
normal random walk with variance 0.005, to update the non-zero elements in a single block. This improved the acceptance rates. 
For $\bB_g$, $\bg(\bx^*_{1,0})$ and $\bD^*_n$, the strategy of block updating using random walk failed. As a remedy we considered the transformation-based MCMC (TMCMC),
recently developed by \ctn{Dutta11}; in particular, we used the additive transformation, which requires much less number of move
types and hence computationally less expensive compared to other, non-additive move types. Briefly, for each of the blocks, 
we generated $\xi\sim N(0,0.05)I(\xi>0)$. Then, for each parameter in the block, we either added or subtracted $\xi$ with equal probability. In other words,
we used the same $\xi$ to update all the parameters in a block, unlike the block random walk proposal. This considerably improved the
acceptance rates. For theoretical and implementation details, see \ctn{Dutta11}. 

We discarded the first 10,000 iterations of the MCMC run as burn-in and stored the subsequent 50,000 iterations for inference. Convergence
is assessed by informal diagnostics, as before. It took an ordinary laptop about 24 hours to implement this multivariate 
experiment.

\subsection{Results of model-fitting}
\label{subsubsec:mult_data_results}

\begin{figure}
\centering
  \subfigure{ \label{fig:x0_1}
   \includegraphics[width=3.5cm,height=3.5cm]{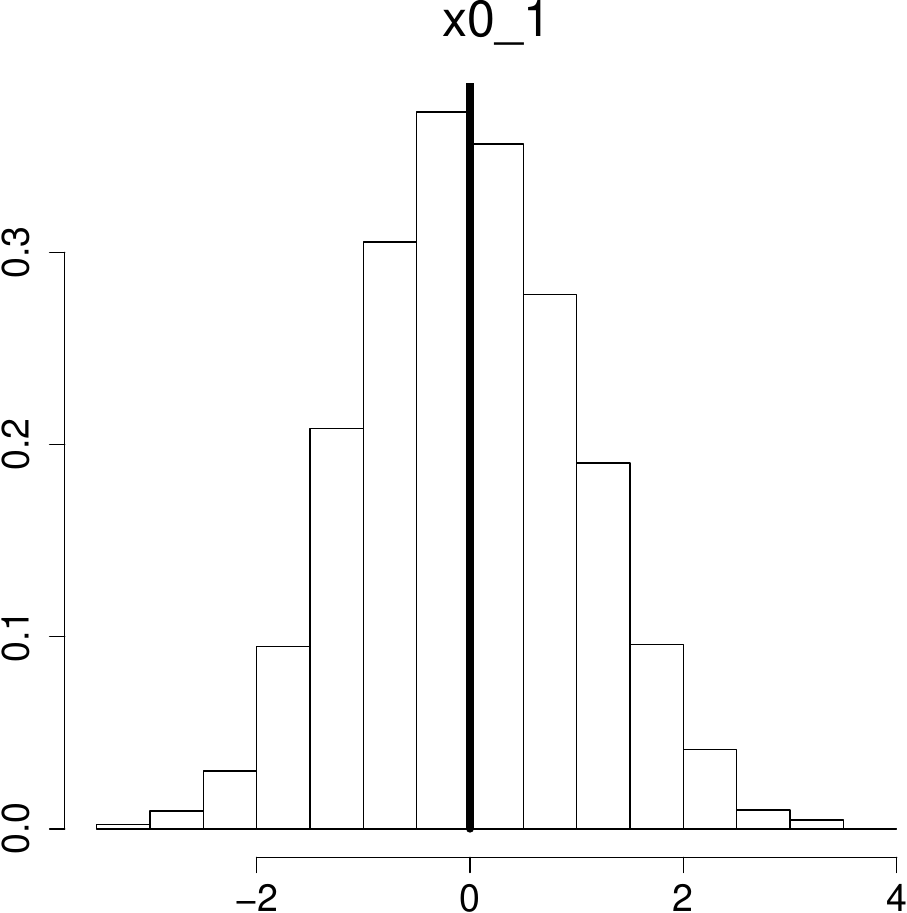}}
   \hspace{2mm}
  \subfigure{ \label{fig:x0_2} 
  \includegraphics[width=3.5cm,height=3.5cm]{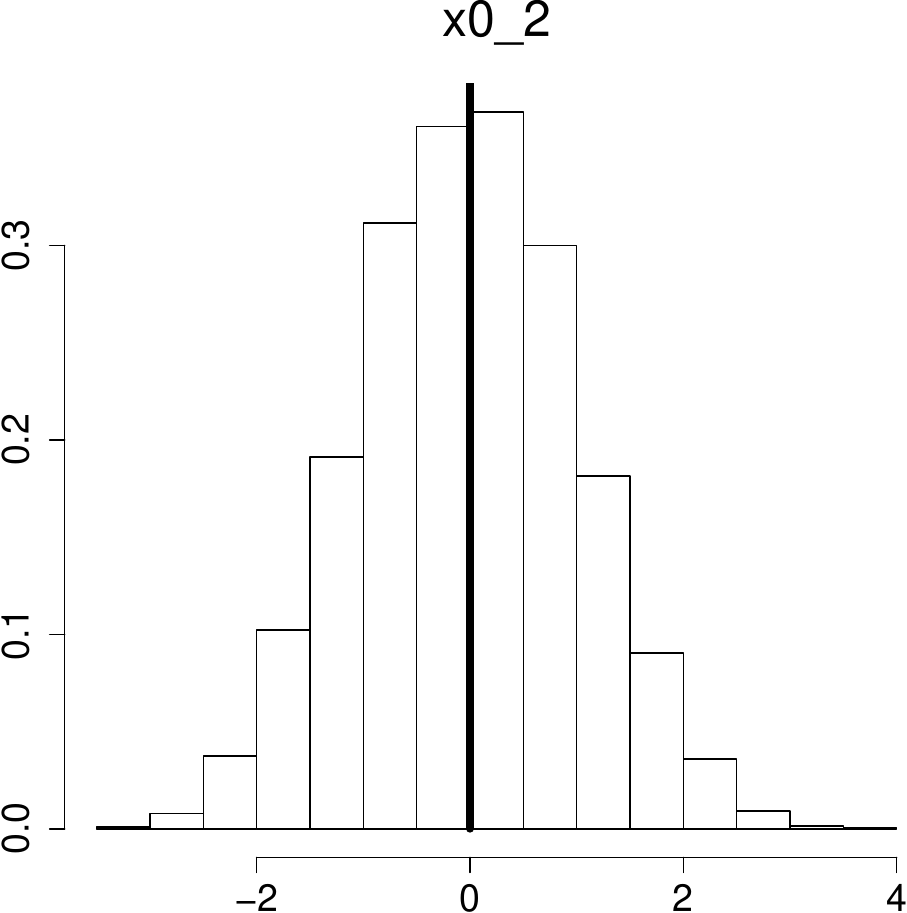}}
   \hspace{2mm}
  \subfigure{ \label{fig:x0_3} 
  \includegraphics[width=3.5cm,height=3.5cm]{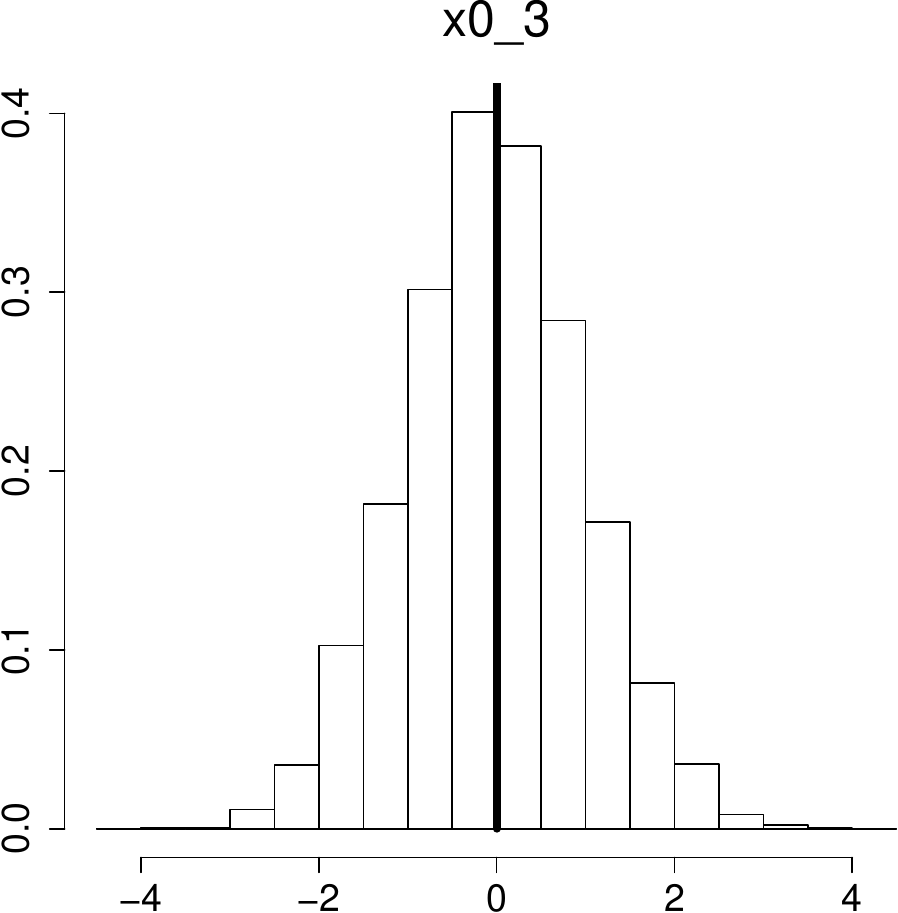}}
  \vspace{2mm}
  \subfigure{ \label{fig:x0_4}
   \includegraphics[width=3.5cm,height=3.5cm]{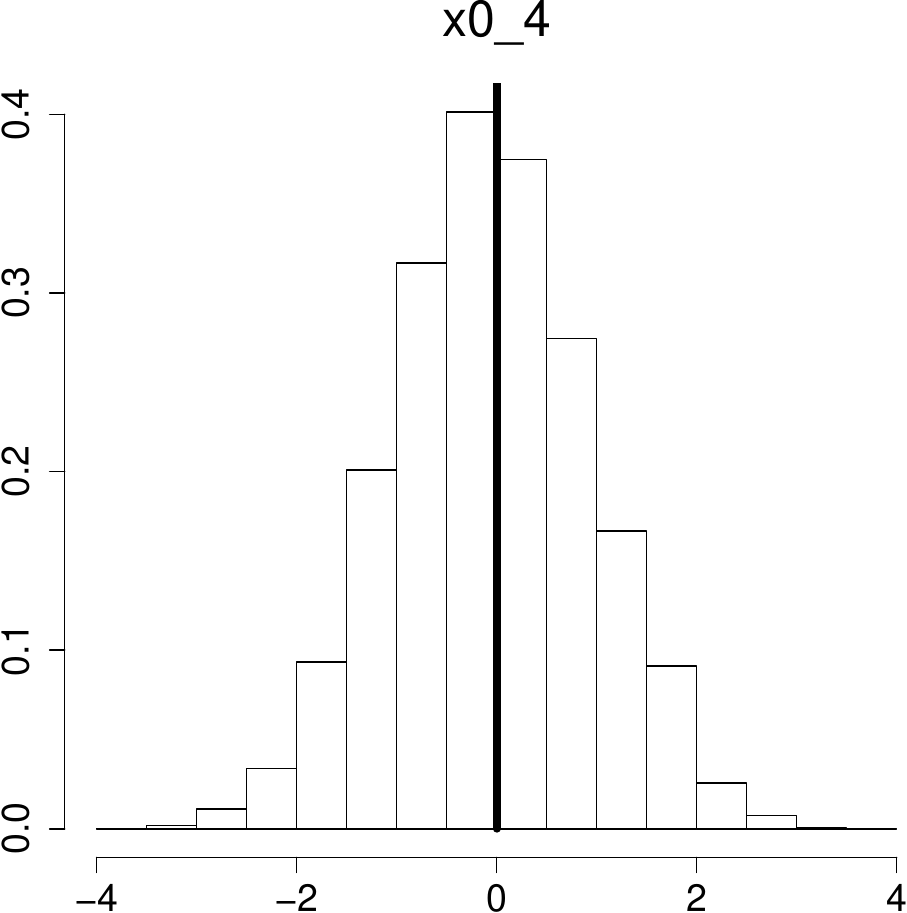}}\\
   \vspace{2mm}
  \subfigure{ \label{fig:forecasted_x1} 
  \includegraphics[width=3.5cm,height=3.5cm]{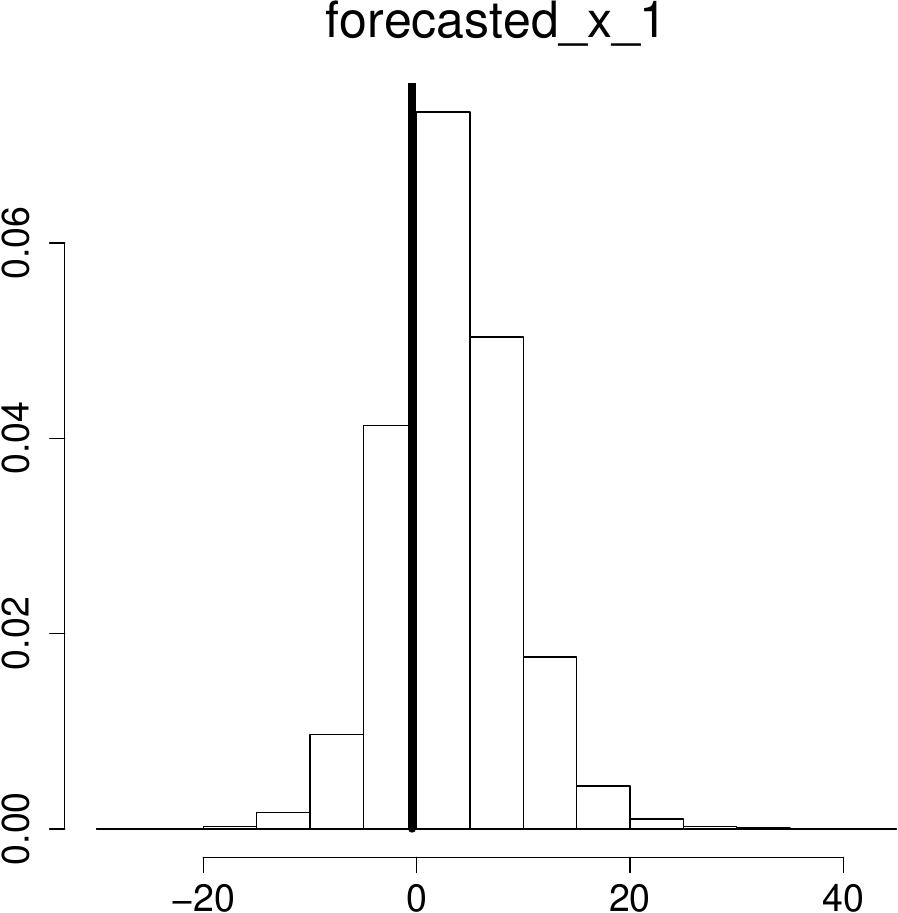}}
   \hspace{2mm}
  \subfigure{ \label{fig:forecasted_x2} 
  \includegraphics[width=3.5cm,height=3.5cm]{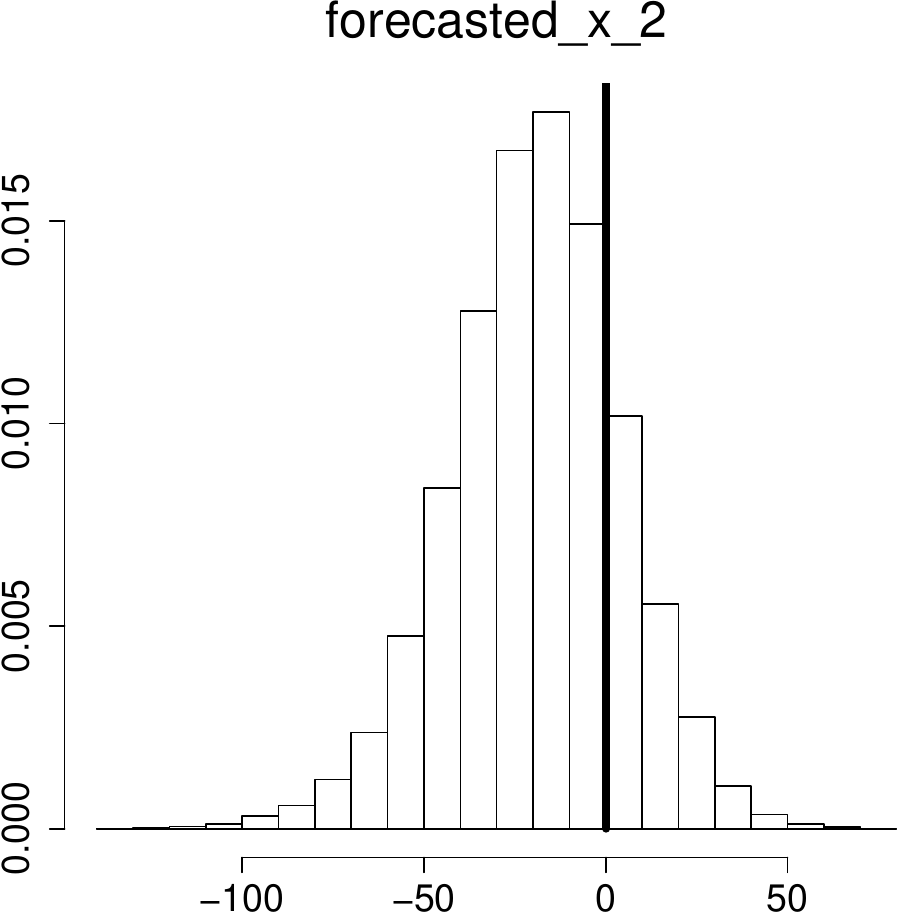}}
  \vspace{2mm}
  \subfigure{ \label{fig:forecasteded_x3}
   \includegraphics[width=3.5cm,height=3.5cm]{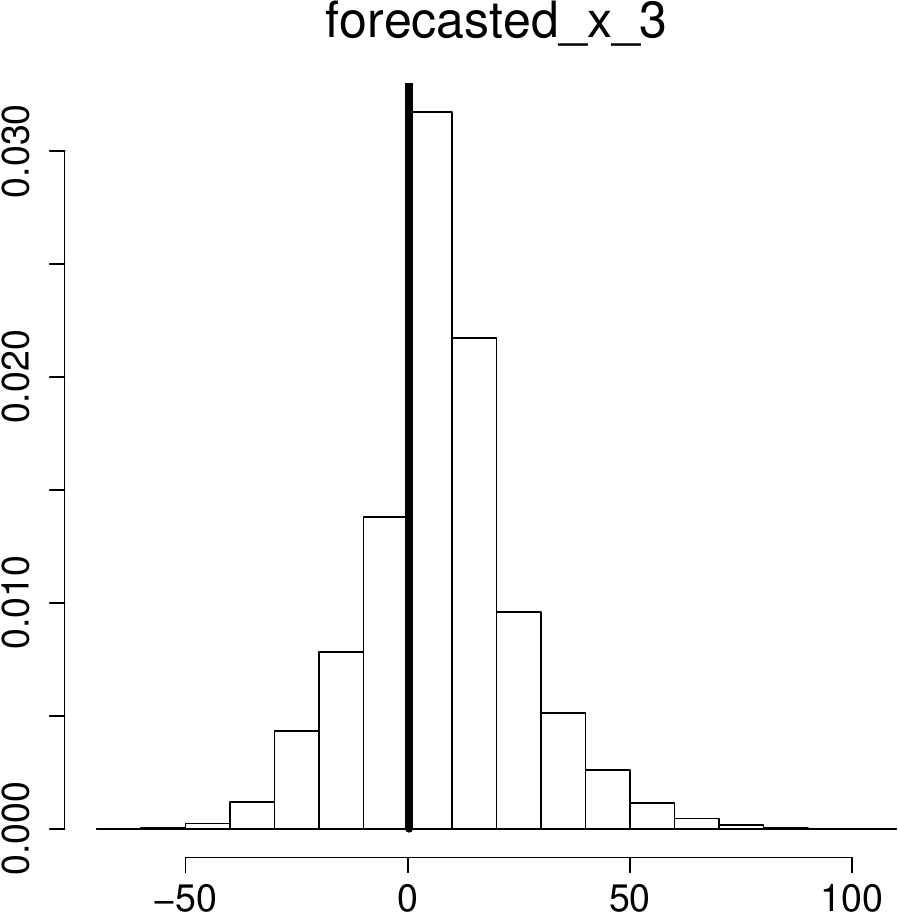}}
   \hspace{2mm}
  \subfigure{ \label{fig:forecasted_x4} 
  \includegraphics[width=3.5cm,height=3.5cm]{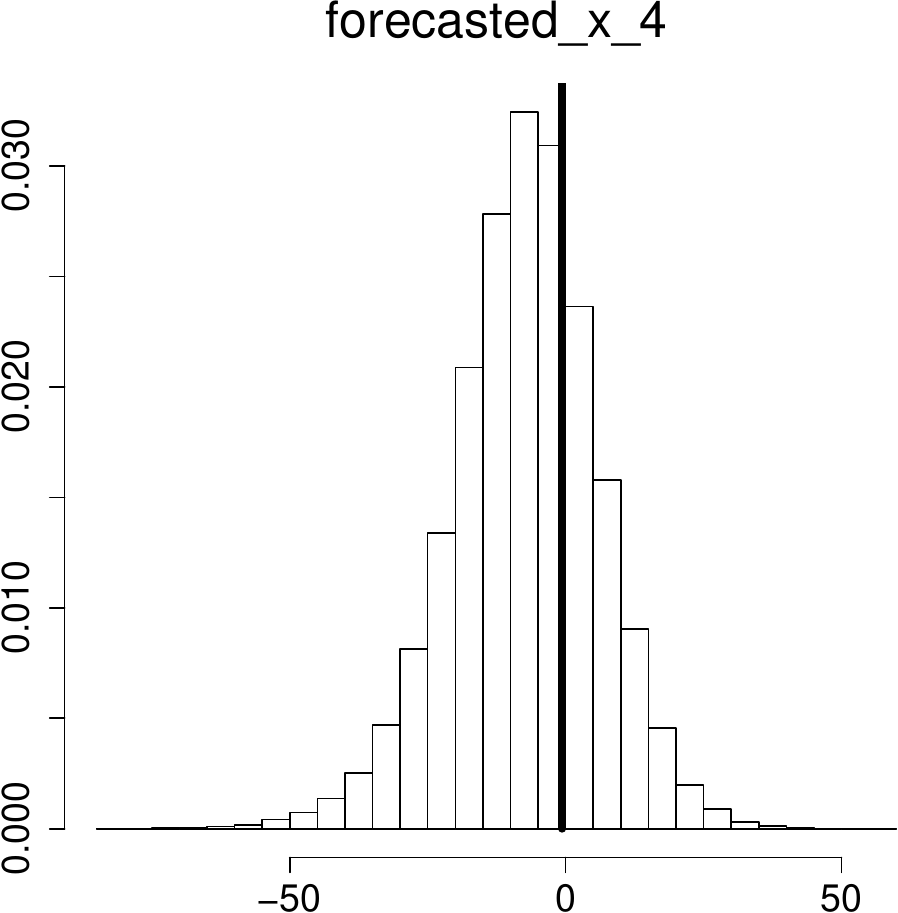}}\\
   \vspace{2mm}
  \subfigure{ \label{fig:forecasted_y1} 
  \includegraphics[width=3.5cm,height=3.5cm]{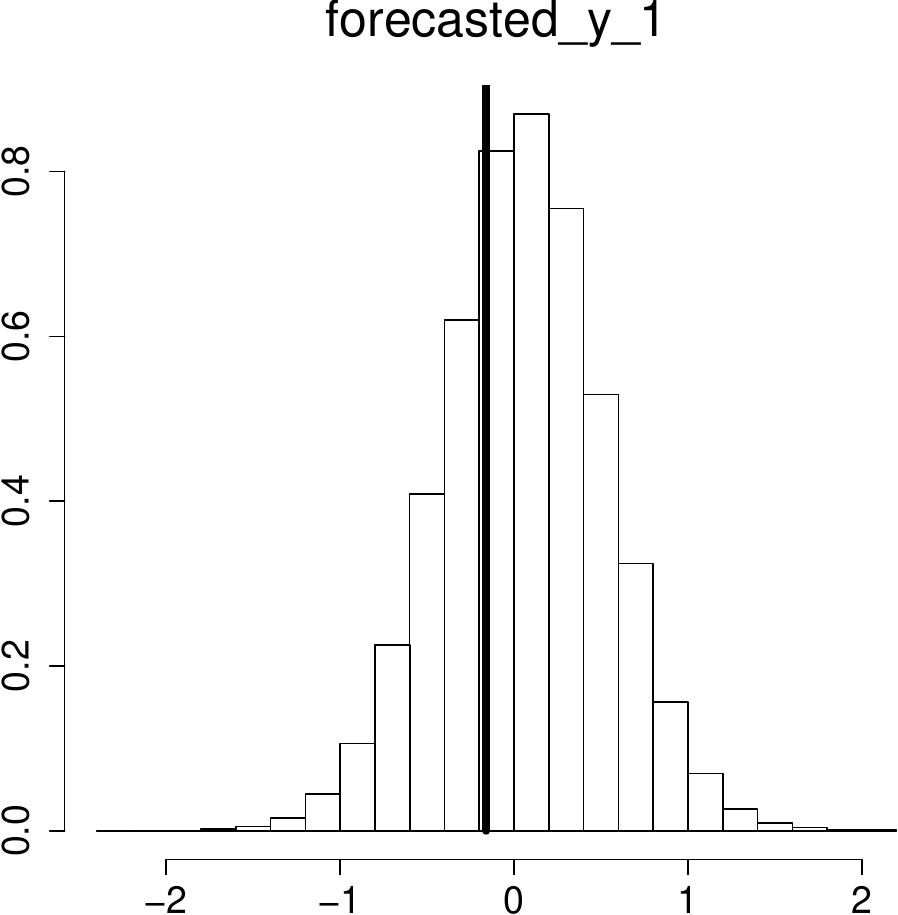}}
   \hspace{2mm}
  \subfigure{ \label{fig:forecasted_y2} 
  \includegraphics[width=3.5cm,height=3.5cm]{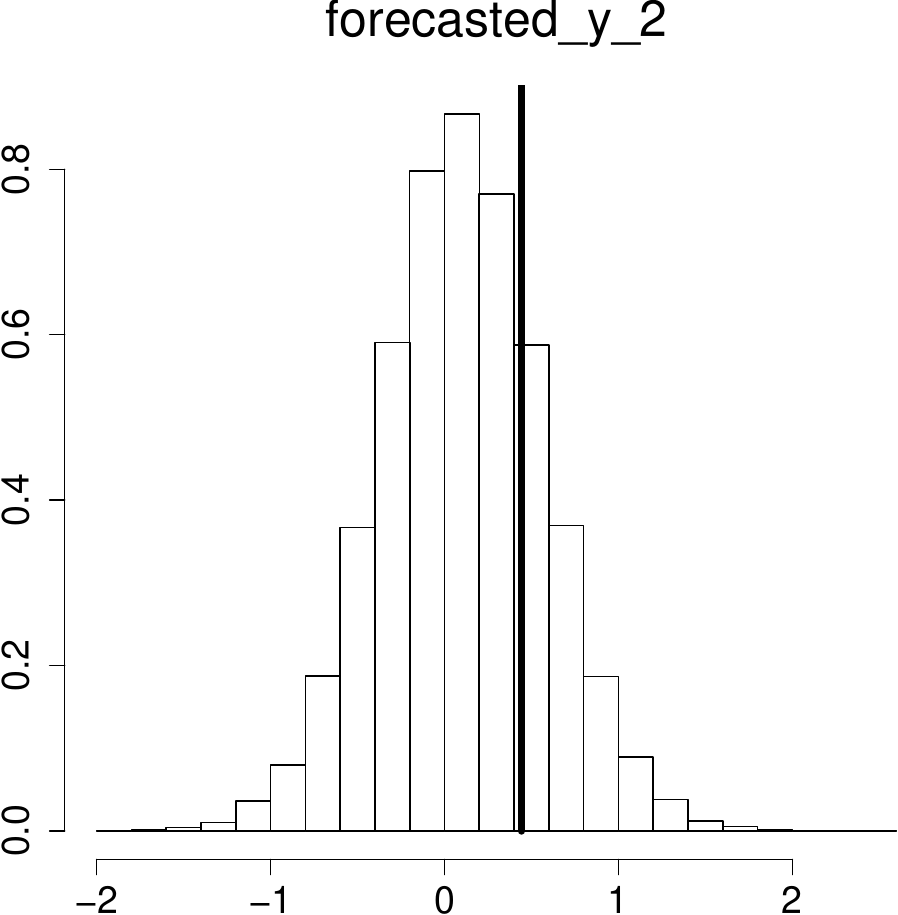}}
  \vspace{2mm}
  \subfigure{ \label{fig:forecasteded_y3}
   \includegraphics[width=3.5cm,height=3.5cm]{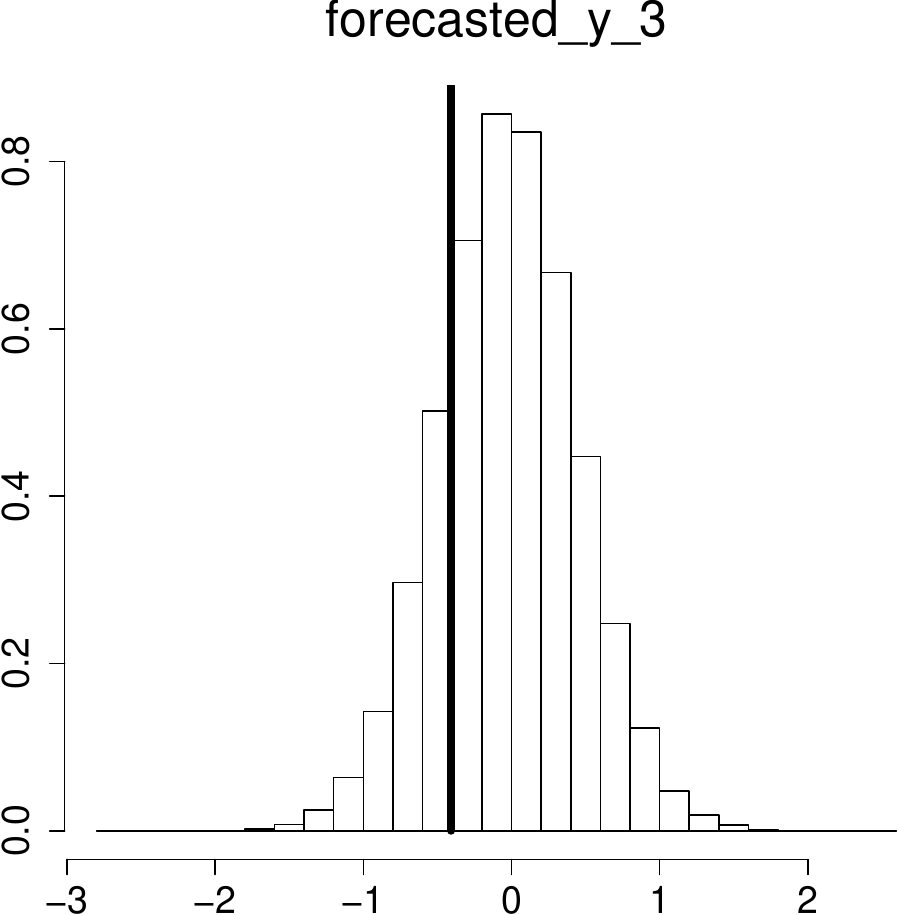}}
   \hspace{2mm}
  \subfigure{ \label{fig:forecasted_y4} 
  \includegraphics[width=3.5cm,height=3.5cm]{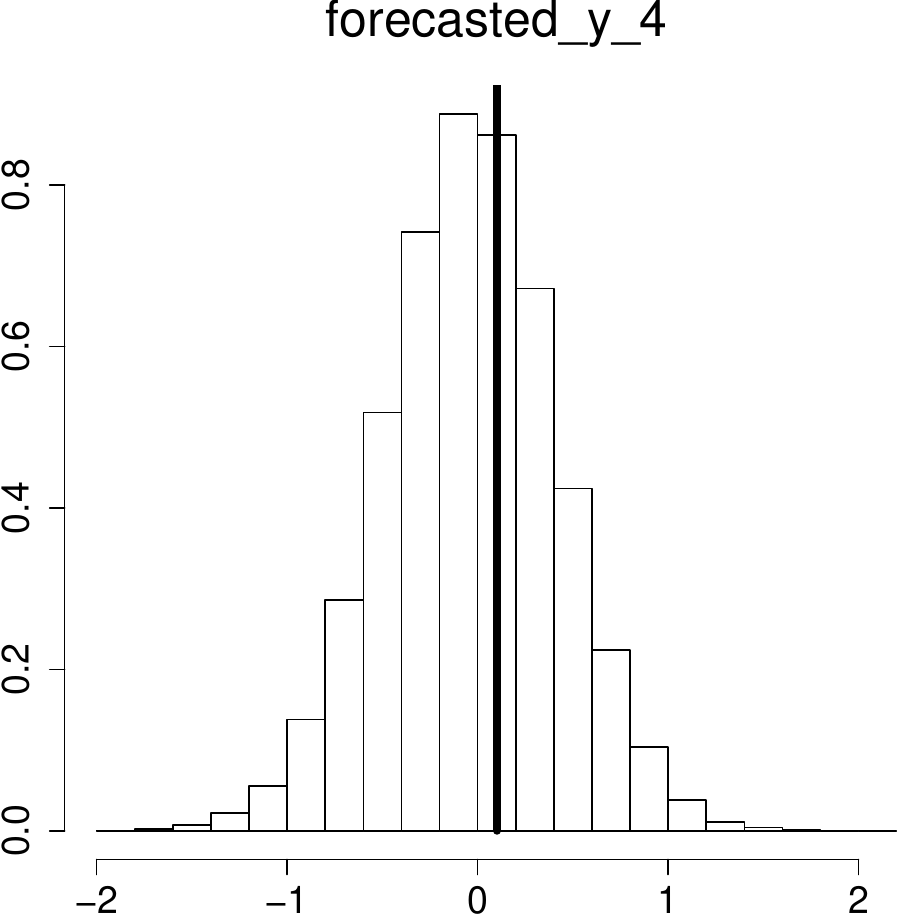}}
 \caption{{\bf Simulation study with data generated from the 4-variate modification of the model of CPS}: 
 Posterior distributions of $\bx_0$, $\bx_{T+1}$ (one-step forecasted $\bx$), and $\by_{T+1}$ (one-step forecasted $\by$). 
 The solid line stands for the true values of the respective parameters.}
 \label{fig:simdata_cps_mult_hist}
 \end{figure}


\begin{figure}
\centering
  \subfigure{ \label{fig:x_time_series_mult1}
   \includegraphics[width=7cm,height=6cm]{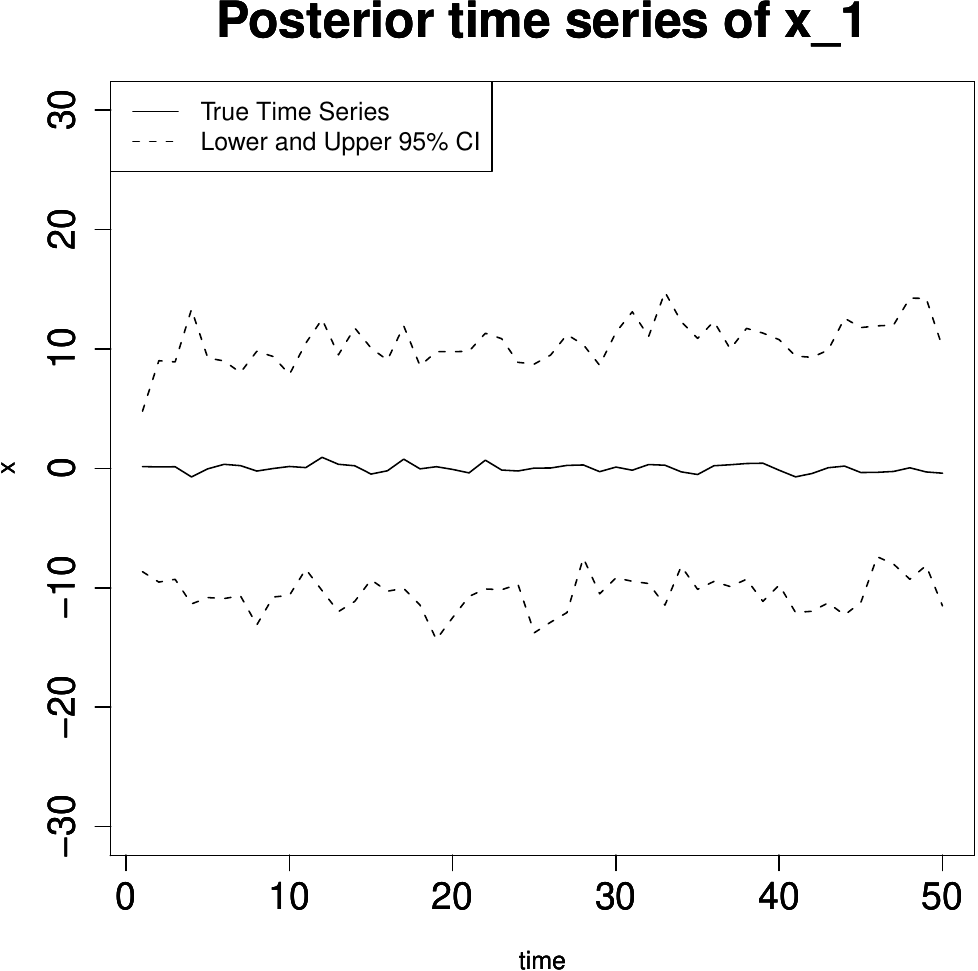}}
   \hspace{2mm}
  \subfigure{ \label{fig:x_time_series_mult2}
   \includegraphics[width=7cm,height=6cm]{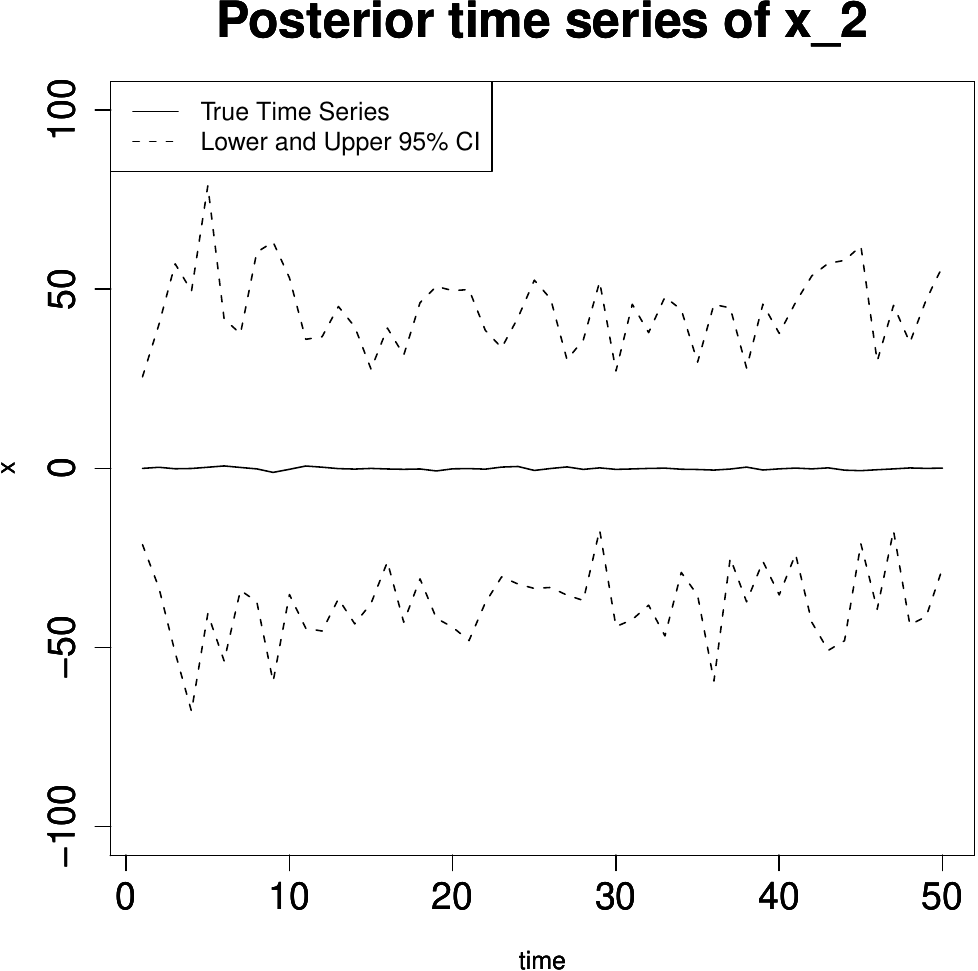}}\\
   \vspace{2mm}
  \subfigure{ \label{fig:x_time_series_mult3}
   \includegraphics[width=7cm,height=6cm]{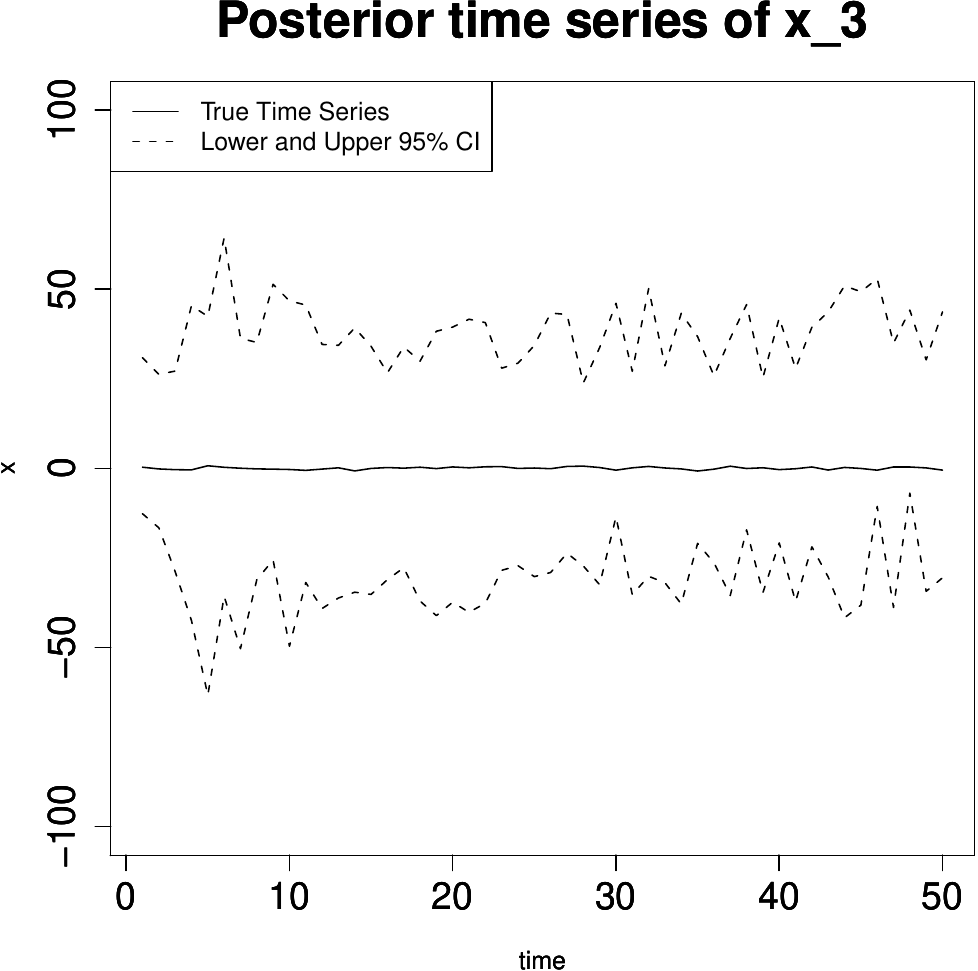}}
   \hspace{2mm}
  \subfigure{ \label{fig:x_time_series_mult4}
   \includegraphics[width=7cm,height=6cm]{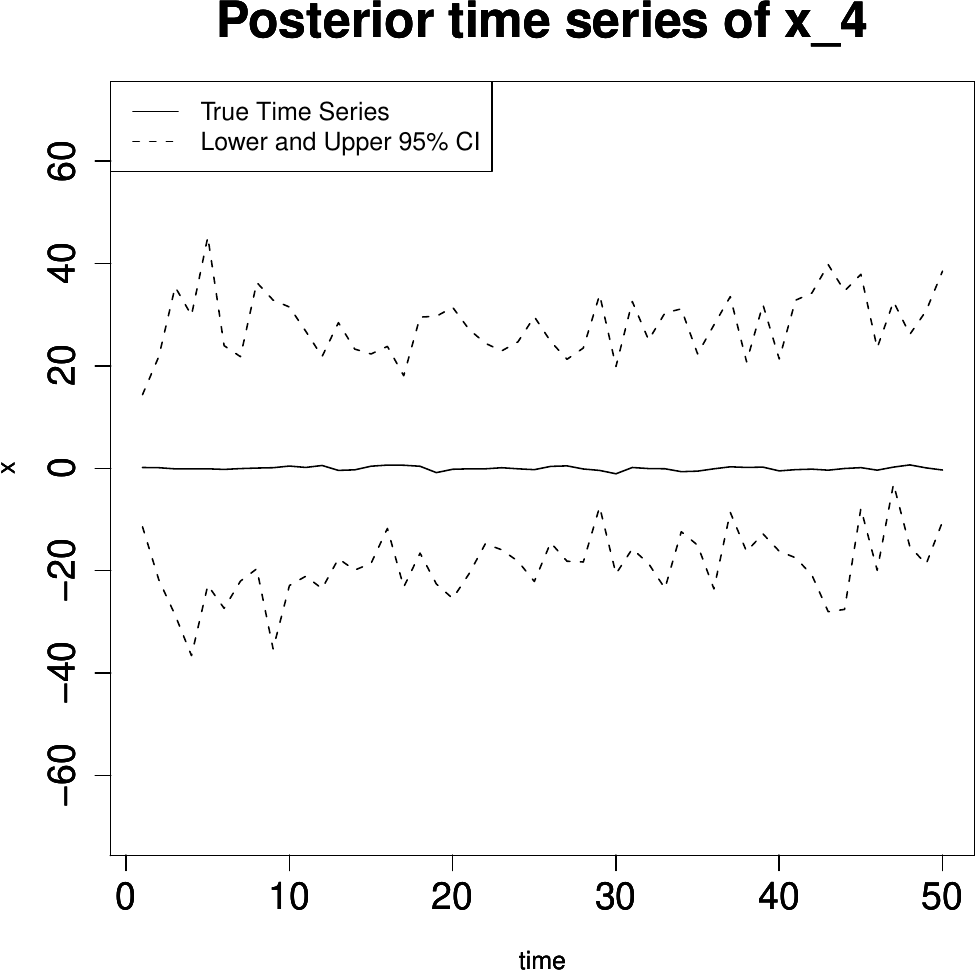}}
 \caption{{\bf Simulation study with data generated from the 4-variate modification of the model of CPS}: 
 95\% highest posterior density credible intervals of the time series $x_{1,j},\ldots,x_{T,j}$; $j=1,2,3,4$. 
 The solid line stands for the true time series.}
 \label{fig:x_time_series_mult}
 \end{figure}

Figures \ref{fig:simdata_cps_mult_hist} and \ref{fig:x_time_series_mult} show that 
our Gaussian process-based nonparametric model performs well in spite of the multidimensional
situation---the true values are well-captured 
by the posterior distributions, and the 
true forecast values are also well-supported by the
corresponding forecast distributions of $\bx_{51}$ and $\by_{51}$. Moreover, each component of the true
4-variate time series fall entirely within the 95\% highest posterior densities of the 
corresponding component of the 4-variate posterior time series. 

\section{Posteriors of the function compositions and comparisons with the true function compositions}

It may be of interest to check how the posterior distributions of the function compositions
at different time points compare with the true function compositions generating the data.
To clarify, for any $x$, let $g^*_0(x)=x$, and let $g^*_t(x)=g(t,g^*_{t-1}(x))$ for $t\geq 1$
be the $t$-step composition of $g(\cdot,\cdot)$. It may be of interest to compare the posterior
of $g^*_t(\cdot)$ with the true $t$-step functional composition of the data-generating evolutionary equation.
Similarly, for $t\geq 1$ it may be of interest to compare the true $t$-step composition of the 
observational equation with the posterior of $f^*_t(\cdot)=f(t,g^*_t(\cdot))$.  

Although technically the comparisons are possible, there is a subtle issue that needs to be understood. 
It is important to note that the observed data
has been generated by a single latent time series, guided by the evolutionary equation, beginning at a single
value $x_0$. Since we consider $x_0$ unknown, we attempt to learn about this unknown through the posterior
distribution of $x_0$. Given the observed data $\bD_T$, the support of the posterior distribution contains
the likely values of $x_0$ that might have given rise to the observed data via the compositions of the
observational and the evolutionary equations. But beyond the support of the posterior of $x_0$ there is no
imformation about the composite functions. Hence, it is reasonable to
compare the true function compositions with $g^*_t(x)$ and $f^*_t(x)$ when $x$ belongs to the support of
the posterior distribution of $x_0$, but not if $x$ is not in the support of the posterior of $x_0$.
Since the support of the posterior of $x_0$ is not likely to be large if the data and the model contain sufficient
information, the composite functions are likely to be close to linear on that relatively small region.
Below we provide details of computing the posteriors of $g^*_t(\cdot)$ and $f^*_t(\cdot)$.

We note that it follows on similar lines as in (16) and (17) of GMRB, that for $t\geq 1$, 
$[g^*_t(x)\mid \bD^*_n,g^*_{t-1}(x),\btheta_g]$
is normal with mean
\[\mu_{g,t}(x)=\bh(t,g^*_{t-1}(x))^\prime \bbeta_g+\bs_{g,\bD^*_n}(t,g^*_{t-1}(x))^\prime \bA_{g,D^*_n}^{-1}(\bD^*_n-\bH_{D^*_n}\bbeta_g)\]
and variance 
\[
\sigma^2_{g,t}(x)=\sigma^2_g\left\{1-\bs_{g,\bD^*_n}(t,g^*_{t-1}(x))^\prime \bA_{g,\bD^*_n}^{-1}\bs_{g,\bD^*_n}(t,g^*_{t-1}(x))\right\}.
\]
Noting that the posterior of $g^*_t(x)$ can be expressed (using conditional independence) as
\[[g^*_t(x)\mid\bD_T]=\int [g^*_t(x)\mid \bD^*_n,g^*_{t-1}(x),\btheta_g][\bD^*_n,g^*_{t-1}(x),\btheta_g\mid\bD_T]
d\bD^*_nd g^*_{t-1}(x)d\btheta_g,\]
given the available posterior samples of $\bD^*_n,g^*_{t-1}(x),\btheta_g$, draws from $[g^*_t(x)\mid \bD^*_n,g^*_{t-1}(x),\btheta_g]$ 
yields posterior samples from $[g^*_t(x)\mid\bD_T]$. Thus, for any $x$, we can compute the posterior distribution of
$g^*_t(x);t=1,2,\ldots$.

Similarly, it follows as in (27) and (28) of GMRB, that for $t\geq 1$, the full conditional of 
$f^*_t(x)=f(t,g^*_t(x))$, given $g^*_t(x)$ is normal, with mean
\[
\mu_{f,t}(x)=\bh(t,g^*_t(x))^\prime \bbeta_f+\bs_{f,\bD_T}(t,g^*_t(x))^\prime \bA_{f,D_T}^{-1}(\bD_T-\bH_{D_T}\bbeta_f)
\]
and variance 
\[
\sigma^2_{f,t}(x)=\sigma^2_f\left\{1-\bs_{f,\bD_T}(t,g^*_t(x))^\prime \bA_{f,\bD_T}^{-1}\bs_{f,\bD_T}(t,g^*_t(x))\right\}.
\]
Following the method discussed for generating posterior samples of $g^*_t(x)$, we can easily generate posterior samples
of $f^*_t(x)$.

The procedure in the multivariate situation is analogous; we only modify the notations $f^*_t(x)$ and $g^*_t(x)$ to
$f^*_{jt}(x)$ and $g^*_{kt}(x)$ to denote the corresponding observational and evolutionary $t$-step 
composite functions for the $j$-th and the $k$-th components of the functions espectively, where $j=1,2,\ldots,p$
and $k=1,\ldots,q$. 

\begin{figure}
\centering
  \subfigure{ \label{fig:f_1_linear}
   \includegraphics[width=5.1cm,height=5.1cm]{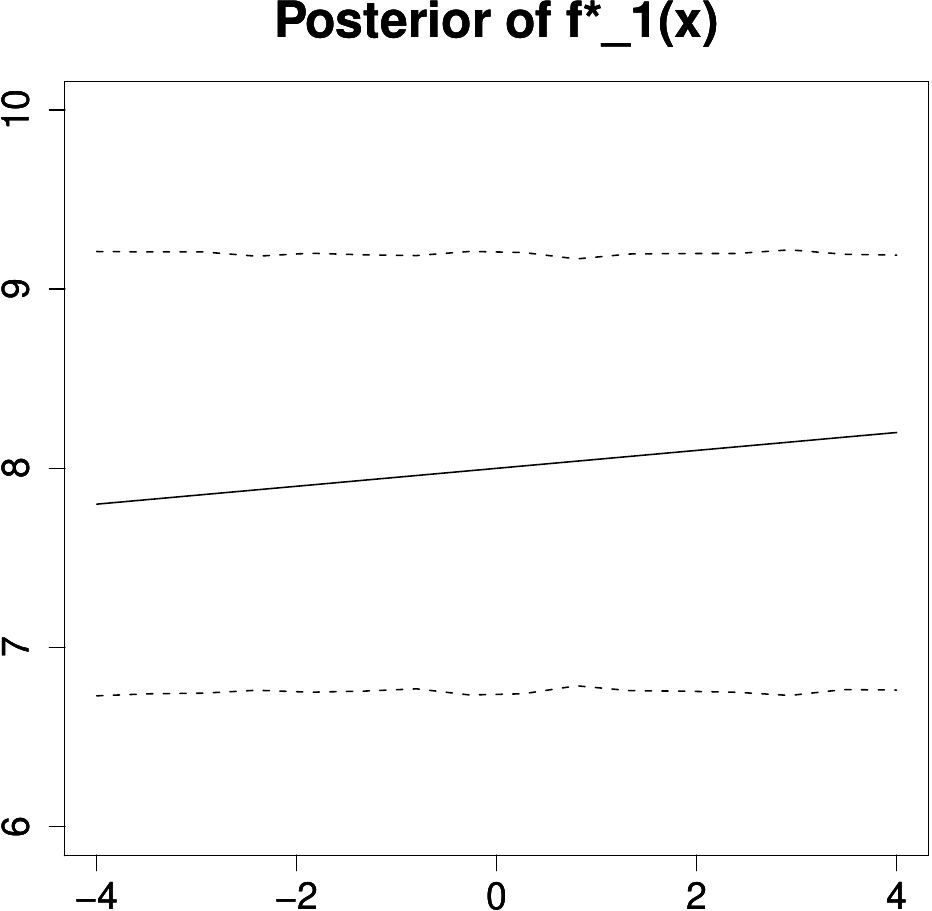}}
   \hspace{2mm}
  \subfigure{ \label{fig:f_2_linear} 
  \includegraphics[width=5.1cm,height=5.1cm]{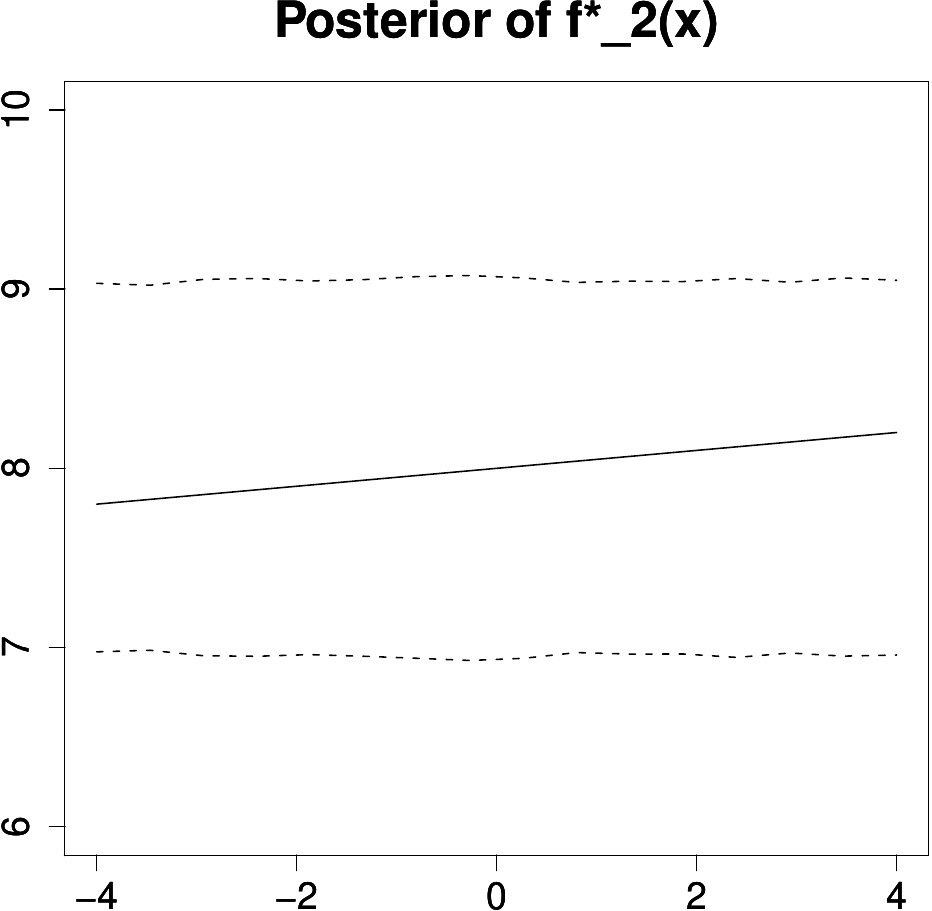}}
   \hspace{2mm}
  \subfigure{ \label{fig:f_3_linear} 
  \includegraphics[width=5.1cm,height=5.1cm]{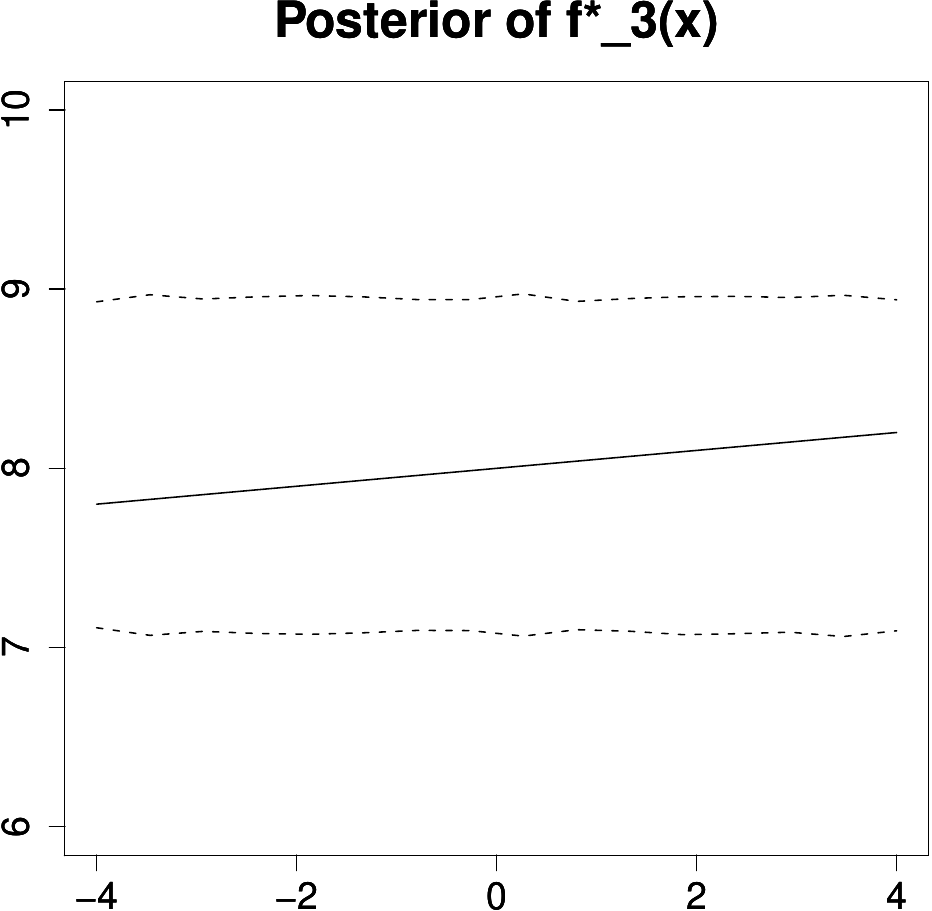}}\\
  \vspace{2mm}
  \subfigure{ \label{fig:g_1_linear}
   \includegraphics[width=5.1cm,height=5.1cm]{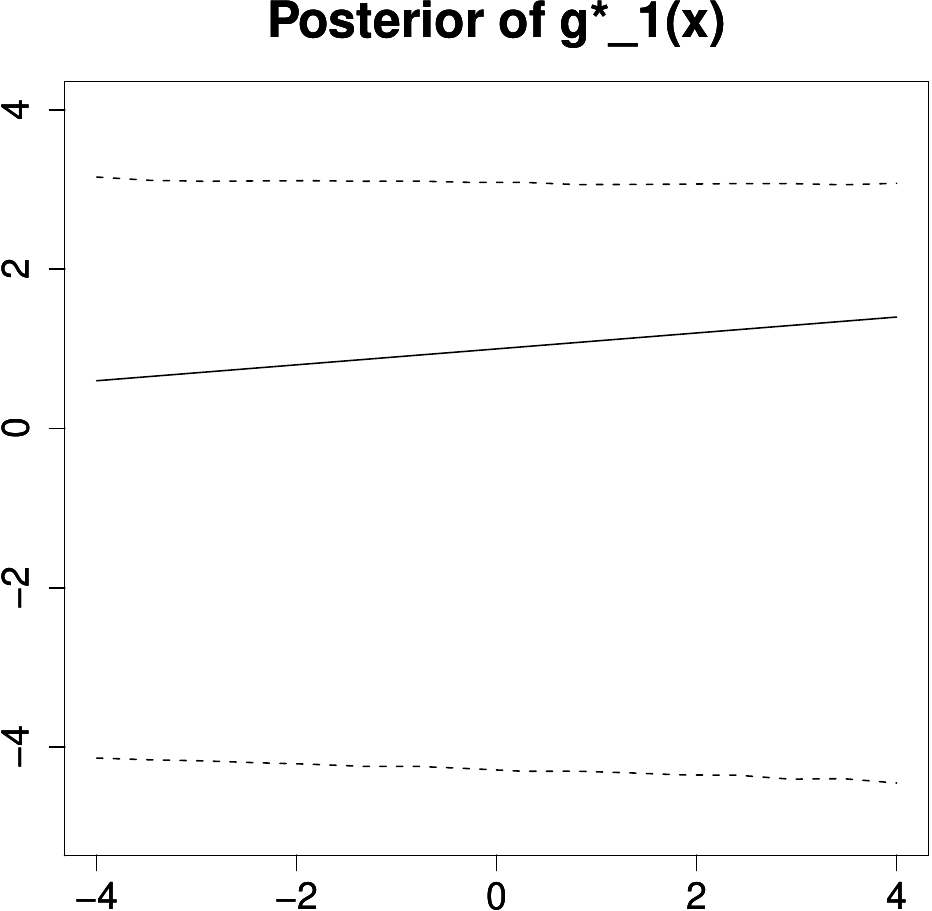}}
   \hspace{2mm}
  \subfigure{ \label{fig:g_2_linear} 
  \includegraphics[width=5.1cm,height=5.1cm]{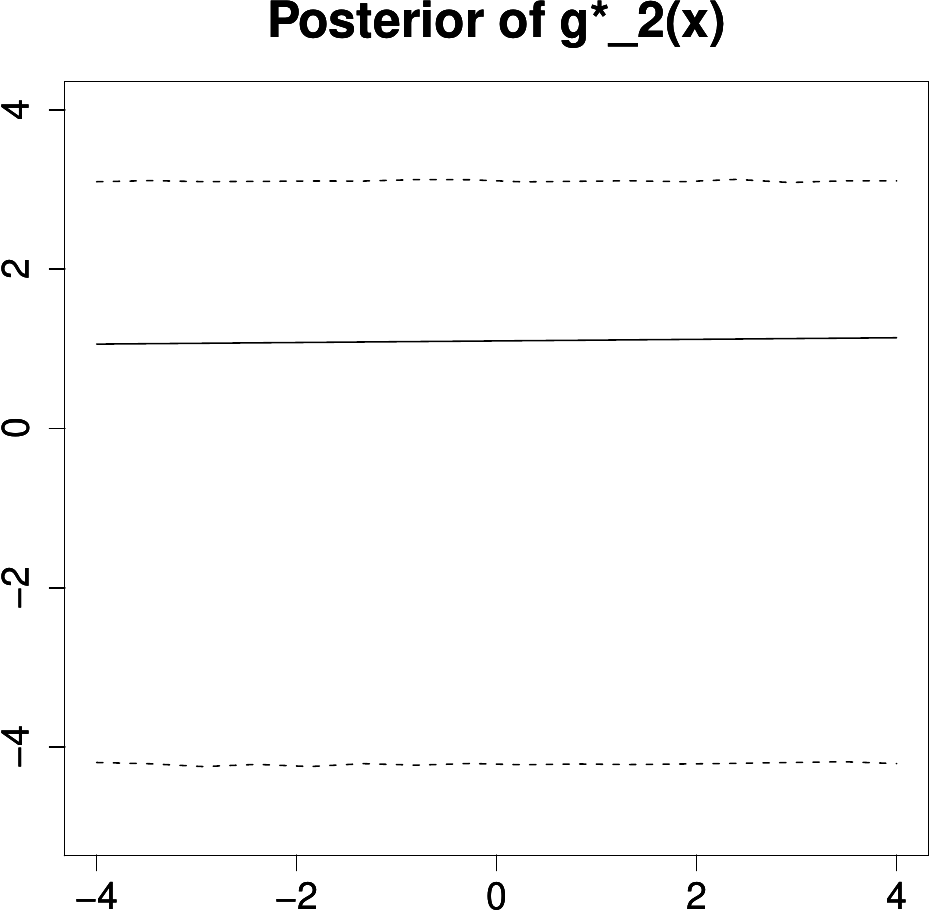}}
   \hspace{2mm}
  \subfigure{ \label{fig:g_3_linear} 
  \includegraphics[width=5.1cm,height=5.1cm]{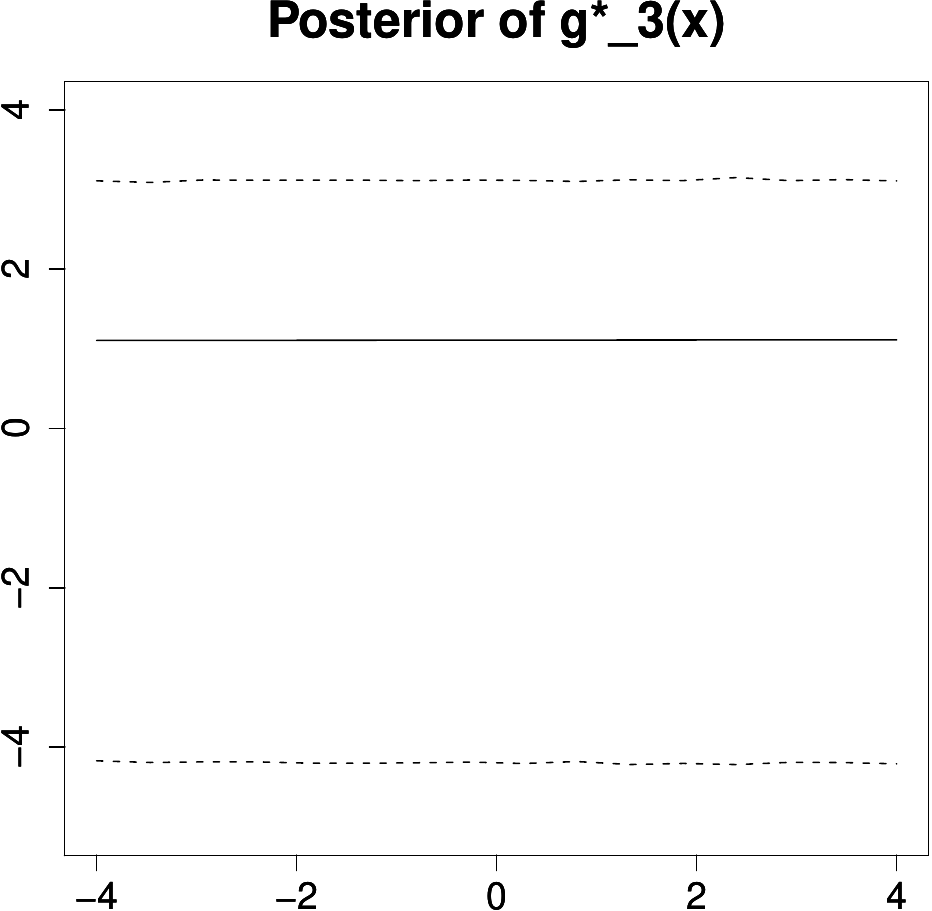}}

 \caption{{\bf Simulation study with data generated from the univariate linear model}: 
 The broken lines denote the 95\% credible regions of the posterior distributions of $f^*_t(x)$ and $g^*_t(x)$ for $t=1,2,3$;
 $x\in [-4,4]$, which is the support of the posterior of $x_0$. The solid lines denote the true composite observational and evolutionary functions.}
 \label{fig:plots_function_linear}
 \end{figure}

\begin{figure}
\centering
  \subfigure{ \label{fig:f_1_cps}
   \includegraphics[width=5.1cm,height=5.1cm]{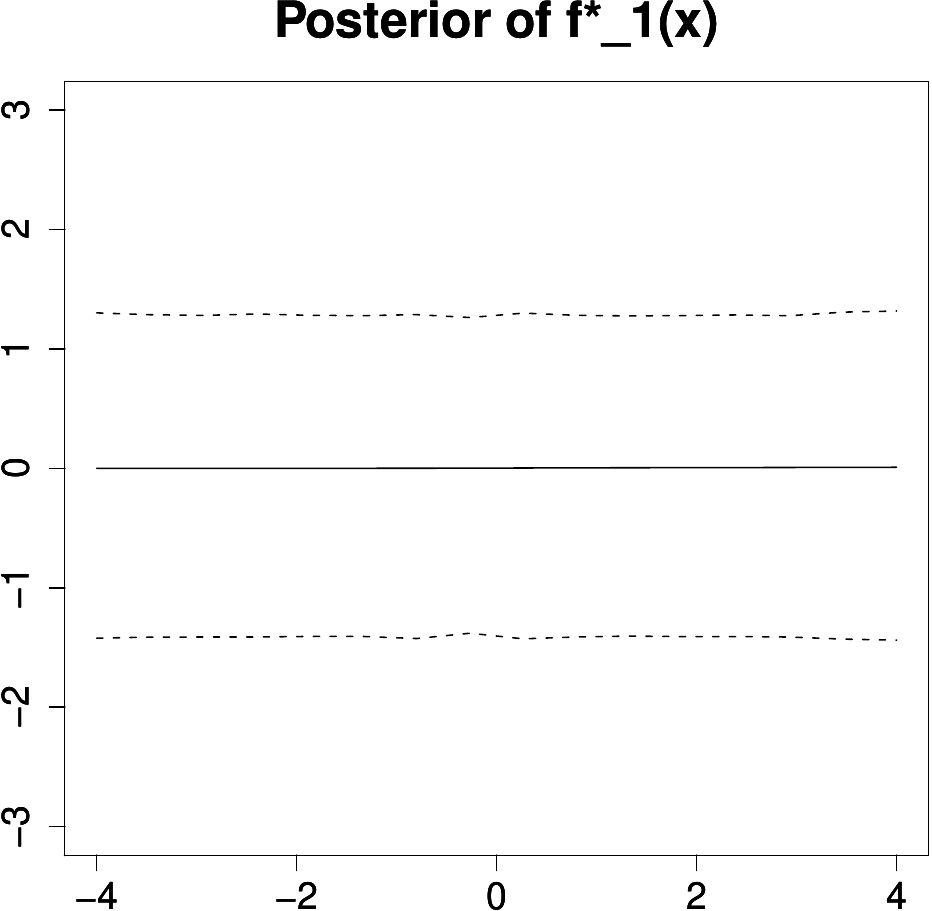}}
   \hspace{2mm}
  \subfigure{ \label{fig:f_2_cps} 
  \includegraphics[width=5.1cm,height=5.1cm]{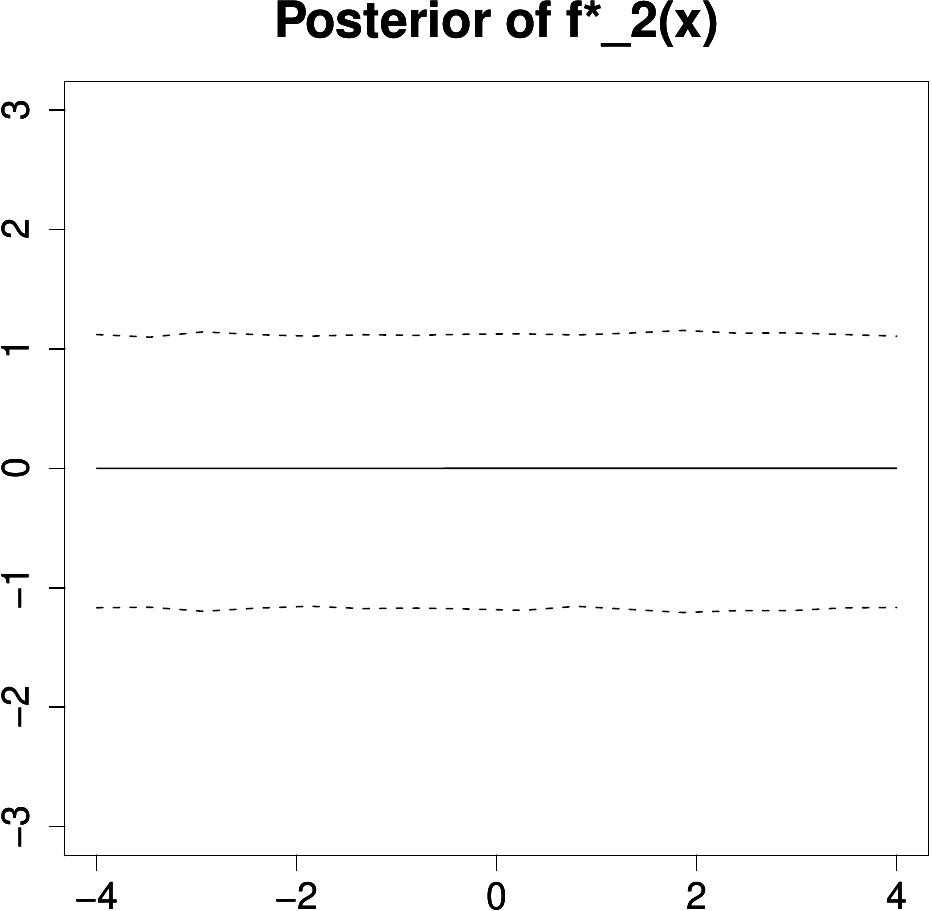}}
   \hspace{2mm}
  \subfigure{ \label{fig:f_3_cps} 
  \includegraphics[width=5.1cm,height=5.1cm]{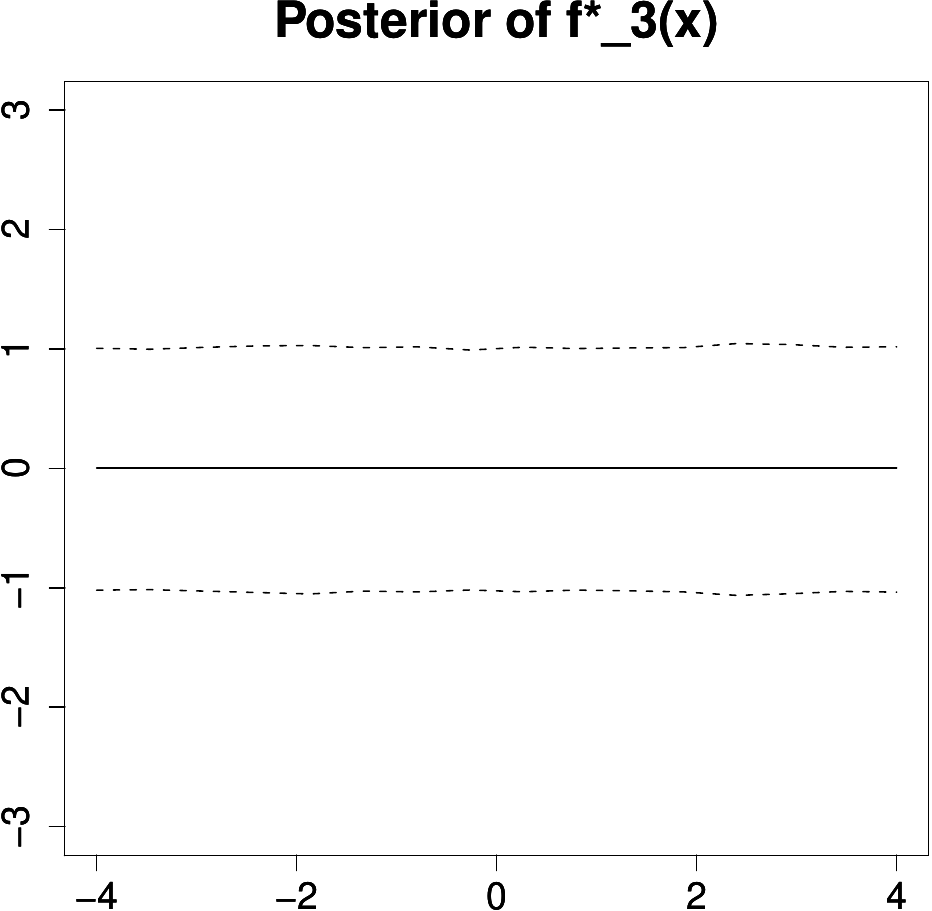}}\\
  \vspace{2mm}
  \subfigure{ \label{fig:g_1_cps}
   \includegraphics[width=5.1cm,height=5.1cm]{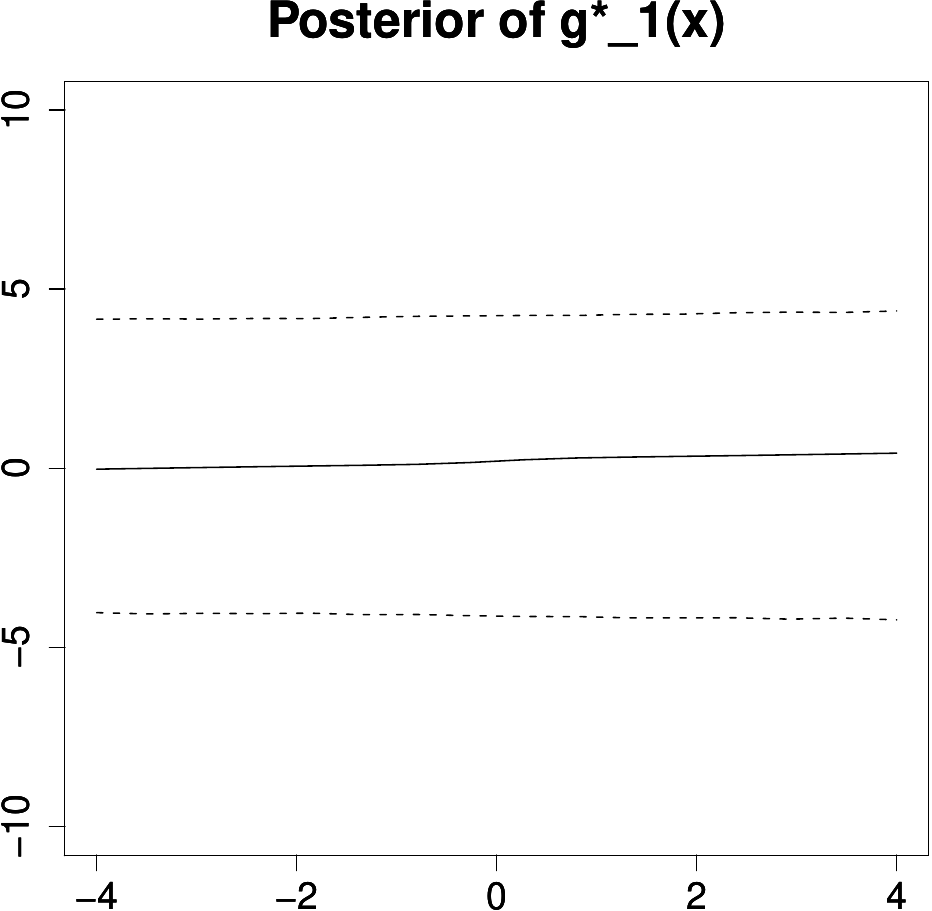}}
   \hspace{2mm}
  \subfigure{ \label{fig:g_2_cps} 
  \includegraphics[width=5.1cm,height=5.1cm]{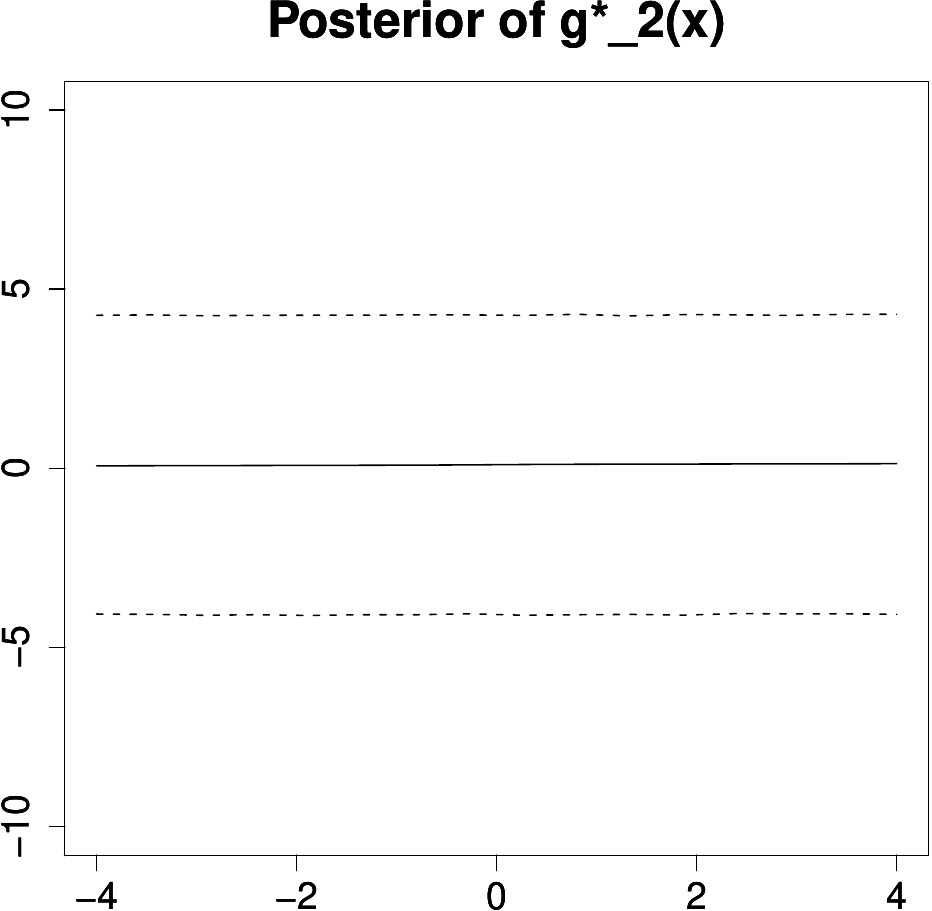}}
   \hspace{2mm}
  \subfigure{ \label{fig:g_3_cps} 
  \includegraphics[width=5.1cm,height=5.1cm]{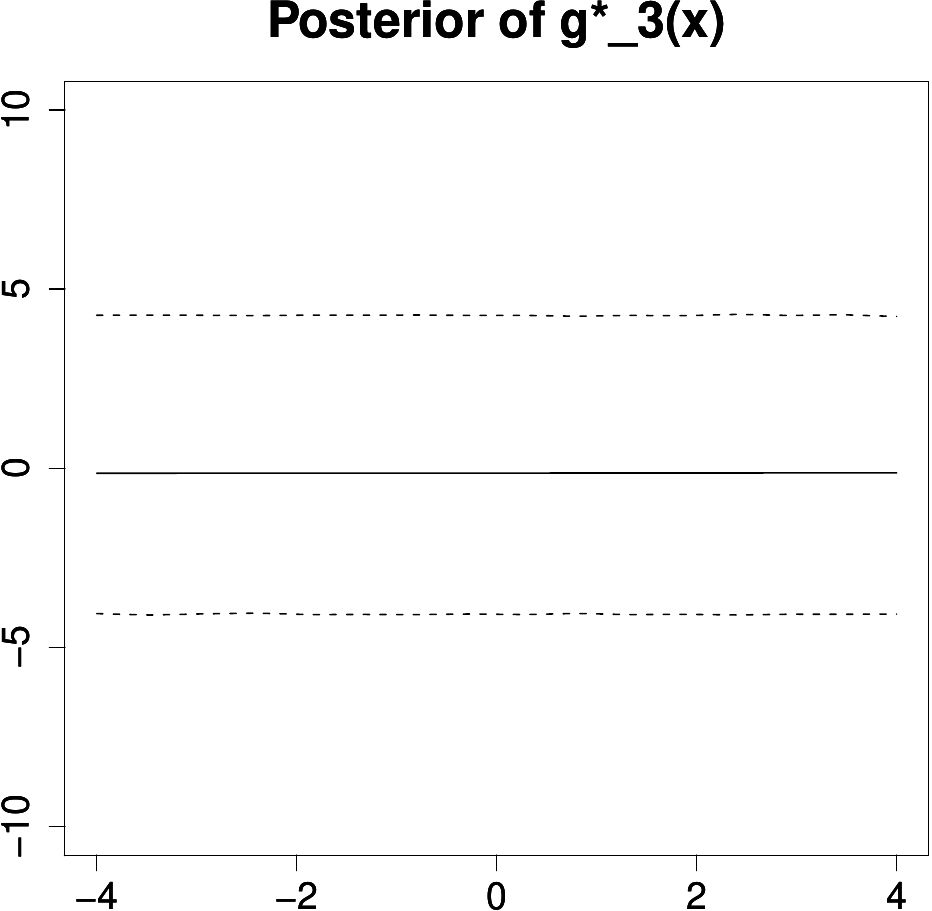}}

 \caption{{\bf Simulation study with data generated from the univariate growth model of CPS}: 
 The broken lines denote the 95\% highest posterior density credible regions of the posterior distributions 
 of $f^*_t(x)$ and $g^*_t(x)$ for $t=1,2,3$;
 $x\in [-4,4]$, which is the support of the posterior of $x_0$. The solid lines denote the true composite 
 observational and evolutionary functions.}
 \label{fig:plots_function_cps}
 \end{figure}

\begin{figure}
\centering
   \includegraphics[width=8.1cm,height=8.1cm]{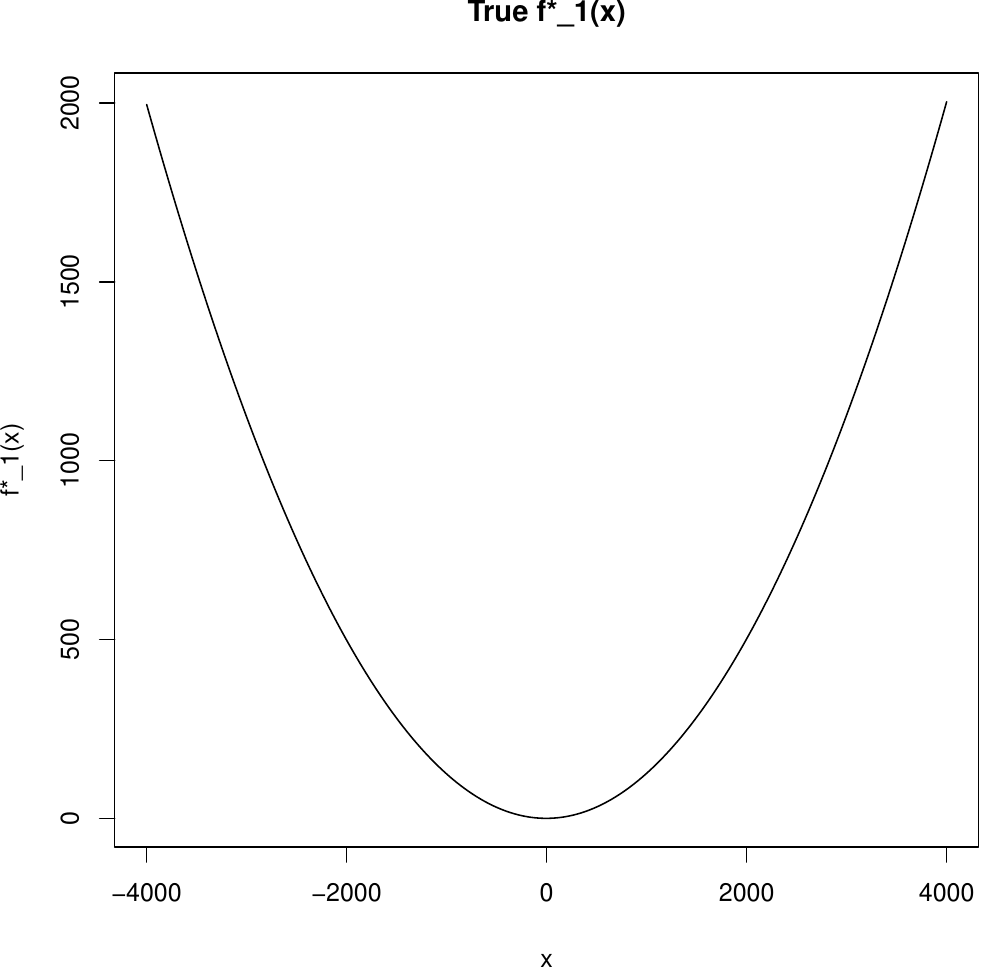}
 \caption{{\bf Simulation study with data generated from the univariate growth model of CPS}: 
Displayed is the true function $f^*_1(\cdot)$ on the much wider domain $[-4000,4000]$.}
 \label{fig:cps_true_f_1}
 \end{figure}

\begin{figure}
\centering
  \subfigure{ \label{fig:f_1_1}
   \includegraphics[width=4.1cm,height=4.1cm]{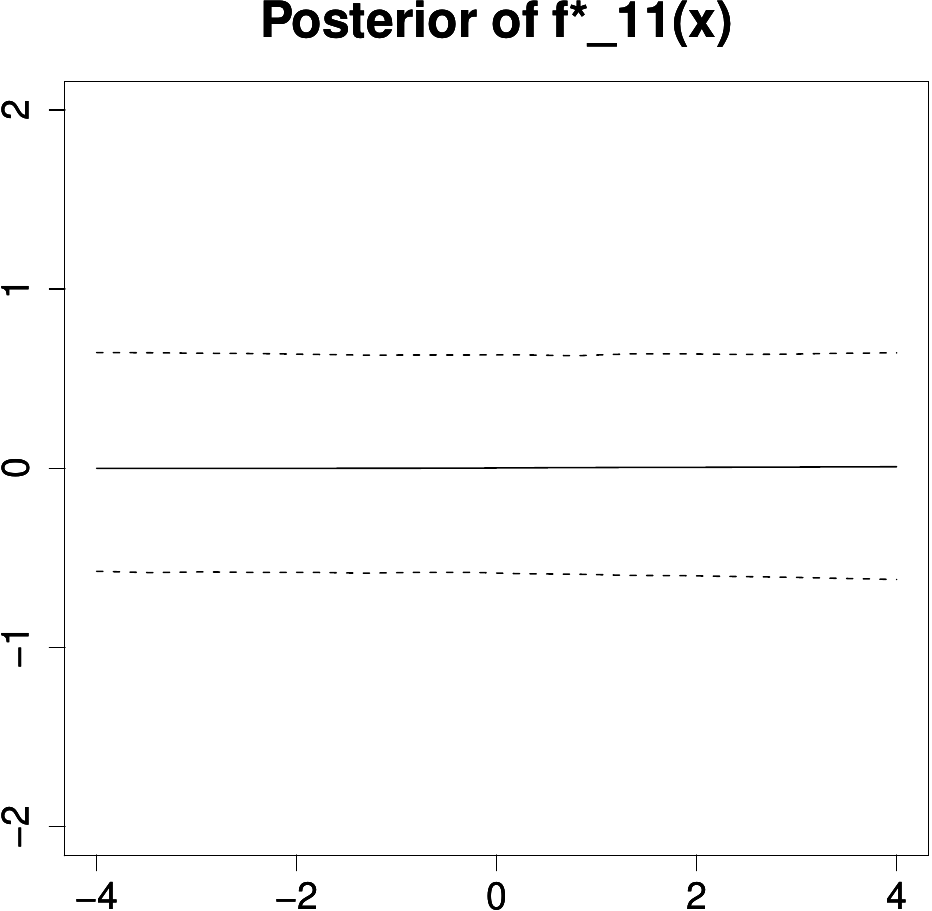}}
   \hspace{2mm}
  \subfigure{ \label{fig:f_2_1} 
  \includegraphics[width=4.1cm,height=4.1cm]{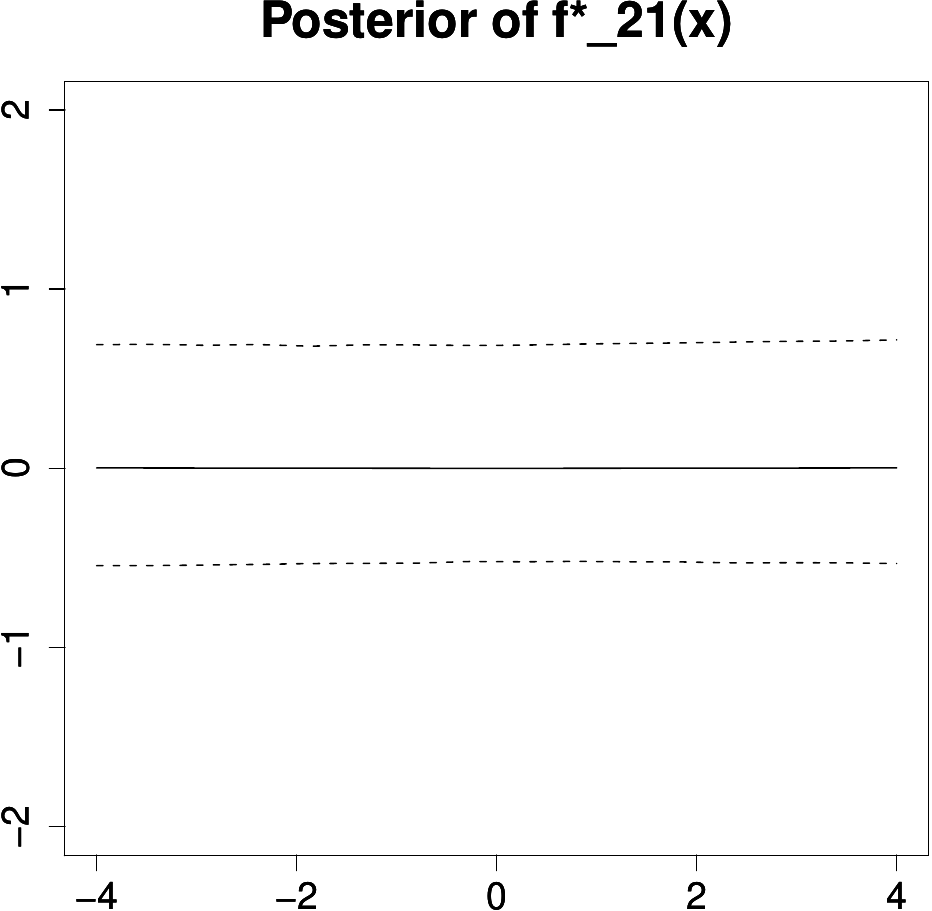}}
   \hspace{2mm}
  \subfigure{ \label{fig:f_3_1} 
  \includegraphics[width=4.1cm,height=4.1cm]{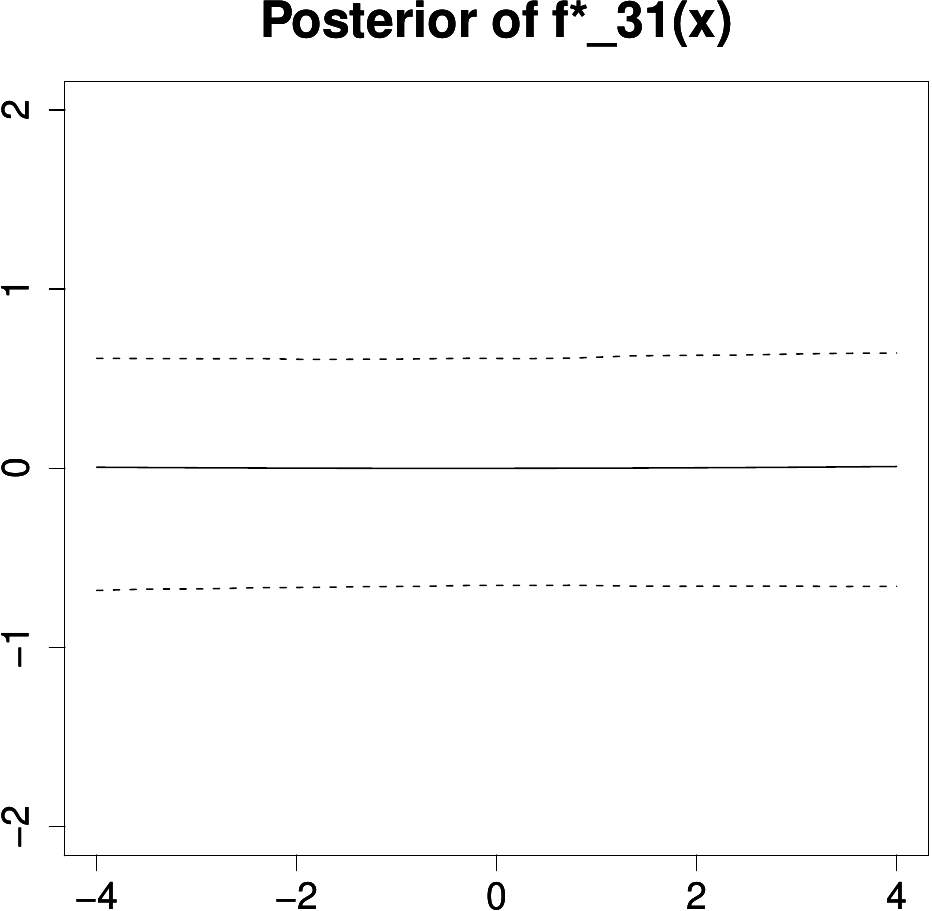}}\\
   \vspace{2mm}
  \subfigure{ \label{fig:f_4_1} 
  \includegraphics[width=4.1cm,height=4.1cm]{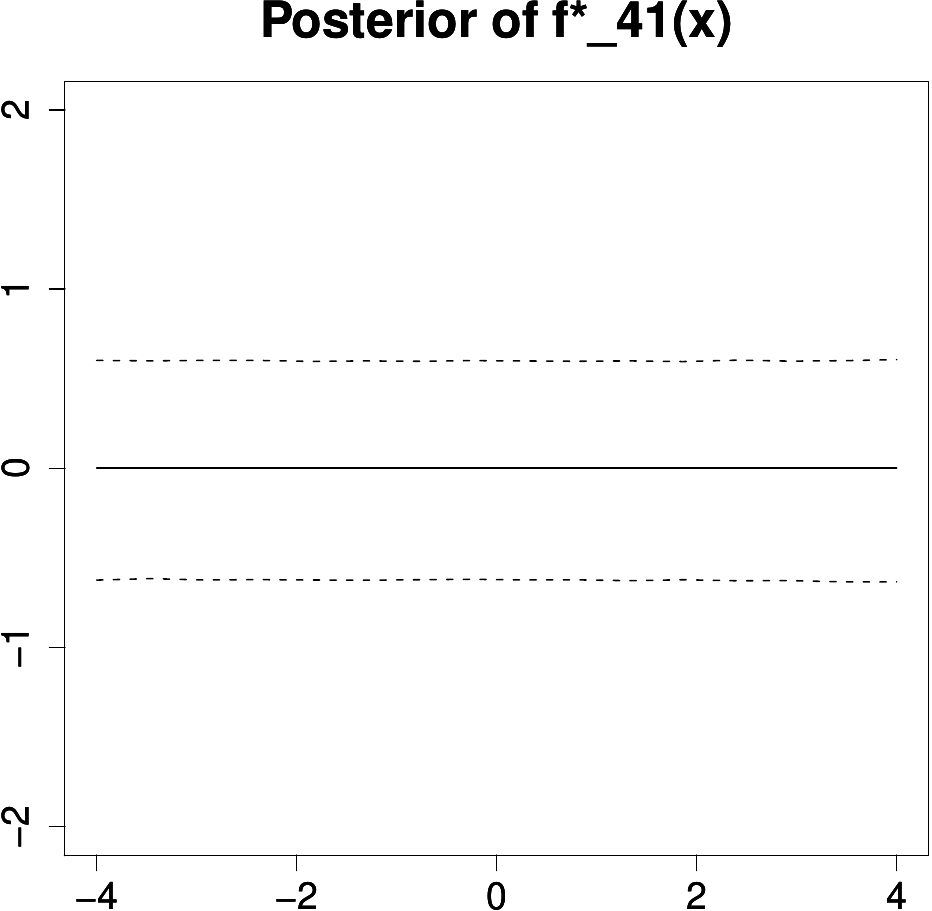}}
  \hspace{2mm}
  \subfigure{ \label{fig:f_1_2}
   \includegraphics[width=4.1cm,height=4.1cm]{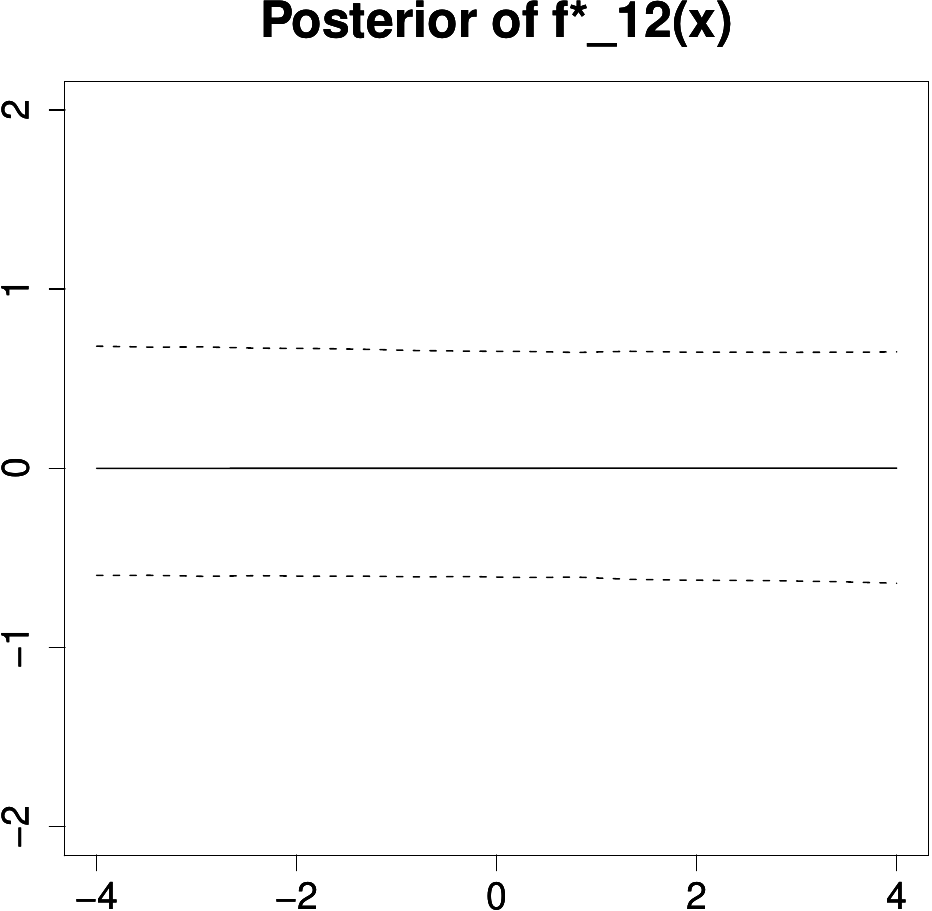}}
   \hspace{2mm}
  \subfigure{ \label{fig:f_2_2} 
  \includegraphics[width=4.1cm,height=4.1cm]{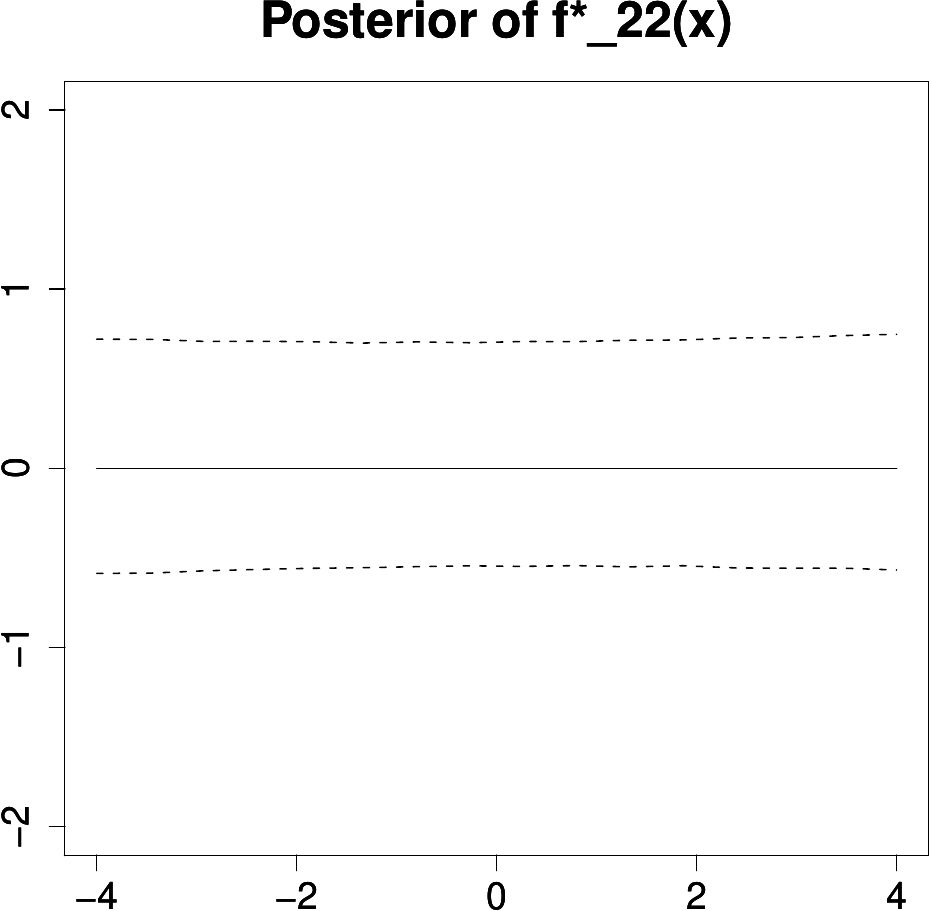}}\\
   \vspace{2mm}
  \subfigure{ \label{fig:f_3_2} 
  \includegraphics[width=4.1cm,height=4.1cm]{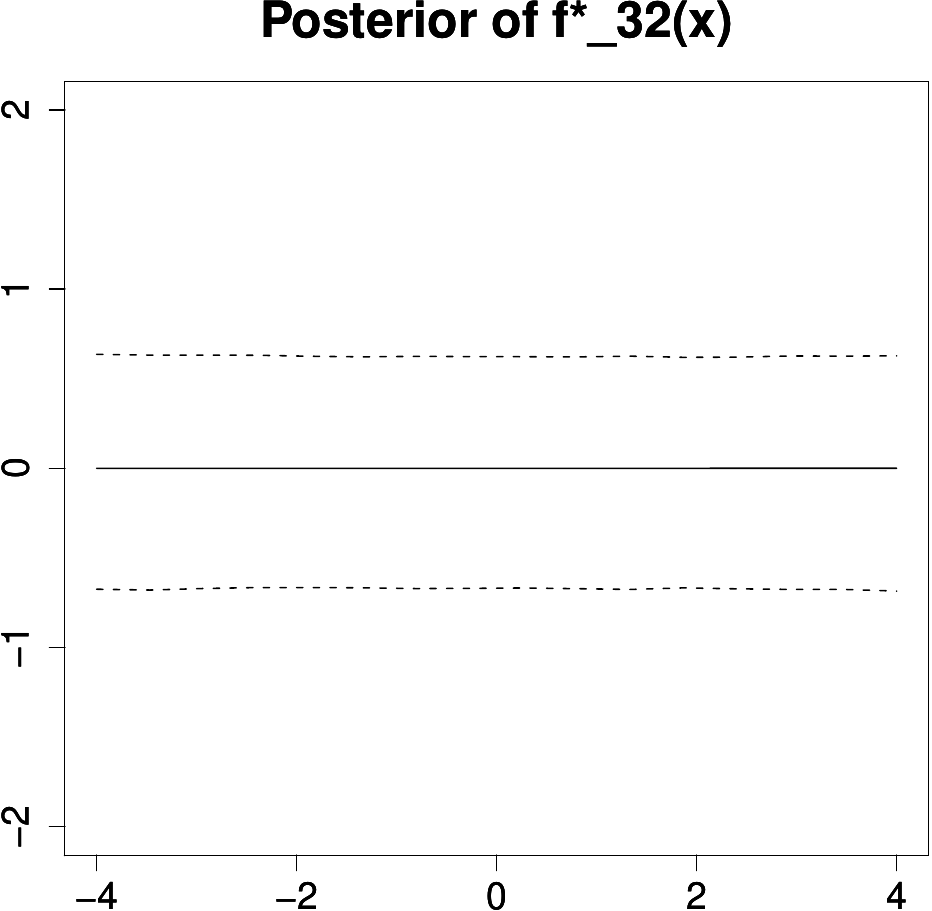}}
   \hspace{2mm}
  \subfigure{ \label{fig:f_4_2} 
  \includegraphics[width=4.1cm,height=4.1cm]{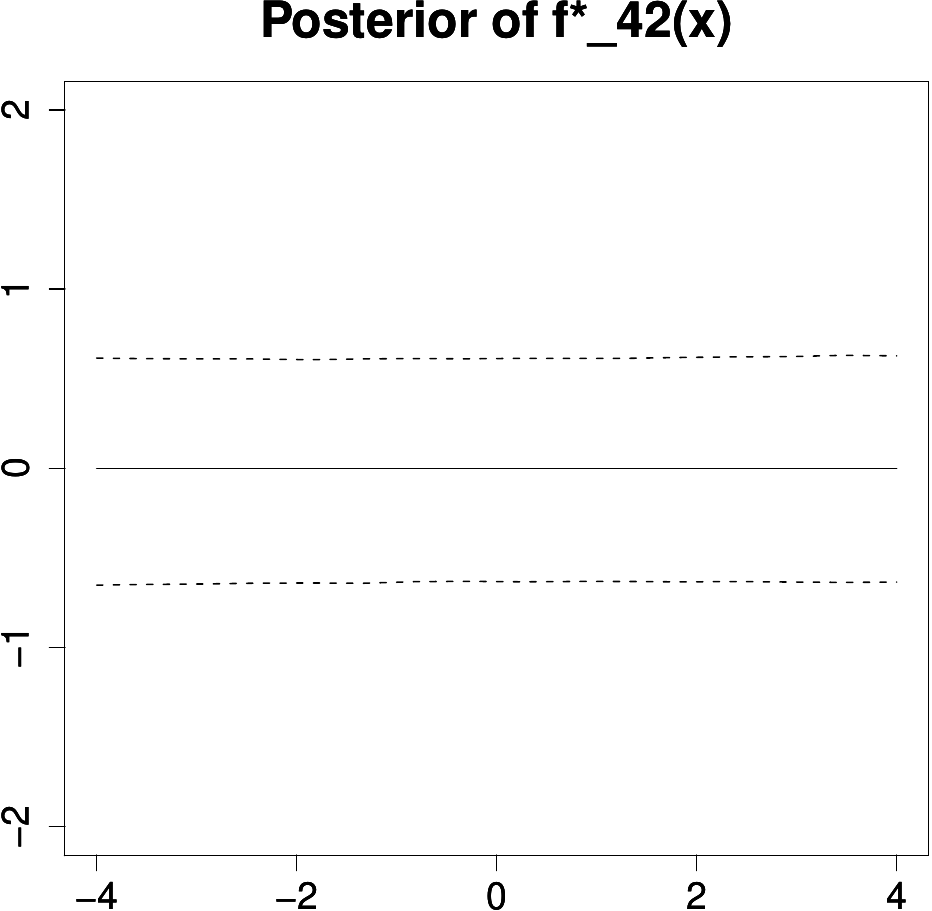}}
  \hspace{2mm}
  \subfigure{ \label{fig:f_1_3}
   \includegraphics[width=4.1cm,height=4.1cm]{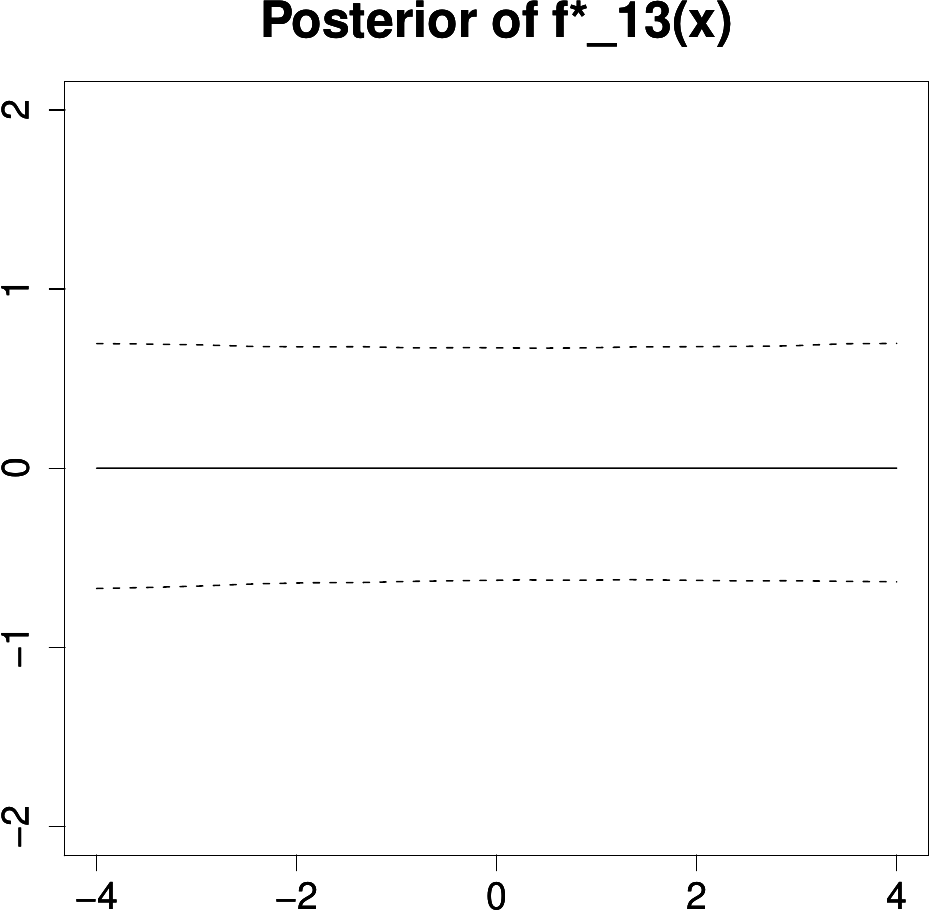}}\\
   \vspace{2mm}
  \subfigure{ \label{fig:f_2_3} 
  \includegraphics[width=4.1cm,height=4.1cm]{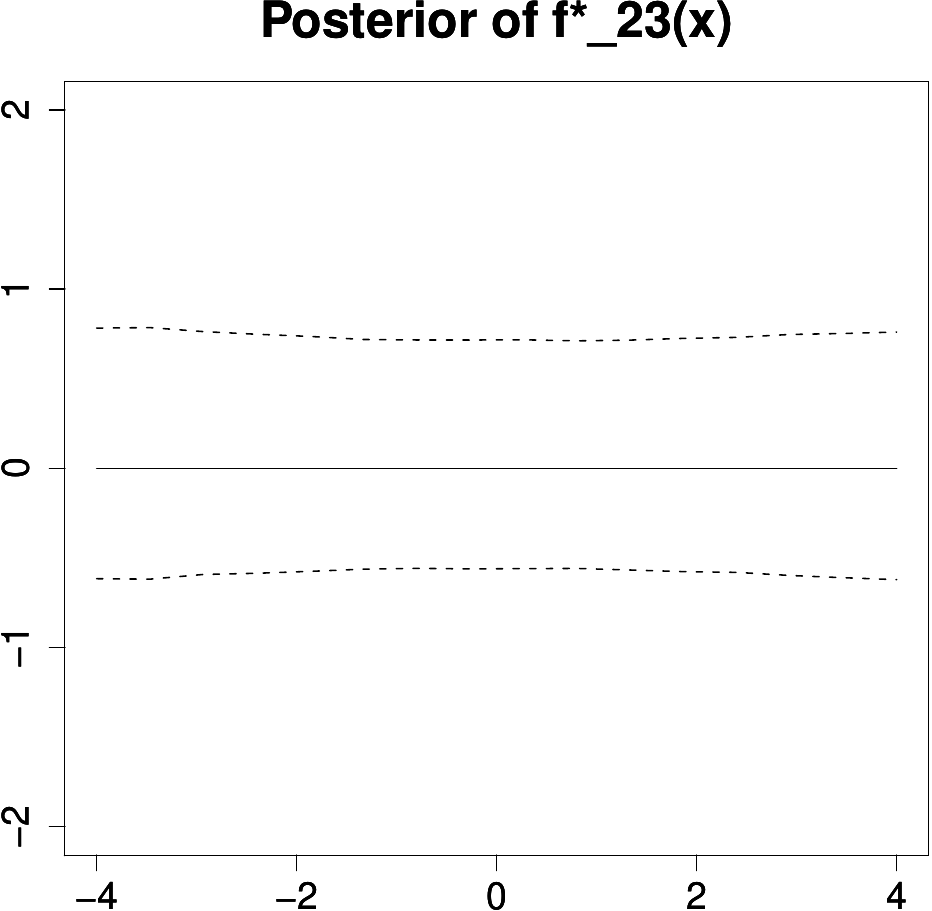}}
   \hspace{2mm}
  \subfigure{ \label{fig:f_3_3} 
  \includegraphics[width=4.1cm,height=4.1cm]{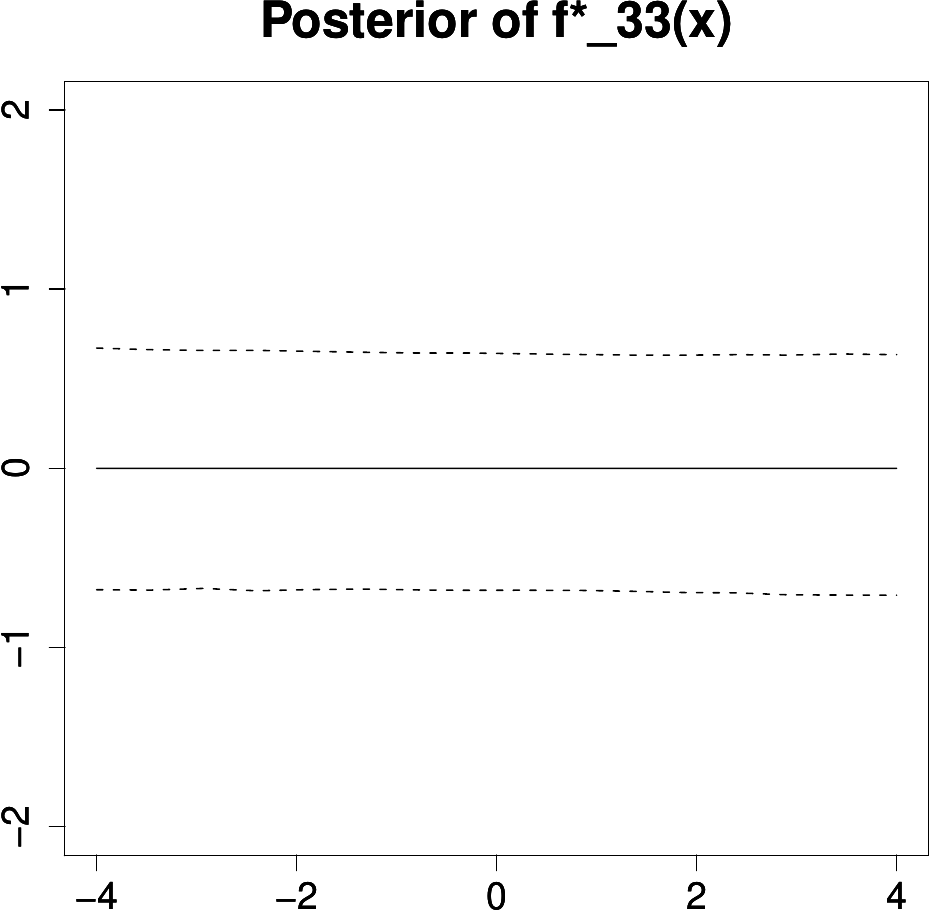}}
   \hspace{2mm}
  \subfigure{ \label{fig:f_4_3} 
  \includegraphics[width=4.1cm,height=4.1cm]{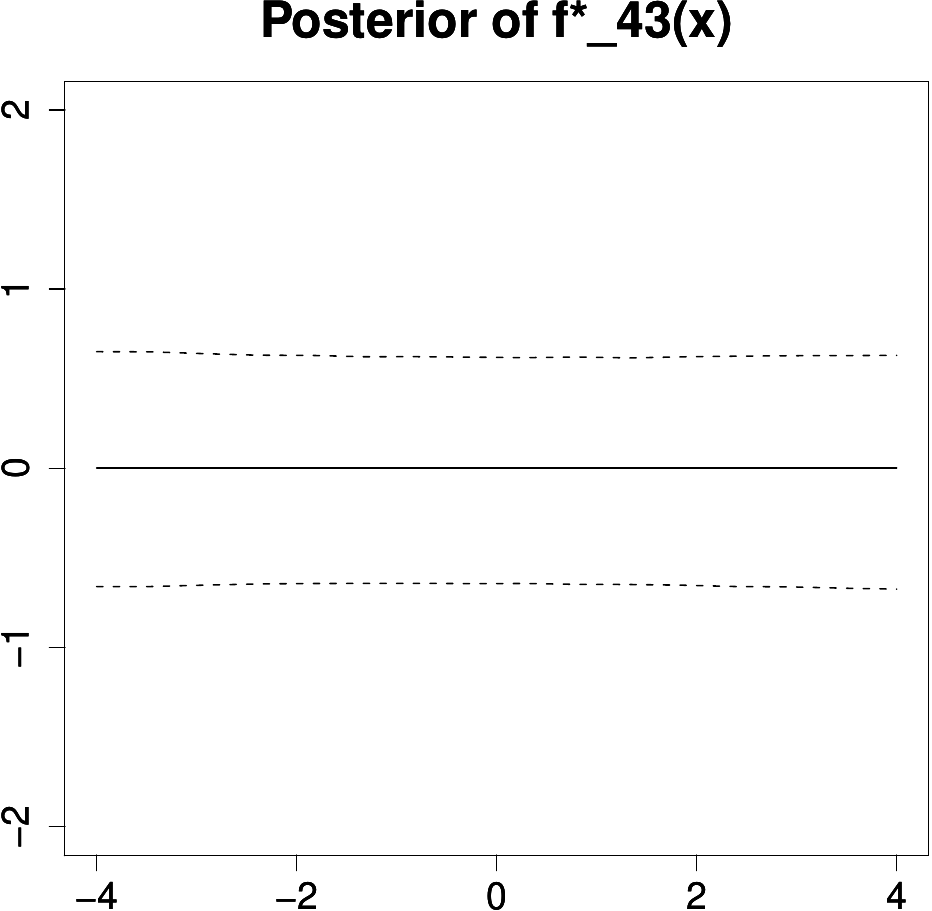}}

 \caption{{\bf Simulation study with data generated from the 4-variate modification of the model of CPS}: 
 The broken lines denote the 95\% highest posterior density credible regions of the posterior distributions 
 of $f^*_{jt}(x)$ for $j=1,2,3,4$ and $t=1,2,3$;
 $x\in [-4,4]$, which is the support of the posterior of $x_0$. The solid lines denote the true composite 
 observational and evolutionary functions.}
 \label{fig:plots_function_mult_cps_f}
 \end{figure}

\begin{figure}
\centering
  \subfigure{ \label{fig:g_1_1}
   \includegraphics[width=4.1cm,height=4.1cm]{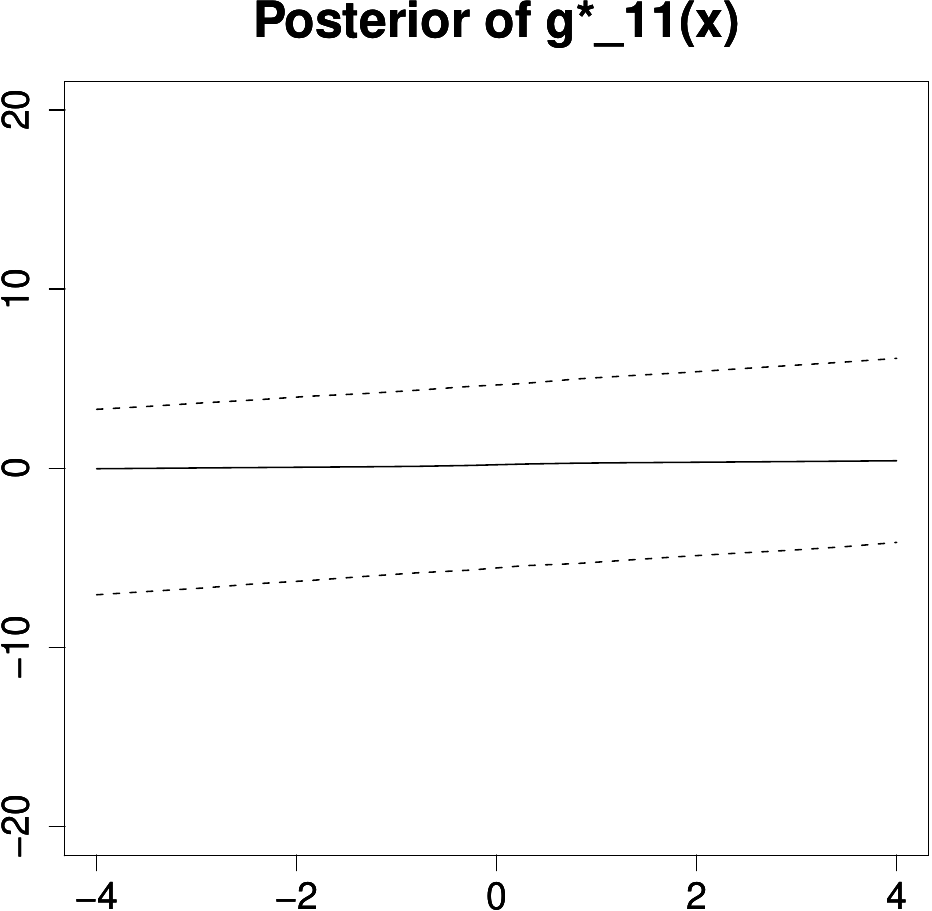}}
   \hspace{2mm}
  \subfigure{ \label{fig:g_2_1} 
  \includegraphics[width=4.1cm,height=4.1cm]{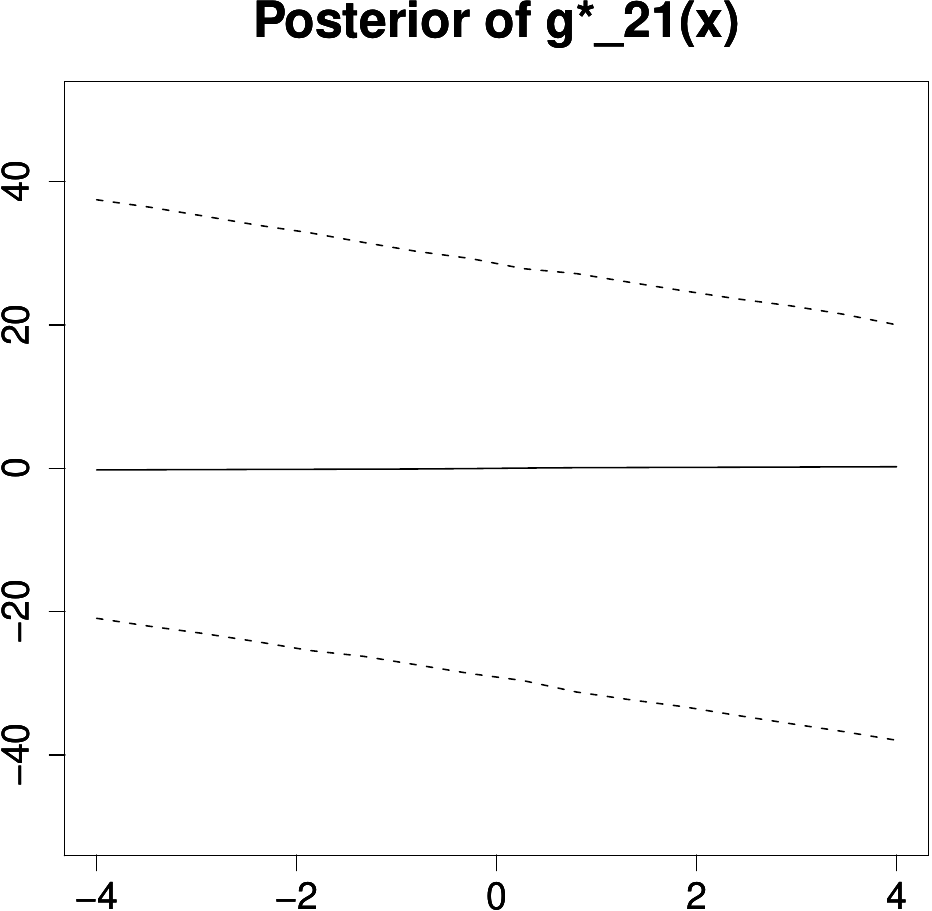}}
   \hspace{2mm}
  \subfigure{ \label{fig:g_3_1} 
  \includegraphics[width=4.1cm,height=4.1cm]{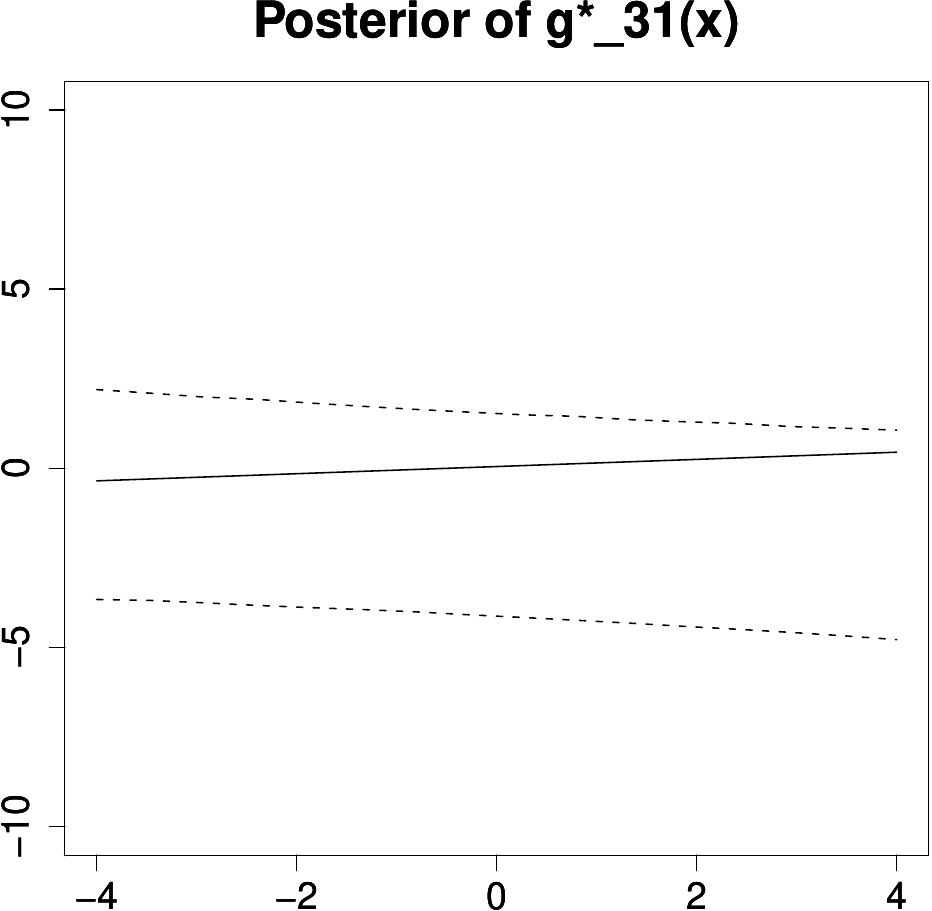}}\\
   \vspace{2mm}
  \subfigure{ \label{fig:g_4_1} 
  \includegraphics[width=4.1cm,height=4.1cm]{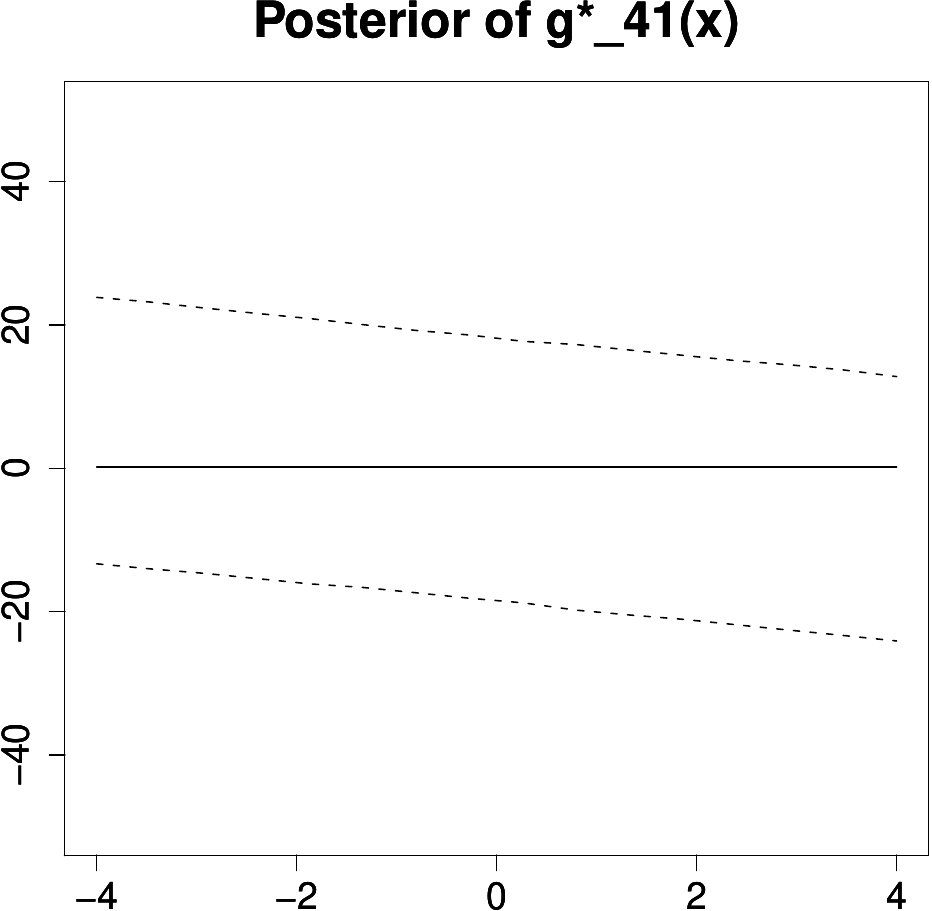}}
  \hspace{2mm}
  \subfigure{ \label{fig:g_1_2}
   \includegraphics[width=4.1cm,height=4.1cm]{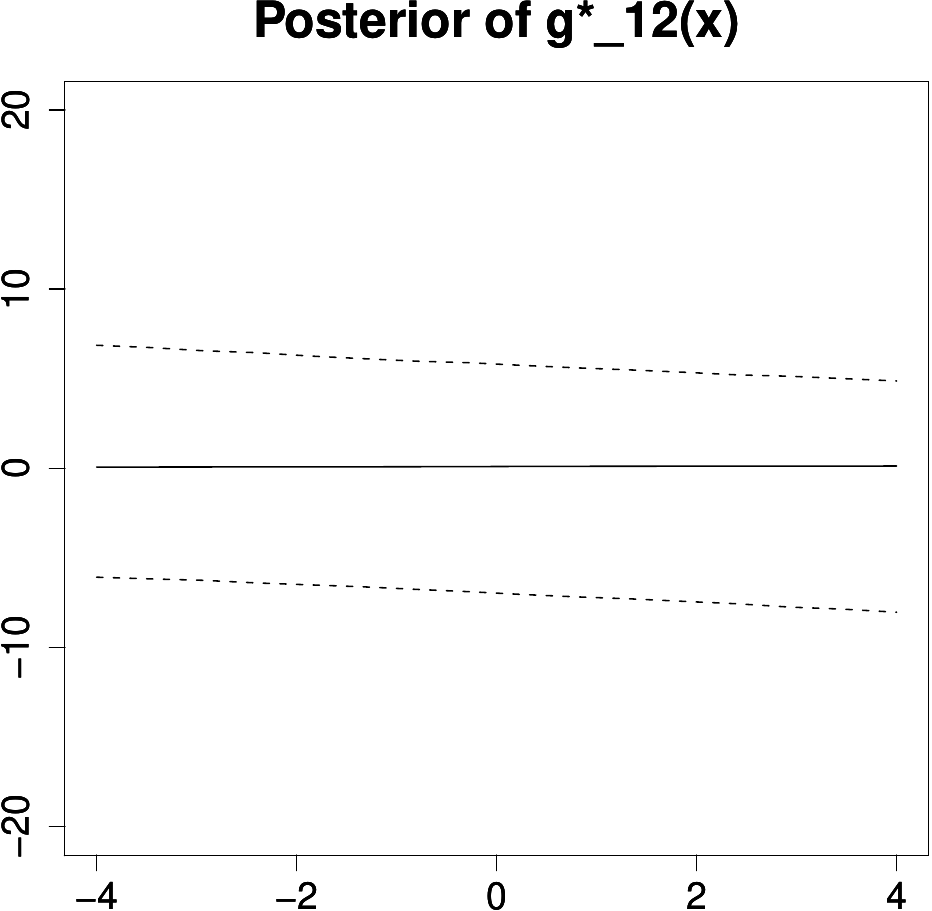}}
   \hspace{2mm}
  \subfigure{ \label{fig:g_2_2} 
  \includegraphics[width=4.1cm,height=4.1cm]{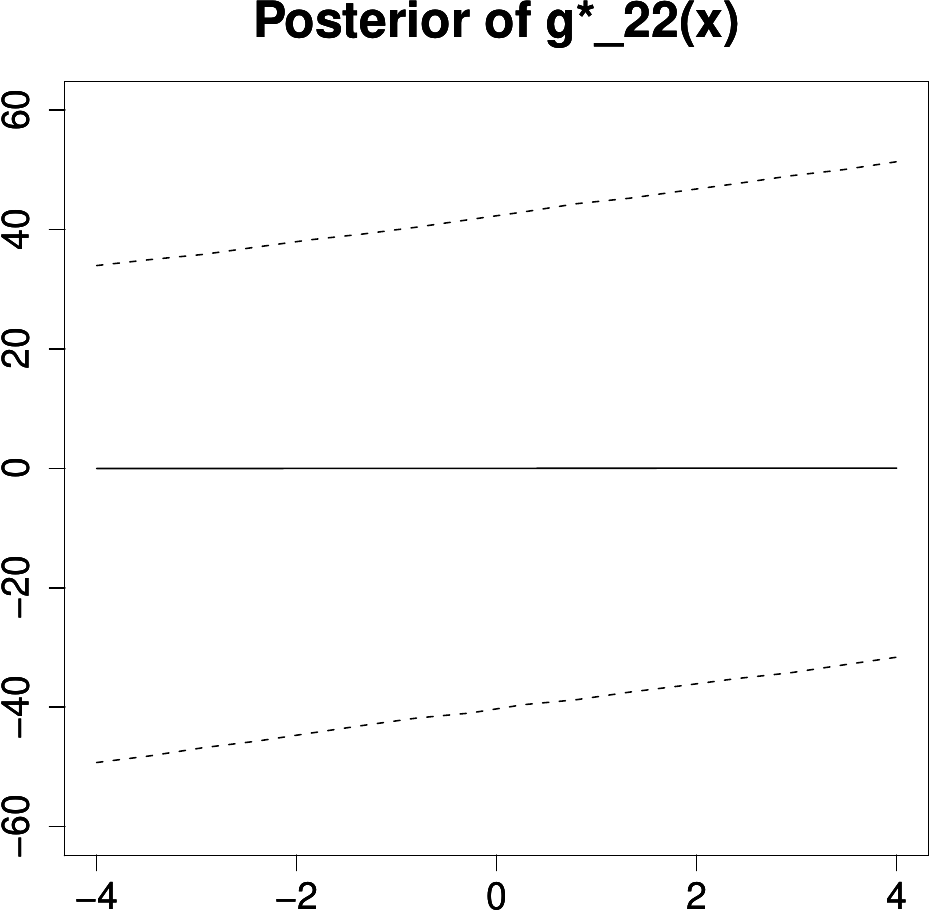}}\\
   \vspace{2mm}
  \subfigure{ \label{fig:g_3_2} 
  \includegraphics[width=4.1cm,height=4.1cm]{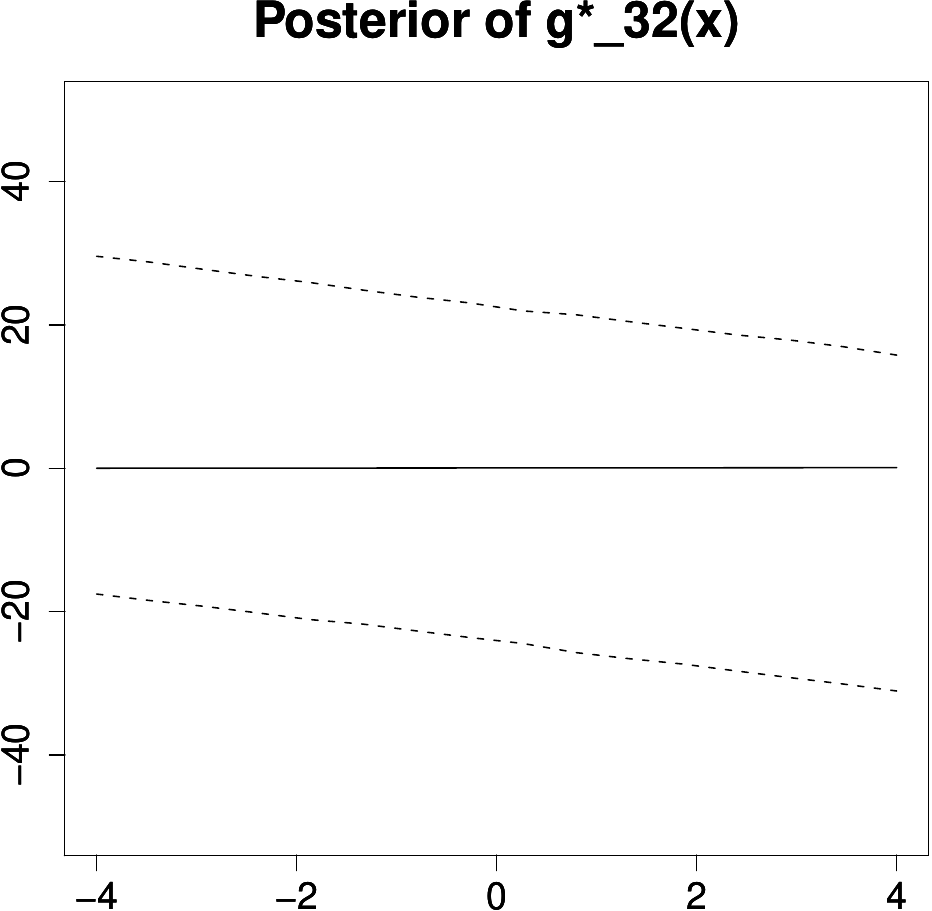}}
   \hspace{2mm}
  \subfigure{ \label{fig:g_4_2} 
  \includegraphics[width=4.1cm,height=4.1cm]{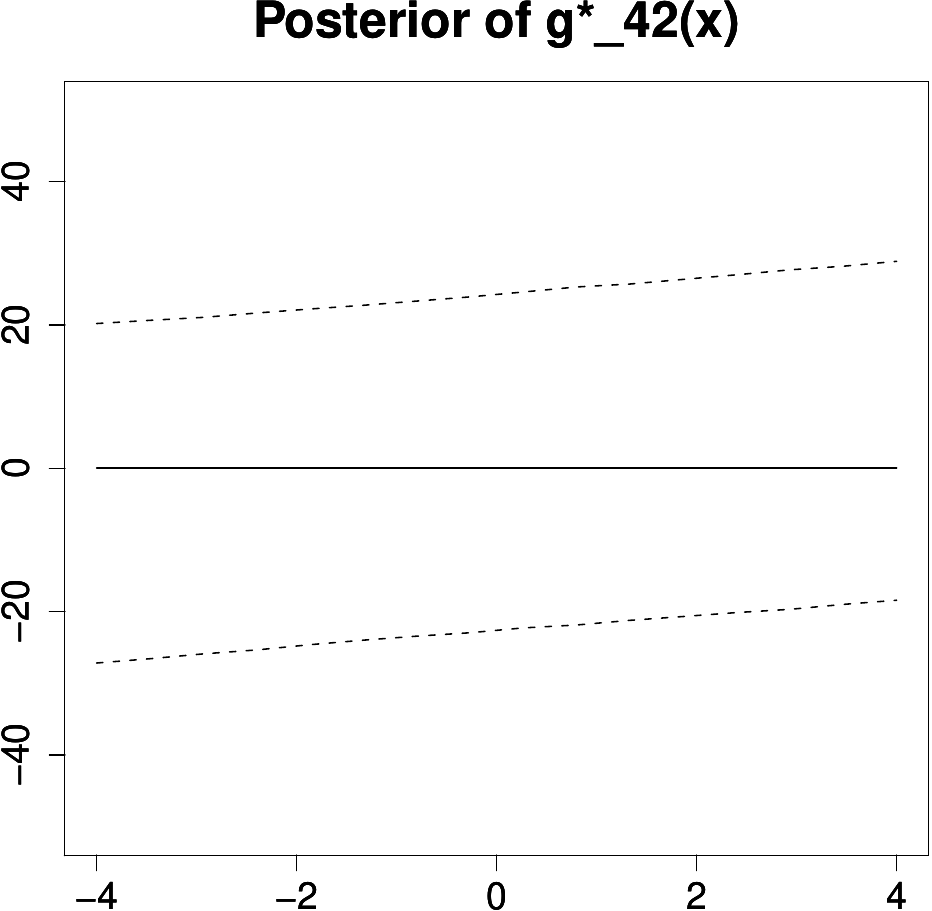}}
  \hspace{2mm}
  \subfigure{ \label{fig:g_1_3}
   \includegraphics[width=4.1cm,height=4.1cm]{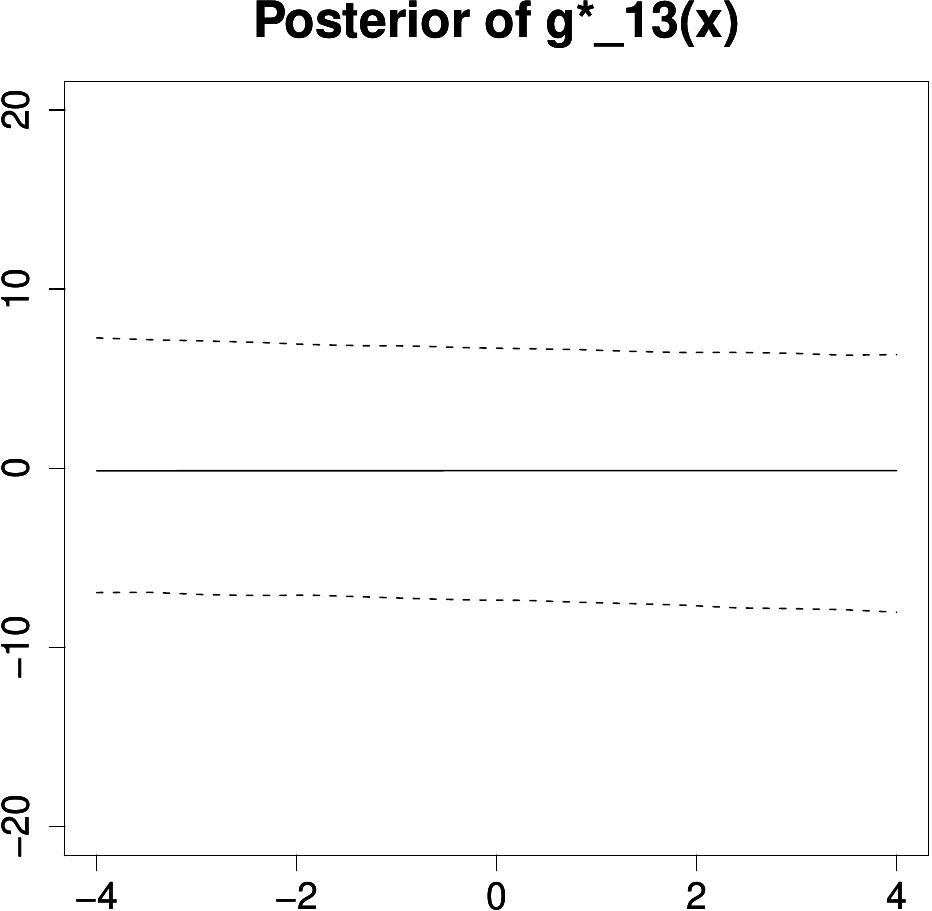}}\\
   \vspace{2mm}
  \subfigure{ \label{fig:g_2_3} 
  \includegraphics[width=4.1cm,height=4.1cm]{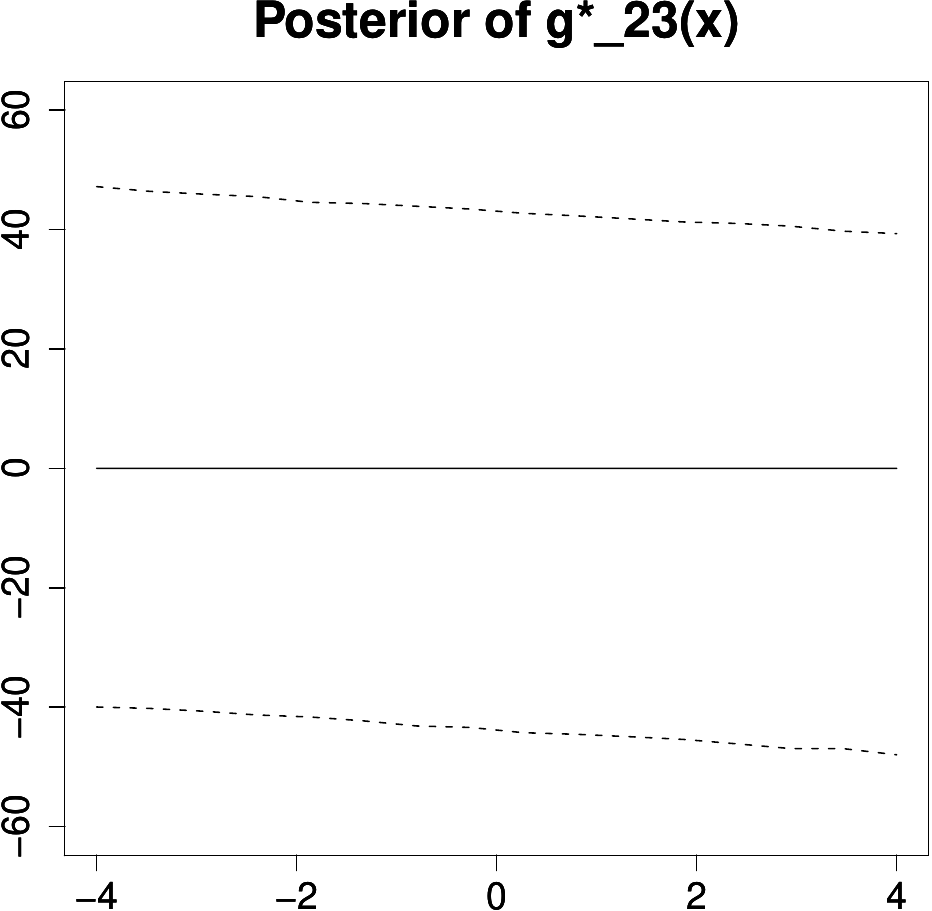}}
   \hspace{2mm}
  \subfigure{ \label{fig:g_3_3} 
  \includegraphics[width=4.1cm,height=4.1cm]{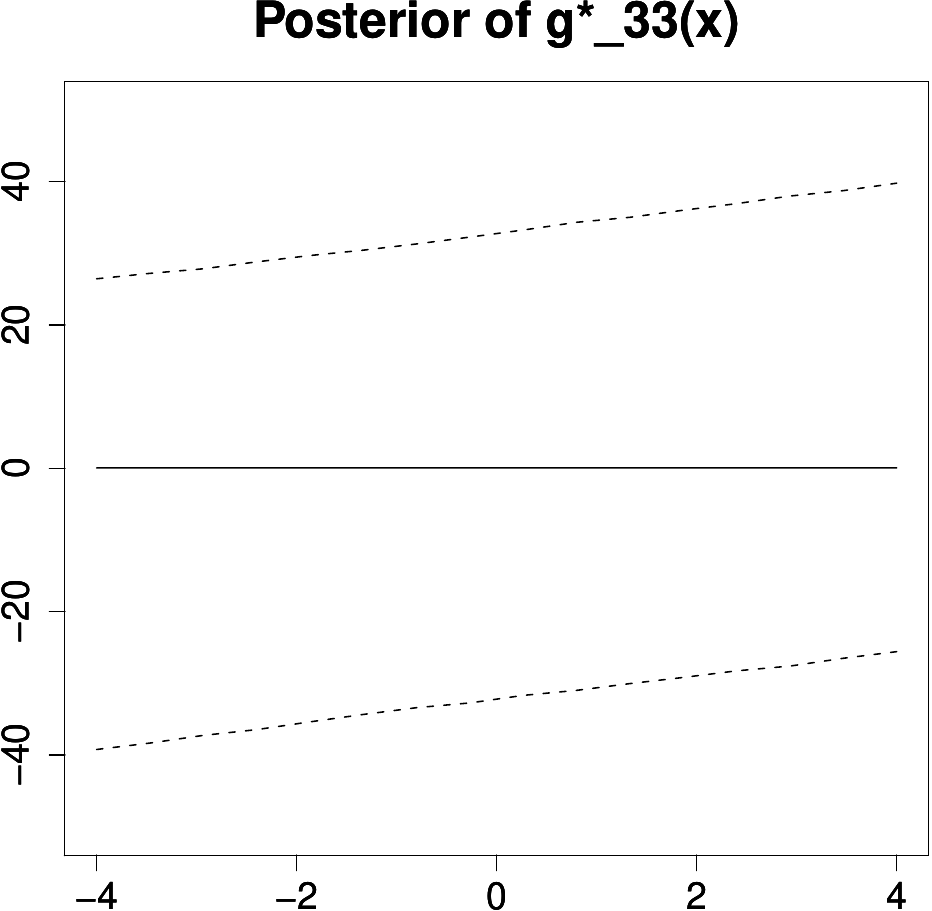}}
   \hspace{2mm}
  \subfigure{ \label{fig:g_4_3} 
  \includegraphics[width=4.1cm,height=4.1cm]{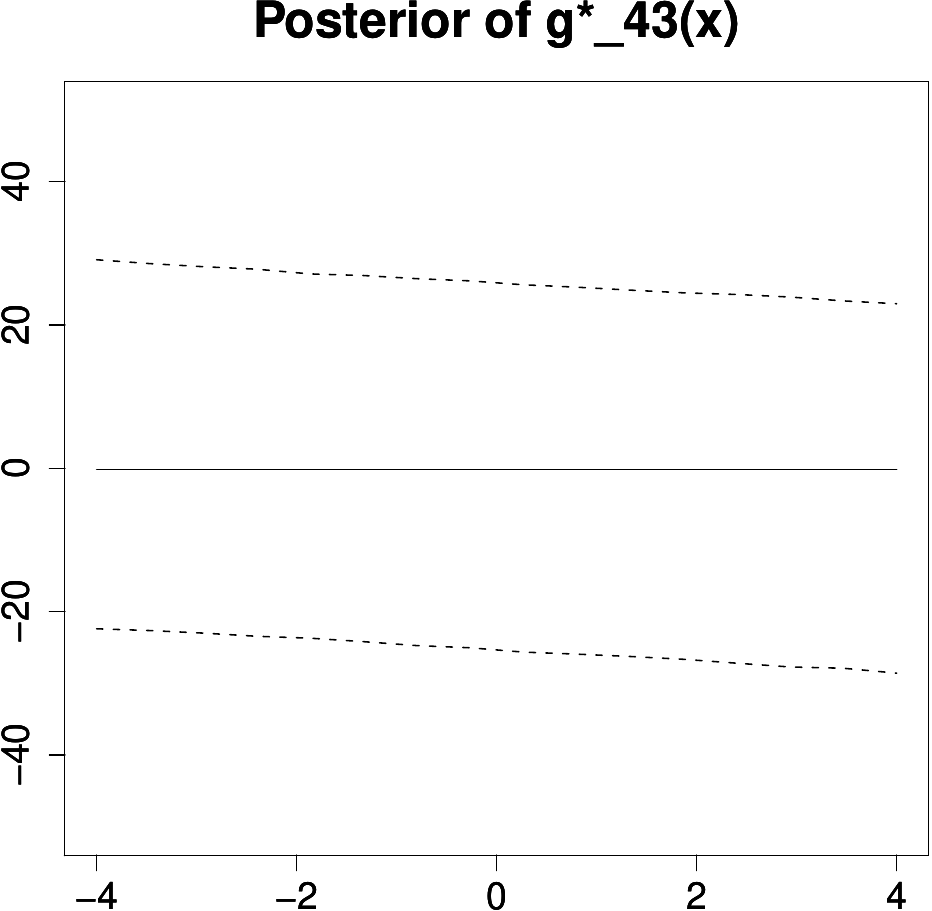}}

 \caption{{\bf Simulation study with data generated from the 4-variate modification of the model of CPS}: 
 The broken lines denote the 95\% highest posterior density credible regions of the posterior distributions 
 of $g^*_{jt}(x)$ for $j=1,2,3,4$ and $t=1,2,3$;
 $x\in [-4,4]$, which is the support of the posterior of $x_0$. The solid lines denote the true composite 
 observational and evolutionary functions.}
 \label{fig:plots_function_mult_cps_g}
 \end{figure}

Figures \ref{fig:plots_function_linear} and \ref{fig:plots_function_cps} display the 95\% highest posterior density
intervals of $f^*_t(x)$ and $g^*_t(x)$ for $t=1,2,3$ and for $x\in [-4,4]$ in the simulation studies concerning the univariate linear model
and the non-linear growth model of CPS respectively. The domain $[-4,4]$ is chosen because the posterior of $x_0$ is supported
on this interval in both the linear and the non-linear cases (see Figures 2 and 5 of GMRB). As already mentioned, the true functions
are almost linear within this short interval. That actually the true functions can be far from linear on a much wider domain is already clear
from the observational and the evolutionary equations in the non-linear example; the true function $f^*_1(\cdot)$
in the case of the non-linear example as shown in Figure \ref{fig:cps_true_f_1} is clearly non-linear over a much wider domain
even though it is close to linear (indeed, almost constant) on $[-4,4]$.
Figures \ref{fig:plots_function_mult_cps_f} and \ref{fig:plots_function_mult_cps_g} show the true functions 
$f^*_{jt}$ and $g^*_{kt}$ for $j,k=1,2,3,4$ and $t=1,2,3$ along with the corresponding 95\% highest posterior density
credible intervals in the multivariate situation. Here also we select the domain of $x$ as $[-4,4]$
because Figure \ref{fig:simdata_cps_mult_hist} shows that {\it a posteriori} each co-ordinate of $x_0$ has support $[-4,4]$.

In all the cases the true, composite, observational and the evolutionary functions fall comfortably within
their respective 95\% highest posterior density credible intervals when $x\in [-4,4]$ is considered. 
We also experimented with $x\in [-30,30]$, but in this case, as already anticipated, in many cases
(not reported here) the true values are excluded from the credible intervals when $x\notin [-4,4]$, 
particularly towards the boundaries of the range.

\newpage

\renewcommand\baselinestretch{1.3}
\normalsize
\bibliographystyle{ECA_jasa}
\bibliography{irmcmc}